%% file: HDR.tex
\PassOptionsToPackage{unicode,bookmarksnumbered=true,linktoc=page}{hyperref}
\PassOptionsToPackage{hyphens}{url}
\PassOptionsToPackage{dvipsnames,svgnames,x11names}{xcolor}
\documentclass[
  english,
  11pt,
  a4paper,
]{memoir}
\usepackage{xcolor}
\usepackage{amsmath,amssymb}
\usepackage{iftex}
\ifPDFTeX
  \usepackage[T1]{fontenc}
  \usepackage[utf8]{inputenc}
  \usepackage{textcomp} % provide euro and other symbols
\else % if luatex or xetex
  \usepackage{unicode-math} % this also loads fontspec
  \defaultfontfeatures{Scale=MatchLowercase}
  \defaultfontfeatures[\rmfamily]{Ligatures=TeX,Scale=1}
\fi
\usepackage[]{libertinus}
\ifPDFTeX\else
\fi
\IfFileExists{upquote.sty}{\usepackage{upquote}}{}
\IfFileExists{microtype.sty}{% use microtype if available
  \usepackage[]{microtype}
  \UseMicrotypeSet[protrusion]{basicmath} % disable protrusion for tt fonts
}{}
\makeatletter
\@ifundefined{KOMAClassName}{% if non-KOMA class
  \IfFileExists{parskip.sty}{%
    \usepackage{parskip}
  }{% else
    \setlength{\parindent}{0pt}
    \setlength{\parskip}{6pt plus 2pt minus 1pt}}
}{% if KOMA class
  \KOMAoptions{parskip=half}}
\makeatother
\usepackage{color}
\usepackage{fancyvrb}

\newcommand{\VERB}{\Verb[commandchars=\\\{\}]}
\DefineVerbatimEnvironment{Highlighting}{Verbatim}{commandchars=\\\{\}}
\newcommand{\DataTypeTok}[1]{\textcolor[rgb]{0.56,0.13,0.00}{#1}}

\usepackage{graphicx}
\makeatletter
\newsavebox\pandoc@box
\newcommand*\pandocbounded[1]{% scales image to fit in text height/width
  \sbox\pandoc@box{#1}%
  \Gscale@div\@tempa{\textheight}{\dimexpr\ht\pandoc@box+\dp\pandoc@box\relax}%
  \Gscale@div\@tempb{\linewidth}{\wd\pandoc@box}%
  \ifdim\@tempb\p@<\@tempa\p@\let\@tempa\@tempb\fi% select the smaller of both
  \ifdim\@tempa\p@<\p@\scalebox{\@tempa}{\usebox\pandoc@box}%
  \else\usebox{\pandoc@box}%
  \fi%
}
\def\fps@figure{htbp}
\makeatother
\ifLuaTeX
\usepackage[bidi=basic,shorthands=off]{babel}
\else
\usepackage[bidi=default,shorthands=off]{babel}
\fi
\ifLuaTeX
  \usepackage{selnolig} % disable illegal ligatures
\fi
\providecommand{\tightlist}{%
  \setlength{\itemsep}{0pt}\setlength{\parskip}{0pt}}
\usepackage[maxnames=10,style=numeric-comp,defernumbers,sorting=citeorder,sortcites,backref]{biblatex}
\usepackage{HDR}

\usepackage{imakeidx}
\makeindex[intoc=true]
\indexsetup{othercode=}
\NewCommandCopy{\oldchap}{\chapter}
\RenewDocumentCommand\chapter{sm}{%
    \IfBooleanTF{#1}%
        {\oldchap*[#2]{#2}}%
        {\oldchap{#2}}%
}

\NewCommandCopy{\oldsec}{\section}
\RenewDocumentCommand\section{sm}{%
    \IfBooleanTF{#1}%
        {\oldsec*{#2\markright{#2}}}%
        {\oldsec{#2}}%
}

\usepackage[intoc]{nomencl}
\makenomenclature
\NewDocumentCommand{\notation}{omD(){}}{%
    \nomenclature[#3]{#2}{#1}%
}

\renewcommand{\nomgroup}[1]{%
    \bigskip
    \ifthenelse{\equal{#1}{D}}{\item[\scshape\large Rings and fields]}{%
    \ifthenelse{\equal{#1}{P}}{\item[\scshape\large Polynomials]}{%
    \ifthenelse{\equal{#1}{C}}{\item[\scshape\large Complexity]}{%
    \ifthenelse{\equal{#1}{V}}{\item[\scshape\large Vectors]}{%
    \ifthenelse{\equal{#1}{W}}{\item[\scshape\large Matrices]}{%
    \ifthenelse{\equal{#1}{A}}{\item[\scshape\large Algorithms]}{%
    \ifthenelse{\equal{#1}{S}}{\item[\scshape\large Sparse polynomials]}{%
            }}}}}}}%
}
\usepackage{bookmark}
\IfFileExists{xurl.sty}{\usepackage{xurl}}{} % add URL line breaks if available
\hypersetup{
  pdftitle={Fast polynomial computations with space constraints},
  pdfauthor={Bruno Grenet},
  pdflang={en},
  colorlinks=true,
  linkcolor={Maroon},
  filecolor={Maroon},
  citecolor={green!50!black},
  urlcolor={Blue},
  pdfcreator={LaTeX via pandoc}}

\title{Fast polynomial computations with space constraints}
\author{Bruno Grenet}
\date{}

\begin{document}
\frontmatter
\maketitle

\mainmatter
\chapter*{Abstract}\label{Section:abstract}
\addcontentsline{toc}{chapter}{Abstract}

The works presented in this habilitation concern the algorithmics of
polynomials. This is a central topic in computer algebra, with numerous
applications both within and outside the field---cryptography,
error-correcting codes, etc. For many problems, extremely efficient
algorithms have been developed since the 1960s. Here, we are interested
in how this efficiency is affected when space constraints are
introduced.

The first part focuses on the time-space complexity of fundamental
polynomial computations---multiplication, division, interpolation,
\ldots{} While \emph{naive} algorithms typically have constant space
complexity, fast algorithms generally require linear space. We develop
algorithms that are both time- and space-efficient. This leads us to
discuss and refine definitions of space complexity for function
computation.

In the second part, the space constraints are put on the inputs and
outputs. Algorithms for polynomials assume in general a \emph{dense}
representation for the polynomials, that is storing the full list of
coefficients. In contrast, we work with \emph{sparse} polynomials, in
which most coefficients vanish. In particular, we describe the first
quasi-linear algorithm for \emph{sparse interpolation}, which plays a
role analogous to the \emph{Fast Fourier Transform} in the sparse
settings. We also explore computationally hard problems concerning
divisibility and factorization of sparse polynomials.

\cleardoublepage
\tableofcontents
\tcblistof[\chapter*]{algs}{List of algorithms}
\addcontentsline{toc}{chapter}{List of algorithms}
\MainMatter

\chapter*{Introduction}\label{Section:introduction}
\addcontentsline{toc}{chapter}{Introduction}

\emph{Computer algebra} is the study of algorithms that manipulate
exactly representable mathematical objects, mostly of an algebraic
nature.\footnote{The names \emph{symbolic computation} or
  \emph{algebraic computing} are almost synonymous.} This includes
computing with integers, rational numbers, and finite fields, as well as
polynomials and matrices over these rings. It is part of the larger
field of \emph{computational mathematics}. It builds on \emph{computer
arithmetic} which is primarily concerned with computing with integers,
and is a close cousin of \emph{numerical computation} which focuses on
computing with approximations of real or complex numbers. Computer
algebra extends from (algebraic) complexity theory on the theoretical
side to implementations on the practical side,\footnote{Widely used
  \emph{computer algebra systems} are available, such as
  \href{http://sagemath.org}{SageMath} or commercial alternatives.} with
strong foundations in various domains of mathematics such as algebra and
number theory.

With a touch of bad faith, one could argue that computer algebra is
actually the origin of the field of algorithmics. The word
\emph{algorithm} comes from the name of the Persian mathematician
Muḥammad ibn Mūsā al-Khwārizmī\,(780--850), while the word
\emph{algebra} comes from the title of his book, \emph{Kitāb
al-mukhtaṣar fī\,ḥisāb al-jabr wa-l-muqābala},\footnote{\emph{The
  Concise Book of Calculation by Restoration and Balancing}. The word
  \emph{al-jabr}, meaning \emph{balancing}, became the modern
  \emph{algebra}.} in which he introduces some of the earliest
algorithms for solving quadratic equations. In fact, other algorithms
describing algebraic computations were described earlier. One example is
Euclid's algorithm, which Knuth refers to as the ``granddaddy of all
algorithms''~\autocite{Knuth1997}.

\begin{figure}
\centering
\includegraphics[width=0.6\linewidth,height=\textheight,keepaspectratio,alt={Pages from al-Khwārizmī's book containing geometrical solutions to two quadratic equations (public domain, via Wikimedia commons).}]{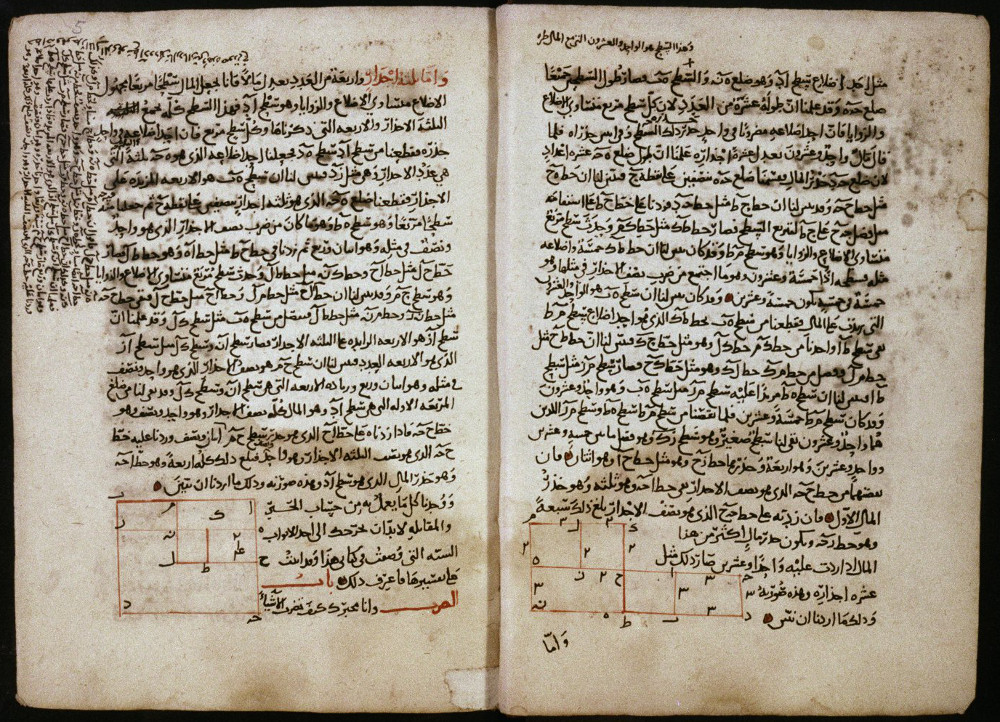}
\caption[Pages from al-Khwārizmī's book containing geometrical solutions
to two quadratic equations (public domain, \emph{via} Wikimedia
commons).]{Pages from al-Khwārizmī's book containing geometrical
solutions to two quadratic equations (public domain, \emph{via}
Wikimedia commons\footnotemark{}).}
\end{figure}
\footnotetext{\url{https://commons.wikimedia.org/wiki/File:Bodleian_MS._Huntington_214_roll332_frame36.jpg}.}

While computer algebra extends beyond polynomial computation, it
undoubtedly represents one of its great successes. Very fast algorithms
have been developed for basic polynomial operations such as
multiplication, Euclidean division and \textsc{gcd} computation, as well
as for more complex problems such as factorization and polynomial system
solving. Implementations of these algorithms in computer algebra systems
or more specialized libraries are used every day to solve practical
problems. Polynomial computations lie at the heart of some related
fields such as cryptography or error-correcting codes, but they are also
essential in seemingly unrelated fields such as robotics.

Although the development of even-faster algorithms and implementations
for polynomial computations is still an active area of research, we take
a step to the side and introduce space constraints to these fast
algorithms. These constraints fall into two categories. In
\hyperref[part:cstspace]{Part~\ref*{part:cstspace}}, we impose some
restrictions on the memory usage of the algorithms. Indeed, faster
algorithms were obtained at the cost of increased memory usage. We
investigate the extent to which we can achieve both time- and
space-efficiency with a single algorithm. In
\hyperref[part:sparse]{Part~\ref*{part:sparse}}, the restrictions
concern the inputs and outputs of the algorithms. Our focus is on sparse
polynomials, which have few terms compared to their degree. Traditional
algorithms usually do not exploit the structure of these polynomials
that benefit from a very compact representation. Consequently, fast
algorithms appear slow with respect to the compact input size. Our goal
is to develop new fast algorithms that take full account of the sparsity
of the inputs and outputs.

The works presented in this document represent most of the research I
conducted since my PhD thesis. The original publications contain more
details, and are referred to throughout the text.\footnote{The
  publications I (co-)authored have labels {[}C⟨\emph{n}⟩{]} (for
  conference publications), {[}J⟨\emph{n}⟩{]} (for journal
  publications), {[}M⟨\emph{n}⟩{]} (for unpublished manuscripts) or
  {[}S⟨\emph{n}⟩{]} (for software). Other references have digit-only
  labels.} Two publications are not presented in the document: one about
a new analysis of Euclid's algorithm~\autocite{GrenetVolkovich2020}, and
a very recent one in the field of
cryptography~\autocite{DumasGalanGrenetMaignanRoche2025}.

\chapter{Preliminaries}\label{chapter:preliminaries}

In this chapter, we introduce the necessary tools from classical
computer algebra.
\hyperref[section:notations]{Section~\ref*{section:notations}} fixes our
(sometimes unusual) notation. Subsequent sections introduce very
classical results in computer algebra that can be found in standard
textbooks~\autocite{vonzurGathenGerhard2013,GeddesCzaporLabahn1992,Pan2001}.
They can be safely skipped by most readers.

\section{Notations}\label{section:notations}

\subsection{Algebra}\label{Section:algebra}

Let \(\mathbb{Z}\) be the ring of integers, \(\mathbb{Q}\) be the field
of rational numbers, \(\mathbb{R}\) the field of real numbers,
\(\mathbb{C}\) the field of complex numbers, and \(\mathbb{F}_q\) be the
finite field with \(q\) elements for a prime power \(q\). We denote an
abstract (commutative) ring (with identity) by the letter
\(\mathsf{R}\). Its group of units is denoted \(\mathsf{R}^\times\).
When it makes sense, we use the notation \(\mathsf{R}_{>0}\) (resp.
\(\mathsf{R}_{\geq 0}\)) to denote the positive (resp. nonnegative)
elements of \(\mathsf{R}\). We denote by \(\mathsf{R}[x]\) the ring of
(univariate) polynomials over \(\mathsf{R}\), by \(\mathsf{R}[x]_{<n}\)
the set of polynomials of degree less than \(n\) and by
\(\mathsf{R}[x_1,\dotsc,x_n]\) the ring of \(n\)-variate polynomials
over \(\mathsf{R}\). The ring of power series over \(\mathsf{R}\) is
\(\mathsf{R}[[x]]\).
\notation[ring of integers\nomrefpage]{\(\mathbb{Z}\)}(dringZ)
\notation[fields of rational, real and complex numbers\nomrefpage]{\(\mathbb{Q}\),
\(\mathbb{R}\), \(\mathbb{C}\)}(dringC)
\notation[finite field with $q$ elements\nomrefpage]{\(\mathbb{F}_q\)}(dringF)
\notation[invertible, positive, nonnegative elements of $\mathsf{R}$\nomrefpage]{\(\mathsf{R}^\times\),
\(\mathsf{R}_{>0}\), \(\mathsf{R}_{\geq 0}\)}(dringzRinv)
\notation[abstract field and ring\nomrefpage]{\(\mathsf{K}\),
\(\mathsf{R}\)}(dringzF)
\notation[rings of univariate and multivariate polynomials over $\mathsf{R}$\nomrefpage]{\(\mathsf{R}[x]\),
\(\mathsf{R}[x_1,\dotsc,x_n]\)}(dringzRpolm)
\notation[set of polynomials of degree $<n$ of over $\mathsf{R}$\nomrefpage]{\(\mathsf{R}[x]_{<n}\)}(dringzRpoln)
\notation[ring of power series over $\mathsf{R}$\nomrefpage]{\(\mathsf{R}[[x]]\)}(dringzRpols)

Vectors are written in bold font, \emph{e.g.} \(\vv\in\mathsf{R}^n\).
Matrices are written in capital letter, \emph{e.g.}
\(M\in\mathsf{R}^{m\times n}\). A polynomial \(f\in\mathsf{R}[x]_{<n}\)
can be identified with its vector of coefficients, written \(\vf\). Due
to this identification, we use the unusual convention that vectors and
matrices are \(0\)-indexed. If \(\vec v\in \mathsf{R}^n\), its entries
are written either \(\vv_{[0]}\), \ldots, \(\vv_{[n-1]}\) or \(v_0\),
\ldots, \(v_{n-1}\). For \(0\leq i<j<n\), \(\vv_{[i,j[}\) is the vector
\((v_i, \dotsc, v_{j-1})\). The entries of
\(M\in \mathsf{R}^{m\times n}\) are denoted \(M_{[i,j]}\) or \(M_{i,j}\)
for \(0\leq i < m\) and \(0\leq j < n\). For a vector
\(\vv\in\mathsf{R}^n\), \(\vv^{\shortleftarrow}\) denotes the vector
defined by \(\vv^{\shortleftarrow}_{[i]} = \vv_{[n-1-i]}\) for
\(0 \leq i < n\). This notation is extended to polynomials and
\(f^{\shortleftarrow}\) denotes the polynomial \(x^{\deg f}f(1/x)\)
whose vector of coefficients is \(\cev f\). The \emph{size} of a
polynomial is the size of its vector of coefficients, that is
\(1+\deg f\)
\notation[vector space and matrix space over $\mathsf{R}$\nomrefpage]{\(\mathsf{R}^n\),
\(\mathsf{R}^{m\times n}\)}(dringzRm)
\notation[$i$th entry of $\vv$, $0 \leq i<n$\nomrefpage]{\(\vv_{[i]}\)
or \(v_i\)}(vvi)
\notation[vector $(v_i, \dotsc, v_{j-1})$\nomrefpage]{\(\vv_{[i,j[}\)}(vvj)
\notation[entry $(i,j)$ of $M$, $0 \leq i<m$, $0 \leq j<n$\nomrefpage]{\(M_{[i,j]}\)
or \(M_{i,j}\)}(wim)
\notation[reversed vector of $\vv$, defined by $\vv^{\shortleftarrow}_{[i]}=\vv_{[n-1-i]}$, $0 \leq i<n$\nomrefpage]{\(\vv^{\shortleftarrow}\)}(vvr)
\notation[reversed polynomial of $f$, defined by $f^{\shortleftarrow}(x) = x^{\deg f}f(1/x)$.\nomrefpage]{\(f^{\shortleftarrow}\)}(par)
\index{polynomial!size}

\subsection{Complexity analyses}\label{Section:complexity-analyses}

We use the Landau notation for asymptotic complexity analyses. Given two
nondecreasing functions \(f\), \(g:\mathbb{R}\to\mathbb{R}_{\geq 0}\),
we write \(f = O(g)\) if there exists \(x_0\) and \(c\) such that
\(f(x) \leq c\cdot g(x)\) for all \(x\geq x_0\). We write \(f = o(g)\)
if for all \(c>0\), there exists \(x_0\) such that
\(f(x) < c\cdot g(x)\) for all \(x\geq x_0\). We also use the notations
\(f = \Omega(g)\) if \(g = O(f)\), \(f = \Theta(g)\) if \(f = O(g)\) and
\(f = \Omega(g)\), and \(f = \omega(g)\) if \(g = o(f)\). We extend the
standard notation \(O(\cdot)\) and write \(f = O^{\widetilde{}}(g)\) if
there exists \(k \in \mathbb{Z}_{>0}\) such that \(f = O(g(\log g)^k)\).
A function is said \emph{quasi-linear in \(x\)} if
\(f = O^{\widetilde{}}(x)\).
\notation[$\exists c, x_0$, $\forall x\geq x_0$, $f(x) \leq c\cdot g(x)$\nomrefpage]{\(f = O(g)\)}(ca)
\notation[$\exists k\geq 0$, $f = O(g \log^k g)$\nomrefpage]{\(f = O^{\widetilde{}}(g)\)}(cc)
\notation[$\forall c$, $\exists x_0$, $\forall x\geq x_0$, $f(x) < c\cdot g(x)$\nomrefpage]{\(f = o(g)\)}(cd)
\notation[$g = O(f)$\nomrefpage]{\(f = \Omega(g)\)}(cb)
\notation[$f=O(g)$ and $g=O(f)$\nomrefpage]{\(f = \Theta(g)\)}(cbb)
\notation[$g = o(f)$\nomrefpage]{\(f = \omega(g)\)}(ce)
\notation[$f = O(g)$ and $g = O(f)$\nomrefpage]{\(f = \Theta(g)\)}(cf)
\index{complexity!quasi-linear}

There are two very natural time complexity analyses in computer algebra.
The \emph{algebraic complexity} of an algorithm over some ring
\(\mathsf{R}\) is the number of ring operations performed by the
algorithm. (Over a field \(\mathsf{K}\), we consider field operations,
including inversions and divisions.) The \emph{bit complexity} of an
algorithm is the number of bit operations it performs. It takes into
account the cost of each ring (or field) operation.
\index{complexity!algebraic} \index{complexity!bit}

We denote by \(\mathsf{Z}(n)\) the (bit) cost of multiplying two
\(n\)-bit integers. A recent celebrated result provides the bound
\(\mathsf{Z}(n) = O(n\log n)\)~\autocite{HarveyvanderHoeven2021}.
Computing a Euclidean division of a \(2n\)-bit integer by an \(n\)-bit
integer has cost \(O(\Z(n))\) and computing the \textsc{gcd} of two
\(n\)-bit integers has cost
\(O(\Z(n)\log n)\)~\autocite{BrentZimmermann2010}. This implies that the
field operations in \(\mathbb{Q}\) have cost \(O(\Z(n)\log n)\) if the
numerators and denominators have at most \(n\) bits. For a prime finite
field \(\mathbb{F}_p\), addition and subtraction have cost
\(O(\log p)\), multiplication has cost
\(O(\Z(\log p)) = O(\log p\log\log p)\) and inversion and divisions have
cost \(O(\Z(\log p)\log\log p) = O(\log p\log^{2}\log p)\). In a
nonprime finite field \(\mathbb{F}_q\), elements are represented by
polynomials, and the cost of basic polynomial operations is presented in
the next sections.
\notation[cost of a product of two $n$-bit integers\nomrefpage]{\(\mathsf{Z}(n) = O(n\log n)\)}(cz)

A \emph{randomized algorithm} is an algorithm that makes some random
choices during its execution. It is called a \emph{Las Vegas algorithm}
if its correctness does not depend on these random choices but its
complexity does. It is called a \emph{Monte Carlo algorithm} if its
complexity does not depend on the random choices but its correctness
does. Finally it is called an \emph{Atlantic City algorithm} if both the
complexity and the correctness depend on the random choices. Note that
is is always possible to turn a Las Vegas algorithm or an Atlantic City
algorithm into a Monte Carlo algorithm. \index{algorithm!Monte Carlo}
\index{algorithm!Atlantic City} \index{algorithm!Las Vegas}

\section{Polynomial products}\label{Section:polynomial-products}

Given two polynomials \(f\in\mathsf{R}[x]_{<m}\) and
\(g\in\mathsf{R}[x]_{<n}\), their product
\(h = f\times g\in\mathsf{R}[x]_{<m+n-1}\) is defined by
\(h_k = \sum_{i+j=k} f_ig_j\) for \(0 \leq k < m+n-1\).

\begin{definition}

A function
\(\M_\mathsf{R} : \mathbb{R}_{\geq 0} \to \mathbb{R}_{\geq 0}\) is a
\emph{polynomial multiplication time} for \(\mathsf{R}\) if

\begin{itemize}
\item
  two polynomials \(f\), \(g\in\mathsf{R}[x]_{<n}\) can be multiplied in
  \(\M_\mathsf{R}(n)\) operations in \(\mathsf{R}\), and
\item
  the function \(n\mapsto \M_\mathsf{R}(n)/n\) is nondecreasing.
\end{itemize}

\end{definition}

\notation[cost of a product of two polynomials in $\mathsf{R}[x]_{<n}$\nomrefpage]{\(\M_\mathsf{R}(n)\)
or \(\M(n)\)}(cm) \index{polynomial!multiplication time}

As long as the context is clear, we shall drop the subscript
\(\mathsf{R}\) and write \(\M(n)\) for \(\M_\mathsf{R}(n)\). Over any
ring \(\mathsf{R}\), the classical polynomial multiplication algorithm
requires \(O(n^{2})\) operations in \(\mathsf{R}\). We shall therefore
always assume that \(\M(n)/n = O(n)\). Better algorithms are also known,
either ring-agnostic or specialized. Over any ring, Karatsuba's
algorithm~\autocite{KaratsubaOfman1963} computes a polynomial product in
\(O(n^{\log 3})\) operations in \(\mathsf{R}\),\footnote{In the whole
  document, \(\log(\cdot)\) denotes the base-\(2\) logarithm, and
  \(\log_b(\cdot)\) denotes the base-\(b\) logarithm.} and its
generalizations known as Toom-Cook's
algorithms~\autocite{Toom1963,Cook1966} compute it in
\(O(n^{\log_{r+1}(2r+1)})\) operations in \(\mathsf{R}\) for every
\(r > 0\). If \(\mathsf{R}\) contains a \(2n\)th principal root of unity
in \(\mathsf{R}\),\footnote{A \emph{principal root of unity} of order
  \(n\) is an element \(\omega\in\mathsf{R}\) such that
  \(\omega^n = 1\), and \(\omega^i-1\in\mathsf{R}^\times\) for
  \(0<i<n\). If \(\mathsf{R}\) is an integral domain, principal roots of
  unity coincide with primitive roots of unity.
  \index{principal root of unity}} FFT-based polynomial multiplication
uses \(O(n\log n)\) operations in
\(\mathsf{R}\)~\autocite{CooleyTukey1965,GentlemanSande1966,Nussbaumer1980}.
In the general case, one can \emph{create} such a root of unity, to get
an algorithm that performs \(O(n\log n\log\log n)\) operations in
\(\mathsf{R}\)~\autocite{CantorKaltofen1991}. For the important case of
a finite field \(\mathbb{F}_q\), the number of operations in
\(\mathbb{F}_q\) may be \(O(n\log n)\) if an appropriate root of unity
exists, and this translates into \(O(n\log n\cdot\log(q)\log\log(q))\)
bit operations by means of fast integer
multiplication~\autocite{HarveyvanderHoeven2022}. This bound is not
attained for all finite fields, and is anyway not the best achievable.
Unconditionally, two polynomials of degree \(<n\) can be multiplied in
\(O(n\log q\log(n\log q)4^{\log^*(n\log q)})\) bit
operations~\autocite{HarveyvanderHoeven2019}, where \(\log^*\) denotes
the iterated logarithm.\footnote{It is defined by \(\log^*(x) = 0\) if
  \(x\leq 1\) and \(\log^*(x) = 1+\log^*(\log x)\) otherwise.} Under
some number-theoretic assumptions, the complexity bound becomes
\(O(n\log q\log(n\log q))\)~\autocite{HarveyvanderHoeven2022}.

Polynomial multiplication is the basis of most algorithms on
polynomials. In many cases, the full result is not required but rather
only part of it. The standard product of polynomials is called the
\emph{full product}. We also define partial products.
\index{polynomial!full product}

\begin{definition}

Let \(f\), \(g\in \mathsf{R}[x]\) of respective sizes \(m\) and \(n\).

\begin{itemize}
\item
  Their \emph{lower product} is the polynomial made of the lower \(m\)
  coefficients of their product, written
  \(\LowProd(f,g) = (f\times g)\bmod x^m\).
\item
  Their \emph{upper product} is the polynomial made of the upper \(n-1\)
  coefficients of their product, written
  \(\UppProd(f,g) = (f\times g)\bquo x^m\).
\item
  If \(m \geq n\), their \emph{middle product} is the polynomial made of
  the central \(m-n+1\) coefficients of their product, written
  \(\MidProd(f,g) = \left((f\times g)\bquo x^{n-1}\right)\bmod x^{m-n+1}\).
\end{itemize}

\end{definition}

\notation[remainder in the Euclidean division of $f$ by $g$\nomrefpage]{\(f\bmod g\)}(pam)
\notation[quotient in the Euclidean division of $f$ by $g$\nomrefpage]{\(f\bquo g\)}(paq)
\notation[lower product of $f$ by $g$\nomrefpage]{\(\LowProd(f,g)\)}(ppl)
\notation[upper product of $f$ by $g$\nomrefpage]{\(\UppProd(f,g)\)}(ppu)
\notation[middle product of $f$ by $g$\nomrefpage]{\(\MidProd(f,g)\)}(ppm)
\index{polynomial!lower product} \index{polynomial!upper product}
\index{polynomial!middle product}

\begin{remark}

As defined, the lower and upper products are not commutative. They
satisfy
\(f\times g = \LowProd(f,g)+x^m\UppProd(f,g) = \LowProd(g,f)+x^n\UppProd(g,f)\).\gobblepar

\end{remark}

We extend the notation \(\M(\cdot)\) and denote by \(\M(m,n)\) the cost
of an algorithm that multiplies two polynomials
\(f\in\mathsf{R}[x]_{<m}\) and \(g\in\mathsf{R}[x]_{<n}\). Assuming
without loss of generality that \(m\geq n\), the product can be
performed using \(\lceil m/n\rceil\) products of size-\(n\) polynomials,
whence \(\M(m,n) \leq \lceil m/n\rceil\M(n)\). The case \(m=n\) is
called a \emph{balanced} (full) product. We also define the balanced
lower and upper products when \(m=n\), and the balanced middle product
when \(m=2n-1\) and the result has size \(n\).

\notation[cost of a product of polynomials of size $m$ and $n$ respectively\nomrefpage]{\(\M(m,n)\)}(cmn)
\index{polynomial!balanced product}

\begin{proposition}[\autocite{HanrotQuerciaZimmermann2004}]

Let \(f\), \(g\in \mathsf{R}[x]\) of respective sizes \(m\) and \(n\).
Then \(\LowProd(f,g)\) can be computed in \(\M(m)\) operations, and
\(\UppProd(f,g)\) in \(\M(n-1)\) operations. The middle product
\(\MidProd(f,g)\) can be computed in \(\M(m,n)\) operations.\gobblepar

\end{proposition}

\section{Other polynomial and power series
computations}\label{Section:other-polynomial-and-power-series-computations}

Let \(\phi\in\mathsf{R}[[x]]\) be some power series. In computations, it
is represented by a \emph{truncation at precision \(n\)}, that is its
value modulo \(x^n\) for some \(n\). If \(f = \phi\bmod x^n\), \(f\) is
said to be a \emph{truncated power series at precision \(n\)}. A
truncated power series at precision \(n\) is thus a size-\(n\)
polynomial. \index{power series truncation}

If \(\phi\), \(\psi\in \mathsf{R}[[x]]\) and \(f = \phi\bmod x^n\),
\(g = \psi\bmod x^n\), the truncation at precision \(n\) of
\(\phi\times\psi\) is \(\LowProd(f,g)\).

\begin{proposition}\label{proposition:seriesprod}

Given the truncations at precision \(n\) of two power series \(\phi\),
\(\psi\in \mathsf{R}[[x]]\), the truncation at precision \(n\) of
\(\phi\times\psi\) can be computed in \(\M(n)\) operations
in~\(\mathsf{R}\).\gobblepar

\end{proposition}

If \(\phi(0)\) is a unit of \(\mathsf{R}\), the series \(\phi\) is
invertible, that is there exists \(\phi^{-1}\in\mathsf{R}[[x]]\) such
that \(\phi\times\phi^{-1} = 1 \in \mathsf{R}[[x]]\). Given a truncation
of \(\phi\) at precision \(n\), the truncation of \(\phi^{-1}\) at
precision \(n\) can be computed by means of Newton iteration.

\begin{proposition}[\autocite{Sieveking1972,Kung1974}]\label{proposition:seriesinv}

Given the truncation at precision \(n\) of a power series
\(\phi\in \mathsf{R}[[x]]\) whose constant coefficient is a unit, the
truncation at precision \(n\) of \(\phi^{-1}\) can be computed in
\(O(\M(n))\) operations in \(\mathsf{R}\).\gobblepar

\end{proposition}

From this, one can also compute a division of power series.

\begin{corollary}\label{corollary:seriesdiv}

Given the truncations at precision \(n\) of two power series \(\phi\),
\(\psi\in \mathsf{R}[[x]]\), where \(\psi(0)\) is a unit, the truncation
at precision \(n\) of \(\phi/\psi\) can be computed in \(O(\M(n))\)
operations in~\(\mathsf{R}\).\gobblepar

\end{corollary}

Power series inversion and division are the basis for fast Euclidean
division of polynomials. Indeed, the quotient \(q = f\bquo g\) can be
computed as a \emph{reversed power series division} (see
\hyperref[section:transposition]{Section~\ref*{section:transposition}}
and \hyperref[figure:EuclideanDiv]{Figure~\ref*{figure:EuclideanDiv}} in
\hyperref[chapter:rorw]{Chapter~\ref*{chapter:rorw}}), and the remainder
\(r = f\bmod g\) is obtained as \(f-gq\).

\begin{proposition}[\autocite{Strassen1973}]\label{proposition:polydiv}

Given two polynomials \(f\) and \(g\in \mathsf{R}[x]\) of respective
sizes \(m+n-1\) and \(n\) such that the leading coefficient of \(g\) is
a unit, the quotient \(q = f\bquo g\) and the remainder \(r = f\bmod g\)
such that \(f = bq+r\) and \(\deg(r)<\deg(g)\) can be computed in
\(O(\M(m)+\M(n))\) operations in \(\mathsf{R}\).\gobblepar

\end{proposition}

Given a size-\(n\) polynomial \(f\in\mathsf{R}[x]\) and \(n\) points
\(\alpha_0\), \ldots, \(\alpha_{n-1}\in\mathsf{R}\), the problem of
\emph{multipoint evaluation} is to compute \(f(\alpha_0)\), \ldots,
\(f(\alpha_{n-1})\). Interpolation does the converse. Given
\(\alpha_0\), \ldots, \(\alpha_{n-1}\in\mathsf{R}\) such that
\(\alpha_i-\alpha_j\) is a unit for all \(i\neq j\), and \(\beta_0\),
\ldots, \(\beta_{n-1}\), the goal is to compute the unique size-\(n\)
polynomial \(f\in\mathsf{R}[x]\) such that \(f(\alpha_i)=\beta_i\) for
\(0\leq i<n\). \index{polynomial!multipoint evaluation}
\index{polynomial!interpolation}

\begin{proposition}[\autocite{BorodinMoenck1974}]\label{proposition:evaluationinterpolation}

Given a size-\(n\) polynomial \(f\in \mathsf{R}\) and a vector
\(\vec \alpha\in \mathsf{R}^n\), the vector
\((f(\alpha_0), \dotsc, f(\alpha_{n-1}))\) can be computed in
\(O(\M(n)\log n)\) operations in \(\mathsf{R}\). Given two vectors
\(\vec \alpha\), \(\vec \beta\in\mathsf{R}^n\) such that
\(\alpha_i-\alpha_j\) is a unit for \(i\neq j\), the unique size-\(n\)
polynomial \(f\) such that \(f(\alpha_i)=\beta_i\) for \(0\leq i<n\) can
be computed in \(O(\M(n)\log n)\) operations in
\(\mathsf{R}\).\gobblepar

\end{proposition}

Note that it makes also sense to define multipoint evaluation for a
size-\(m\) polynomial and \(n\) points when \(m\neq n\). If \(m < n\),
one can perform \(\lceil n/m\rceil\) multipoint evaluations in size
\(m\), in \(O(\frac{n}{m}\M(m)\log m)=O(\M(n)\log m)\) operations in
\(\mathsf{R}\). If \(m < n\), \(f\) is first reduced modulo
\(\prod_i (x-\alpha_i)\) (which is to be computed anyway by the
algorithm) and multipoint evaluation is applied to the size-\(n\)
remainder, for a total of \(O(\M(m-n) + \M(n)\log n)\) operations in
\(\mathsf{R}\). When the vector \(\vec \alpha\) has some structure, the
computation can usually be sped up~\autocite{BostanSchost2005}. In
particular if it is a geometric progression, the complexities drop to
\(O(\M(n))\)~\autocite{Bluestein1970,BostanSchost2005}.

Given two monic polynomials \(f\) and \(g\), their greatest common
divisor (\textsc{gcd}) can be computed with a fast variant of Euclid's
algorithm. This also provides Bézout coefficients \(u\) and \(v\) such
that \(uf+vg = \textsc{gcd}(f,g)\).

\begin{proposition}[\autocite{BrentGustavsonYun1980}]\label{proposition:gcd}

Given two monic polynomials \(f\), \(g\in\mathsf{R}[x]\) of respective
size \(m\) and \(n\) where \(m\geq n\), their \textsc{gcd} and the
corresponding Bézout coefficients can be computed in
\(O(\M(m-n) + \M(n)\log n)\) operations in \(\mathsf{R}\).\gobblepar

\end{proposition}

\begin{remark}

In all the previous complexities, the logarithmic factors disappear in
the terms \(\M(n)\log(n)\) as soon as
\(\M(n) = \Omega(n^{1+\varepsilon})\) for some \(\varepsilon>0\). This
is the case if the underlying multiplication algorithm is for instance
the naive one (\(\M(n) =O(n^{2})\)) or Karatsuba's (\(O(n^{\log 3})\)).

In practice, the constants of the leading terms in the complexities play
a major role for the efficiency of the algorithms. For better constants
than the original algorithms, we refer to more recent
works~\autocite{BostanLecerfSchost2003,HanrotQuerciaZimmermann2004,BostanSchost2005,vanderHoeven2025}.\gobblepar

\end{remark}

Equipped with these results, we obtain the bit costs of the field
operations in a finite field \(\mathbb{F}_q\) where \(q = p^s\) for some
prime number \(p\). Since each element of \(\mathbb{F}_q\) is
represented by a polynomial in \(\mathbb{F}_p[x]_{<s}\), addition and
subtraction use \(s\) operations in \(\mathbb{F}_p\), that is
\(O(s\Z(\log p))\) bit operations. Multiplication can be computed in
\(O(\M_{\mathbb{F}_p}(s))\) operations in \(\mathbb{F}_p\), that is
\(O^{\widetilde{}}(s\log p)\) bit operations. Under some
number-theoretic assumptions, the complexity becomes
\(O(s\log p\log(s\log p))\)~\autocite{HarveyvanderHoeven2022}. Inversion
and division are computed in \(O(\M_{\mathbb{F}_p}(s)\log s)\)
operations in \(\mathbb{F}_p\), that is \(O^{\widetilde{}}(s\log p)\)
bit operations.

\section{Linear recurrent
sequences}\label{Section:linear-recurrent-sequences}

\emph{Linear recurrent sequences}, also known as
\emph{constant-recursive} or \emph{C-finite sequences}, are a very
useful tool in computer algebra. For a sequence
\((u_n)_{n\geq 0}\in\mathsf{R}^\mathbb{N}\), its \emph{generating
series} is the power series \(\phi = \sum_{n\geq 0} u_n x^n\).

\begin{definition}

A sequence \((u_n)_{n\geq 0} \in \mathsf{R}^\mathbb{N}\) is \emph{linear
recurrent} if there exist \(a_0\), \ldots, \(a_{k-1}\) such that for all
\(n\geq 0\), \[u_{n+k} = \sum_{j=0}^{k-1} a_j\cdot u_{n+j}.\] The
polynomial \(p = x^k-\sum_{j=0} a_j x^j\) is a \emph{characteristic
polynomial} of \((u_n)_n\). The \emph{minimal polynomial} of \((u_n)_n\)
is its least-degree characteristic polynomial. The \emph{order} of
\((u_n)_n\) is the degree of its minimal polynomial.\gobblepar

\end{definition}

\index{linear recurrent sequence} \index{minimal polynomial}

The following proposition states some fundamental equivalent
characterizations of a linear recurrent sequence. We state it only in
the simple situation where the minimal polynomial splits and is
square-free\footnote{A degree-\(d\) polynomial \emph{splits} over
  \(\mathsf{R}\) if it has \(d\) roots in \(\mathsf{R}\), counting with
  multiplicities. It is square-free if its roots are pairwise distinct.}
over \(\mathsf{R}\). We make the assumption in this document that the
minimal polynomials of all the linear recurrent sequences split and are
square-free in \(\mathsf{R}\). Recall that the polynomial
\(p^{\shortleftarrow}\) is defined as \(x^{\deg p}p(1/x)\).

\begin{proposition}\label{proposition:linrec}

Let \((u_n)_{n\geq 0}\in \mathsf{R}^\mathbb{N}\), and
\(\phi = \sum_{n\geq 0} u_n x^n\) its generating function. The following
statements are equivalent:

\begin{enumerate}[label=(\roman*)]

\item

\((u_n)_{n\geq 0}\) is linear recurrent with minimal polynomial
\(p = \prod_{i=0}^{k-1} (x-\rho_i)\) with pairwise distinct roots
\(\rho_0\), \ldots, \(\rho_{k-1}\in \mathsf{R}\);
\label{proposition:linrec:minpoly}

\item

\(\phi = q/p^{\shortleftarrow}\) for some polynomial
\(q\in\mathsf{R}[x]\) of degree \(<k\); \label{proposition:linrec:rat}

\item

\(u_n = \sum_{i=0}^{k-1} \lambda_i \rho_i^n\) for some \(\lambda_0\),
\ldots, \(\lambda_{k-1}\) that only depend on the initial conditions
\(u_0\), \ldots, \(u_{k-1}\). \label{proposition:linrec:expsum}

\end{enumerate}\gobblepar

\end{proposition}

Similar and further equivalences can be given in more general
settings~\autocite{EverestvanderPoortenShparlinskiWard2003}. An
algorithmic view of this proposition gives different representations for
a linear recurrent sequence.

\begin{proposition}\label{proposition:lrsrepr}

A linear recurrence sequence \((u_n)_{n\geq 0}\) of order at most \(k\)
is completely determined by any of the following data:

\begin{enumerate}[label=(\roman*)]

\item

\(2k\) initial terms \(u_0\), \ldots, \(u_{2k-1}\);
\label{proposition:lrsrepr:init}

\item

its minimal polynomial \(p\), together with the \(k\) initial terms
\(u_0\), \ldots, \(u_{k-1}\); \label{proposition:lrsrepr:minpoly}

\item

a rational representation \(q/p^{\shortleftarrow}\) of its generating
series \(\phi\); \label{proposition:lrsrepr:rat}

\item

the vectors \(\vec \lambda\) and \(\vec \rho\) such that
\(u_n = \sum_{i=0}^{k-1} \lambda_i \rho_i^n\) for \(n\geq 0\).
\label{proposition:lrsrepr:expsum}

\end{enumerate}\gobblepar

\end{proposition}

\begin{figure}

\input{linrec.tikz}
\caption{Conversion between representations of linear recurrent sequences}
\label{figure:linrec}

\end{figure}
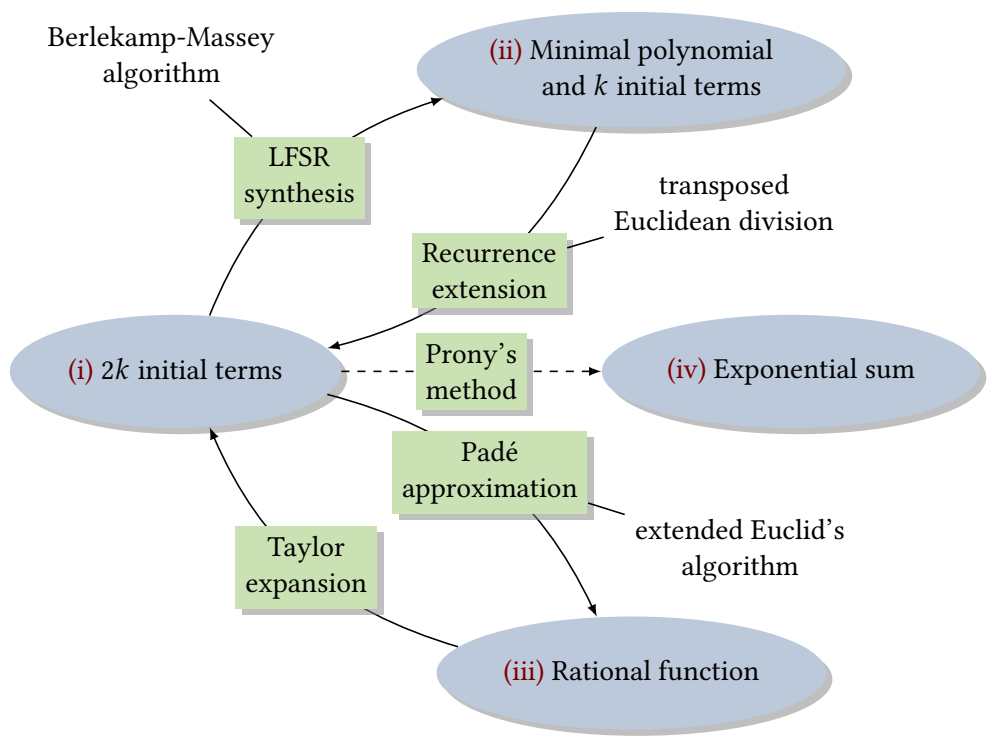

Note that in all cases, the representation consists of \(2k\) elements
from \(\mathsf{R}\). Conversions between these different representations
are known under various names in the literature, and associated to some
classical algorithms. These are represented in
\hyperref[figure:linrec]{Figure~\ref*{figure:linrec}}.

Both representations \ref{proposition:lrsrepr:minpoly} and
\ref{proposition:lrsrepr:rat} contain the minimal polynomial \(p\). The
conversion between \(2k\) initial terms of \((u_n)_n\) and each of these
two representations give rise to two families of algorithms. In
information theory, the conversion \ref{proposition:lrsrepr:init} →
\ref{proposition:lrsrepr:minpoly} is known as \emph{LFSR
synthesis}.\footnote{LFSR stands for Linear-Feedback Shift Register.} It
is computed using Berlekamp-Massey's
algorithm~\autocite{Berlekamp1968,Berlekamp2015,Massey1969} in quadratic
time and serves as building block for decoding algorithms of some
error-correcting codes related to BCH codes. A fast variant of this
algorithm has complexity \(O(\M(k)\log k)\). The other direction is
known as \emph{recurrence extension} and can be computed as a
\emph{transposed} Euclidean division~\autocite{Shoup1991}.

Originating in complex analysis, a Padé approximant is a rational
function approximation of a power series. The conversion
\ref{proposition:lrsrepr:init} → \ref{proposition:lrsrepr:rat} is
exactly a Padé approximant computation, viewing the \(2k\) initial terms
as the truncated power series \(\phi\bmod x^{2k}\). The extended
Euclidean algorithm can be used to compute it in quadratic time. Using
fast Euclidean algorithm
(\hyperref[proposition:gcd]{Proposition~\ref*{proposition:gcd}}), the
complexity drops to \(O(\M(k)\log k)\). The other direction is
\emph{Taylor expansion} and can be computed using Euclidean division.

It has been noticed that Berlekamp-Massey's algorithm and the Euclidean
algorithm are in a sense dual of each other~\autocite{Dornstetter1987}.
Using any of the two fast variants provides the following result.

\begin{proposition}\label{proposition:minpoly}

Given \(u_0\), \ldots, \(u_{2k-1}\in\mathsf{R}\), the minimal polynomial
of \((u_i)_{0\leq i<2k}\) can be computed with \(O(\M(k)\log k)\)
operations in \(\mathsf{R}\).\gobblepar

\end{proposition}

Finally, the conversion \ref{proposition:lrsrepr:init} →
\ref{proposition:lrsrepr:expsum} is known as \emph{Prony's method} in
numerical analysis, and the conversion is made by first computing the
minimal polynomial, that is using either the conversion
\ref{proposition:lrsrepr:init} → \ref{proposition:lrsrepr:minpoly} or
\ref{proposition:lrsrepr:init} → \ref{proposition:lrsrepr:rat} first.
More details on this method are given in
\hyperref[section:SparseInterpolationBB]{Section~\ref*{section:SparseInterpolationBB}}.

\section{Polynomial operations as structured linear
algebra}\label{Section:polynomial-operations-as-structured-linear-algebra}

Most polynomial and power series computations presented so far are
bilinear maps. Therefore, one can fix any of the inputs to get a linear
map that has a matrix representation. The corresponding matrices have
structures that we present now. Note that all these structures can be
encompassed in the more general framework of \emph{low displacement
rank}~\autocite{KailathKungMorf1979}. Pan has written a thorough
treatment of the links between polynomial computations and structured
matrix computations~\autocite{Pan2001}.

Recall that a polynomial \(f\in\mathsf{R}[x]_{<n}\) is identified with
its vector of coefficients \(\vf\in\mathsf{R}^n\). Conversely, a vector
\(\vv\in\mathsf{R}^n\) can be viewed as a polynomial
\(v\in\mathsf{R}[x]_{<n}\). The polynomial \(f^{\shortleftarrow}\) is
defined by \(f^{\shortleftarrow}(x)=x^{\deg f}f(1/x)\) and its vector of
coefficients \(\cev f\) is defined by
\(f^{\shortleftarrow}_i=f_{\deg(f)-i}\).

We first consider polynomial products. They correspond to Toeplitz
matrix-vector products.

\begin{definition}

Let \(\vec \alpha\in \mathsf{R}^{m+n-1}\). The \emph{Toeplitz matrix}
\(T_{m,n}(\vec \alpha)\) is the \(m\times n\) matrix
\[T = \begin{pmatrix}
  \alpha_{m-1} & \alpha_m & \dotsc & \alpha_{m+n-2}\\
  \alpha_{m-2} & \alpha_{m-1} & \dotsc & \alpha_{m+n-3} \\
  \vdots&\vdots&&\vdots\\
  \alpha_1 & \alpha_2 & \dotsc & \alpha_n\\
  \alpha_0 & \alpha_1 & \dotsc & \alpha_{n-1}
  \end{pmatrix}\] defined by \(T_{i,j} = \alpha_{m-1+j-i}\) for
\(0 \leq i < m\) and \(0 \leq j < n\). If \(m = n\), we denote the
square matrix \(T_{m,m}(\vec \alpha)\) by \(T_m(\vec \alpha)\).

If \(\vec \alpha\in\mathsf{R}^{m}\), the \emph{lower} and \emph{upper
triangular Toeplitz matrices} are the \(m\times m\) matrices
\(L_m(\vec \alpha) = T_m(\vec \alpha\Vert\vec 0_{m-1})\) and
\(U_m(\vec \alpha) = T_m(\vec 0_{m-1}\Vert\vec \alpha)\) where
\(\vec 0_{m-1}\in\mathsf{R}^{m-1}\) is the all-zero vector, and
\(\cdot\Vert\cdot\) denotes concatenation.\gobblepar

\end{definition}

\notation[Toeplitz $m\times n$ and $m\times m$ matrices built on $\vec \alpha$\nomrefpage]{\(T_{m,n}(\vec \alpha)\),
\(T_m(\vec \alpha)\)}(wxT)
\notation[lower and upper triangular Toeplitz matrices built on $\vec \alpha$\nomrefpage]{\(L_m(\vec \alpha)\),
\(U_m(\vec \alpha)\)}(wxTs)
\notation[concatenation of two vectors\nomrefpage]{\(\vu\Vert\vv\)}(vvc)
\notation[all-zero vector of length $m$\nomrefpage]{\(\vec 0_m\)}(vvz)
\index{matrix!Toeplitz}

\begin{proposition}

Let \(f\), \(g\in \mathsf{R}[x]\) of respective sizes \(m\) and \(n\).
Then

\begin{itemize}
\item
  \(h = f\times g\) is equivalent to
  \(\vh = T_{m+n-2,n}(\vec 0_{n-2}\Vert\cev f\Vert\vec 0_{n-1}) \cdot \vg\);
\item
  \(h = \MidProd(f,g)\) is equivalent to
  \(\vh = T_{m,n}(\cev f)\cdot\vg\);
\item
  \(h = \LowProd(f,g)\) is equivalent to \(\vh = L_m(\cev f)\cdot\vg\);
\item
  \(h = \UppProd(f,g)\) is equivalent to
  \(\vh = U_{n-1}(\cev f_{[m-n+1,m[})\cdot\vg\).
\end{itemize}

\end{proposition}

As a result, Toeplitz matrix-vector products can be computed in
\(O(\M(m,n))\) operations in \(\mathsf{R}\). Lower triangular Toeplitz
system solving corresponds to power series division and can be computed
in \(O(\M(n))\) operations in \(\mathsf{R}\).

A special case of Toeplitz matrix is the case of a circulant matrix.

\begin{definition}

Let \(\vec \alpha\in \mathsf{R}^m\). The \emph{circulant matrix}
\(C_m(\vec \alpha)\) is the \(m\times m\) matrix \[C = \begin{pmatrix}
\alpha_0 & \alpha_1 & \alpha_2 & \dotsc & \alpha_{m-2} & \alpha_{m-1} \\
\alpha_{m-1} & \alpha_0 & \alpha_1 & \dotsc & \alpha_{m-3} & \alpha_{m-2} \\
\vdots & \vdots & \vdots && \vdots & \vdots\\
\alpha_1 & \alpha_2 & \alpha_3 & \dotsc & \alpha_{m-1} & \alpha_0
\end{pmatrix}\] defined by \(C_{i,j} = \alpha_{(j-i)\bmod m}\) for
\(0 \leq i, j < m\).

For \(\lambda\in \mathsf{R}\), the \emph{\(\lambda\)-circulant matrix}
\(C^\lambda_m(\vec \alpha)\) is the \(m\times m\) matrix
\[C^\lambda = \begin{pmatrix}
\alpha_0 & \alpha_1 & \alpha_2 & \dotsc & \alpha_{m-2} & \alpha_{m-1} \\
\lambda\cdot\alpha_{m-1} & \alpha_0 & \alpha_1 & \dotsc & \alpha_{m-3} & \alpha_{m-2} \\
\vdots & \vdots & \vdots && \vdots & \vdots\\
\lambda\cdot\alpha_1 & \lambda\cdot\alpha_2 & \lambda\cdot\alpha_3 & \dotsc & \lambda\cdot\alpha_{m-1} & \alpha_0
\end{pmatrix}\] defined by \(C^\lambda_{i,j} = \alpha_{j-i}\) for
\(0 \leq i \leq j < m\) and
\(C^\lambda_{i,j} = \lambda\cdot\alpha_{m-i+j}\) for
\(0\leq j<i<m\).\gobblepar

\end{definition}

\notation[circulant and $\lambda$-circulant $m\times m$ matrices with first row $\vec \alpha$\nomrefpage]{\(C_m(\vec \alpha)\),
\(C_m^\lambda(\vec \alpha)\)}(wxC) \index{matrix!circulant}

These matrices correspond to \emph{polynomial convolutions}.

\index{polynomial!convolution}

\begin{proposition}

Let \(f\), \(g\in \mathsf{R}[x]_{<n}\) and \(\lambda\in \mathsf{R}\).
Then \(h = f\times g \bmod x^n-\lambda\) is equivalent to
\(\vh = C_n^\lambda(\vec f)^\transp\cdot\vg\). Both problems can be
computed in \(\M(n)\) operations in \(\mathsf{R}\).\gobblepar

\end{proposition}

We now turn to evaluation and interpolation, that correspond to
Vandermonde matrices.

\begin{definition}

Let \(\vec \alpha\in \mathsf{R}^m\). The \emph{Vandermonde matrix}
\(V_{m,n}(\vec \alpha)\) is the \(m\times n\) matrix
\[V = \begin{pmatrix}
1 & \alpha_0 & \alpha_0^{2} & \dotsc & \alpha_0^{n-1} \\
1 & \alpha_1 & \alpha_1^{2} & \dotsc & \alpha_1^{n-1} \\
\vdots&\vdots&\vdots&&\vdots\\
1 & \alpha_{m-2} & \alpha_{m-2}^{2} & \dotsc & \alpha_{m-2}^{n-1} \\
1 & \alpha_{m-1} & \alpha_{m-1}^{2} & \dotsc & \alpha_{m-1}^{n-1}
\end{pmatrix}\] defined by \(V_{i,j} = \alpha_i^j\) for \(0 \leq i < m\)
and \(0 \leq j < n\). If \(m = n\), we denote the square matrix
\(V_{m,m}(\vec \alpha)\) by \(V_m(\vec \alpha)\).

If \(\alpha\in R\), a special case of Vandermonde matrix is the
\emph{Fourier matrix}\footnote{When \(\alpha\) has order \(m\), it is
  also known as the \emph{DFT matrix}, for \emph{discrete Fourier
  transform}.}
\(F_m(\alpha) = V_m(1,\alpha,\alpha^{2},\dotsc,\alpha^{m-1})\).

\end{definition}

\notation[Vandermonde $m\times n$ and $m\times m$ matrices with second column $\vec \alpha$\nomrefpage]{\(V_{m,n}(\vec \alpha)\),
\(V_m(\vec \alpha)\)}(wxV)
\notation[DFT matrix $V_m(1,\alpha,\dotsc,\alpha^{m-1})$\nomrefpage]{\(F_m(\alpha)\)}(wxVsF)
\index{matrix!Vandermonde} \index{matrix!Fourier}

Note that a Fourier matrix and its transpose have both the Vandermonde
structure.

\begin{proposition}

Let \(f\in \mathsf{R}[x]\) of size \(n\),
\(\vec \alpha\in \mathsf{R}^m\) and
\(\ve = (f(\alpha_0),\dotsc,f(\alpha_{m-1})) \in\mathsf{R}^m\). Then
\(\ve = V_{m,n}(\vec \alpha)\cdot\vec f\). Conversely, if
\(\vec \alpha\), \(\vec \beta\in \mathsf{R}^m\), the vector of
coefficients of the unique size-\(m\) polynomial such that
\(f(\alpha_i) = \beta_i\), \(0\leq i<m\), is
\(\vf = V_m(\vec \alpha)^{-1}\cdot\vec \beta\).\gobblepar

\end{proposition}

A consequence of these equivalences is that (square) Vandermonde
matrix-vector products and Vandermonde system solving can be computed in
\(O(\M(n)\log n)\) operations in \(\mathsf{R}\). In the case of a
Fourier matrix, both can be computed with \(O(\M(n))\) operations.

\section{Straight-line programs and arithmetic
circuits}\label{section:SLP}

Straight-line programs are a representation of polynomials by programs
of evaluation that have neither loop nor test. They have been very
successfully used in computer algebra for instance for polynomial
factorization~\autocite{Kaltofen2003} or for polynomial system
solving~\autocite{GiustiHeintzMoraisMorgensternPardo1998}. We shall need
them as inputs of some sparse interpolation algorithm, \emph{cf.}
\hyperref[chapter:sparseinterpolation]{Chapter~\ref*{chapter:sparseinterpolation}}.
And the transposition principle presented in
\hyperref[section:transposition]{Section~\ref*{section:transposition}}
can be phrased with \emph{linear} straight-line programs.

\begin{definition}\label{definition:SLP}

A \emph{straight-line program} (SLP) \(\mathcal{S}\) with \(n\)
variables \(x_1\), \ldots, \(x_n\) over some ring \(\mathsf{R}\) is a
list of \(\ell\) instructions. The \(i\)th instruction,
\(0\leq i<\ell\), is of the form \(r_i \gets u \star v\) where \(r_i\)
is a \emph{register}, \(u\) (resp. \(v\)) is either a variable, a
constant from \(\mathsf{R}\) or a register \(r_j\), \(j < i\), and
\(\star \in \{+,-,\times\}\). Over a field \(\mathsf{K}\), an SLP
\emph{with divisions} allows divisions, that is
\(\star\in\{+,-,\times,/\}\).

The output of an SLP is a tuple \((r_{i_0}, \dotsc, r_{i_m})\) of
\emph{output registers}. A single-output SLP has \(r_{\ell-1}\) as
unique output register.

The result of an SLP (resp. an SLP with divisions) on inputs
\(\alpha_1\), \ldots, \(\alpha_n\in\mathsf{R}\) (resp. \(\mathsf{K}\))
is the \(m\) values of its output registers when each variable \(x_j\)
is replaced by the value \(\alpha_j\), and each instruction is executed
in order by interpreting the operation \(\star\) by the corresponding
operation in \(\mathsf{R}\) (resp. \(\mathsf{K}\)).\gobblepar

\end{definition}

\index{straight-line program} \index{program!straight-line}
\index{SLP|see{straight-line program}}

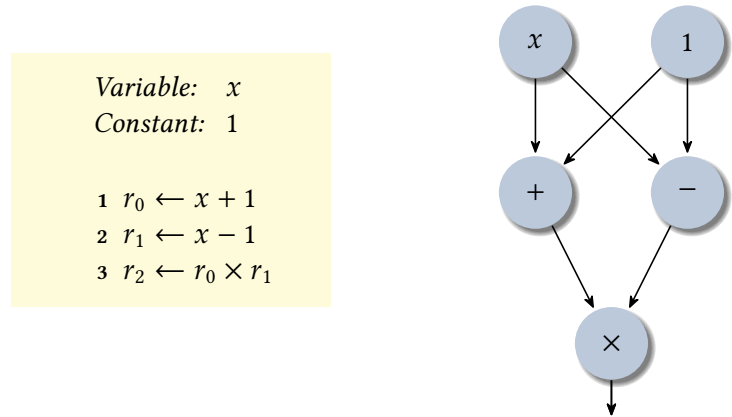
\begin{figure}

\hfill
\begin{tcolorbox}[mybox=jaune!25,width=.3\textwidth,nobeforeafter,breakable=false]
\begin{algo}
\begin{description}
\item[Variable:] $x$
\item[Constant:] $1$
\end{description}
\begin{enumerate}
\item $r_0 \gets x + 1$
\item $r_1 \gets x - 1$
\item $r_2 \gets r_0 \times r_1$
\end{enumerate}
\end{algo}
\end{tcolorbox}\hfill
\input{circuit.tikz}
\hfill\null
\caption{Straight-line program for the polynomial $x^2-1$ (left) and its arithmetic circuit representation (right).}
\label{figure:slp}

\end{figure}

A single-output straight-line program computes a polynomial function of
its inputs (or a rational function if it allows divisions). It admits a
graphical representation called an \emph{arithmetic circuit}. It is
formally a directed acyclic graph with one vertex per variable, one
vertex per constant, and one vertex per instruction. The vertices
associated to the variables and the constant have in-degree \(0\). The
vertex associated to an instruction \(r_i \gets u\star v\) has two
incoming arcs, one from the vertex associated to \(u\) and one from the
vertex associated to \(v\). An example is given in
\hyperref[figure:slp]{Figure~\ref*{figure:slp}}.

\index{arithmetic circuit}

\section{Transposition principle and
reversion}\label{section:transposition}

The \emph{transposition
principle}~\autocite{Fiduccia1972,Fiduccia1973,KaminskiKirkpatrickBshouty1988,Shoup1994}
relates, for any matrix \(M\in\mathsf{R}^{m\times n}\), the cost of
computing \(M\cdot\vv\) for \(\vv\in\mathsf{R}^n\) and the cost of
\(M^\transp\cdot\vw\) for \(\vw\in\mathsf{R}^m\), or equivalently to
compute \(\vw\cdot M\). (Here and thereafter, we adopt the convention
that a vector is understood as a column vector in a matrix-vector
product while it is understood as a row vector in a vector-matrix
product.) It can be formalized with the use of \emph{directed acyclic
graphs} or \emph{linear programs}~\autocite{BostanLecerfSchost2003}. A
linear program is a variant of a straight-line program, with only linear
operations.

\begin{definition}\label{definition:linearprogram}

A \emph{linear program} with \(n\) inputs over \(\mathsf{R}\) is a
straight-line program where each instruction is of the form
\(r_i \gets \pm u \pm v\) or \(r_i \gets \lambda\cdot u\) where \(u\)
(resp. \(v\)) is either a variable, a constant or a register \(r_j\),
\(j < i\), and \(\lambda \in \mathsf{R}\).\gobblepar

\end{definition}

\index{program!linear}

A linear program computes a linear mapping. A linear program for a
matrix \(M\) is a linear program that, on input \(\vv\in\mathsf{R}^n\),
computes \(\vw = M\cdot\vv\). The transposition principle is a program
transformation technique to obtain a linear program for the transposed
of a matrix.

\begin{proposition}\label{proposition:transposition}

Any linear program of size \(\ell\) for \(M\in\mathsf{R}^{m\times n}\)
can be turned into a linear program of size \(\ell+m-n\) for the
transpose matrix \(M^\transp\).\gobblepar

\end{proposition}

\notation[transposed of the matrix $M$\nomrefpage]{\(M^\transp\)}(wt)
\index{transposition principle}

Examples of transposition are very common in polynomial computation. For
instance, the transposed version of (full) polynomial multiplication is
the middle product. The lower and upper products are (almost\footnote{In
  our definition, the matrix of the lower product has a nonzero
  diagonal, contrary to the matrix of the upper product.}) transposed of
each other. Some of the fastest algorithms have been found by means of
the transposition principle, first designing a fast algorithm for the
transposed problem~\autocite{BostanLecerfSchost2003,BostanSchost2005}.

We shall use another program transformation technique, coined
\emph{reversion}. Given a linear program for a matrix
\(M\in\mathsf{R}^{m\times n}\), we can obtain a same size linear program
for the matrix \(M^{\shortleftarrow}\in\mathsf{R}^{m\times n}\) defined
by \(M^{\shortleftarrow}_{i,j} = M_{m-1-i,n-1-j}\). The computation
\(\vw \gets M^{\shortleftarrow}\cdot\vv\) is equivalent to
\(\cev w \gets M\cdot\cev v\). The transformation is extremely simple:
reverse the order of the input registers (\(r_{-n+1}\) becomes \(r_0\),
\(r_{-n+2}\) becomes \(r_{-1}\), etc.) and of the output registers (the
output tuple becomes \((i_{m-1}, \dotsc, i_0)\)). With this
transformation, we can for instance define the \emph{reversed power
series division} that takes as inputs \(f = \phi\bmod x^n\) and
\(g = \psi\bmod x^n\) such that the leading coefficient of \(g\) is a
unit, and returns \(h\) such that
\(h^{\shortleftarrow} = f^{\shortleftarrow}/g^{\shortleftarrow} \bmod x^n\).
This is a basic operation in the fast Euclidean algorithm, \emph{cf.}
\hyperref[figure:EuclideanDiv]{Figure~\ref*{figure:EuclideanDiv}} on
page \pageref{figure:EuclideanDiv}.

\begin{proposition}\label{proposition:reversion}

Any linear program of size \(\ell\) for \(M\in\mathsf{R}^{m\times n}\)
can be turned into a linear program of the same size for the reversed
matrix \(M^{\shortleftarrow}\).\gobblepar

\end{proposition}

\notation[reversed of the matrix $M$\nomrefpage]{\(M^{\shortleftarrow}\)}(wr)
\index{algorithm!reversed} \index{reversed power series division}

\begin{remark}

Propositions \ref{proposition:transposition} and
\ref{proposition:reversion} look superficially very similar. Yet the
transposition principle requires a proof, even if not very involved, and
some questions about it remain open~\autocite{Kaltofen2000}. The
reversion on the other hand is the simple remark that one can reverse
the order of the indices in a program that manipulates vectors or
arrays.\gobblepar

\end{remark}

\part{Time- and space-efficient polynomial
computations}\label{part:cstspace}

\chapter*{Summary}\label{Section:summary}
\addcontentsline{toc}{chapter}{Summary}

As presented in
\hyperref[chapter:preliminaries]{Chapter~\ref*{chapter:preliminaries}},
the complexity of many polynomial operations has been reduced from
quadratic for the classical algorithms to subquadratic and even
quasi-linear time. This is due to faster algorithms for polynomial
multiplication on the one hand, and algorithmic reductions from many
operations to multiplication with the smallest possible overhead on the
other hand.

Nevertheless, the price to pay for these faster algorithms is an
increase in the space complexity. The quadratic polynomial
multiplication algorithm is easily seen to require no extra space. Other
quadratic algorithms can be implemented without extra space as
well~\autocite{Monagan1993}. But fast multiplication algorithms or other
fast polynomial algorithms require at least a linear amount of space,
and sometimes even more. As a simple example, consider Karatsuba's
multiplication algorithm~\autocite{KaratsubaOfman1963}. To compute
\(f\times g\), the two polynomials are written as \(f_0+x^mf_1\) and
\(g_0+x^mg_1\), and their product is written
\(f_0g_0 + x^m((f_0+f_1)(g_0+g_1)-f_0g_0-f_1g_1) + x^{2m}f_1g_1\). Using
three half-sized recursive calls decreases the complexity of
\(O(n^{\log 3})\). But in terms of space complexity, a linear amount of
space is required to store the intermediate results \(f_0+f_1\),
\(g_0+g_1\), \(f_0g_0\), \(f_1g_1\) and \((f_0+f_1)(g_0+g_1)\).

The goal of this part is to present some simultaneously fast and
constant-space algorithms for many basic polynomial operations. In
\hyperref[chapter:models]{Chapter~\ref*{chapter:models}}, we first
discuss the models of computation. There are some subtleties to properly
define time-space complexity classes for functions, and we argue that
the standard complexity classes are not very well suited for this case.
We also exhibit the links with traditional complexity theory, and in
particular explain that our \emph{constant space} more or less
corresponds to the traditional \emph{logarithmic space}. Chapters
\ref{chapter:rorw} and \ref{chapter:rwrw} present two series of
incomparable results. The first one is in a more restricted model. The
second one uses a more permissive model but provides stronger results,
focusing on \emph{cumulative} operations such as \(c \pe a\times b\).
Finally, \hyperref[chapter:automatic]{Chapter~\ref*{chapter:automatic}}
presents a partial automatization of the results of
\hyperref[chapter:rwrw]{Chapter~\ref*{chapter:rwrw}}, that is algorithms
to produce constant-space variants of standard algorithms with the same
asymptotic running times. We also apply these techniques to
linear-algebraic problems.

This is based on a series of works with Pascal Giorgi (U. Montpellier)
and Daniel S. Roche (U.S. Naval
Academy)~\autocite{GiorgiGrenetRoche2019,GiorgiGrenetRoche2020} and with
Jean-Guillaume Dumas (U. Grenoble
Alpes)~\autocite{DumasGrenet2024,DumasGrenet2024a,DumasGrenet2025}.

\section*{Notations and
conventions}\label{Section:notations-and-conventions}
\addcontentsline{toc}{section}{Notations and conventions}

In this part, we consider polynomials over an abstract ring
\(\mathsf{R}\). We assume that it is an integral domain, although
several results still hold in more general settings.

The coefficient of degree \(i\) of a polynomial \(f\) is denoted
\(f_{[i]}\). We keep non-bracketed subscripts such as \(f_0\) or \(f_1\)
to denote parts of the polynomial \(f\). The size-\((j-i)\) polynomial
made of terms of degree \(i\) to \(j-1\) is either denoted \(f_{[i,j[}\)
or \([f]_i^j\). Using the more mathematical notation \(f\bmod x^j\) and
\(f\bquo x^i\), we have the equality
\[f_{[i,j[} = [f]_i^j = (f\bmod x^j) \bquo x^i = (f\bquo x^i)\bmod x^{j-i} = \sum_{d=0}^{j-i-1} f_{[d+i]} x^d.\]
Our algorithms make use of the two program transformations presented in
\hyperref[section:transposition]{Section~\ref*{section:transposition}},
namely \emph{transposition} and \emph{reversion}. The \emph{reverse} of
a polynomial \(f\) is \(f^{\shortleftarrow} = x^{\deg f} f(1/x)\).
Combining both notations, \(f^{\shortleftarrow}_{[i,j[}\) denotes the
polynomial \(\sum_{d=0}^{j-i-1} f_{[j-1-d]} x^d\).
\notation[coefficient of degree $i$ of $f$\nomrefpage]{\(f_{[i]}\)}(pi)
\notation[polynomial $\sum_{d=i}^{j-1} f_{[d]}x^{d-i}$\nomrefpage]{\(f_{[i,j[}\)
or \([f]_i^j\)}(pii)
\notation[polynomial $\sum_{d=i}^{j-1} f_{[d]}x^{j-d-1}$\nomrefpage]{\(f\gets_{[i,j[}\)}(pit)

As mentioned earlier, we describe cumulative algorithms. Therefore, our
basic operations are not only assignments but also \emph{compound
assignments} or \emph{fused operations}. We use the notation \(x\se v\)
for the assignment of the value \(v\) to \(x\), \(x \pe v\) for
\(x \se x+v\), \(x \me v\) for \(x\se x-v\), \(x \fe v\) for
\(x\se x\times v\) and \(x \de v\) for \(x\se x/v\). We also extend the
latter to \(x \fe v\bmod m\) for \(x\se (x\times v)\bmod m\) and
\(x \de v\bmod m\) for \(x\se(x/v)\bmod m\).
\notation[affectation of the value $v$ to the variable $a$\nomrefpage]{\(a \gets v\)
or \(a\se v\)}(alg)
\notation[compound affectations $a \se a+v$, $a\se a-v$\nomrefpage]{\(a\pe v\),
\(a\me v\)}(alga)
\notation[compound affectations $a\se a\times v$, $a \se a/v$\nomrefpage]{\(a\fe v\),
\(a\de v\)}(algb)
\notation[compound affectation $a \se (a\times v)\bmod w$\nomrefpage]{\(a\fe v\bmod w\)}(algc)
\notation[compound affectation $a \se (a/v)\bmod w$\nomrefpage]{\(a \de v\bmod w\)}(algd)

Several algorithms that we describe have complexity \(O(\M(n)\log n)\).
Actually, the extra logarithmic factor occurs when \(\M(n)\) is
quasi-linear. Otherwise, the complexity is actually \(O(\M(n))\). We
introduce the notation \(\M^*(n)\) for these complexities. Formally,
\(\M^*(n) = O(\M(n)\log(n))\), or \(O(\M(n))\) if
\(\M(n) = \Omega(n^{1+\varepsilon})\) for some \(\varepsilon>0\).
\notation[$O(\M(n)\log n)$ or $O(\M(n))$ if $\M(n) = \Omega(n^{1+\varepsilon})$ for some $\varepsilon>0$\nomrefpage]{\(\M^*(n)\)}(cms)

Finally, the literature on constant-space algorithms, although small,
uses a very diverse and inconsistent vocabulary to describe the
properties of these algorithms.\footnote{I plead guilty.} Below are the
terms used in this document:

\begin{itemize}
\tightlist
\item
  \emph{constant-space algorithm}: an algorithm that uses \(O(1)\) extra
  space, beyond its input(s) and output(s);
\item
  \emph{in-place algorithm}: an algorithm that replaces (part of) its
  inputs by the output;
\item
  \emph{cumulative algorithm}: an algorithm that adds its output to
  (part of) its inputs.
\end{itemize}

An \emph{in-place} algorithm may use \emph{constant space}, but this is
not required. A \emph{linear-space in-place algorithm} replaces its
input by its output, using a work space of linear size. A
\emph{cumulative} algorithm is a special case of an in-place algorithm.

\index{algorithm!constant space} \index{algorithm!in-place}
\index{algorithm!cumulative}

\chapter{The computational models -- and why they do
matter}\label{chapter:models}

To express our algorithms and analyze their time-space complexities, a
proper model of computation must be defined. Straight-line programs
suffer two limitations. They handle fixed-size inputs, and there is no
meaningful notion of space complexity attached to them. In
\hyperref[section:algalg]{Section~\ref*{section:algalg}}, we introduce
our models of computation based on the \emph{algebraic Random Access
Machine (RAM)}~\autocite{AhoHopcroftUllman1974,PreparataShamos1990}.
While standard computational complexity theory defines space complexity
using machines with read-only inputs and write-only
outputs~\autocite{AroraBarak2009}, we argue in
\hyperref[section:space]{Section~\ref*{section:space}} that this model
is not suitable for the algebraic computations we are interested in, and
we define several models of space complexity.
\hyperref[section:callstack]{Section~\ref*{section:callstack}} discusses
space complexity of recursive algorithms, in particular the role of the
call stack. Finally, we relate our definitions to standard space
complexity classes in
\hyperref[section:cstspace-comparisons]{Section~\ref*{section:cstspace-comparisons}}.

The title of this chapter is borrowed from~\autocite[Chapter
1]{AroraBarak2009}: \emph{The computational model---and why it doesn't
matter}.

\section{Algebraic algorithms and their models of
computation}\label{section:algalg}

We first define our objects of study, namely algebraic problems.

\begin{definition}\label{definition:AlgebraicProblem}

A \emph{fixed-size algebraic problem} over some ring \(\mathsf{R}\) is a
mapping \(\pi:\mathsf{R}^n \to \mathsf{R}^n\). An \emph{algebraic
problem} over \(\mathsf{R}\) is a family \((\pi_n)_{n\geq 0}\) of
fixed-size algebraic problem, where
\(\pi_n:\mathsf{R}^n\to\mathsf{R}^n\).\gobblepar

\end{definition}

\index{algebraic problem} \index{algebraic problem!fixed-size}
\index{fixed-size!algebraic problem}

We view an algebraic problem as a rewriting process. Given
\(\vv\in \mathsf{R}^n\), the goal is to replace \(\vv\) by \(\pi(\vv)\).
In particular, an algorithm that computes an algebraic problem \(\pi\)
is given as inputs the entries of \(\vv\) in some registers, and must
replace the value \(\vv_{[i]}\) in the \(i\)th register by
\(\pi(\vv)_{[i]}\).

In our definition of an algebraic problem, there is \emph{a priori} no
distinction between inputs and outputs. For the example of polynomial
multiplication, we would like to say that \(f\) and \(g\) are the inputs
and \(h\) the output. We define a notion of inputs and outputs that can
be used informally.

\begin{definition}\label{definition:InputOutput}

Let \(\pi:\mathsf{R}^n\to\mathsf{R}^n\) be a fixed-size algebraic
problem.

\begin{itemize}
\item
  The \emph{inputs} of \(\pi\) are the indices \(i\in\{0,\dotsc,n-1\}\)
  such that \(\pi(\vv)\) \emph{depends on} \(\vv_{[i]}\), that is such
  that there exists \(\vv\), \(\vec{v'}\in \mathsf{R}^n\) such that
  \(\vv_{[i]}\neq\vv'_{[i]}\) but \(\vv_{[j]} = \vv'_{[j]}\) for
  \(i \neq j\), and \(\pi(\vv) \neq \pi(\vec{v'})\).
\item
  The \emph{outputs} of \(\pi\) are the indices \(i\in\{0,\dotsc,n-1\}\)
  such that there exists \(\vv\in\mathsf{R}^n\) where
  \(\pi(\vv)_{[i]} \neq \vv_{[i]}\).
\end{itemize}

\end{definition}

In many cases, it is more natural that an algebraic problem operates on
tuple of vectors. For instance, a polynomial multiplication
\(h\se f\times g\) operates on the triple of coefficient vectors
\((\vf,\vg,\vh)\). Viewing the triple as one long vector or three
smaller size vectors is equivalent. The definition of an algebraic
problem is general enough to encompass several situations. Classical
functions from computer science have an input and a (separate) output. A
computation such as \(y \se f(x)\) where the input is \(x\) and the
output is \(y\) is represented by \((x,y)\mapsto(x,f(x))\) in the model.
Clearly, \(x\) is the input, and \(y = f(x)\) is the output. But the
model also allows cumulative computations such as \(y \pe f(x)\),
represented by \((x,y)\mapsto(x,y+f(x))\). In such a case, \(x\) is
still an input, and \(y\) is both an input and an output. Finally, one
can also have in-place computations such as \(x \se f(x)\), modeled as
\(x\mapsto f(x)\) where \(x\) is both the input and output. Without
further precision, an algebraic problem is thus an in-place problem.
Note that cumulative problems are a special case of in-place problems.

An SLP computes a fixed-size algebraic problem. To be able to define
space complexity, we refine
\hyperref[definition:SLP]{Definition~\ref*{definition:SLP}}.

\begin{definition}\label{definition:algprogram}

A \emph{fixed-size algebraic program} over \(\mathsf{R}\) has \(n\)
input-output registers \(r_0\), \ldots, \(r_{n-1}\), \(s\) temporary
registers \(t_0\), \ldots, \(t_{s-1}\) and \(\ell\) instructions of the
form \(u \se v\star w\) where \(u\), \(v\),
\(w\in\{r_0,\dotsc,r_{n-1},t_0,\dotsc,t_{s-1}\}\cup \mathsf{R}\)
(\(u\notin \mathsf{R}\)) and \(\star\in\{+,-,\times\}\).

It computes a fixed-size algebraic problem \(\pi\) if, given the
initialization of its input-output registers \(r_i\) to \(v_i\),
\(0\leq i<n\) and of its temporary registers to \(0\), the final value
of the input-output registers is \(\pi(\vv)_{[i]}\),
\(0\leq i<n\).\gobblepar

\end{definition}

\index{fixed-size!algebraic program}
\index{program!fixed-size algebraic}

From a fixed-size algebraic program, we can build an SLP of the same
length by replacing the \(i\)-th instruction \(u\se v\star w\) by
\(r_i\gets u\star w\) and the subsequent uses of \(u\) on the
right-hand-side of an instruction by \(r_i\). To handle inputs of any
size, one can use families of fixed-size algebraic programs. This
defines a \emph{nonuniform} model of computation. Although it is very
much adapted to algebraic complexity theory and in particular to proving
lower bounds~\autocite{Burgisser2000}, it is much less so to design
algorithms and prove upper bounds that reflect the practice of
programming. Instead of adding an outside uniformity requirement, we
prefer work directly with a uniform model of computation, namely the
algebraic RAM.

An algebraic RAM is parameterized by some ring \(\mathsf{R}\). It is
made of \emph{algebraic registers}, each containing a element from
\(\mathsf{R}\), and \emph{pointer registers} that store integers. We
provide one possible definition. Other definitions are possible,
\emph{cf.} for instance~\autocite{Seiller2024,ChanusMazzaRogers2025} for
recent formalizations.

\begin{definition}\label{definition:algebraicRAM}

An \emph{algebraic RAM} over a ring \(\mathsf{R}\) has algebraic
registers \((r_i)_{i\geq 0}\) storing elements of \(\mathsf{R}\), and
pointer registers \((p_i)_{i\geq 0}\) storing nonnegative integers. It
is controlled by a list of numbered instructions of one of the following
forms:

\begin{description}
\item[algebraic instructions]
\(r_a \se r_b\star r_c\), \(r_a\se r_b\) or \(r_a \se \lambda\) where
\(a\), \(b\) and \(c\) are either integer constants or pointer registers
(\emph{indirect addressing}), \(\lambda\in\mathsf{R}\), and
\(\star\in\{+,-,\times\}\);
\item[pointer instructions]
\(p_a \se p_b\star p_c\), \(p_a\se p_b\) or \(p_a \se m\) where \(a\),
\(b\) and \(c\) are either integer constants or pointer registers
(\emph{indirect addressing}), \(m\in\mathbb{Z}_{\geq 0}\), and
\(\star\in\{+,-,\times,/,\scalebox{.9}{\%}\}\);
\item[branching]
\textsc{if} \(r_a = 0\) \textsc{then} \textsc{goto} \(\ell\), where
\(a\) is either an integer constant or a pointer register and \(\ell\)
is an instruction number.
\end{description}

On input \(\vv\in\mathsf{R}^n\), \(v_i\) is stored in \(r_i\) for
\(0\leq i<n\), \(p_0\) contains \(n\), and the other registers are
initialized to \(0\). The instructions are executed in order, but if a
\textsc{goto} is encountered and the corresponding register contains
\(0\). Algebraic instructions are interpreted in \(\mathsf{R}\) and
pointer instructions in \(\mathbb{Z}_{\geq 0}\): \(a-b\) is actually
\(\max(0,a-b)\), \(a/b = a\bquo b\) and
\(a\mathbin{\scalebox{.9}{\%}}b = a\bmod b\). The computation stops when
the last instruction is executed (if no \textsc{goto} is applied). The
result is the content of the registers \(r_0\), \ldots, \(r_{n-1}\).

An algebraic RAM is \emph{honest} if for any input size \(n\), the
values of the pointer registers are \(O(\log n)\) during the
computation.\gobblepar

\end{definition}

\index{Random Access Machine (RAM)!algebraic} \index{register}

There is no limitation in our definition on the number of registers
used, or the magnitudes of the integers within the pointer registers. An
\emph{honest} algebraic RAM prevents any \emph{cheating} and corresponds
to the \emph{transdichotomous model}~\autocite{FredmanWillard1993} where
the word size of the pointers is large enough to write the size of the
inputs and outputs in \(O(1)\) pointers. In the rest of this document,
all algebraic RAMs are honest. Actually, we do not write the algorithms
with the formal syntax of algebraic RAMs, rather in a standard
pseudocode using conditional statements, loops and function calls
(including recursive calls). A translation from the former to the latter
is a classical exercise.

\section{Space complexity of algebraic algorithms}\label{section:space}

The standard model for space complexity in computational complexity
theory assumes that the inputs are read-only and the outputs
write-only~\autocite{AroraBarak2009}. The space complexity of an
algorithm is the number of extra registers required by the algorithm,
not counting the inputs and outputs. We depart from this model for
several reasons. On the theoretical side, a time-space quadratic lower
bound is known for polynomial multiplication in this
model~\autocite{Abrahamson1986}. This means that fast multiplication
algorithms, thereby fast algorithms on polynomials more generally,
require a polynomial amount of extra space. On the practical side, a
programmer allows some memory for the output. There is no good reason to
forbid the use of this space as work space. Our definition of algebraic
problems as a rewriting process makes the traditional space complexity
model inoperative. \emph{It's not a bug, it's a feature.}

Our goal is to analyze the time- and space-complexity of algorithms for
algebraic problems. There are some subtleties in defining the space
complexity of a function, especially for low-space algorithms. The
common feature of all possible definitions is that only the extra space
is counted, not the inputs nor the outputs. In our definition of an
algebraic problem, this means that the \emph{algebraic space} of an
algorithm computing some problem \(\pi:\mathsf{R}^n\to\mathsf{R}^n\) is
the number of algebraic registers used by the algorithm, \emph{in
addition to the \(n\) input/output registers}. We also define the
\emph{pointer space} of an algorithm as the number of pointers it uses.

\begin{definition}\label{definition:SpaceComplexity}

The \emph{space complexity} of a fixed-size algebraic program is the
number of temporary registers it uses.

An algebraic RAM computing some algebraic problem \((\pi_n)_{n\geq 0}\)
has \emph{algebraic space complexity} \(s_a(n)\) if, on any input of
size \(n\), the only algebraic registers modified by the machine are
\(r_0\), \ldots, \(r_{n+s_a(n)-1}\). It has \emph{pointer space
complexity} \(s_p(n)\) if the only modified pointer registers are
\(p_0\), \ldots, \(p_{s_p(n)-1}\).\gobblepar

\end{definition}

\index{space complexity} \index{space complexity}

\begin{remark}

We shall always assume in the definition of an algebraic problem
\(\pi_n:\mathsf{R}^n\to\mathsf{R}^n\) that each of the \(n\) entries is
either an input, an output, or both in the sense of
\hyperref[definition:InputOutput]{Definition~\ref*{definition:InputOutput}}.
Indeed, an entry that is neither an input nor an output plays no role in
the computation and is only a placeholder. Prohibiting these prevents
cheating when defining space complexity.\gobblepar

\end{remark}

Our definition of space complexity does not put any restriction on the
use of the input and output registers of the machine. This allows to
define richer notions of space complexity, based on read-write
permissions for the registers. We define three \emph{permission models}
for algebraic RAMs. They are based on the notions of \emph{input-only}
registers, that are inputs but not outputs, and \emph{output-only}
registers.

\begin{definition}

An algebraic RAM has permissions

\begin{itemize}
\item
  \texttt{ro/wo} if it never writes in an input-only register, and never
  reads from an output-only register;
\item
  \texttt{ro/rw} if it never writes in an input-only register;
\item
  \texttt{rw/rw} if it has no read nor write restriction.
\end{itemize}

\end{definition}

\index{permission models}
\index{rowo@\texttt{ro/wo}|see{permission models}}
\index{rorw@\texttt{ro/rw}|see{permission models}}
\index{rwrw@\texttt{rw/rw}|see{permission models}}

The standard model \texttt{ro/wo} has one theoretically attractive
feature. Consider two functions \(f:\mathsf{R}^\ell\to\mathsf{R}^m\) and
\(g:\mathsf{R}^m\to\mathsf{R}^n\) and their composition
\(h = g\circ f:\mathsf{R}^\ell\to\mathsf{R}^n\). (In our model, this
would be \(\pi:(\vu,\vv)\mapsto(\vu,f(\vu))\),
\(\rho:(\vv,\vw)\mapsto(\vv,g(\vv))\) and
\(\rho\circ\pi:(\vu,\vw)\mapsto(\vu,g\circ f(\vu))\) where
\(\vu\in \mathsf{R}^\ell\), \(\vv\in \mathsf{R}^m\) and
\(\vw\in \mathsf{R}^n\).) If \(f\) and \(g\) can both be computed in
space \(s(n)\) for some \(s\), then \(g\circ f\) can also be computed in
the same space. To avoid storing the intermediate result
\(\vv = f(\vu)\), the technique is to recompute each entry of \(\vv\)
when it is needed during the computation of \(g\)~\autocite[Chapter
4]{AroraBarak2009}. Unfortunately, this \emph{composition theorem} does
not hold in the time-space settings. Since entries of \(f(\vu)\) must be
(in general) computed several times, the time complexity of the
composition of the two algorithms is not the sum of their original time
complexities. This makes this model much less attractive for time-space
complexity considerations.

The model \texttt{ro/rw} corresponds fairly closely to practice. Once
the output space has been allocated, the programmer can use it as she
wants. On the other hand, it is quite common to declare inputs (that are
\emph{input-only}) as constant using keywords such as
\VERB|\DataTypeTok{const}| in \textsc{c} or \textsc{c++}, making an
input read-only. This is in particular useful for parallel programming
if there are parallel accesses to the inputs.

The model \texttt{rw/rw} is also natural in practice, especially in the
\emph{rewriting} view of algebraic problems. For instance for matrix
computations, it is possible to replace a matrix \(M\) by its \(LU\)
decomposition, where \(L\) is stored in the lower triangular part of
\(M\) and \(U\) in the upper triangular part. For such computations, the
inputs must obviously not be declared constant since they are modified
by definition. The drawback of this model is to make parallel accesses
to the inputs more complex.

\section{Call stack and tail recursion}\label{section:callstack}

\index{call stack} \index{tail recursive call}

The model of algebraic RAM is able to simulate recursive calls. This
requires to use a call stack. In our model the call stack will be made
of pointers only. In most cases, the call stack for an algorithm
operating on a size-\(n\) vector will be of size \(O(\log n)\).
Therefore, several of our algorithms use \(O(1)\) algebraic registers
and \(O(\log n)\) pointers for the call stack.

A special case of recursive algorithms are tail recursive algorithms
where the only recursive call is the last instruction of the algorithm.
In such a case, the call stack is not required. This means that a tail
recursive algorithm using \(O(1)\) pointers can be simulated by loops
still using \(O(1)\) pointers. The situation can be generalized using
\emph{tail recursion modulo cons}~\autocite{FriedmanWise1975} and its
generalizations such as \emph{tail recursion modulo
context}~\autocite{LeijenLorenzen2023}, or \emph{continuation-passing
style}~\autocite{Reynolds1972}. In our case, we shall not need such
involved programming techniques. We only need to generalize tail
recursion to the following situation. Consider an algebraic problem
\((\vv,\vw)\mapsto(\vv,f(\vv,\vw))\) and a recursive algebraic algorithm
\(\mathcal{A}\) that makes only one recursive call. We assume that after
the recursive call, \(\mathcal{A}\) only operates on \(\vv\). Then, one
can define another algorithm \(\mathcal{A}'\), tail recursive, that
simply ignores the post-treatment on \(\vv\). Then \(\mathcal{A}'\)
computes some problem \((\vv,\vw)\mapsto(\vx,f(\vv,\vw))\). And the
post-treatment on \(\vv\) is some algorithm \(\mathcal{A}"\) that
computes \(\vx\mapsto\vv\). Since \(\mathcal{A}'\) is tail recursive, it
can be simulated by an algebraic RAM without call stack, and the
sequential application of \(\mathcal{A}'\) and \(\mathcal{A}"\)
simulates \(\mathcal{A}\). Therefore, \(\mathcal{A}\) can be simulated
an algebraic RAM without call stack.

\section{Comparisons with standard space complexity
theory}\label{section:cstspace-comparisons}

The main difference in our model is the relaxation on the inputs and
outputs that are not read-only and write-only respectively. The more
traditional model suffers from quadratic lower bounds as explained, but
is also completely irrelevant for cumulative or in-place computations.
The second difference is that we focus on time-space complexity classes.
Our ultimate goal is algorithms that have quasi-linear time complexity,
ideally \(O(\M(n))\), and use a constant number of algebraic and pointer
registers. This could be phrased in the settings of \emph{fine-grained}
complexity theory~\autocite{VassilevskaWilliams2019}, and more
specifically in terms of fine-grained time-space complexity
classes~\autocite{LincolnVassilevskaWilliamsWangWilliams2016}.

Let us focus on space complexity. To compare our results with standard
complexity classes such as \(\mathsf{L}\), we need to get back to bit
complexity. Since algebraic algorithms naturally manipulate two kinds of
data (pointers and algebraic elements), they do not nicely fit within
the model. Yet consider an algorithm that takes as inputs \(n\) elements
from a finite field \(\mathbb{F}_p\). The input bit size is
\(O(n\log p)\), and a logarithmic space means an extra space of size
\(O(\log(n\log p)) = O(\log n+\log\log p)\). Depending on the relative
magnitudes of \(n\) and \(p\), a same algorithm could be considered in
\(\mathsf{L}\) or not. We can distinguish two regimes:

\begin{itemize}
\tightlist
\item
  If \(\log p = \omega(\log n)\), it is not even possible to store a
  single algebraic element beyond the inputs and outputs. On the other
  hand, it is possible to store up to \(\log_n p\) pointers. In this
  regime, details on the authorized algebraic operations in the model
  impact the space complexity. A cumulative product of ring elements
  \(c \pe a\times b\) can be performed in logarithmic space if some kind
  of \emph{fused multiply-add} is in the set of operations of the
  machine. But if only additions and multiplications are allowed, the
  intermediate result \(t = a\times b\) must be stored, and the
  algorithm does not run in logarithmic space.
\item
  If \(\log p = O(\log n)\), only a constant number of pointers can be
  stored, but it is possible to store a larger number of algebraic
  registers (at most \(\log_p n\)).
\end{itemize}

Therefore, with the caveat of very large fields, one can consider that
an algorithm that uses a constant number of registers of both kinds has
a logarithmic space complexity. Nevertheless, as the above discussion
shows, classical complexity classes are not the best option to study
algebraic algorithms.

Another comparison can be made with two space complexity models that
have been recently highlighted by the spectacular result that
\(\textsf{TIME}(t(n)) \subseteq \textsf{SPACE}(\sqrt{t(n)\log n})\)~\autocite{Williams2025}.
Both models use the multi-tape Turing machine. The first model, coined
\emph{global storage model}, is due to
Goldreich~\autocite{Goldreich2008}. It is a Turing machine that has a
\emph{global tape} where both the input and the output are written (as
well as oracle queries), and one or several \emph{local tapes} that
serve as work space. This model is very close to the model
\texttt{rw/rw} we defined. The second model is known as \emph{catalytic
computation}~\autocite{BuhrmanCleveKouckyLoffSpeelman2014}. In this
model, a \emph{catalytic tape} is given that initially contains some
data. The Turing machine is allowed to write on this catalytic tape, but
it must be ultimately restored in its initial state. In our model
\texttt{rw/rw}, this corresponds to adding a \emph{dummy} input
\(\vec c\) that is neither an input nor an input. A problem \(\pi\)
becomes \(\pi^*:(u,c)\mapsto(\pi(u),c)\).

\chapter{\texorpdfstring{Algorithms in the \texttt{ro/rw}
model}{Algorithms in the ro/rw model}}\label{chapter:rorw}

The main line of work for space-efficient polynomial computations is the
investigation of low-space polynomial multiplication algorithms in the
\texttt{ro/rw} model. The exact space complexity of Karatsuba's
algorithm is analyzed by Maeder when one preallocates all the necessary
memory once~\autocite{Maeder1993}. Then, an unpublished note by Thomé
shows how to implement the algorithm using exactly \(n\) extra algebraic
registers~\autocite{Thome2002a}. Roche describes several low-space
multiplication algorithms: a variant of Karatsuba's algorithm that uses
only constant algebraic space, and a constant-space FFT-based algorithm
for polynomials of power-of-two size~\autocite{Roche2009,Roche2011}.
Together with Harvey, they extend this latest result to any
size~\autocite{HarveyRoche2010}. Low-space Toom-Cook algorithms have
been investigated in the context of polynomials over
\(\mathbb{F}_2\)~\autocite{SuFan2012}.

In our work with Pascal Giorgi and Daniel S. Roche, we first provide
generic (\emph{algorithm-agnostic}) reductions for polynomial
multiplication, proving that any linear-space multiplication algorithm
has a constant-space variant with the same asymptotic time complexity.
We also investigate \emph{subproducts} such as lower and upper products
or middle product. In particular, we highlight the links between the
different problems in a \emph{fine-grained time-space} framework. In a
second work, we extend the results to other classical computer algebra
operations such as power series inversion, Euclidean division,
multipoint evaluation and interpolation. We refer to the original
publications~\autocite{GiorgiGrenetRoche2019,GiorgiGrenetRoche2020} for
detailed analyses of the implied \emph{hidden} constants in the time
complexities.

The general idea of our algorithms is to use the free space in the
output space as work space. Since the size of this space decreases over
time while new coefficients of the result are computed, we need to adapt
the standard algorithms to take this decrease into account. We design
(tail) recursive algorithms that compute fewer coefficients at each
recursive call to keep the space complexity constant.

For recursive algorithms, it is customary to assume that the input size
is a power of two, or at least even. This is classically ensured by
\emph{padding} the input with zeroes if necessary. In the context of
constant-space algorithms, this is not possible. Nevertheless, we can
always use \emph{fake padding} on the inputs. While accessing a
nonexisting index in an array usually results in an error, we only need
a data structure implementation where the error is replaced by returning
\(0\). (This is easily implementable by catching exceptions in most
programming languages.) Note though that fake padding cannot be used on
the output in our context, since we use the output space as work space.

\index{fake padding}

\section{Generic reductions for polynomial
multiplication}\label{Section:generic-reductions-for-polynomial-multiplication}

In this section, we describe \emph{reductions} from any linear-space
multiplication algorithm to a constant-space variant with close time
complexity. We consider the standard full product of two polynomials
\(f\), \(g\in \mathsf{R}[x]\), as well as the middle, lower and upper
products. We then unravel the links that exist between low-space
algorithms for all these variants.

\subsection{Full product}\label{Section:full-product}

\begin{figure}

\centering
\input{semicumulativeprod.tikz}

\caption{\hyperref[algorithm:SemiCumulativeProduct]{Algorithm~\ref*{algorithm:SemiCumulativeProduct}}
as a Toeplitz matrix-vector product. The first step is \(f_0g\)
(purple), then \(f_1g_0\) (red) and finally a tail recursive call
\(f_1g_1\) (blue). \emph{Fake padding} is used to handle the shaded
parts.} \label{figure:SemiCumulativeProduct}\gobblepar

\end{figure}
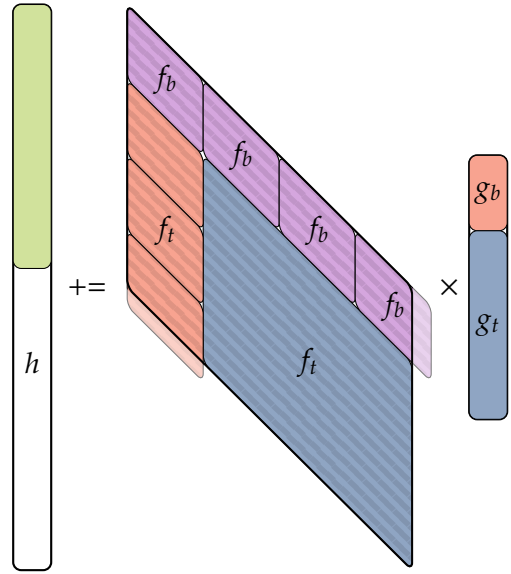

The reduction starts with a full product algorithm that, given two
size-\(n\) polynomials \(f\) and \(g\), computes their product
\(f\times g\) in \(O(\M(n))\) operations using \(\leq cn\) extra
registers. Our goal is to obtain a constant-space algorithm that
computes \(h \se f\times g\). Writing \(f = f_b+x^kf_t\) and
\(g=g_b+x^kg_t\) where \(f_b\) and \(g_b\) have size \(k\) for some
\(k < n\), \[h = f_bg + x^kf_tg_b + x^{2k} f_tg_t.\] The strategy is to
compute \(f_bg\) and \(f_tg_b\) using several calls to the linear-space
multiplication algorithm and \(f_tg_t\) by a (tail) recursive call. Yet
\(f_bg+x^kf_tg_b\) must be written in \(h_{[0,n+k-1[}\) and
\(x^{2k}f_tg_t\) in \(h_{[2k,2n-1[}\). Since they overlap, the recursive
call is not possible. Therefore, we generalize the problem to a
\emph{semi-cumulative full product}, that is \(h \pe f\times g\) where
\(h\bquo x^{n-1} = 0\). The recursive call becomes
\(h_{[2k,2n-1[} \pe f_t\times g_t\), where \(f_t\) and \(g_t\) have size
\(n-k\), and \(h\bquo x^{n-k-1} = 0\), hence it is legitimate. It
remains to set \(k\) so that the computations of \(f_bg\) and \(f_tg_b\)
can be performed in constant space, using the free space in the output
to store intermediate results. The algorithm is illustrated as a
Toeplitz matrix-vector product in
\hyperref[figure:SemiCumulativeProduct]{Figure~\ref*{figure:SemiCumulativeProduct}}.

\index{algorithm!semi-cumulative}
\index{polynomial!product!full!semi-cumulative}

Since \(f_bg\) can be computed using \(\lceil n/k\rceil\) calls to the
linear-space multiplication algorithm in size \(k\), this computation
requires \(ck+2k-1\) extra space. The same holds for \(f_tg_b\). And the
free space in \(h\) is \(h_{[n+k-1,2n-1[}\), of size \(n-k\). Therefore,
\(k\) must satisfy \((c+2)k-1 \leq n-k\), that is
\(k \leq (n+1)/(c+3)\). The formal description is given as
\hyperref[algorithm:SemiCumulativeProduct]{Algorithm~\ref*{algorithm:SemiCumulativeProduct}}.

\begin{algorithm}[SemiCumulativeProduct]

\begin{description}
\item[Inputs:]
\(f\), \(g\in \mathsf{R}[x]\) of size \(n\) \comment{\emph{read-only}}

\(h\in\mathsf{R}[x]\) of size \(2n-1\) such that \(h\bquo x^{n-1}=0\)
\comment{\emph{read-write}}
\item[Output:]
\(h \pe f\times g\)
\item[Required:]
full product algorithm with space complexity \(\leq cn\)
\item[Notations:]
\(k = \lfloor\frac{n+1}{c+3}\rfloor\) and \(\ell = \lceil n/k\rceil\)

write \(f = f_b+x^kf_t\) and \(g = g_b+x^kg_t\)
\end{description}

\begin{enumerate}

\item

if \(k = 0\): \(h \pe fg\) \comment{constant space}

\item

\(h_{[0,n+k-1[} \pe f_b\times g\) \comment{\(\ell\) products, free:
\(h_{[n+k-1,2n[}\)}

\item

\(h_{[k,n+k-1[} \pe f_t\times g_b\) \comment{\(\ell-1\) products, free:
\(h_{[n+k-1,2n[}\)}

\item

\(h_{[2k,2n[} \pe f_t\times g_t\) \comment{tail recursive call}

\end{enumerate}\gobblepar

\end{algorithm}

\begin{theorem}[\autocite{GiorgiGrenetRoche2019}]

\hyperref[algorithm:SemiCumulativeProduct]{Algorithm~\ref*{algorithm:SemiCumulativeProduct}} (\textsc{SemiCumulativeProduct})
is correct, requires \(O(1)\) extra space, and performs \(O(\M(n))\)
operations. \gobblepar

\end{theorem}

\subsection{Lower and upper
products}\label{Section:lower-and-upper-products}

We are now given lower and upper product algorithms that take as inputs
two size-\(n\) polynomials \(f\), \(g\in \mathsf{R}[x]\) and return
\((f\times g)\bmod x^n\) and \((f\times g)\bquo x^n\) respectively, both
in \(O(\M(n))\) operations and using \(\leq cn\) extra space. We
describe a constant-space algorithm to compute
\(h = (f\times g)\bmod x^n\). The reversed algorithm
(\hyperref[section:transposition]{Section~\ref*{section:transposition}})
provides a constant-space upper algorithm.

\begin{figure}

\centering
\input{lowerprod.tikz}

\caption{\hyperref[algorithm:LowerProduct]{Algorithm~\ref*{algorithm:LowerProduct}}
as a lower triangular Toeplitz matrix-vector product. The first steps
correspond to the bottom strip covered by triangular Toeplitz matrices.
The tail recursive call corresponds to the top triangular part.}
\label{figure:LowerProduct}\gobblepar

\end{figure}
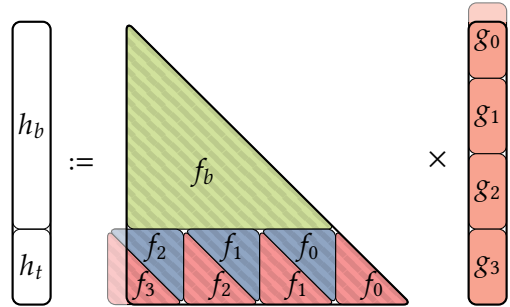

First write \(f = f_b+x^{n-k}f_t\), \(g = g_b+x^{n-k} g_t\) and
\(h = h_b+x^{n-k}h_t\) for some \(k\). Since
\(h_b = f_b\times g_b\bmod x^{n-k}\), it can be computed by a tail
recursive call. We focus on computing \(h_t\). An illustration as a
triangular Toeplitz matrix-vector product is given in
\hyperref[figure:LowerProduct]{Figure~\ref*{figure:LowerProduct}}.

Let \(\ell = \lceil n/k\rceil\) and \(r = k\ell-n\). Write now \(f\) and
\(g\) as a sum of size-\(k\) polynomials, namely
\(f = \sum_{i=0}^{\ell-1} f_i x^{ki}\) and
\(g = g_0+\sum_{j=1}^{\ell-1} g_jx^{kj-r}\) where \(f_0\), \ldots,
\(f_{\ell-2}\), \(g_1\), \ldots{} \(g_{\ell-1}\) have size \(k\), and
\(f_{\ell-1}\) and \(g_0\) have size \(k-r\). Their product can be
expanded as
\[f\times g = \sum_{i=0}^{\ell-1}\sum_{j=0}^{\ell-1} f_ig_jx^{k(i+j)-r_{[j\neq 0]}}\]
where \(r_{[j\neq 0]} = r\) if \(j\neq 0\) and \(0\) otherwise. We want
to compute \(h_1 = [f\times g]_{n-k}^n\). The degrees of nonzero terms
of \(f_ig_jx^{k(i+j)-r_{[j\neq 0]}}\) are between
\(k(i+j)-r_{[j\neq 0]}\) and \(k(i+j+2)-r_{[j\neq 0]}-2\). Therefore, if
\(k(i+j)-r_{[j\neq 0]} \geq n\) or \(k(i+j+2)-r_{[j\neq 0]}-2 < n-k\),
that is if \(i+j\geq\ell\) or \(i+j<\ell-2\), this summand plays no role
in the computation of \(h_1\). In other words,
\[h_1 = \left[x^{n-k}\cdot\sum_{i=0}^{\ell-1} f_ig_{\ell-1-i} + x^{n-2k}\cdot\sum_{i=0}^{\ell-2} f_ig_{\ell-2-i}\right]_{n-k}^n.\]
All nonzero terms in the first sum have degree \(\geq n-k\), whence we
need to compute \(\sum_i f_ig_{\ell-1-i}\bmod x^k\) using \(\ell\) lower
products. The second sum has degree at most \(n-2\). Therefore, we
simply need to compute \(\sum_{i=0}^{\ell-2}f_ig_{\ell-2-i}\bquo x^k\)
using \(\ell-1\) upper products.

Finally, to get the algorithm, we have to check that the linear-space
lower and upper multiplication algorithms have enough free space in the
output space. The results of these calls are written in \(h_1\) of size
\(k\), therefore \(n-k\) free registers are available in the output
space. To compute the two sums, we need \(k\) registers to store
intermediate results in addition to the \(ck\) registers required by the
linear-space algorithms. Therefore, the algorithm works as long as
\((c+1)k \leq n-k\), or \(k \leq n/(c+2)\). The formal description is
given as
\hyperref[algorithm:LowerProduct]{Algorithm~\ref*{algorithm:LowerProduct}}.

\begin{algorithm}[LowerProduct]

\begin{description}
\item[Inputs:]
\(f\), \(g\in \mathsf{R}[x]\) of size \(n\) \comment{\emph{read-only}}
\item[Output:]
\(h \se f\times g\bmod x^n\) \comment{\emph{read-write}}
\item[Required:]
lower and upper products algorithms with space complexity \(\leq cn\)
\item[Notations:]
\(k = \lfloor\frac{n}{c+3}\rfloor\), \(\ell = \lceil n/k\rceil\) and
\(r = k\ell-n\)

write \(f = \sum_{i=0}^{\ell-1} f_ix^{ki}\) and
\(g = g_0+\sum_{j=1}^{\ell-1} g_j x^{kj-r}\)
\end{description}

\begin{enumerate}

\item

if \(n < c+2\): \(h \se f\times g\bmod x^n\) \comment{constant space}

\item

for \(i = 0\) to \(\ell-1\): \label{line:lowprod:lowmul}

\item

~~~\(h_{[n-k,n[} \pe f_i\times g_{\ell-1-i}\bmod x^k\) \comment{lower
product, free: \(h_{[0,n-k[}\)}

\item

for \(i = 0\) to \(\ell-2\):

\item

~~~\(h_{[n-k,n[} \pe f_i\times g_{\ell-2-i}\bquo x^k\) \comment{upper
product, free: \(h_{[0,n-k[}\)} \label{line:lowprod:uppmul}

\item

\(h_{[0,n-k[} \se f_{[0,n-k[}\times g_{[0,n-k[}\bmod x^{n-k}\)
\comment{tail recursive call}

\end{enumerate}\gobblepar

\end{algorithm}

\begin{theorem}[\autocite{GiorgiGrenetRoche2019}]

\hyperref[algorithm:LowerProduct]{Algorithm~\ref*{algorithm:LowerProduct}} (\textsc{LowerProduct})
is correct, requires \(O(1)\) extra space, and performs \(O(\M(n))\)
operations. \gobblepar

\end{theorem}

As for
\hyperref[algorithm:SemiCumulativeProduct]{Algorithm~\ref*{algorithm:SemiCumulativeProduct}} (\textsc{SemiCumulativeProduct}),
the polynomial \(h\) may contain some data initially. Assume for
instance that the top \(n-s\) coefficients are nonzero for some \(s\).
The first step is to compute \(h_{[s,n[} \pe (f\times g)_{[s,n[}\).
Similarly to the computation made in Lines \ref{line:lowprod:lowmul} to
\ref{line:lowprod:uppmul}, this can be computed using lower and upper
products. Using the constant-space \textsc{LowerProduct} and its
reversed algorithm \textsc{UpperProduct}, one can compute parts of the
results (of size \(n-s\)) in the free space of \(h\), and then add it to
\(h_{[s,n[}\). As long as \(s \geq n/2\), this strategy works. Actually,
the same strategy adapts when \(s < n/2\) by computing \((fg)_{[s,n[}\)
by chunks of size \(s\). Then, the computation
\(h_{[0,s[} = fg\bmod x^s\) is another call to \textsc{LowerProduct}.

\begin{corollary}[unpublished]\label{corollary:SemiCumulativeLowerProduct}

Given \(f\), \(g\), \(h\in\mathsf{R}[x]\) of size \(n\) such that
\(h\bmod x^s = 0\) for some \(s > 0\), one can compute
\(h \pe fg\bmod x^n\) with \(O(1)\) extra space and \(O(\M(n))\)
operations if \(s \geq n/2\), and \(O((\frac{n}{s})^{2}\M(s))\)
otherwise.\gobblepar

\end{corollary}

Note that \((\frac{n}{s})^{2}\M(s) = O(\frac{n}{s}\M(n))\). If further
\(s \geq \alpha n\) for some constant \(\alpha\),
\(O((\frac{n}{s})^{2}\M(s)) = O(\frac{1}{\alpha}\M(n)) = O(\M(n))\).
Therefore, if the amount of free space is a constant fraction of \(n\),
the algorithm has the same asymptotic complexity as
\hyperref[algorithm:LowerProduct]{Algorithm~\ref*{algorithm:LowerProduct}} (\textsc{LowerProduct}).

Later, we shall need a special case of this algorithm, when \(g\) has
itself size \(s\). In this variant, small values of \(s\) make the
computation easier.

\begin{corollary}[unpublished]\label{corollary:SemiCumulativeLowerProduct2}

Given \(f\), \(g\), \(h\in\mathsf{R}[x]\) of respective sizes \(n\),
\(s\) and \(n\) such that \(s\leq n\) and \(h\bmod x^s = 0\), one can
compute \(h \pe fg\bmod x^n\) with \(O(1)\) extra space and \(O(\M(n))\)
operations.\gobblepar

\end{corollary}

\subsection{Middle product}\label{Section:middle-product}

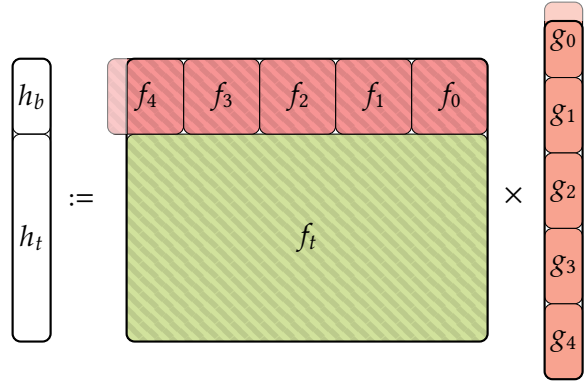
\begin{figure}

\centering
\input{middleprod.tikz}

\caption{\hyperref[algorithm:MiddleProduct]{Algorithm~\ref*{algorithm:MiddleProduct}}
as a rectangular Toeplitz matrix-vector product. The first step
corresponds to the top red strip and the tail recursive call to the
bottom green strip.} \label{figure:MiddleProduct}\gobblepar

\end{figure}

As a starting point of the reduction, we use a \emph{balanced} middle
product algorithm that computes the middle product of
\(f\in\mathsf{R}[x]_{<2n-1}\) and \(g\in\mathsf{R}[x]_{<n}\), in
\(O(\M(n))\) operations and using \(\leq cn\) extra space. We use it to
design a constant-space algorithm for the more general \emph{unbalanced}
middle product that takes as inputs \(f\in\mathsf{R}[x]_{<m+n-1}\) and
\(g\in\mathsf{R}[x]_{<n}\) returns their middle product
\([f\times g]_{n-1}^{m+n-1}\). The generalization is required to set up
the recursion. An illustration as a triangular Toeplitz matrix-vector
product is given in
\hyperref[figure:MiddleProduct]{Figure~\ref*{figure:MiddleProduct}}.

Let \(h = [f\times g]_{n-1}^{m+n-1}\) and write \(h = h_b + x^k h_t\)
for some \(k < m\). Then
\(h_b = [f\times g]_{n-1}^{n-1+k} = [f_b\times g]_{n-1}^{n-1+k}\) where
\(f_b=f_{[0,n-1+k[}\) and
\(h_1 = [f\times g]_{n-1+k}^{m+n-1} = [f_t\times g]_{n-1}^{m-k+n-1}\)
where \(f_t = f_{[k,m+n-1[}\). These are two middle products, in size
\((n+k-1,n)\) and \((m+n-k-1,n)\) respectively. The algorithm computes
\(h_b\) using \(\lceil n/k\rceil\) balanced middle products with extra
space \((c+1)k\), and \(h_t\) with a tail recursive call. As long as
\((c+1)k \leq m-k\), that is \(k \leq m/(c+2)\), the algorithm is in
constant space. Since the recursive call is in size \((m+n-k-1,n)\) and
does not decrease the size of \(g\), the complexity becomes
\(O(\M^*(n))\) (in the balanced case \(m=n\)). The formal description is
given as
\hyperref[algorithm:MiddleProduct]{Algorithm~\ref*{algorithm:MiddleProduct}}.

\begin{algorithm}[MiddleProduct]

\begin{description}
\item[Inputs:]
\(f\), \(g\in \mathsf{R}[x]\) of respective sizes \(m+n-1\) and \(n\)
\comment{\emph{read-only}}
\item[Output:]
\(h \se [f\times g]_{n-1}^{m+n-1}\) \comment{\emph{read-write}}
\item[Required:]
\emph{Balanced} middle product algorithm with space complexity
\(\leq cn\)
\item[Notation:]
\(k = \lfloor\frac{m}{c+2}\rfloor\)
\end{description}

\begin{enumerate}

\item

if \(m < c+2\): \(h = [f\times g]_n^{m+n-1}\) \comment{constant space}

\item

\(h_{[0,k[} \se [f_{[0,n-1+k[}\times g]_{n-1}^{n-1+k}\)
\comment{\(\lceil\frac{n}{k}\rceil\) middle products, free:
\(h_{[k,m[}\)}

\item

\(h_{[k,m[} \se [f_{[k,m+n-1[}\times g]_{n-1}^{m-k+n-1}\) \comment{tail
recursive call}

\end{enumerate}\gobblepar

\end{algorithm}

\begin{theorem}[\autocite{GiorgiGrenetRoche2019}]

\hyperref[algorithm:MiddleProduct]{Algorithm~\ref*{algorithm:MiddleProduct}} (\textsc{MiddleProduct})
is correct, requires \(O(1)\) extra space, and performs
\(O(\frac{\nu}{\mu}\M^*(\mu))\) operations, where \(\mu=\min(m,n)\) and
\(\nu=\max(m,n)\).\gobblepar

\end{theorem}

\subsection{Time-space preserving reductions between polynomial
products}\label{Section:time-space-preserving-reductions-between-polynomial-products}

We have proved that for the full, upper, and middle products, any
linear-space algorithm can be turned into a constant-space variant, with
a limited increase in time complexity. We now consider the relationships
between these problems, in terms of time-space complexity.

\begin{definition}

A problem \(A\) is \emph{time-space reducible} to a problem \(B\),
denoted \(A \leq_\textsf{TISP} B\), if there exists an algorithm for
\(A\) that, given access to an algorithm for \(B\) that runs in time
\(t(n)\) and space \(s(n)\), runs in time \(O(t(n))\) and space
\(O(s(n))\). We write \(A \equiv_\textsf{TISP} B\) if
\(A\leq_\textsf{TISP} B\) and \(B\leq_\textsf{TISP} A\).\gobblepar

\end{definition}

\notation[time-space preserving reducibility and equivalence\nomrefpage]{\(\leq_\textsf{TISP}\)
and \(\equiv_\textsf{TISP}\)}(czz)

To compare the problems, we define them formally.

\begin{definition}

Let \(f\), \(g\), \(h\), \(\ell\) and \(u\in\mathsf{R}[x]\) where \(f\),
\(g\) and \(\ell\) have size \(n\), \(h\) has size \(2n-1\) and \(u\)
has size \(n-1\). We define the following problems by the computation
they perform:

\begin{itemize}
\item
  \(\FullProd\): \(h \se f\times g\);
\item
  \(\FullProd^+\): \(h \pe f\times g\), assuming that
  \(h\bquo x^n = 0\);
\item
  \(\LowProd\): \(\ell \se f\times g\bmod x^n\);
\item
  \(\UppProd\): \(u \se f\times g\bquo x^n\);
\item
  \(\MidProd\): \(f \se [h\times g]_{n-1}^{2n-1}\).
\end{itemize}

\end{definition}

By reversal, lower and upper products are equivalent, that is
\(\LowProd\equiv_\textsf{TISP}\UppProd\):
\[(f \times g) \bquo x^n = \left((f\bquo x)^{\shortleftarrow} \times (g\bquo x)^{\shortleftarrow} \bmod x^{n-1}\right)^{\shortleftarrow}.\]

Since \(f\times g = (fg)\bmod x^n + x^n \left[(fg)\bquo x^n\right]\),
the previous equivalence implies
\(\FullProd \leq_\textsf{TISP} \LowProd\). For \(\FullProd^+\), one can
first compute \((fg)\bmod x^n\) in the free (upper) part of \(h\), add
it to the lower part, and finally compute \((fg)\bquo x^n\) in the upper
part. Therefore, \(\FullProd^+ \leq_\textsf{TISP} \LowProd\).

Conversely, given an algorithm for \(\FullProd^+\), write
\(f = f_0+x^{\lfloor n/2\rfloor}f_1\) and
\(g = g_0+x^{\lfloor n/2\rfloor}g_1\). Then
\((fg)\bmod x^n = f_0g_0 + x^{\lfloor n/2\rfloor}((f_0g_1 + f_1g_0)\bmod x^{\lceil n/2\rceil})\).
This computation reduces to \(\FullProd^+\): First compute
\(h = f_0g_1\) since \(h\) is free; Erase the upper coefficients of
\(h\) and compute \(h\pe f_1g_0\); Finally move the lower coefficients
to the upper part of \(h\) and compute \(h \pe f_0g_0\). This shows
\(\LowProd \leq_\textsf{TISP} \FullProd^+\), whence the equivalence.

Finally, writing
\((fg)\bmod x^n = [(x^{n-1}\cdot f)\times g]_{n-1}^{2n-1}\) proves
\(\LowProd \leq_\textsf{TISP} \MidProd\) using \emph{fake padding}.

As a result, we obtain the following reductions and equivalences.

\begin{theorem}[\autocite{GiorgiGrenetRoche2019}]

There exist time-space reductions between full, lower, upper and middle
products as depicted below:
\[\FullProd \leq_\textsf{TISP} \FullProd^+ \equiv_{\textsf{TISP}} \LowProd \equiv_\textsf{TISP} \UppProd \leq_\textsf{TISP} \MidProd.\]\gobblepar

\end{theorem}

\section{Power series inversion and Euclidean
division}\label{Section:power-series-inversion-and-euclidean-division}

The goal of this part is to prove that the fast Euclidean division
algorithm can be made constant-space while preserving the same
asymptotic time complexity \(O(\M(n))\). To this end, we first
investigate the case of power series inversion and division which serve
as building blocks.

\subsection{Power series inversion and
division}\label{Section:power-series-inversion-and-division}

Our goal is to adapt Newton iteration for power series inversion to make
it work in constant space. Let \(\phi\in \mathsf{R}[[x]]\) be an
invertible power series and \(\psi\in \mathsf{R}[[x]]\) be its inverse.
We assume that \(f = \phi\bmod x^n\) is its truncation at precision
\(n\), and we aim to compute \(g = \psi\bmod x^n\). Assume that at some
point, \(g = \psi\bmod x^k\) for some \(k \leq n/2\). Then one step of
Newton iteration updates \(g\) as
\[g \se g + (1-gf)\cdot g \bmod x^{2k}\] so that the new value of \(g\)
satisfies \(g = \psi\bmod x^{2k}\)~\autocite{vonzurGathenGerhard2013}.
Since \(gf = 1\bmod x^k\) by assumption, the lower part of \((1-gf)\) is
zero. For efficiency reasons, and in particular space efficiency, the
update can be computed as
\[g \se g - x^k\times\left( \left[(f^*\bmod x^{2k-1})\times g\right]_{k-1}^{2k-1} \times g \right)\bmod x^k\]
where \(f^* = f\bquo x\). That is, \(g\) is updated using a middle
product followed by a lower product. If \(\psi\) is to be computed at
precision \(n\), \(g\) is stored as a vector of size \(n\). When \(k\)
is small enough, we can use the free space \(g_{[2k,n[}\) to compute the
middle and lower products, without extra space. But when \(k\)
approaches \(n/2\), no free space is available anymore. The solution is
to slow down the computation and compute fewer coefficients than what a
standard step would do. This is summarized in the following lemma.

\begin{lemma}\label{lemma:NewtonItGen}

Let \(\phi\in \mathsf{R} [[x]]\) invertible,
\(f = \phi\bmod x^{k+\ell}\) and \(g\) be the inverse of \(\phi\) at
precision \(k\). Then
\[g - x^k\times\left( \left[f^*\times g\right]_{k-1}^{k+\ell-1} \times g \right)\bmod x^\ell\]
is the inverse of \(\phi\) at precision \(k+\ell\), where
\(f^* = f\bquo x\).\gobblepar

\end{lemma}

The algorithm adjusts the value of \(\ell\) to keep enough free space.
It requires \(\ell\) registers for the result of the middle product,
\(\ell\) other registers for intermediate results since the middle
product is unbalanced, and \(c\ell\) registers as extra space for the
middle product computation itself. The same holds for the lower product.
Together, the condition on \(\ell\) is \((c+2)\ell \leq n-k\). We remark
that the formula of
\hyperref[lemma:NewtonItGen]{Lemma~\ref*{lemma:NewtonItGen}} computes
\(\ell\) new coefficients of the inverse by using \(k+\ell-1\)
coefficients of \(f\). The number of coefficients keeps increasing
during the algorithm, even if \(\ell\) itself decreases. As for the
constant-space middle product, the time complexity increases to
\(O(\M^*(n))\).

\begin{algorithm}[Inversion]

\begin{description}
\item[Inputs:]
\(f = \phi\bmod x^n\) such that \(\phi_0\) is a unit
\comment{\emph{read-only}}
\item[Output:]
\(g \se \psi\bmod x^n\) such that \(\phi\times\psi = 1\) in
\(\mathsf{R}[[x]]\) \comment{\emph{read-write}}
\item[Required:]
Middle and lower product algorithms with space complexity \(\leq cn\)
\end{description}

\begin{enumerate}

\item

\(g_{[0]} \se f_{[0]}^{-1}\)

\item

let \(k = 1\) and \(\ell = 1\)

\item

while \(\ell > 0\):

\item

~~~~\(g_{[n-\ell,n[} \se [f_{[1,k+\ell[}\times g_{[0,k[}]_{k-1}^{k+\ell-1}\)
\comment{middle product, free: \(g_{[k,n-\ell[}\)}

\item

~~~~\(g_{[k,k+\ell[} \se -(g_{[0,\ell[} \times g_{[n-\ell,n[})\bmod x^\ell\)
\comment{lower product, free: \(g_{[k+\ell,n-\ell[}\)}

\item

~~~~update \(k = k+\ell\) and
\(\ell = \min(k,\lfloor\frac{n-k}{c+2}\rfloor)\)

\item

\(g_{[k,n[} \se -g_{[0,n-k[}\times[f_{[1,n[}\times g_{[0,k[}]_{k-1}^{n-1} \bmod x^{n-k}\)
\comment{constant space}

\end{enumerate}\gobblepar

\end{algorithm}

\begin{theorem}[\autocite{GiorgiGrenetRoche2019}]\label{theorem:CstSpaceInversion}

\hyperref[algorithm:Inversion]{Algorithm~\ref*{algorithm:Inversion}} (\textsc{Inversion})
is correct, requires \(O(1)\) extra space, and performs \(O(\M^*(n))\)
operations. \gobblepar

\end{theorem}

The division \(\phi/\psi\) of two power series \(\phi\),
\(\psi\in \mathsf{R} [[x]]\) can be computed as \(\phi\times\psi^{-1}\).
The drawback is that it requires to store the intermediate result
\(\psi^{-1}\). Karp and Markstein's trick incorporates the
multiplication into the final step of Newton iteration when computing
\(\psi^{-1}\)~\autocite{KarpMarkstein1997}. Since the last iteration is
replaced in our constant-space variant by several iterations, we need to
incorporate the multiplication into several iterations. The following
lemma is a generalization of their method, that can also be seen as a
generalization of the standard Newton iteration for \(\phi/\psi\).

\begin{lemma}\label{lemma:NewtonItDivision}

Let \(\phi\), \(\psi\in \mathsf{R} [[x]]\), \(\psi\) invertible,
\(f = \phi\bmod x^{k+\ell}\), \(g=\psi\bmod x^{k+\ell}\),
\(g^* =g\bquo x\), and \(h = \phi/\psi\bmod x^k\). Then
\[h^* = h + x^k\cdot\left( (g^{-1}\bmod x^\ell) \times \left(f\bquo x^k-[g^*\times h]_{k-1}^{k+\ell-1}\right)\right)\bmod x^\ell\]
satisfies \(h^* = \phi/\psi\bmod x^{k+\ell}\).\gobblepar

\end{lemma}

The algorithm has the same structure as
\hyperref[algorithm:Inversion]{Algorithm~\ref*{algorithm:Inversion}} (\textsc{Inversion}),
using the free output space as work space. Note that \(g^{-1}\) is
needed during the whole algorithm, but with less and less precision
since \(\ell\) has to decrease. Therefore, we can compute it once at the
beginning, and store it in the reversed order as the last coefficients
of \(h\) and progressively overwrite the unneeded coefficients.

\begin{algorithm}[Division]

\begin{description}
\item[Inputs:]
\(f = \phi\bmod x^n\), \(g = \psi\bmod x^n\) such that \(\psi_0\) is a
unit \comment{\emph{read-only}}
\item[Output:]
\(h \se \phi/\psi\bmod x^n\) \comment{\emph{read-write}}
\item[Required:]
middle and lower product, and power series inversion algorithms with
space complexity \(\leq cn\)
\end{description}

\begin{enumerate}

\item

let \(k = \lfloor\frac{n}{c+2}\rfloor\) and
\(\ell = \lfloor\frac{n-k}{c+3}\rfloor\)

\item

\(h_{[n-k,n[}^{\shortleftarrow} \se g_{[0,k[}^{-1} \bmod x^k\)
\comment{inversion, free: \(h_{[0,n-k[}\)}

\item

\(h_{[0,k[} \se (f_{[0,k[} \times h_{[n-k,n[}^{\shortleftarrow})\bmod x^k\)
\comment{lower product, free: \(h_{[k,n-k[}\)}

\item

while \(\ell > 0\):

\item

~~~~\(h_{[n-2\ell,n-\ell[} \se -[g_{[1,k+\ell[}\times h_{[0,k[}]_{k-1}^{k+\ell-1}\)
\comment{middle product, free: \(h_{[k,n-2\ell[}\)}

\item

~~~~\(h_{[n-2\ell,n-\ell[} \pe f_{[k,k+\ell[}\)

\item

~~~~\(h_{[k,k+\ell[} \se (h_{[n-2\ell,n-\ell[}\times h_{[n-\ell,n[}^{\shortleftarrow})\bmod x^\ell\)
\comment{lower product, free: \(h_{[k+\ell,n-2\ell[}\)}

\item

~~~~update \(k = k+\ell\) and \(\ell = \lfloor\frac{n-k}{c+3}\rfloor\)

\item

\(h_{[k,n[} \se (f_{[k,n[}-[g_{[1,n[}\times h_{[0,k[}]_{k-1}^{n-1}) \times h_{[k,n[}^{\shortleftarrow} \bmod x^{n-k}\)
\comment{constant space}

\end{enumerate}\gobblepar

\end{algorithm}

\begin{theorem}[\autocite{GiorgiGrenetRoche2019}]\label{theorem:Division}

\hyperref[algorithm:Division]{Algorithm~\ref*{algorithm:Division}} (\textsc{Division})
is correct, requires \(O(1)\) extra space, and performs \(O(\M^*(n))\)
operations. \gobblepar

\end{theorem}

Constant-space power series division is the main ingredient of a
constant-space Euclidean division of polynomials. In this use, the
dividend \(\phi\) is actually an intermediate result. Therefore, it can
be overwritten during the computation.
\hyperref[lemma:NewtonItDivision]{Lemma~\ref*{lemma:NewtonItDivision}}
shows than once the first \(k\) coefficients of \(\phi/\psi\) have been
computed, the first \(k\) coefficients of \(\phi\) are not needed
anymore. The idea is then to actually replace \(\phi\bmod x^n\) by
\(\phi/\psi\bmod x^n\), using an extra linear space. As a result, the
complexity is back to \(O(\M(n))\).

We describe the algorithm in full generality when the extra space has
any size \(s\). At each iteration,
\(\ell = \lfloor\frac{s}{c+3}\rfloor\) new coefficients are computed.
The algorithm thus uses \(O(n/s)\) iterations of cost
\(O(\frac{n}{s}\M(s))\).

\begin{algorithm}[SmallSpaceInPlaceDivision]

\begin{description}
\item[Inputs:]
\(f = \phi\bmod x^n\) \comment{\emph{read-write}}

\(g = \psi\bmod x^n\) such that \(\psi_0\) is a unit
\comment{\emph{read-only}}
\item[Output:]
\(f \se \phi/\psi\bmod x^n\)
\item[Required:]
middle product, lower product and inversion algorithms with space
complexity \(\leq cn\), and an extra space \(t\) of size \(s\)
\end{description}

\begin{enumerate}

\item

let \(k = 0\) and \(\ell = \lfloor\frac{s}{c+3}\rfloor\)

\item

\(t_{[0,\ell[} \se g_{[0,\ell[}^{-1}\bmod x^\ell\) \comment{inversion,
free: \(t_{[\ell,s[}\)}

\item

\(t_{[\ell,2\ell[} \se (f_{[0,\ell[} \times t_{[0,\ell[})\bmod x^\ell\)
\comment{lower product, free: \(t_{[2\ell,s[}\)}

\item

\(f_{[0,\ell[} \se t_{[\ell,2\ell[}\)

\item

while \(k < n\):

\item

~~~~update \(k = k+\ell\) and
\(\ell = \min(\lfloor\frac{s}{c+3}\rfloor, n-k)\)

\item

~~~~\(t_{[\ell,2\ell[} \se -[g_{[1,k+\ell[}\times f_{[0,k[}]_{k-1}^{k+\ell-1}\)
\comment{middle product, free: \(t_{[2\ell,s[}\)}

\item

~~~~\(t_{[\ell,2\ell[} \pe f_{[k,k+\ell[}\)

\item

~~~~\(f_{[k,k+\ell[} \se (t_{[\ell,2\ell[}\times t_{[0,\ell[})\bmod x^\ell\)
\comment{lower product, free: \(t_{[2\ell,s[}\)}

\end{enumerate}\gobblepar

\end{algorithm}

\begin{theorem}[\autocite{GiorgiGrenetRoche2019}]\label{theorem:SmallSpaceInPlaceDivision}

\hyperref[algorithm:SmallSpaceInPlaceDivision]{Algorithm~\ref*{algorithm:SmallSpaceInPlaceDivision}} (\textsc{SmallSpaceInPlaceDivision})
is correct, requires \(s\) extra space, and performs
\(O((\frac{n}{s})^{2}\M(s))\) operations in \(\mathsf{R}\).\gobblepar

\end{theorem}

If \(s = \alpha n\) for some constant \(\alpha\), the complexity becomes
\(O(\M(n))\). For \(s = 1\) on the other hand, the complexity is
\(O(n^{2})\) and the algorithm corresponds to the naive algorithm, which
requires no extra space.

\subsection{Euclidean division}\label{Section:euclidean-division}

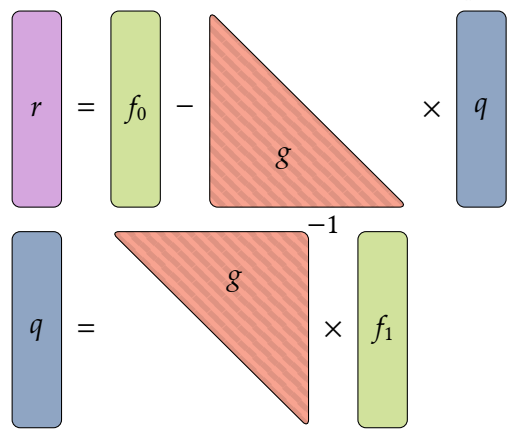
\begin{figure}

\centering{}

\subtop[Start with the equation $f = gq+r$ that defines the Euclidean division.]{\input{euclideandiv1.tikz}}

\medskip

\subtop[Split it into two equations $f_0 = (gq)\bmod x^n+r$ and $f_1 = (gq)\bquo x^n$.]{\input{euclideandiv2.tikz}}
\medskip

\subtop[Solve the two equations to get $r = f_0 - gq\bmod x^n$ and $q^{\shortleftarrow} = (g^{\shortleftarrow}/f_1^{\shortleftarrow})\bmod x^n$.]{\input{euclideandiv3.tikz}}

\caption{Derivation of the fast Euclidean division algorithm \emph{via}
linear algebra. In the last part, an upper product must be inverted.
This corresponds to a reversed power series division. The remainder is
computed with a lower product.} \label{figure:EuclideanDiv}\gobblepar

\end{figure}

Given \(f\in\mathsf{R}[x]_{<m+n-1}\) and \(g\in \mathsf{R}[x]_{<n}\)
whose leading coefficient is a unit, the goal is to compute a size-\(m\)
quotient \(q\) and a size-\((n-1)\) remainder \(r\) such that
\(f = gq+r\). We also consider the computations of the sole quotient or
the sole remainder. Fast algorithms for Euclidean division start by
computing the quotient \(q\). Since
\(f\bquo x^{n-1} = (g\times q) \bquo x^{n-1}\) is an upper product,
\(q\) can be computed as the inverse of an upper product, namely a
\emph{reversed} power series division
\(q \se \left(f^{\shortleftarrow} / g^{\shortleftarrow} \bmod x^m\right)^{\shortleftarrow}\).
The remainder is computed as \(r \se f-gq\).
\hyperref[figure:EuclideanDiv]{Figure~\ref*{figure:EuclideanDiv}} is a
linear-algebraic presentation of this idea for \(m=n\).

As a first remark, computing the quotient only without any space for the
remainder is equivalent to power series division.

If \(m < n-1\), the size-\(m\) quotient can be computed using the free
space of the remainder as the work space, in time \(O(\M(m))\). Then,
\(r\) is computed as \(r = (f-gq)\bmod x^{n-1}\) with a constant-space
lower product \(gq\bmod x^{n-1}\) in time \(O(\M(n))\), for a total
complexity \(O(\M(m)+\M(n))\). If the same algorithm is used to compute
the remainder only, the space for the quotient counts as extra space.

Let us focus on the case \(m \geq n-1\) where the best complexity
\(O(\frac{m}{n} \M(n))\) relies on the long division algorithm that
computes the quotient block by block. For the first blocks, there is
enough free space in \(q\) to use non-constant-space algorithms. The
difficulty is to compute the last block of the quotient. It actually
boils down to the case \(m = n-1\). In this situation,
\(f = f_0 + x^{n-1} f_1\) has size \(2n-1\) and \(g\) has size \(n\).
Hence, we need to compute the \(n\) coefficients of \(q\) and \((n-1)\)
coefficients of \(r\). To compute \(q\), we rely on
\hyperref[algorithm:SmallSpaceInPlaceDivision]{Algorithm~\ref*{algorithm:SmallSpaceInPlaceDivision}} (\textsc{SmallSpaceInPlaceDivision}).
Since
\(q^{\shortleftarrow} = f_1^{\shortleftarrow} / g^{\shortleftarrow}\bmod x^n\),
we set \(q\se f_1\), then
\(q^{\shortleftarrow} \de g^{\shortleftarrow} \bmod x^{n-1}\) using the
free space of \(r\) as extra space. Finally, the remainder is computed
using a constant-space lower product.

In the presentation below, for simplicity, we use the same method to
compute each block of the quotient even though only the last division
needs to be compute in place. Replacing these in-place divisions by
linear-space divisions would improve the overall complexity by a
constant.

\begin{algorithm}[EuclideanDivision]

\begin{description}
\item[Inputs:]
\(f\) of size \(m+n-1\), \(g\) of size \(n\), \(m \geq n-1\)
\comment{\emph{read-only}}
\item[Outputs:]
\(q\) of size \(m\) and \(r\) of size \(n-1\) such that \(f = gq+r\)
\comment{\emph{read-write}}
\item[Required:]
constant-space lower product algorithm

in-place reversed division algorithm with space complexity \(<n\)
\item[Notations:]
\(k = \lfloor\frac{m}{n}\rfloor\), \(\ell = m\bmod n\)

write \(q = \sum_{j=0}^k q_j x^{jn}\),
\(f = \sum_{j=0}^{k+1} f_j x^{jn}\)

let \(g_{\smash *} = g_{[0,n-1[}\), \(q_{j*} = q_{j[0,n-1[}\) and
\(q_j^* = q_{j[1,n[}\)
\end{description}

\begin{enumerate}

\item

\(q_k \se f_{k+1}\)

\item

\(q_k^{\shortleftarrow} \de g_{[n-\ell,n[}^{\shortleftarrow} \bmod x^\ell\)
\comment{reversed division, free: \(r\), \(q_{[0,kn[}\)}

\item

\(q_{k-1}^* \se -(g_{[0,\ell[}\times q_k)\bmod x^\ell\) \comment{lower
product, free: \(r\), \(q_{[0,(k-1)n[}\)}

\item

\(q_{k-1} \pe f_k\)

\item

for \(j = k-1\) to \(1\):

\item

~~~~\(q_j^{\shortleftarrow} \de g^{\shortleftarrow} \bmod x^n\)
\comment{reversed division, free: \(r\), \(q_{[0,jn[}\)}

\item

~~~~\(q_{j-1}^* \se -(g_{\smash *}\times q_{j*})\bmod x^{n-1}\)
\comment{lower product, free: \(r\), \(q_{[0,(j-1)n[}\)}

\item

~~~~\(q_{j-1} \pe f_j\)

\item

\(q_0^{\shortleftarrow} \de g^{\shortleftarrow}\bmod x^n\)
\comment{reversed division, free: \(r\)}

\item

\(r \se -(g_{\smash *}\times q_{0*})\bmod x^{n-1}\) \comment{lower
product with no work space}

\item

\(r \pe f_{[0,n-1[}\)

\end{enumerate}\gobblepar

\end{algorithm}

\begin{theorem}[\autocite{GiorgiGrenetRoche2019}]\label{theorem:CstSpaceEuclideanDivision}

\hyperref[algorithm:EuclideanDivision]{Algorithm~\ref*{algorithm:EuclideanDivision}} (\textsc{EuclideanDivision})
is correct, requires \(O(1)\) extra space, and performs
\(O(\frac{m}{n} \M(n))\) operations in \(\mathsf{R}\).\gobblepar

\end{theorem}

If only the remainder is to be computed, one can actually forget about
the coefficients of the quotient during the computation. This means that
only one block of the quotient needs to fit in the extra space. Using
the small-space in-place reversed division and the semi-cumulative lower
product, any space \(s < n\) is sufficient to compute the remainder. (In
the presentation, we assume \(s \leq n-1\) to be able to store blocks of
size \(s\) within the remainder space.)

\begin{algorithm}[SmallSpaceRemainder]

\begin{description}
\item[Inputs:]
\(f\) of size \(m+n-1\), \(g\) of size \(n\), \(m \geq n-1\)
\comment{\emph{read-only}}
\item[Outputs:]
\(r\) of size \(n-1\) such that \(r = f\bmod g\)
\comment{\emph{read-write}}
\item[Required:]
constant-space semi-cumulative lower product algorithm

in-place reversed division algorithm with space complexity \(\leq n-1\)

extra space \(t\) of size \(s\leq n-1\)
\item[Notations:]
\(k = \lfloor\frac{m}{s}\rfloor\), \(\ell = m\bmod s\)

\(f_j = f_{[n-1+(j-1)s,n-1+js[}\) for \(0\leq j\leq k\) and
\(f_{k+1} = f_{[m+n-1-\ell,m+n-1[}\)

\(r^* = r_{[n-1-s,n-1[}\), \(g^* = g_{[n-s,n[}\) and
\(g_{\smash *} = g_{[0,n-1[}\)
\end{description}

\begin{enumerate}

\item

\(t_{[0,\ell[} \se f_{k+1}\)

\item

\(t^{\shortleftarrow}_{[0,\ell[} \de g_{[n-\ell,n[}^{\shortleftarrow}\bmod x^\ell\)
\comment{reversed division, free: \(r\)}

\item

\(r \se (g_{\smash *}\times t)\bmod x^{n-1}\) \comment{lower product in
constant space}

\item

\(t \se r^* + f_k\)

\item

for \(j = k-1\) to \(0\):

\item

~~~~\(r_{[s,n[} \se r_{[0,n-1-s[}\) \comment{right shift}

\item

~~~~\(t^{\shortleftarrow} \de (g^*)^{\shortleftarrow} \bmod x^s\)
\comment{reversed division, free: \(r_{[0,s[}\)}

\item

~~~~\(r \me (g_{\smash *}\times t)\bmod x^{n-1}\) \comment{lower product
in constant space}

\item

~~~~\(t \se r^* + f_j\)

\item

\(r \se t\)

\item

\(r_{[0,n-1-s[} \pe f_{[0,n-1-s[}\)

\end{enumerate}\gobblepar

\end{algorithm}

\begin{theorem}[unpublished]

\hyperref[algorithm:SmallSpaceRemainder]{Algorithm~\ref*{algorithm:SmallSpaceRemainder}} (\textsc{SmallSpaceRemainder})
is correct, requires an extra space of size \(s\), and performs
\(O(\frac{m}{s}\M(n))\) operations.\gobblepar

\end{theorem}

For \(s = \Theta(1)\), the complexity bound becomes \(O(m\M(n))\).
Actually, a slightly finer analysis shows that the complexity is
\(O(mn)\). One can notice that this is the standard long division
algorithm, that indeed requires no extra space. For \(s = \Theta(n)\),
the complexity is \(O(\frac{m}{n}\M(n))\) as for the linear-space
algorithm.

\section{Multi-point evaluation and
interpolation}\label{Section:multi-point-evaluation-and-interpolation}

The standard and fastest algorithms for multipoint evaluation and
interpolation do not use only a linear amount of extra space, but rather
an extra space of size \(n\log n+O(n)\) to store the so-called
subproduct tree. Von zur Gathen and Shoup noticed that the space
complexity of multipoint evaluation can be reduced to linear with the
same asymptotic time
complexity~\autocite{vonzurGathenShoup1992,vonzurGathenShoup1992a}. For
instance consider the evaluation of a size-\(n\) polynomial \(f\) on
\(n\) points. By grouping the points as \(\log(n)\) groups of
\(n/\log(n)\) points, one can perform \(\log(n)\) multipoint evaluations
that each require \(n+O(n/\log n)\) extra space. The complexity is
\(O(\M(\frac{n}{\log n}\log(\frac{n}{\log n})\log n) = O(\M(n)\log n)\).
Once a linear-space algorithm is known for multipoint evaluation,
obtaining a constant-space version uses the same ideas as before. First
evaluate \(f\) on \(k < n\) points using the free output space as work
space, and recur. We obtain the following result.

\begin{theorem}[\autocite{GiorgiGrenetRoche2019}]

Let \(f\in \mathsf{R}[x]\) of size \(n\) and \(a_0\), \ldots,
\(a_{n-1}\) in \(\mathsf{R}\). One can compute \(f(a_0)\), \ldots,
\(f(a_{n-1})\) in \(O(\M(n)\log(n))\) operations using \(O(1)\) extra
registers.\gobblepar

\end{theorem}

Our approach may appear a bit cumbersome. We fix a number \(k\) of
points on which to perform a partial multipoint evaluation. Then, in
order for this multipoint evaluation to be computable in linear space,
this set of points itself is split into \(\log(k)\) blocks of
\(k/\log(k)\) points. It is of course possible to directly choose a
value of \(k\) so that there is enough free space to perform the
standard multipoint evaluation of space complexity \(k\log(k)+O(k)\).
This probably improves slightly the complexity by a constant factor, but
makes the analysis painful.

The main difficulty lies in interpolation. Let \((a_0,b_0)\), \ldots,
\((a_{n-1},b_{n-1})\) be \(n\) pairs of evaluations with invertible
differences \(a_i-a_j\), \(i\neq j\). The goal is to compute the unique
size-\(n\) polynomial \(f\in\mathsf{R}[x]\) such that \(f(a_i)=b_i\) for
\(0\leq i<n\), with a constant-space algorithm. The first ingredient is
to provide a variant of polynomial interpolation that computes
\(f\bmod x^k\) using \(O(k)\) extra space. For simplicity, let us assume
that \(k\) divides \(n\). For \(1 \leq i \leq n/k\), let
\(m_i= \prod_{j=k(i-1)}^{ki-1}
(x-a_j)\) and \(s_i=m/m_i\) where \(m = \prod_{i=1}^n (x-a_i)\). Note
that \(s_i =
\prod_{j\neq i} m_j\). Lagrange interpolation formula can be written by
block as
\[f(x) =\sum_{i=1}^{n/k} \sum_{j=k(i-1)}^{ki-1} b_j \cdot \frac{m_j(x)}{m_j(a_j)} = m(x)\sum_{i=1}^{n/k}
\frac{n_i(x)}{m_i(x)} = \sum_{i=1}^{n/k}  n_i(x)s_i(x)\] for some
size-\(k\) polynomials \(n_1\), \ldots, \(n_{n/k}\). Therefore, we need
to compute
\[f \bmod x^k= \sum_{i=1}^{n/k} n_i\cdot(s_i \bmod x^k) \bmod x^k.\]

One can observe that for \(k(i-1)\leq j<ki\),
\(f(a_j) = n_i(a_j)s_i(a_j)\). Since \(m_i(a_j) = 0\),
\(s_i(a_j) = (s_i\bmod m_i)(a_j)\). Therefore, \(n_i\) is the unique
size-\(k\) polynomial satisfying
\(n_i(a_j)=b_j / (s_i \bmod m_i)(a_j)\). It can be computed using
evaluation and interpolation: First compute \(s_i \bmod m_i\); Evaluate
it at the \(a_j\)'s; Perform \(k\) divisions to get each \(n_i(a_j)\);
Finally interpolate \(n_i\).

The second ingredient is to generalize the previous approach when some
initial coefficients of \(f\) are known. Writing \(f = g + x^s h\) where
\(g\) is known, we want to compute \(h\bmod x^k\) from some evaluations
of \(f\). Since \(f = g+x^sh\), we can write
\(h(a_j)=(f(a_j)-g(a_j))/a_j^s\). Therefore, the algorithm sketched for
\(f\bmod x^k\) can be generalized for \(h\bmod x^k\), by computing the
evaluations of \(h\) using multipoint evaluation and fast
exponentiation.
\hyperref[algorithm:PartialInterpolation]{Algorithm~\ref*{algorithm:PartialInterpolation}}
describes this approach.

\begin{algorithm}[PartialInterpolation]

\begin{description}
\item[Inputs:]
\(g\in \mathsf{R}[x]\) of size \(s\), \((a_i,b_i)_{0\leq i<n-s}\) with
\(a_i-a_j\) invertible for \(i\neq j\), an integer \(k \leq n-s\)
\comment{\emph{read-only}}
\item[Output:]
\(h\bmod x^k\) where \(f = g+x^s\cdot h\) is the unique size-\(n\)
polynomial such that \(f(a_i)=b_i\), \(0\leq i<n-s\)
\comment{\emph{read-write}}
\item[Required:]
linear-space \textsc{mp-evaluation} and \textsc{interpolation}

constant-space full product and Euclidean division
\end{description}

\begin{enumerate}

\item

for \(i = 1\) to \((n-s)/k\)\,:

\item

~~~~\(m_i \se \prod_{j=k(i-1)}^{ki-1} (x-a_j)\)

\item

~~~~\(s_i^{(k)} \se 1\); \(s_i^{(m)} \se 1\)

\item

~~~~for \(j=1\) to \((n-s)/k\), \(j \neq i\):

\item

~~~~~~~~\(m_k \se \prod_{\ell=k(j-1)}^{kj-1} (x-a_\ell)\)

\item

~~~~~~~~\(s_i^{(k)} \se s_i^{(k)} \times m_j\bmod x^k\)

\item

~~~~~~~~\(s_i^{(m)} \se s_i^{(m)} \times m_j\bmod m_i\)

\item

~~~~\(g^{(m)} \gets g\bmod m_i\)

\item

~~~~\((c_0, \dotsc, c_{k-1}) \se \textsc{mp-evaluation}(s_i^{(m)}, (a_{k(i-1)}, \dotsc, a_{ki-1}))\)

\item

~~~~\((d_0, \dotsc, d_{k-1}) \se \textsc{mp-evaluation}(s_i^{(m)}, (a_{k(i-1)}, \dotsc, a_{ki-1}))\)

\item

~~~~for \(j=0\) to \(k-1\):
\(c_j \se (b_{j+k(i-1)}-d_j)/(a_{j+k(i-1)}^s c_j)\)

\item

~~~~\(n_i \se \textsc{interpolation}((a_{j+k(i-1)},c_j)_{0\leq j<k})\)

\item

~~~~\(h \pe n_i \times s_i^{(k)}\bmod x^k\)

\end{enumerate}\gobblepar

\end{algorithm}

\begin{lemma}

\hyperref[algorithm:PartialInterpolation]{Algorithm~\ref*{algorithm:PartialInterpolation}} (\textsc{PartialInterpolation})
is correct, requires \(O(k)\) extra space, and has complexity
\(O(\frac{(n-s)^{2}}{k^{2}}\M(k)\log k)\).\gobblepar

\end{lemma}

From this algorithm, a constant-space interpolation algorithm directly
follows. We start with \(h = 0\), and set \(k\) so that the extra space
required by
\hyperref[algorithm:PartialInterpolation]{Algorithm~\ref*{algorithm:PartialInterpolation}} (\textsc{PartialInterpolation})
fits within the free output space of size \(n-k\). We progressively
compute new coefficients of the interpolant \(f\). Altogether, we obtain
the following result.

\begin{theorem}[\autocite{GiorgiGrenetRoche2019}]

Given \(n\) pairs \((a_i,b_i)_{0\leq i<n}\) where \(a_i-a_j\) is
invertible for \(i\neq j\), the unique size-\(n\) interpolant \(f\) such
that \(f(a_i) = b_i\) for \(0\leq i<n\) can be computed in
\(O(\M(n)\log n)\) operations and \(O(1)\) extra space.\gobblepar

\end{theorem}

The constant in the time complexity of this algorithm is quite large.
While the fastest quasi-linear-space algorithm runs in time
\(\frac{5}{2}\M(n)\log(n)+O(n)\)~\autocite{BostanLecerfSchost2003}, our
algorithm is estimated to run in time \(105\M(n)\log(n) + O(n)\). As
mentioned earlier, an approach to reduce the constant could be to bypass
the linear-space multipoint evaluation and interpolation algorithms, and
to directly work with the fastest known algorithm.

\begin{openproblem}

What are the smallest values of \(c_e\) and \(c_i\) such that multipoint
evaluation and interpolation can be computed in constant space and time
complexity \(c_e\M(n)\log(n)+O(n)\) and \(c_i\M(n)\log(n) + O(n)\)
respectively?\gobblepar

\end{openproblem}

\chapter{\texorpdfstring{Algorithms in the \texttt{rw/rw}
model}{Algorithms in the rw/rw model}}\label{chapter:rwrw}

We have seen constant-space variants of several polynomial algorithms,
including Euclidean division. But for the case of computing the
remainder only, the best we could achieve is (small) linear space. In
this chapter, we use the more permissive model \texttt{rw/rw} to
investigate this problem. To achieve the result, we must revisit the
whole chain of algorithms: full, lower, upper and middle products, power
series inversion and division, and finally Euclidean division. In
particular, we design algorithms for \emph{cumulative} and
\emph{in-place} variants of these operations. As a result, we can also
perform more general operations such as a cumulative modular product,
that is \(r \pe f\times g\bmod h\).

In contrast with the algorithms in the \texttt{ro/rw} model where the
output was initially free, we start here with a full output. Therefore,
the techniques must be completely different and cannot rely on the
output space to serve as work space. One of our main techniques is to
use pre- and post-additions on the outputs to \emph{distribute} an
intermediate result at different places. Assume that we need to compute
\(y \pe f(x)\) and \(z\pe f(x)\) for some \(f\). A standard algorithm
would compute \(f(x)\) once and store it into a temporary space, before
adding it to \(y\) and \(z\). In our case, assuming we know an algorithm
to perform \(y\pe f(x)\), we pre-subtract \(y\) from \(z\) (\(z\me y\)),
compute \(y\pe f(x)\), and post-add \(y\) to \(z\) (\(y\pe z\)). More
generally, our techniques can be viewed as transforming some algorithms
to their \emph{reversible} counterparts where some or all of the
algorithm can be \emph{undone}.

This kind of transformations can be applied automatically on any
\emph{bilinear} algorithm to derive a constant-space variant of it, at
least in a nonuniform model. In many cases, the transformation can be
applied uniformly to produce a constant-space uniform algorithm.
Actually, the space is not fully constant. For recursive algorithms,
there exists \emph{a priori} a call stack of (usually) logarithmic size,
that amounts to \(O(\log n)\) pointers.

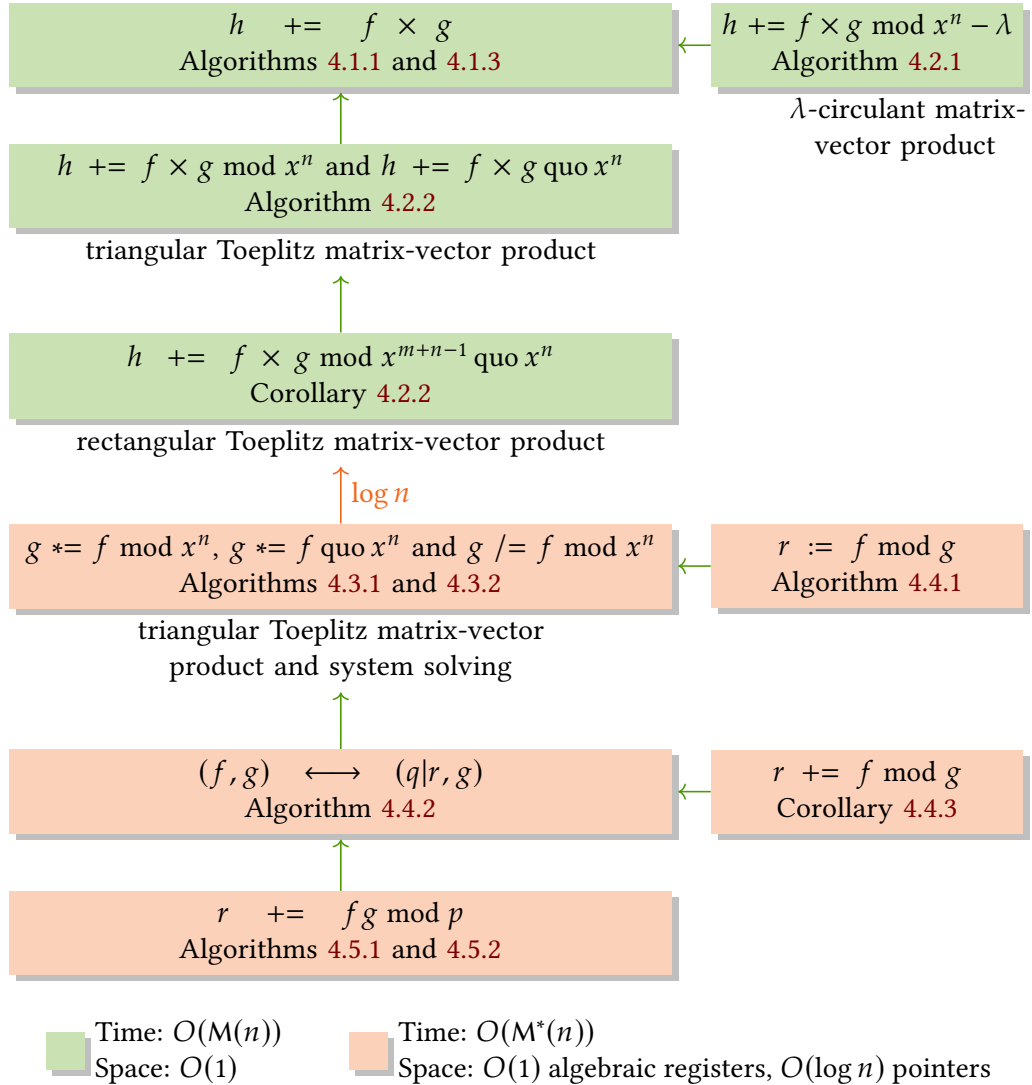
\begin{figure}

\centering
\input{reductions.tikz}

\caption{Summary of the reductions of the chapter.}
\label{figure:reductions}\gobblepar

\end{figure}

In this chapter, we describe algorithms that ultimately lead to a fast
remainder algorithm and a fast modular multiplication algorithm with
constant algebraic space. The series of reductions is summarized in
\hyperref[figure:reductions]{Figure~\ref*{figure:reductions}} on
p.~\pageref{figure:reductions}. Some algorithms for full products can be
obtained automatically, and the ones that we present are optimized
variants. The technique for such an automatization is postponed to
\hyperref[chapter:automatic]{Chapter~\ref*{chapter:automatic}}, where
some consequences in linear algebra are also presented.

\section{Cumulative full products}\label{section:cumul-full-products}

As usual in computer algebra, the building block is polynomial
multiplication. The goal is to perform \(h \pe f\times g\) in constant
space. Contrary to what we had in the \texttt{ro/rw} model, we do not
provide a generic reduction from a standard multiplication algorithm,
but rather revisit the main polynomial multiplication algorithms.

\subsection{Karatsuba's algorithm}\label{Section:karatsubas-algorithm}

Let \(f\) and \(g\) of size \(n\), and \(h = f\times g\). Write
\(f = f_0+x^m f_1\) and \(g = g_0+x^mg_1\). Karatsuba's algorithm
expresses the product \(h  = f\times g\) as
\[h = f_0g_0+x^{2m} f_1g_1 + x^m(f_0g_0+f_1g_1-(f_0-f_1)\times(g_0-g_1))\]
which leads to a recursive algorithm with three recursive calls, and a
running time \(O(n^{\log 3})\)~\autocite{KaratsubaOfman1963}.

\begin{figure}
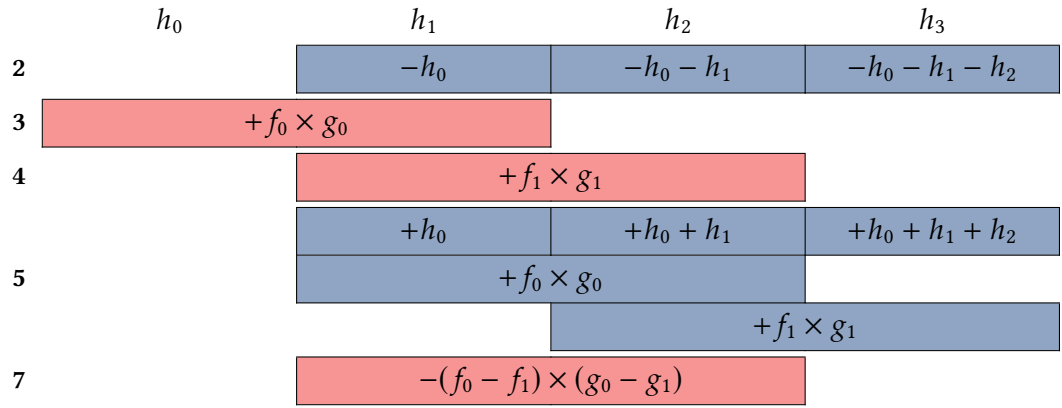


\(\begin{tblr}{
    width=\linewidth,
    hline{2} = {3-Z}{solid},
    hline{3} = {1}{3-Z}{solid},
    hline{3} = {2}{2-3}{solid},
    hline{4} = {1}{2-3}{solid},
    hline{4} = {2}{3-4}{solid},
    hline{5} = {1}{3-4}{solid},
    hline{5} = {2}{3-Z}{solid},
    hline{6,7} = {3-Z}{solid},
    hline{8} = {1}{4-5}{solid},
    hline{8} = {2}{3-4}{solid},
    hline{9} = {1}{3-4}{solid},
    vline{2} = {3}{solid},
    vline{3,5} = {2,4-6,8}{solid},
    vline{4} = {2-Y}{solid},
    vline{3,5} = {2,4-6,8}{solid},
    vline{6} = {2,5,7}{solid},
    colspec = {rX[c]X[c]X[c]X[c]},
    column{1} = {mode=text,font=\small\bfseries},
    cell{2,5}{3-Z} = {bleute},
    cell{3}{2} = {c=2}{c,rougete},
    cell{4,8}{3} = {c=2}{c,rougete},
    cell{6}{3} = {c=2}{c,bleute},
    cell{7}{4} = {c=2}{c,bleute},
    cell{5}{1} = {r=3}{m},
}
  & h_0 & h_1 & h_2 & h_3 \\
2 &     & -h_0& -h_0-h_1 & -h_0 -h_1 -h_2\\
3 & + f_0\times g_0 \\
4 & & + f_1\times g_1\\
5 & & +h_0 & +h_0+h_1 & +h_0+h_1+h_2\\
  & & + f_0 \times g_0 \\
  & & & + f_1 \times g_1 \\
7 & & - (f_0-f_1) \times (g_0-g_1)
\end{tblr}\)

\caption{Data movements in
\hyperref[algorithm:CumulativeKaratsuba]{Algorithm~\ref*{algorithm:CumulativeKaratsuba}},
where \(h_0 = h_{[0,m[}\), \(h_1 = h_{[m,2m[}\), \(h_2 = h_{[2m,3m[}\)
and \(h_3 = h_{[3m,2n-1[}\). In blue, the pre- and post-additions; in
red, the results of the recursive calls.}

\label{figure:CumulativeKaratsuba}\gobblepar

\end{figure}

To make it cumulative and constant-space, the first remark is that it is
easy to compute \(f_0-f_1\) and \(g_0-g_1\) in the input space, since it
can be easily undone. The main challenge is to compute only once
\(f_0g_0\) and \(f_1g_1\) and \emph{distribute} them at two distinct
locations.
\hyperref[algorithm:CumulativeKaratsuba]{Algorithm~\ref*{algorithm:CumulativeKaratsuba}}
uses pre- and post-additions to this end. Its data movements are
represented in
\hyperref[figure:CumulativeKaratsuba]{Figure~\ref*{figure:CumulativeKaratsuba}}.

\begin{algorithm}[CumulativeKaratsuba]

\begin{description}
\item[Inputs:]
\(f\), \(g\in\mathsf{R}[x]\) of size \(n\), \(h\in\mathsf{R}[x]\) of
size \(2n-1\) \comment{\emph{read-write}}
\item[Output:]
\(h \pe f\times g\)
\item[Notations:]
\(m = \lceil n/2\rceil\) and write \(f = f_0 + x^mf_1\),
\(g = g_0+x^mg_1\)
\end{description}

\begin{enumerate}

\item

if \(n < 2\): \(h \pe f\times g\) \comment{constant space}

\item

\(h_{[m,2m[} \me h_{[0,m[}\)\,; \(h_{[2m,3m[} \me h_{[m,2m[}\)\,;
\(h_{[3m,2n-1[} \me h_{[2m,2n-1-m[}\)

\item

\(h_{[0,2m-1[} \pe f_0 \times g_0\) \comment{recursive call}

\item

\(h_{[m,3m-1[} \pe f_1 \times g_1\) \comment{recursive call}

\item

\(h_{[3m,2n-1[} \pe h_{[2m,2n-1-m[}\)\,;
\(h_{[2m,3m[} \pe h_{[m,2m[}\)\,; \(h_{[m,2m[} \pe h_{[0,m[}\)

\item

\(f_0 \me f_1\)\,; \(g_0 \me g_1\)

\item

\(h_{[m,3m-1[} \me f_0\times g_0\) \comment{recursive call}

\item

\(f_0 \pe f_1\)\,; \(g_0 \pe g_1\)

\end{enumerate}\gobblepar

\end{algorithm}

\begin{theorem}[\autocite{DumasGrenet2024a,DumasGrenet2025}]

\hyperref[algorithm:CumulativeKaratsuba]{Algorithm~\ref*{algorithm:CumulativeKaratsuba}} (\textsc{CumulativeKaratsuba})
is correct, requires \(O(n^{\log 3})\) operations, and only uses a call
stack of \(O(\log n)\) pointers as extra space.\gobblepar

\end{theorem}

The same algorithm can be extended to the case where \(f\) and \(g\) do
not have the same size.

\begin{corollary}

Given three polynomials \(f\), \(g\) and \(h\) of respective sizes
\(m\), \(n\) and \(m+n-1\), \(m\geq n\), one can compute
\(h \pe f\times g\) in \(O(mn^{\log 3-1})\) operations, using a call
stack of \(O(\log n)\) pointers.\gobblepar

\end{corollary}

\subsection{FFT/TFT-based
algorithm}\label{Section:ffttft-based-algorithm}

The FFT-based algorithms are well-suited to perform a cumulative full
product. Given \(f\) and \(g\) of size \(n\) and \(h\) of size \(2n-1\),
one can compute three size-\((2n-1)\) FFTs of \(f\), \(g\) and \(h\) to
get \(\vec{\hat f}\), \(\vec{\hat g}\) and \(\vec{\hat h}\) such that
\(\hat f_i = f(\omega^i)\) where \(\omega\) is a \((2n-1)\)st principal
root of unity, and similarly for \(\vec{\hat g}\) and \(\vec{\hat h}\).
Then, \(\hat h_i \pe \hat f_i\hat g_i\) for \(0\leq i<2n\) followed by
an inverse FFT provides \(h \pe fg\).

To make this algorithm constant-space, we can use the fact that the
original FFT algorithm works \emph{in place}, that is a size-\(n\)
polynomial \(f\) is replaced by its size-\(n\) FFT, in time
\(O(n\log n)\)~\autocite{CooleyTukey1965}. This \emph{a priori} only
covers cases where \(n\) is a power of two, but it has been shown that
the truncated Fourier transform (TFT)~\autocite{vanderHoeven2004} and
its inverse can also be computed in
place~\autocite{HarveyRoche2010,Roche2011,Arnold2013,Coxon2022}. Yet,
one difficulty remains. A size-\((2n-1)\) TFT of \(f\) and \(g\) is
needed while they only have size \(n\). The solution is to compute these
TFTs in several steps, restoring the initial \(f\) and \(g\) between two
steps. Note that without permutation, a \emph{decimation-in-frequency}
FFT algorithm computes the \emph{bit-reversed} DFT of the input
polynomial, that is
\((f(\omega^{[0]_k}), f(\omega^{[1]_k}), \dotsc, f(\omega^{[2^k-1]_k}))\),
where \([i]_k =\sum_{j=0}^{k-1} d_j 2^{k-j-1}\) is the length-\(k\)
bit-reversal of \(i = \sum_{j=0}^{k-1} d_j 2^j\), \(d_j\in\{0,1\}\).

\notation[length-$k$ bit-reversal of $i\in{0,\dotsc,2^{k-1}}$\nomrefpage]{\([i]_k\)}(aa)

\begin{algorithm}[PartialFourierTransform]

\begin{description}
\item[Inputs:]
\(f\) of size \(n\), two integers \(k\) and \(\ell\) such that
\(2^\ell \leq n\) \comment{\emph{read-write}}

a principal \(2^p\)th root of unity \(\omega\), with
\((k+1)2^\ell \leq 2^p\)
\item[Output:]
the first \(2^\ell\) coefficients of \(f\) replaced by
\(f(\omega^{[k\cdot 2^\ell+i]_p})\), \(0 \leq i < 2^\ell\)
\end{description}

\begin{enumerate}

\item

for \(i = 0\) to \(n-1\): \(f_{[i]} \fe \omega^{i[k\cdot 2^\ell]_p}\)

\item

for \(i = 2^\ell\) to \(n-1\): \(f_{[i-2^\ell]} \pe f_i\)

\item

\(f_{[0,2^\ell[} \se \textsc{FFT}(f_{[0,2^\ell[}, \omega^{2^{p-\ell}})\)

\end{enumerate}\gobblepar

\end{algorithm}

This algorithm is easily inverted by undoing each operation. From this
partial Fourier transform and its inverse denoted
\(\textsc{PartialFourierTransform}^{-1}\), we can build a cumulative
FFT-based multiplication algorithm.

\begin{algorithm}[CumulativeFFTMultiplication]

\begin{description}
\item[Inputs:]
\(f\), \(g\), \(h\) of size \(m\), \(n\) and \(m+n-1\) respectively,
\(m\leq n\) \comment{\emph{read-write}}

a principal \(2^p\)th root of unity \(\omega\), where
\(p = \lceil\log(m+n-1)\rceil\)
\item[Output:]
\(h \pe f\times g\)
\end{description}

\begin{enumerate}

\item

\(h \se \textsc{TFT}(h, \omega)\) \comment{\(h\) replaced by
\((h(1),h(\omega), \dotsc, h(\omega^{m+n-2})\)}

\item[]

let \(r = m+n-1\)

\item

while \(r > 0\):

\item

~~~~~~let \(\ell = \lfloor\log\min(r,m)\rfloor\) and
\(t = \lfloor\log\min(r,n)\rfloor-\ell\)

\item

~~~~~~let \(k = m+n-1-r\)

\item

~~~~~~\(g \se \textsc{PartialFourierTransform}_{k,\ell+t}(g, \omega)\)

\item

~~~~~~for \(s=0\) to \(2^t-1\):

\item

~~~~~~~~~~\(f \se \textsc{PartialFourierTransform}_{s+k\cdot 2^t,\ell}(f,\omega)\)

\item

~~~~~~~~~~for \(i = 0\) to \(2^\ell-1\):
\(h_{[i+(k\cdot 2^t+s)\cdot 2^\ell]} \pe a_{[i]} b_{[i+s\cdot 2^\ell]}\)

\item

~~~~~~~~~~\(f \se \textsc{PartialFourierTransform}^{-1}_{s+k\cdot 2^t,\ell}(f,\omega)\)

\item

~~~~~~\(g \se \textsc{PartialFourierTransform}^{-1}_{k,\ell+t}(g,\omega)\)

\item

~~~~~~let \(r = r-2^{\ell+t}\)

\end{enumerate}\gobblepar

\end{algorithm}

\begin{theorem}[\autocite{DumasGrenet2024a,DumasGrenet2025}]

\hyperref[algorithm:CumulativeFFTMultiplication]{Algorithm~\ref*{algorithm:CumulativeFFTMultiplication}}
is correct, uses \(O(n\log n)\) operations, and uses no extra
space.\gobblepar

\end{theorem}

\section{Cumulative convolutions and short
products}\label{Section:cumulative-convolutions-and-short-products}

From a cumulative constant-space full product algorithm, we can actually
derive cumulative convolutions \(h \pe f\times g\bmod x^n-\lambda\),
including the short product (\(\lambda = 0\)). We first assume that
\(\lambda\neq 0\). Assume for simplicity that \(n\) is even, and
\(\lambda = 1\). Writing \(f = f_0+x^mf_1\) and \(g = g_0+x^mg_1\) for
\(m = n/2\), \(fg = f_0g_0 + x^m(f_0g_1+f_1g_0) + x^nf_1g_1\). Computing
modulo \(x^n-1\), \(x^nf_1g_1 \equiv f_1g_1\). Similarly,
\(f_0g_1 \equiv (f_0g_1)_{[m,n[} + x^m (f_0g_1)_{[0,m[}\) (and the same
holds for \(f_1g_0\)). Therefore, given a cumulative full product
algorithm, we can easily compute \(fg\bmod x^n-1\) by writing these
subproducts in the right registers.

\begin{figure}
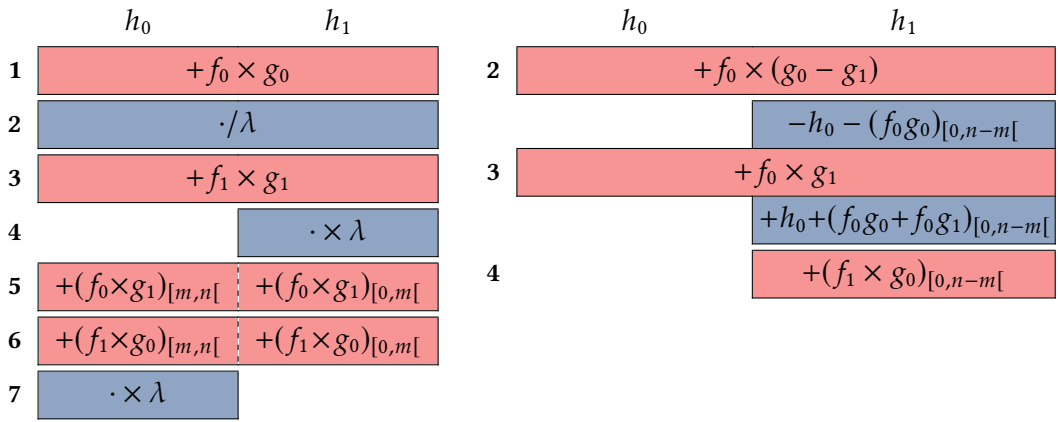


\(\begin{tblr}[baseline=t]{
    width=.42\linewidth,
    colspec = {rX[c]X[c]},
    column{1} = {mode=text,font=\small\bfseries},
    hline{2} = {2-Z}{solid},
    hline{9} = {2}{solid},
    hline{2-5,7,8} = {1}{2-Z}{solid},
    hline{6} = {1}{3}{solid},
    hline{3,4,6,7} = {2}{2-Z}{solid},
    hline{5} = {2}{3}{solid},
    hline{8} = {2}{2}{solid},
    vline{2} = {2-4,6-8}{solid},
    vline{3} = {5,8}{solid},
    vline{3} = {6,7}{dashed},
    vline{4} = {2-7}{solid},
    cell{2,4}{2} = {c=2}{c,rougete},
    cell{6,7}{2,3} = {c,rougete},
    cell{3}{2} = {c=2}{c,bleute},
    cell{5}{3} = {bleute},
    cell{8}{2} = {bleute},
}
  & h_0 & h_1 \\
1 & + f_0\times g_0\\
2 & \cdot / \lambda \\
3 & + f_1\times g_1\\
4 & & \cdot\times \lambda \\
5 & +(f_0 \times g_1)_{[m,n[} & + (f_0\times g_1)_{[0,m[} \\
6 & +(f_1 \times g_0)_{[m,n[} & + (f_1\times g_0)_{[0,m[} \\
7 & \cdot\times\lambda
\end{tblr}\) \hfill \(\begin{tblr}[baseline=t]{
    width=.55\linewidth,
    colspec = {rX[c,3]X[c,4]},
    column{1} = {mode=text,font=\small\bfseries},
    hline{2,4,5} = {2-Z}{solid},
    hline{7} = {3}{solid},
    hline{3} = {1}{2-Z}{solid},
    hline{6} = {1}{3}{solid},
    hline{3,6} = {2}{3}{solid},
    vline{2} = {2,4}{solid},
    vline{3} = {3,5,6}{solid},
    vline{4} = {2-Z}{solid},
    cell{2,4}{2} = {c=2}{c,rougete},
    cell{6}{3} = {c,rougete},
    cell{3,5}{3} = {bleute},
    cell{3}{1} = {r=3}{m},
}
  & h_0 & h_1 \\
2 & + f_0\times(g_0-g_1)\\
3 & & -h_0-(f_0g_0)_{[0,n-m[}\\
  & + f_0\times g_1\\
  & & \!\!\mathmbox{+h_0{+}(f_0g_0{+}f_0g_1)_{[0,n-m[}} \\
4 & & + (f_1\times g_0)_{[0,n-m[} \\
\end{tblr}\)

\caption{Illustration of Algorithms
\ref{algorithm:CumulativeConvolution} (left) and
\ref{algorithm:CumulativeLowerProduct} (right). Pre- and post-operations
are in blue, and products in red.}

\label{figure:Convolutions}\gobblepar

\end{figure}

\hyperref[algorithm:CumulativeConvolution]{Algorithm~\ref*{algorithm:CumulativeConvolution}}
generalizes the approach to any unit \(\lambda\), and any \(n\). It is
illustrated in
\hyperref[figure:Convolutions]{Figure~\ref*{figure:Convolutions}}. Note
that a Karatsuba-like approach decreases the number of calls to the full
product subroutine to three instead of four~\autocite{DumasGrenet2025}.

\begin{algorithm}[CumulativeConvolution]

\begin{description}
\item[Inputs:]
\(f\), \(g\), \(h\) of size \(n\), \(\lambda\in \mathsf{R}^\times\)
\comment{\emph{read-write}}
\item[Output:]
\(h \pe f\times g\bmod x^n-\lambda\)
\item[Required:]
a cumulative constant-space full product algorithm
\item[Notations:]
\(m = \lceil n/2\rceil\) and \(s = n\bmod 2\)

write \(f = f_0+x^mf_1\) and \(g=g_0+x^mg_1\)
\end{description}

\begin{enumerate}

\item

\(h_{[0,2m-1[} \pe f_0\times g_0\)

\item

\(h \de \lambda\)

\item

\(h_{[s,n-1[} \pe f_1\times g_1\)

\item

\(h_{[m,n[} \fe \lambda\)

\item

\(h_{[m,n[}\Vert h_{[0,m[} \pe f_0\times g_1\)
\comment{\(\cdot\Vert\cdot\) denotes concatenation}

\item

\(h_{[m,n[}\Vert h_{[0,m[} \pe f_1\times g_0\)

\item

\(h_{[0,m[} \fe \lambda\)

\end{enumerate}\gobblepar

\end{algorithm}

The case \(\lambda = 0\) is to be treated differently. The subproduct
\(f_1\times g_1\) is unneeded. But the lower parts of the two cross
products \(f_0g_1\) and \(f_1g_0\) must be computed. This could be done
using recursive calls, but the algorithm would make two recursive calls.
This would add an extra logarithmic factor to the complexity when
\(\M(n)\) is quasi-linear, and would require a call stack.

Write \(f = f_0+x^mf_1\) and \(g = g_0+x^mg_1\). Our goal is to compute
\(h \pe f\times g \bmod x^n\). We can expand
\(f\times g\bmod x^n = f_0g_0 + x^m (f_0g_1+f_1g_0\bmod x^{n-m})\). The
first product \(f_0g_0\) is a full product, but \(f_0g_1\) and
\(f_1g_0\) must be recursive calls. If we rewrite the expansion in a
slightly unnatural way as
\[f\times g\bmod x^n = (f_0(g_0-g_1) + f_0g_1) + x^m (f_0g_1 + f_1g_0\bmod x^{n-m})\]
we can make only one recursive call to compute \(f_1g_0\bmod x^{n-m}\),
and compute the full product \(f_0g_1\) before distributing it.

This is
\hyperref[algorithm:CumulativeLowerProduct]{Algorithm~\ref*{algorithm:CumulativeLowerProduct}},
illustrated in
\hyperref[figure:Convolutions]{Figure~\ref*{figure:Convolutions}}. We
can actually improve the leading constant in the complexity of this
algorithm by using different formulas~\autocite{DumasGrenet2025}.

\begin{algorithm}[CumulativeLowerProduct]

\begin{description}
\item[Inputs:]
\(f\), \(g\), \(h\) of size \(n\) \comment{\emph{read-write}}
\item[Output:]
\(h \pe f\times g\bmod x^n\)
\item[Required:]
a cumulative constant-space full product algorithm
\item[Notations:]
\(m = \lceil n/2\rceil\)

write \(f = f_0+x^m f_1\) and \(g = g_0+x^mg_1\) and
\(g_0^* = g_0\bmod x^{n-m}\)
\end{description}

\begin{enumerate}

\item

if \(n < 2\): \(h \pe f\times g\bmod x^n\) \comment{constant space}

\item

\(g_0 \me g_1\); \(h_{[0,n[} \pe f_0\times g_0\); \(g_0 \pe g_1\)
\comment{\(f_0(g_0-g_1)\)}

\item

\(h_{[m,n[} \me h_{[0,n-m[}\); \(h_{[0,n[} \pe f_0 \times g_1\);
\(h_{[m,n[} \pe h_{[0,n-m[}\) \comment{\(f_0g_1\)}

\item

\(h_{[m,n[} \pe f_1\times g_0^*\bmod x^{n-m}\) \comment{tail recursive
call}

\end{enumerate}\gobblepar

\end{algorithm}

\begin{theorem}[\autocite{DumasGrenet2024,DumasGrenet2025}]

\hyperref[algorithm:CumulativeLowerProduct]{Algorithm~\ref*{algorithm:CumulativeLowerProduct}} (\textsc{CumulativeLowerProduct})
is correct, requires no extra space, and runs in time
\(O(\M(n))\).\gobblepar

\end{theorem}

As a consequence of this algorithm, we can actually compute any
\emph{slice} of a polynomial product, in constant space with
accumulation. The first remark is that we can compute an upper product
with the same algorithm, used in reversed mode. Then, for a size-\(m\)
polynomial \(f\) and a size-\(n\) polynomial \(g\), the \emph{slice}
\([f\cdot g]_s^t\) for \(0\leq s < t \leq m+n\) can be decomposed as a
sum of lower and upper products. The nature of cumulative algorithms
make their sequential composition straightforward.

\begin{corollary}\label{corollary:CumulativeSlice}

Let \(f\), \(g\), \(h\) of respective sizes \(m\), \(n\), and \(r\)
where \(0 < r <m+n\), and \(s\) such that \(0 \leq s < m+n-r\). Then
\(h \pe [f\cdot g]_s^{r+s}\) can be computed with no extra space in time
\(O(\M(r))\).\gobblepar

\end{corollary}

\section{In-place power series
computations}\label{Section:in-place-power-series-computations}

We consider two power series \(\phi\) and \(\psi\), and the operations
of multiplication \(\phi\times\psi\) and division \(\phi/\psi\). By
\emph{in-place} we mean that we aim to replace (the truncation of)
\(\phi\) by (the truncation of) \(\phi\times\psi\) or \(\phi/\psi\). The
first operation corresponds to a polynomial lower product.

Let \(f = \phi\bmod x^n\), \(g = \psi\bmod x^n\) and
\(h=\phi\times\psi\bmod x^n\). Write \(f = f_0+x^kf_1\),
\(g = g_0+x^kg_1\) and \(h = h_0+x^k h_1\) where
\(k = \lceil n/2\rceil\). Then \(h_0 = f_0\times g_0\bmod x^k\), and
\(h_1 = [f_0\times g]_k^n + (f_1\times g_0\bmod x^{n-k})\). This yields
the following algorithm that replaces \(f\) by \(h\), by first computing
\(h_1\). Due to the two recursive calls needed to compute \(h_0\) and
\(h_1\), the complexity becomes \(O(\M^*(n))\).

\begin{algorithm}[InPlaceLowerProduct]

\begin{description}
\item[Inputs:]
\(f = \phi\bmod x^n\) and \(g = \psi\bmod x^n\)
\comment{\emph{read-write}}
\item[Output:]
\(f = \phi\times\psi\bmod x^n\)
\item[Notation:]
\(k = \lceil n/2\rceil\)
\end{description}

\begin{enumerate}

\item

if \(n < 2\): \(f \fe g\bmod x^n\) \comment{constant space}

\item

\(f_{[k,n[} \fe g_{[0,n-k[} \bmod x^{n-k}\) \comment{recursive call}

\item

\(f_{[k,n[} \pe [f_{[0,k[} \times g_{[1,n[}]_k^n\)
\comment{\hyperref[corollary:CumulativeSlice]{Corollary~\ref*{corollary:CumulativeSlice}}}

\item

\(f_{[0,k[} \fe g_{[0,k[} \bmod x^k\) \comment{recursive call}

\end{enumerate}\gobblepar

\end{algorithm}

We now turn to power series division. A solution to compute
\(\varphi/\psi\) in place could be to first invert \(\psi\) in place,
and then to multiply \(\varphi\) by \(\psi^{-1}\) in place (replacing
one or the other of the series). Unfortunately, it is not clear that
in-place inversion is possible. Instead, the solution is to directly use
Newton iteration with \(\varphi/\psi\). With the same notations as
before, the algorithm is: \(h_0 \se f_0/g_0\bmod x^k\);
\(h_1 \se (f_1 - [g\times h_0]_{k}^n) / g_0 \bmod x^{n-k}\). This is
exactly the inverse algorithm of
\hyperref[algorithm:LowerProduct]{Algorithm~\ref*{algorithm:LowerProduct}} (\textsc{LowerProduct}).

\begin{algorithm}[InPlaceDivision]

\begin{description}
\item[Inputs:]
\(f = \phi\bmod x^n\) and \(g = \psi\bmod x^n\)
\comment{\emph{read-write}}
\item[Output:]
\(f = \phi/\psi\bmod x^n\)
\item[Notation:]
\(k = \lceil n/2\rceil\)
\end{description}

\begin{enumerate}

\item

if \(n < 2\): \(f \de g\bmod x^n\) \comment{constant space}

\item

\(f_{[0,k[} \de g_{[0,k[} \bmod x^k\) \comment{recursive call}

\item

\(f_{[k,n[} \me [f_{[0,k[} \times g_{[1,n[}]_k^n\)
\comment{\hyperref[corollary:CumulativeSlice]{Corollary~\ref*{corollary:CumulativeSlice}}}

\item

\(f_{[k,n[} \de g_{[0,n-k[} \bmod x^{n-k}\) \comment{recursive call}

\end{enumerate}\gobblepar

\end{algorithm}

\begin{theorem}[\autocite{DumasGrenet2024,DumasGrenet2025}]

\hyperref[algorithm:InPlaceLowerProduct]{Algorithm~\ref*{algorithm:InPlaceLowerProduct}} (\textsc{InPlaceLowerProduct})
and
\hyperref[algorithm:InPlaceDivision]{Algorithm~\ref*{algorithm:InPlaceDivision}} (\textsc{InPlaceDivision})
are correct, require \(O(\M^*(n))\) operations, and only use a call
stack of \(O(\log n)\) pointers as extra space.\gobblepar

\end{theorem}

The second algorithm will be used mainly in \emph{reversed mode} in the
sense of
\hyperref[proposition:reversion]{Proposition~\ref*{proposition:reversion}},
that is to compute
\(f^{\shortleftarrow} \de g^{\shortleftarrow} \bmod x^n\) where
\(f = \phi\bmod x^n\) and \(g = \psi\bmod x^n\). This algorithm is
called a \emph{reversed power series division algorithm}, or in short
\emph{reversed division}.

As mentioned above, this approach does not compute the inversion in
place. It is not even clear that there exists an algorithm, even a slow
one, for this task.

\begin{openproblem}

Given \(f = \phi\bmod x^n\), invertible, is it possible to compute
\(f \se \phi^{-1}\bmod x^n\) in place?\gobblepar

\end{openproblem}

\section{Remainder computation}\label{Section:remainder-computation}

To get a constant-space remainder algorithm, the idea is to start from
\hyperref[algorithm:SmallSpaceRemainder]{Algorithm~\ref*{algorithm:SmallSpaceRemainder}} (\textsc{SmallSpaceRemainder})
and replace the basic operations by their in-place variants. As a
result, the algorithm has complexity \(O(\frac{m}{n}\M^*(n))\) rather
than \(O(\frac{m}{n}\M(n))\) due to the use of in-place building blocks
that have complexity \(O(\M^*(n))\).

\begin{algorithm}[Remainder]

\begin{description}
\item[Inputs:]
\(f\) of size \(m+n-1\), \(g\) of size \(n\), \(m \geq n-1\)
\comment{\emph{read-write}}
\item[Output:]
\(r\) of size \(n-1\) such that \(r = f\bmod g\)
\item[Required:]
In-place lower product and reversed division algorithms
\item[Notations:]
\(k = \lfloor m/n\rfloor\), \(\ell = m\bmod n\), \(g^* = g\bquo x\) and
\(g_{\smash *} = g\bmod x^{n-1}\)
\end{description}

\begin{enumerate}

\item

\(r \se f_{[m+n-1-\ell,m+n-1[}\)

\item

for \(j = k-1\) to \(1\):

\item

~~~~\(r^{\shortleftarrow} \de (g^*)^{\shortleftarrow} \bmod x^{n-1}\)
\comment{in-place reversed division}

\item

~~~~\(r \fe -g_{\smash *}\bmod x^{n-1}\) \comment{in-place lower
product}

\item

~~~~\(r \pe f_{[jn-1,(j+1)n-1[}\)

\item

\(r^{\shortleftarrow} \de (g^*)^{\shortleftarrow}\bmod x^{n-1}\)
\comment{in-place reversed division}

\item

\(r \fe -g_{\smash *}\bmod x^{n-1}\) \comment{in-place lower product}

\item

\(r \pe f_{[0,n-1[}\)

\end{enumerate}\gobblepar

\end{algorithm}

\begin{theorem}[\autocite{DumasGrenet2024,DumasGrenet2025}]

\hyperref[algorithm:Remainder]{Algorithm~\ref*{algorithm:Remainder}} (\textsc{Remainder})
is correct, requires \(O(\frac{m}{n}\M^*(n))\) operations, and only uses
a call stack of \(O(\log n)\) pointers as extra space.\gobblepar

\end{theorem}

As a variant of the preceding algorithm, it is actually possible to
\emph{replace} the input \(f\) by the quotient and the remainder, still
using no extra space. An important remark is that the obtained algorithm
is \emph{reversible}, that is it can be inverted to restore \(f\) from
the quotient and the remainder.

\begin{algorithm}[InPlaceEuclideanDivision]

\begin{description}
\item[Inputs:]
\(f\) of size \(m+n-1\), \(g\) of size \(n\) \comment{\emph{read-write}}
\item[Output:]
\(f_{[0,n[} \se r\) and \(f_{[n,n+m-1[} \se q\) s.t. \(f = bq+r\),
\(\deg q <m\), \(\deg r <n-1\)
\item[Required:]
Cumulative lower product and in-place reversed division algorithms
\item[Notations:]
\(k = \lfloor m/n\rfloor\), \(\ell = m\bmod n\), \(g^* = g\bmod x\) and
\(g_{\smash *} = g\bmod x^{n-1}\)

write \(f = \sum_{i=0}^{k-1} f_i x^{ni}\), with \(\deg(f_{k-1}) = \ell\)
\end{description}

\begin{enumerate}

\item

\(f_{k-1}^{\shortleftarrow} \de g_{[n-\ell,n[}^{\shortleftarrow}\bmod x^\ell\)
\comment{in-place reversed division}

\item

\(f_{k-2} \me g_{\smash *}\times f_{k-1}\bmod x^n\) \comment{cumulative
lower product}

\item

for \(i=k-2\) to \(1\):

\item

~~~~\(f_i^{\shortleftarrow} \de (g^*)^{\shortleftarrow}\bmod x^n\)
\comment{in-place reversed division}

\item

~~~~\(f_{i-1} \me g_{\smash *}\times f_i\bmod x^n\)\comment{cumulative
lower product}

\end{enumerate}\gobblepar

\end{algorithm}

\begin{theorem}[\autocite{DumasGrenet2024a,DumasGrenet2025}]

\hyperref[algorithm:InPlaceEuclideanDivision]{Algorithm~\ref*{algorithm:InPlaceEuclideanDivision}} (\textsc{InPlaceEuclideanDivision})
is correct, requires \(O(\frac{m}{n}\M^*(n))\) operations, and uses only
a call stack of \(O(\log n)\) pointers as extra space. Further, it is
reversible and the computation can be undone within the same time and
space complexity bounds.\gobblepar

\end{theorem}

A consequence of the reversibility is that one can use this algorithm to
perform some cumulative remainder computation \(r \pe f\bmod g\). First,
replace \(f\) by the remainder and quotient \([f\bmod g,f\bquo g]\);
Then compute \(r \pe f\bmod g\); Finally, restore \(f\).

\begin{corollary}\label{corollary:CumulativeRemainder}

Given \(f\) of size \(m+n-1\), \(g\) of size \(n\) and \(r\) of size
\(n-1\), one can compute \(r \pe f\bmod g\) in \(O(\frac{m}{n}\M^*(n))\)
operations, with a call stack of \(O(\log n)\) pointers.\gobblepar

\end{corollary}

\section{Modular product}\label{Section:modular-product}

Our last goal is to extend the previous results to the modular product.
More precisely, given \(f\), \(g\), \(p\) and \(r\), the goal is to
compute \(r \pe f\times g\bmod p\). One example of such a computation is
the multiplication is finite field extensions, where \(p\) defines the
extension. To stick with this example, let us assume that \(f\), \(g\)
and \(r\) have the same size \(n\) and \(p\) has size \(n+1\).
\index{polynomial!modular product}

Let \(h = f\times g\) of size \(2n-1\), and write \(h = h_0+x^n h_1\)
where \(h_0\) has size \(n\) and \(h_1\) has size \(n-1\). Then
\(h\bmod p =h_0 + (x^n h_1\bmod p)\). Adding \(h_0\) to \(r\) is easy
and only requires a constant-space cumulative lower product algorithm.
To compute \(x^nh_1\bmod p\), we first compute the size-\((n-1)\)
quotient as
\(q^{\shortleftarrow} = h_1^{\shortleftarrow}/ (p^*)^{\shortleftarrow}\bmod x^{n-1}\)
where \(p^* = p\bquo x^{2}\). Then we get \(x^nh_1\bmod p\) as
\((x^nh_1-q\times p)\bmod x^n = -(q\times p)\bmod x^n\). Since \(h\) is
known only as \(f\times g\), we first have to compute \(h_1\) into the
higher-degree \(n-1\) coefficients \(f^*\) of \(f\) as
\(f^* \fe g^* \bquo x^{n-1}\) using an in-place upper product. This
computation can be undone as
\((f^*)^{\shortleftarrow} \de (g^*)^{\shortleftarrow}\bmod x^{n-1}\)
using an in-place reversed division.

\begin{algorithm}[ModularMultiplication]

\begin{description}
\item[Inputs:]
\(f\), \(g\), \(r\) of size \(n\), \(p\) of size \(n+1\)
\comment{\emph{read-write}}
\item[Output:]
\(r \pe f\times g\bmod p\)
\item[Required:]
Cumulative lower product algorithm

In-place upper product and reversed division algorithms
\item[Notations:]
\(f^* =f_{[1,n[}\), \(g^* = g_{[1,n[}\), \(p^* = p_{[2,n+1[}\),
\(p_{\smash *} = p_{[0,n[}\)

\(f^{\shortleftarrow}_{\smash *} = f_{[1,n[}^{\shortleftarrow}\),
\(g^{\shortleftarrow}_{\smash *} = g_{[1,n[}^{\shortleftarrow}\),
\(p^{\shortleftarrow}_{\smash *} = p_{[2,n+1[}^{\shortleftarrow}\),
\end{description}

\begin{enumerate}

\item

\(r \pe f\times g \bmod x^n\) \comment{\(r\pe h_0\), cumulative lower
product}

\item

\(f^* \fe g^* \bquo x^n\) \comment{in-place upper product}
\label{step:uppmul}

\item

\(f_{\smash *}^{\shortleftarrow} \de p_{\smash *}^{\shortleftarrow}\bmod x^{n-1}\)
\comment{in-place reversed division} \label{step:revdiv}

\item

\(r \me f^*\times p_{\smash *}\bmod x^{n-1}\)
\comment{\(r \pe x^nh_1\bmod p\), cumulative lower product}

\item

\(f^* \fe p^* \bquo x^n\) \comment{undo \ref{step:revdiv}, in-place
upper product}

\item

\(f_{\smash *}^{\shortleftarrow} \de g_{\smash *}^{\shortleftarrow} \bmod x^{n-1}\)
\comment{undo \ref{step:uppmul}, in-place reversed division}

\end{enumerate}\gobblepar

\end{algorithm}

\begin{theorem}[\autocite{DumasGrenet2024,DumasGrenet2025}]

\hyperref[algorithm:ModularMultiplication]{Algorithm~\ref*{algorithm:ModularMultiplication}} (\textsc{ModularMultiplication})
is correct, requires \(O(\M^*(n))\) operations, and only uses a call
stack of size \(O(\log n)\) as extra space.\gobblepar

\end{theorem}

Although the example of the multiplication in finite field extension
assumes that \(f\), \(g\) and \(r\) have size \(n\) and \(p\) has size
\(n+1\), the reversible in-place Euclidean division algorithm makes it
possible to extend the result to any sizes.

\begin{algorithm}[ModularMultiplicationAllSizes]

\begin{description}
\item[Inputs:]
\(f\), \(g\), \(r\) and \(p\) of respective sizes \(\ell\), \(m\), \(n\)
and \(n+1\) \comment{\emph{read-write}}
\item[Output:]
\(r \pe f\times g\bmod p\)
\item[Required:]
In-place Euclidean division algorithm and its inverse

Cumulative modular multiplication algorithm
\item[Notations:]
\(q_f = f\bquo p\), \(r_f = f\bmod p\), \(q_g = g\bquo p\),
\(r_g = g\bmod p\)
\end{description}

\begin{enumerate}

\item

\(f \se [r_f,q_f]\) \comment{only if \(\ell > n\), in-place Euclidean
division} \label{step:quoremf}

\item

\(g \se [r_g,q_g]\) \comment{only if \(m > n\), in-place Euclidean
division} \label{step:quoremg}

\item

\(r \pe f_{[\ell-n,\ell[} \times g_{[m-n,n[} \bmod p\)
\comment{cumulative modular multiplication}

\item

\(g \se q_g \times p + r_g\) \comment{undo \ref{step:quoremg} if
\(m > n\), inverted Euclidean division}

\item

\(f \se q_f \times p + r_f\) \comment{undo \ref{step:quoremf} if
\(\ell>n\), inverted Euclidean division}

\end{enumerate}\gobblepar

\end{algorithm}

\begin{theorem}[\autocite{DumasGrenet2024a,DumasGrenet2025}]

\hyperref[algorithm:ModularMultiplicationAllSizes]{Algorithm~\ref*{algorithm:ModularMultiplicationAllSizes}} (\textsc{ModularMultiplicationAllSizes})
is correct, requires \(O(\frac{\ell+m}{n}\M^*(n))\) operations, and only
uses a call stack of size \(O(\log n)\) as extra space.\gobblepar

\end{theorem}

For the case of multiplication in finite field extensions, the modulus
is often chosen to be sparse. Our algorithms can be analyzed in this
case and the complexities refined~\autocite{DumasGrenet2025}.

\chapter{\texorpdfstring{The automatic approach for algorithms in the
\texttt{rw/rw}
model}{The automatic approach for algorithms in the rw/rw model}}\label{chapter:automatic}

The algorithms described in
\hyperref[section:cumul-full-products]{Section~\ref*{section:cumul-full-products}}
have been obtained semi-automatically. More precisely, they are improved
version of algorithms that were obtained automatically. This chapter is
devoted to this automatization. We present algorithms that produce
constant-space variants of standard algorithms with similar complexity.
It uses the framework of bilinear algorithms, that can be formalized by
a direct extension of linear programs
(\hyperref[definition:linearprogram]{Definition~\ref*{definition:linearprogram}})
to \emph{bilinear programs}. A \textsc{c++} implementation of these
algorithms and (heuristic) optimizations is available in the
\href{http://github.com/jgdumas/plinopt}{\textsc{PLinOpt}}
library~\autocite{DumasGrenetPernetSedoglavic2024}.

\hyperref[section:cstspace-linalg]{Section~\ref*{section:cstspace-linalg}}
applies this automatization to linear-algebraic problems.

\section{The general framework}\label{section:framework}

\begin{definition}\label{definition:bilinear-algorithm}

A \emph{cumulative bilinear algorithm} is given by a triple \((A,B,C)\)
where \(A\in\mathsf{R}^{t\times m}\), \(B\in\mathsf{R}^{t\times n}\) and
\(C\in\mathsf{R}^{s\times t}\) such that neither \(A\) nor \(B\) nor
\(C\) contains an all-zero row. Given \(\vec x\in \mathsf{R}^m\),
\(\vec y\in \mathsf{R}^n\) and \(\vec z\in \mathsf{R}^s\), it computes
\(\vec z \pe C((A\vec x)\odot(B\vec y))\) where \(\odot\) is the
component-wise product, that is
\[z_k \pe \sum_{\ell=0}^{t-1} C_{k,\ell}\left[\left(\sum_{i=0}^{m-1} A_{\ell,i}x_i\right)\cdot\left(\sum_{j=0}^{n-1} B_{\ell,j}y_j\right)\right]\]
for \(0\leq k < s\).\gobblepar

\end{definition}

\index{algorithm!bilinear}

From the representation \((A,B,C)\) of a cumulative bilinear algorithm,
an actual sequence of operations to compute
\(\vz \pe C((A\vx)\odot(B\vy))\) is easily obtained. Yet, it requires
some extra space to store \(A\vx\) and \(B\vy\), and then their
component-wise product. Using techniques of pre- and post-additions as
used in \hyperref[chapter:rwrw]{Chapter~\ref*{chapter:rwrw}}, we can
obtain a constant space algorithm. To avoid any unneeded technicalities,
we assume that the input matrices have entries in a field.
\hyperref[algorithm:ConstantSpaceBilinear]{Algorithm~\ref*{algorithm:ConstantSpaceBilinear}}
is implemented as the tool
\href{https://github.com/jgdumas/plinopt/blob/main/src/inplacer.cpp}{\texttt{inplacer}}
of \textsc{PLinOpt}~\autocite{DumasGrenetPernetSedoglavic2024}.

\begin{algorithm}[ConstantSpaceBilinear]

\begin{description}
\item[Inputs:]
\(A\in\mathsf{K}^{t\times m}\), \(B\in\mathsf{K}^{t\times n}\),
\(C\in\mathsf{K}^{s\times t}\) \comment{\emph{read-only}}
\(\vec x\in\mathsf{K}^m\), \(\vec y\in\mathsf{K}^n\),
\(\vec z\in \mathsf{K}^s\) \comment{\emph{read-write}}
\item[Output:]
\(\vec z \pe C((A\vec x)\odot(B\vec y))\)
\end{description}

\begin{enumerate}

\item

for \(u = 0\) to \(t-1\):

\item

~~~~\(x_i \se \sum_{\ell=0}^m A_{u,\ell}x_\ell\) for some \(i\) s.t.
\(A_{u,i} \neq 0\)

\item

~~~~\(y_j \se \sum_{\ell=0}^n B_{u,\ell}y_\ell\) for some \(j\) s.t.
\(B_{u,j} \neq 0\)

\item

~~~~\(z_k \de C_{k,u}\) for some \(k\) s.t. \(C_{k,u} \neq 0\)
\label{line:zkdiv}

\item

~~~~\(z_\ell \me C_{\ell,u}z_k\) for each \(\ell\neq k\) s.t.
\(C_{\ell,u}\neq 0\)

\item

~~~~\(z_k \pe x_i\cdot y_j\)

\item

~~~~\(z_\ell \pe C_{\ell,u}z_k\) for each \(\ell\neq k\) s.t.
\(C_{\ell,u}\neq 0\)

\item

~~~~\(z_k \fe C_{k,u}\) \label{line:zkmul}

\item

~~~~\(y_j \se (y_j-\sum_{\ell\neq j} B_{u,\ell}y_\ell)/B_{u,j}\)

\item

~~~~\(x_i \se (x_i-\sum_{\ell\neq i} A_{u,\ell}x_\ell)/A_{u,i}\)

\end{enumerate}\gobblepar

\end{algorithm}

A cumulative bilinear algorithm has three kinds of operations:
additions, multiplications and \emph{scalar} multiplications where one
multiplicand is a constant from \(A\), \(B\) or \(C\). Let \(\sigma(M)\)
and \(\tau(M)\) denote the number of nonzero entries and the number of
entries \(\notin\{0,1,-1\}\) in \(M\), respectively.

\begin{theorem}[\autocite{DumasGrenet2024a,DumasGrenet2025}]\label{theorem:ConstantSpaceBilinear}

Given a cumulative bilinear algorithm
\((A,B,C)\in\mathsf{R}^{t\times m}\times\mathsf{R}^{t\times n}\times\mathsf{R}^{s\times t}\),
\hyperref[algorithm:ConstantSpaceBilinear]{Algorithm~\ref*{algorithm:ConstantSpaceBilinear}} (\textsc{ConstantSpaceBilinear})
computes \(\vec z \pe C((A\vec x)\odot(B\vec y))\) \emph{in constant
space} using \(t\) cumulative products \(z_k \pe x_iy_j\),
\(2(\sigma(A)+\sigma(B)+\sigma(C))-5t\) additions, and
\(2(\tau(A)+\tau(B)+\tau(C))\) in-place scalar multiplications, without
any further copy.\gobblepar

\end{theorem}

Without space constraint, a cumulative bilinear algorithm performs
\((\sigma(A)-m)+(\sigma(B)-n)+(\sigma(C)-t)+s\) additions, \(t\)
multiplications and \((\tau(A)+\tau(B)+\tau(C))\) scalar
multiplications. The number of operations is less than doubled by the
constant-space variant. While the number of operations without space
constraint is exact, the constant-space algorithm can be optimized by a
clever scheduling of its operations.

Note that this result also applies to \emph{recursive} algorithm, where
the products \(z_k \pe x_iy_j\) are replaced by recursive calls. This is
in particular the case for matrix multiplication, as shown in
\hyperref[section:cstspace-linalg]{Section~\ref*{section:cstspace-linalg}}.
The same technique cannot be directly applied to (recursive) polynomial
multiplication algorithms such as Karatsuba's algorithm. The problem is
that the output size of a recursive call is not the same as the input
size.

If we consider the product \emph{à la} Karatsuba of two size-\(2\)
polynomials \(f = f_0+f_1x\) and \(g = g_0+g_1x\), the result
\(h = h_0+h_1x+h_2x^{2}\) can be computed as \(h_0 = f_0g_0\),
\(h_1 = f_0g_1-f_1g_0\) and \(h_2 = f_0g_0+f_1g_1-(f_0-f_1)(f_0-g_1)\).
Therefore, we obtain the three matrices
\[A = B = \begin{pmatrix} 1&0\\0&1\\1&-1\end{pmatrix}\text{ and } C = \begin{pmatrix} 1&0&0\\1&1&-1\\0&1&0\end{pmatrix}.\]
\hyperref[algorithm:ConstantSpaceBilinear]{Algorithm~\ref*{algorithm:ConstantSpaceBilinear}} (\textsc{ConstantSpaceBilinear})
produces a Karatsuba-like constant-space algorithm with only three
multiplications for the product of two size-\(2\) polynomials. The
problem is the extension to a recursive algorithm. Assume that \(f\) and
\(g\) have size \(n = 2k\). We need to replace the three products by
recursive calls to size-\(k\) polynomial products. But the result of the
recursive calls is a size-\((2k-1)\) polynomial, so that there is some
overlap between for instance \(h_0\) and \(h_1\).

To formalize the extension, we consider a two-dimensional version of
bilinear algorithms.

\begin{definition}\label{definition:2Dbilinear-algorithm}

Let \(\circ:\mathsf{R}\times\mathsf{R} \to \mathsf{R} ^{2}\) be an
operator, and
\(\circledcirc:\mathsf{R}^t \times \mathsf{R}^t \to \mathsf{R}^{t+1}\)
be its component-wise extension defined by
\(\vec u\circledcirc\vec v = \vec w\) where
\(w_{i} = (u_{i-1}\circ v_{i-1})_{[1]} + (u_i\circ v_i)_{[0]}\) for
\(1\leq i<t\), \(w_0 = (u_0\circ v_0)_{[0]}\) and
\(w_t = (u_{t-1}\circ v_{t-1})_{[1]}\). A \emph{2D-cumulative bilinear
algorithm} is given by a \((A,B,C)\) where
\(A\in\mathsf{R}^{t\times m}\), \(B\in\mathsf{R}^{t\times n}\) and
\(C\in\mathsf{R}^{s\times(t\mathbf{+1})}\). Given
\(\vec x\in \mathsf{R}^m\), \(\vec y\in \mathsf{R}^n\) and
\(\vec z\in \mathsf{R}^s\), it computes
\(\vec z \pe C((A\vec x)\circledcirc(B\vec y))\).\gobblepar

\end{definition}

\hyperref[algorithm:ConstantSpaceBilinear]{Algorithm~\ref*{algorithm:ConstantSpaceBilinear}} (\textsc{ConstantSpaceBilinear})
can be extended to 2D-bilinear algorithms. The operations on the inputs
remain the same. Assume that some result
\(\pi_u = \left(\sum_i A_{u,i}x_i\right)\circ\left(\sum_j B_{u,j}y_j\right)\)
has to be distributed in two places:
\(\smat[z_k\\z_{k+1}] \pe C_{k,u}\pi_u\) and
\(\smat[z_\ell\\z_{\ell+1}] \pe C_{\ell,u}\pi_u\). If \(\ell > k+1\),
there is no overlap and we can proceed as in the standard bilinear case.
We simply have to replace for instance \(z_k \de C_{k,u}\) by
\(\smat[z_k\\z_{k+1}] \de C_{k,u}\), that is \(z_k \de C_{k,u}\) and
\(z_{k+1} \de C_{k,u}\). Other operations on \(\vec z\) can be dealt
with similarly. But if \(\ell = k+1\), there is an overlap. This forces
to be careful in the order of the computations. Finally, we replace
\hyperref[line:zkdiv]{Lines~\ref*{line:zkdiv}} to~\ref{line:zkmul} by
the following ones.

\noindent\gobblepar

\begin{tcolorbox}[nobeforeafter,mybox=jaune!25]

\begin{algo}

\begin{enumerate}[label=\arabic*.,start=4]

\item

\(z_k \de C_{k,u}\) for some \(k\) s.t. \(C_{k,u} \neq 0\)

\item

\(z_\ell \me C_{\ell,u}z_k\) for each \(\ell\neq k\) s.t.
\(C_{\ell,u} \neq 0\)

\item

\(z_{k+1} \de C_{k,u}\)

\item

\(z_{\ell+1} \me C_{\ell,u}z_{k+1}\) for each \(\ell \neq k\) s.t.
\(C_{\ell,u}\neq 0\)

\item

\(\smat[z_k\\z_{k+1}] \pe x_i\circ y_j\)

\item

\(z_{\ell+1} \pe C_{\ell,u}z_{k+1}\) for each \(\ell \neq k\) s.t.
\(C_{\ell,u}\neq 0\)

\item

\(z_{k+1} \fe C_{k,u}\)

\item

\(z_{\ell} \pe C_{\ell,u}z_k\) for each \(\ell \neq k\) s.t.
\(C_{\ell,u}\neq 0\)

\item

\(z_k \fe C_{k,u}\)

\end{enumerate}\gobblepar

\end{algo}\gobblepar

\end{tcolorbox}

This is implemented as the tool
\href{https://github.com/jgdumas/plinopt/blob/main/src/trilplacer.cpp}{\texttt{trilplacer}}
of \textsc{PLinOpt}~\autocite{DumasGrenetPernetSedoglavic2024}.

This algorithm can be applied to any recursive polynomial multiplication
algorithm. We note that to exactly fits within the framework, one has to
define the operator \(u_i\circ v_j = (0,u_i\cdot v_j)\), so that
\(\vec u\circledcirc\vec v\) corresponds to a polynomial product
extended with a zero as leading coefficient since a polynomial product
does not double the size of the input.

\hyperref[algorithm:CumulativeKaratsuba]{Algorithm~\ref*{algorithm:CumulativeKaratsuba}} (\textsc{CumulativeKaratsuba})
can almost directly be obtained from the three matrices \(A\), \(B\) and
\(C\) using this algorithm. The actual algorithm removes some useless
computations and in particular deals with the useless leading zeroes.
The same technique can be used to get a constant-space variant of
Toom-Cook algorithms.

\section{Application to linear algebra}\label{section:cstspace-linalg}

In this section, we investigate the implications of
\hyperref[theorem:ConstantSpaceBilinear]{Theorem~\ref*{theorem:ConstantSpaceBilinear}}
for linear-algebraic computations. First, the theorem can be applied to
matrix multiplication. Let us start with a \(2\times 2\) cumulative
matrix multiplication \(Z \pe XY\). Strassen-Winograd algorithm computes
\(Z \pe XY\) using seven products rather than eight for a naive
algorithm. The recursive extension of this algorithm to \(n\times n\)
matrices has time complexity \(O(n^{\log 7})\) and requires
\(\Theta(n^{2})\) temporary registers. The most space-efficient variants
of this algorithm reduce the number of temporary registers to
\(n^{2}\)~\autocite{Huss-LedermanJacobsonTsaoTurnbullJohnson1996} or
even \(\frac{2}{3}n^{2}\)~\autocite{BoyerDumasPernetZhou2009}. These
results were in the \texttt{ro/rw} model. Using
\hyperref[theorem:ConstantSpaceBilinear]{Theorem~\ref*{theorem:ConstantSpaceBilinear}},
we obtain a constant-space variant of this algorithm in the
\texttt{rw/rw} model. To fit within the framework of
\hyperref[definition:bilinear-algorithm]{Definition~\ref*{definition:bilinear-algorithm}},
a matrix \(X = \smat(x_{00}&x_{01}\\x_{10}&x_{11})\) is viewed as the
vector \(\vec x = (x_{00},x_{01},x_{10},x_{11})\). The resulting
algorithm can be slightly optimized \emph{by hand} to obtain an
\emph{optimal} variant.

\begin{algorithm}[ConstantSpaceStrassenWinograd]

\begin{description}
\item[Inputs:]
\(X\), \(Y\), \(Z\in \mathsf{K}^{n\times n}\) for some \(n = 2^k\)
\comment{\emph{read-write}}
\item[Output:]
\(Z \pe X\cdot Y\)
\item[Notations:]
\(X = \smat(X_{00}&X_{01}\\X_{10}&X_{11})\),
\(Y = \smat(Y_{00}&Y_{01}\\Y_{10}&Y_{11})\),
\(Z = \smat(Z_{00}&Z_{01}\\Z_{10}&Z_{11})\) with blocks of size
\(\frac{n}{2}\)
\end{description}

\begin{enumerate}

\item

if \(n = 1\): return \(z_{00} \pe x_{00}y_{00}\)

\item

\(X_{10} \me X_{00}\); \(Y_{01} \me Y_{11}\); \(Z_{10} \me Z_{11}\)

\item

\(Z_{11} \pe X_{10} \cdot Y_{01}\) \comment{recursive call}

\item

\(X_{10} \pe X_{11}\); \(Y_{01} \me Y_{00}\); \(Z_{01} \me Z_{11}\)

\item

\(Z_{11} \me X_{10} \cdot Y_{01}\) \comment{recursive call}

\item

\(Z_{00} \me Z_{11}\)

\item

\(Z_{11} \pe X_{00} \cdot Y_{00}\) \comment{recursive call}

\item

\(Z_{00} \pe Z_{11}\); \(Y_{01} \pe Y_{10}\); \(Z_{10} \pe Z_{11}\)

\item

\(Z_{10} \pe X_{11} \cdot Y_{01}\) \comment{recursive call}

\item

\(Y_{01} \pe Y_{11}\); \(Y_{01} \me Y_{10}\); \(X_{10} \me X_{01}\)

\item

\(Z_{01} \me X_{10} \cdot Y_{11}\) \comment{recursive call}

\item

\(X_{10} \pe X_{01}\); \(X_{10} \pe X_{00}\)

\item

\(Z_{11} \pe X_{10} \cdot Y_{01}\) \comment{recursive call}

\item

\(Z_{01} \pe Z_{11}\); \(Y_{01} \pe Y_{00}\); \(X_{10} \me X_{11}\)

\item

\(Z_{00} \pe X_{01} \cdot Y_{10}\) \comment{recursive call}

\end{enumerate}\gobblepar

\end{algorithm}

\begin{theorem}[\autocite{DumasGrenet2024a,DumasGrenet2025}]\label{theorem:ConstantSpaceStrassenWinograd}

\hyperref[algorithm:ConstantSpaceStrassenWinograd]{Algorithm~\ref*{algorithm:ConstantSpaceStrassenWinograd}} (\textsc{ConstantSpaceStrassenWinograd})
is correct. If \(n\) is a power of two, it uses \(8n^{\log 7}+O(n)\)
operations and only uses a call stack of \(O(\log n)\) pointers as extra
space. Furthermore, the constant \(8\) is the best achievable for
algorithms performing \(O(n^{\log 7})\) operations with the same space
complexity.\gobblepar

\end{theorem}

It is well known that many linear-algebraic computations reduce to
matrix multiplications. In particular, these reductions have been made
space-efficient
in~\autocite{DumasGiorgiPernet2008,DumasPernetSultan2013}. The only
extra space required is due to the underlying matrix multiplications.
Plugging our constant-space Strassen-Winograd algorithm provides fast
constant-space algorithms for these tasks.

\begin{corollary}\label{corollary:ConstantSpaceLinearAlgebra}

Let \(M\), \(T \in \mathsf{K}^{m\times m}\) and \(A\),
\(B\in\mathsf{K}^{m\times n}\) where \(T\) is upper triangular, and let
\(\omega = \log 7\).

\begin{enumerate}[label=\roman*.]

\item

The following task can be computed in time \(O(m^{\omega-1} n)\) using a
call stack of size \(O(\log n)\) as only extra space:

\begin{itemize}
\item
  \(A \se T\cdot A\) (TRMM);
\item
  \(A \se T^{-1}\cdot A\) (TRSM);
\item
  \(M \pe \alpha A\cdot A^\transp\) (SYRK);
\item
  \(\textsf{Low}(M) \pe \textsf{Low}(A\cdot B^\transp + B\cdot A^\transp)\)
  (SYR2K), where \(\textsf{Low}(M)\) denotes the lower triangular part
  of \(M\).
\end{itemize}

\item

The following tasks can be computed in time \(O(m^{\omega})\) using a
call stack of size \(O(\log n)\) as only extra space:

\begin{itemize}
\item
  \(T \se T^{-1}\) (INVT);
\item
  \(N \pe M^{2}\) (SQUARE).
\end{itemize}

\item

The following tasks can be computed in time \(O(m^{\omega})\), using
\(O(n)\) pointers for storing permutations:\footnote{The PLUQ
  decomposition of \(A\) is \(A = P\smat[L\\M]\smat[U&V]Q\) where \(P\)
  and \(Q\) are \(n\times n\) permutation matrices, and \(L\) (resp.
  \(U\)) is an \(r\times r\) lower (resp. upper) triangular matrix. Due
  to the dimensions, the four matrices \(L\), \(M\), \(U\) and \(V\) can
  be stored in \(A\) as \(\smat[L\backslash U & V\\M&0]\). The
  permutation matrices \(P\) and \(Q\) may be stored in the bottom right
  square of dimensions \((n-r)\times(n-r)\), only if \(r\) is small
  enough. Otherwise, one needs \(O(n)\) pointers to store these
  permutations. But if the base ring is large enough (more than \(mn\)
  elements), it is actually possible to store them inside \(A\). A
  permutation is required only when a null pivot is encountered, and
  this zero can be replaced by some indices. The same applies for the
  inverse. More details in~\autocite[Remark 18]{DumasGrenet2025}.}

\begin{itemize}
\item
  \(A \se \smat[L\backslash U&V\\M&0]\) (PLUQ), where \(L\backslash U\)
  denotes the square matrix with lower part \(L\) and upper part \(U\);
\item
  \(M \se M^{-1}\) (INV).
\end{itemize}

\end{enumerate}\gobblepar

\end{corollary}

In these results, the asymptotic time complexity remains the same as for
the original algorithms. In terms of space complexity, \(O(\log n)\)
pointers are needed in the call stack. The naive matrix multiplication
algorithm only requires \(O(1)\) pointers. For polynomial
multiplication, FFT-based algorithms do without call stack.

\begin{openproblem}

Given \(X\), \(Y\), \(Z\in\mathsf{K}^{n\times n}\), is it possible to
compute \(Z \pe X\cdot Y\) in \(O(n^{\log 7})\) operations, with no
extra space?\gobblepar

\end{openproblem}

\chapter{Conclusions and
perspectives}\label{Section:conclusions-and-perspectives}

We have shown that many polynomial computations, hence many structured
linear-algebraic computations, as well as many unstructured
linear-algebraic computations, admit time- and space-efficient
algorithms. (Our results on polynomials can be to some extent
generalized to matrices with low displacement
rank~\autocite{DumasGrenet2025}.) These results hold in different
models. As mentioned earlier, the traditional space complexity model
(\texttt{ro/wo} in our language) is unsuitable to time-space complexity
analysis. The arguably most natural relaxation of this model is
\texttt{ro/rw} where the inputs remain read-only. (Even for cumulative
operations such as \(c \pe a\times b\), it is natural to ask for \(a\)
and \(b\) to be read-only.) Nevertheless, some operations such as
computing the remainder in a Euclidean division of polynomials have
still no fast and constant-space algorithm in this model. We
investigated an even more relaxed model, \texttt{rw/rw}, where we showed
that any bilinear operations (and actually more general operations too)
can be performed both fast and in constant \emph{algebraic} space. This
means that these algorithms may still require a call stack of
logarithmic size to store some pointers. In this model, we also provide
some \emph{in-place} algorithms where the inputs are overwritten by the
output.

\section{Further constant-space algebraic
algorithms}\label{Section:further-constant-space-algebraic-algorithms}

This work on time- and space-efficient algorithms in computer algebra is
relatively new and many questions remain open. A first set of open
questions is to improve the current algorithms: remove the need for a
call stack when there is one, transfer some algorithms from the model
\texttt{rw/rw} to the model \texttt{ro/rw}, reduce the hidden constants
in the complexities, \ldots{} Theoretical studies on the relative power
of these models would also be of great interest. For instance, quadratic
lower bounds for polynomial multiplications in the model
\texttt{ro/wo}~\autocite{Abrahamson1986} proved this model unsuitable
for time-space complexity. Is it possible to prove some similar lower
bounds for the model \texttt{ro/rw} which would show the necessity of
the model \texttt{rw/rw}? In the model \texttt{rw/rw}, the special case
of in-place computations of the form \(x \se f(x)\) where the input is
replaced by the output even raises some computability questions. We have
shown for instance that power series multiplication or division can be
computed in place, one of the inputs being replaced by the output. For
division, the dividend is replaced but it is not clear whether the same
result holds with the divisor being replaced instead. It is also open
how to perform power series inversion (where there is only one input) in
place. Maybe the simplest question about in-place computations was
raised by Roche~\autocite{Roche2009}, about polynomial multiplication.
Note that the product of two monic polynomials of respective degrees
\(m\) and \(n\) has degree \(m+n\). Therefore, if one represents a monic
polynomial by the vector of its non-leading coefficients, this problem
is an algebraic problem \(\pi:\mathsf{R}^{m+n}\to\mathsf{R}^{m+n}\).

\begin{openproblem}

Is it possible to replace two monic polynomials \(f\) and \(g\) by their
product, without any extra space? Or at least with a sublinear amount of
space?\gobblepar

\end{openproblem}

Note that this question may well receive different answers depending on
the exact model of computation: single-tape or multi-tape Turing
machine, algebraic RAM, \ldots{} To phrase it in the traditional
complexity-theoretic framework, one may restrict to
\(\mathsf{R} = \mathbb{F}_2\). This becomes a question about a
particular function \(p:\{0,1\}^* \to \{0,1\}^*\) which is
length-preserving.

\section{Space-preserving transposition
principle}\label{Section:space-preserving-transposition-principle}

As mentioned in
\hyperref[section:transposition]{Section~\ref*{section:transposition}},
the transposition principle relates the time complexities of a linear
program and its transposed. Relating their time-space complexities is
raised as an open problem by Kaltofen~\autocite[Open Problem
6]{Kaltofen2000}. To have a meaningful notion of space complexity, this
can be phrased in the model of fixed-size algebraic program
(\hyperref[definition:algprogram]{Definition~\ref*{definition:algprogram}}),
restricted to linear operations. A first answer to Kaltofen's question
was given by Bostan, Lecerf and
Schost~\autocite{BostanLecerfSchost2003}. In their model, they prove
that the number of operations and the number of registers are exactly
preserved by transposition. But their model not only gives read-write
permissions on the inputs but also allows their destruction, that is
their programs compute \((\vv,\vw)\mapsto(\vx,M\cdot\vw)\) where \(\vx\)
can be anything. By contrast,
\hyperref[theorem:ConstantSpaceBilinear]{Theorem~\ref*{theorem:ConstantSpaceBilinear}}
shows that the cumulative problem
\((\vv,\vw)\mapsto(\vv,\vw+M\cdot\vv)\) can be computed using no extra
register while asymptotically preserving the number of operations, in
the model \texttt{rw/rw}. By transposition, the same applies to the
transposed problem. These different and incomparable results suggest an
updated and generalized version of Kaltofen's question.

\begin{openproblem}

Given a fixed-size linear algebraic program to compute
\((\vv,\vw)\mapsto(\vv,M\cdot\vv)\) with \(t\) instructions using \(s\)
registers, what are the feasible pairs \((S,T)\) such that there exists
a fixed-size linear algebraic program for
\((\vw,\vv)\mapsto(\vw,M^\transp\cdot\vv)\) with \(T\) instructions
using \(S\) registers?\gobblepar

\end{openproblem}

The problem is not completely defined. The answer may depend on the
permission model. It can be asked in the two models \texttt{ro/rw} and
\texttt{rw/rw}, and even in a mixed case where the source program has
\texttt{ro/rw} permissions while the target program has \texttt{rw/rw}
permissions.

\section{Beyond computer algebra}\label{Section:beyond-computer-algebra}

The questions investigated in this part extend beyond computer algebra.
More generally, the question is about the time-space complexity for some
function \(f:\Sigma^*\to\Sigma^*\) over some finite alphabet \(\Sigma\).
As we have seen for algebraic computations, the permission models play a
role. And the relative powers of these models remain to be understood. A
close recent line of inquiry is catalytic
computation~\autocite{BuhrmanCleveKouckyLoffSpeelman2014}, as presented
in
\hyperref[section:cstspace-comparisons]{Section~\ref*{section:cstspace-comparisons}}.
For instance, a recent result~\autocite{CookLiMertzPyne2025} showed that
if one algorithm computes some function \(f\) in polynomial time and
another one computes \(f\) in catalytic logarithmic space, then there
exists a single algorithm to compute \(f\) in both polynomial time and
catalytic logarithmic space. The questions we investigated in this part
are in some sense \emph{fine-grained} versions of this result (but in a
different model). Relevant to computer algebra would be to extend our
results to the more classical catalytic model to determine if one can
replace the \texttt{rw} permissions on the input by some catalytic
space. Relevant to complexity theory would be to define and study
complexity classes corresponding to catalytic logarithmic space, but
where the catalytic space is actually restricted to the input and/or
output.

\section{Practical aspects}\label{Section:practical-aspects}

\begin{figure}
\centering
\pandocbounded{\includegraphics[keepaspectratio,alt={Timing comparisons between constant-space Strassen-Winograd algorithm and reference FFLAS-FFPACK implementation (with cubic implementation as reference).}]{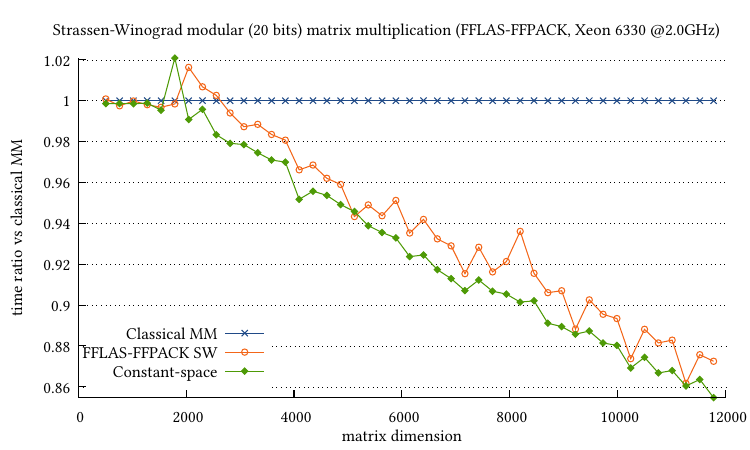}}
\caption{Timing comparisons between constant-space Strassen-Winograd
algorithm and reference FFLAS-FFPACK implementation (with cubic
implementation as reference).}\label{figure:StrassenTimings}
\end{figure}

Although the presentation of our results is mainly theoretical, the
relevance to practical computations is a central question. It is yet too
early to provide definite answers, but some preliminary experimental
results show that at least some of our algorithms
from~\autocite{DumasGrenet2024,DumasGrenet2024a,DumasGrenet2025} are in
practice at least as fast as standard algorithms that require some extra
space. For instance, Clément Pernet made an implementation\footnote{Available
  in branch \texttt{strassen-inplace} of the git repository
  \url{https://github.com/linbox-team/fflas-ffpack}.} of the
constant-space Strassen-Winograd algorithm
(\hyperref[algorithm:ConstantSpaceStrassenWinograd]{Algorithm~\ref*{algorithm:ConstantSpaceStrassenWinograd}})
in the FFLAS-FFPACK
library~\autocite{DumasGiorgiPernet2008,TheFflas-FfpackGroup2019}.
\hyperref[figure:StrassenTimings]{Figure~\ref*{figure:StrassenTimings}}
provides a comparison of this implementation with the reference
implementation of Strassen-Winograd algorithm in FFLAS-FFPACK that
already tries to minimize memory allocations. The two implementations
have similar running times. The constant-space implementation seems to
exhibit a more stable behavior although no explanation is currently
known for this phenomenon.

\begin{figure}
\centering
\pandocbounded{\includegraphics[keepaspectratio,alt={Timing comparisons between constant-space Karatsuba algorithm and the original algorithm, either with pre-allocated stack or with on-the-fly allocations}]{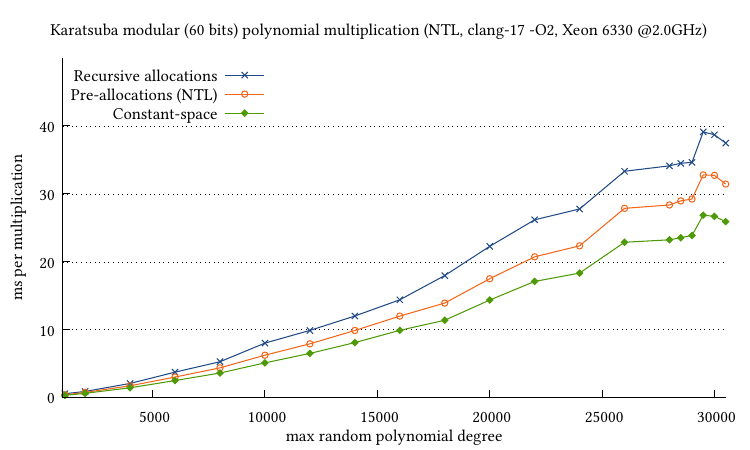}}
\caption{Timing comparisons between constant-space Karatsuba algorithm
and the original algorithm, either with pre-allocated stack or with
on-the-fly allocations}\label{figure:KaraTimings}
\end{figure}

Similarly,
\hyperref[figure:KaraTimings]{Figure~\ref*{figure:KaraTimings}} provides
a comparison between a recursive implementation that allocates some
memory at each level of recursion, the implementation in
NTL~\autocite{Shoup2021} that carefully allocates only once all the
necessary memory, and the constant-space variant
(\hyperref[algorithm:CumulativeKaratsuba]{Algorithm~\ref*{algorithm:CumulativeKaratsuba}}).\footnote{These
  implementations are due to Jean-Guillaume Dumas.} The constant-space
variant is here the fastest. Nevertheless, these results are not fully
reproducible. Using a different compiler (GCC instead of Clang) or a
different processor produces different results. But on-the-fly
allocations are always slower, and in all cases the constant-space
implementation is at most 10\% slower than the NTL implementation.
Further investigation is needed to understand the dependency on the
compiler and the processor architecture.

As a prudent conclusion, it seems that the constant-space variants at
least partially offset the increase in the number of operations with
fewer memory allocations. A more in-depth study is required to draw more
definitive conclusions. Practical gains can probably be expected from
algorithms in linear algebra, since memory allocation can play a
significant role.

\section{Constant-space quantum
algorithms}\label{Section:constant-space-quantum-algorithms}

On a more theoretical side, the algorithms we described in the model
\texttt{rw/rw} are reversible~\autocite{Lecerf1963,Bennett1973}. One
application of reversible computation is quantum computing where, by
definition, each computation must be
reversible~\autocite{NielsenChuang2010}. It can be shown that our
reversible algorithms can indeed be phrased in the formalism of quantum
computation.\footnote{This has been proved by Lucas Ottow for the case
  of Karatsuba's algorithm, during an internship co-supervised by Pascal
  Giorgi and myself. This result is unpublished.} In these settings,
space complexity corresponds to auxiliary (or \emph{ancilla}) qubits.
Reducing the number of such additional qubits is of utmost importance
due to the difficulty to build systems with many qubits. A quantum
computing version of
\hyperref[algorithm:ModularMultiplication]{Algorithm~\ref*{algorithm:ModularMultiplication}} (\textsc{ModularMultiplication})
provides for instance a multiplication algorithm for finite field
extensions that use no extra qubits while still maintaining fast
asymptotics. (The algorithm itself can use either Karatsuba's algorithm
or FFT-based multiplication as building block.) This improves some known
results in the field of quantum
computing~\autocite{KepleySteinwandt2015,ParentRoettelerMosca2018,Gidney2019,vanHoof2020,Kahanamoku-MeyerYao2024}.
Quite recently, some efforts have been made to reduce the space
complexity of some important quantum algorithms. As an example, Regev
has proposed a new quantum factoring algorithm~\autocite{Regev2025} that
has a lower (time) complexity than Shor's algorithm~\autocite{Shor1997},
and some subsequent works reduce the number of qubits required to
implement
it~\autocite{ChevignardFouqueSchrottenloher2024,RagavanVaikuntanathan2024}.
It would be interesting to investigate to what extent our technique may
improve the time-space complexity of this algorithm.

\part{Sparse polynomial computations}\label{part:sparse}

\chapter*{Summary}\label{Section:summary-1}
\addcontentsline{toc}{chapter}{Summary}

Classical computer algebra algorithms for polynomial manipulations
assume that the input in given in \emph{dense representation}, that is
by the vector of its coefficients. For a degree-\(d\) polynomial
\(f\in\mathsf{R}[x]\), the representation consists of \((d+1)\) elements
from \(\mathsf{R}\). This representation is not adapted when most of the
coefficients of \(f\) are zero, that is for \emph{sparse polynomials}.
Instead, one can use a \emph{sparse representation} where only the
nonzero terms are represented, by pairs \((c,e)\) where \(c\) is the
coefficient and \(e\) the exponent. If the polynomial has \(t\) nonzero
terms and degree \(d\), the representation requires \(t\) elements from
\(\mathsf{R}\) and \(t\log d\) bits to represent the exponents. Compared
to the dense representation, the sparse representation may be much more
compact for polynomials with few nonzero terms.

Since the representation is more compact, applying standard algorithms
to sparsely represented polynomials is inefficient. An algorithm with
quasi-linear complexity in the dense representation may have exponential
cost in the sparse representation. The point is that standard algorithms
do not exploit the sparsity of their inputs. Some algorithms were
developed early on for sparse
polynomials~\autocite{Johnson1974,Zippel1979}. But early hardness
results~\autocite{Plaisted1977,Plaisted1978,Plaisted1984} have probably
hindered the research on fast algorithms for sparse polynomials during
several decades. More recently, sparse polynomial arithmetic has been
the subject of renewed
interest~\autocite{NahshonShpilka2025,NahshonShpilka2024,MonaganHuang2024,HuangGao2023,HuangGao2025,HuMonagan2021,Nakos2020,GiesbrechtHuangSchost2020,ArnoldRoche2015,vanderHoevenLecerf2013,MonaganPearce2011a,MonaganPearce2009,KaltofenKoiran2005,KaltofenKoiran2006},
see also the very nice survey of Roche~\autocite{Roche2018}. Despite
these works and in contrast with dense polynomial arithmetic, no sparse
polynomial multiplication algorithm with subquadratic complexity in the
general case was described before 2021.

A central tool to develop algorithms for sparse polynomials is
\emph{sparse interpolation.} In very generic terms, the goal is to
reconstruct a polynomial \(f\) in sparse (explicit) representation from
some implicit representation. This is also known in algebraic complexity
as \emph{sparse polynomial reconstruction} or \emph{sparse polynomial
learning}~\autocite{ShpilkaYehudayoff2010}. Since the seminal work of
Ben-Or and Tiwari~\autocite{Ben-OrTiwari1988}, this problem has
attracted considerable attention in computer
algebra~\autocite{AlonMansour1995,ArnoldGiesbrechtRoche2013,ArnoldGiesbrechtRoche2014,ArnoldGiesbrechtRoche2016,ArnoldRoche2014,BlaserJindal2014,BorodinTiwari1991,ClausenDressGrabmeierKarpinski1991,GiesbrechtLabahnLee2006,GiesbrechtLabahnLee2009,GrigorievKarpinski1993,GrigorievKarpinskiSinger1990,Huang2019,Huang2021,Huang2023,HuangGao2019,HuangGao2020,HuangRao1999,ImamogluKaltofenYang2018,JavadiMonagan2010,KaltofenLakshmanWiley1990,KaltofenLee2003,KaltofenLeeLobo2000,KaltofenLakshman1988,KaltofenYang2021,KhochtaliRocheTian2015,LakshmanSaunders1995,Mansour1995,MuraoFujise1996,vanderHoevenLecerf2014,vanderHoevenLecerf2024,vanderHoevenLecerf2025,Werther1994}.
But until recently, these attempts felt short of providing a
quasi-linear time sparse interpolation algorithm.

In
\hyperref[chapter:sparseinterpolation]{Chapter~\ref*{chapter:sparseinterpolation}},
we present the main approaches for sparse interpolation, as well as a
new (dense) polynomial root-finding algorithm that is useful in this
context. In
\hyperref[chapter:sparseinterpolationZZ]{Chapter~\ref*{chapter:sparseinterpolationZZ}},
we present the first, and only to this date, quasi-linear algorithm for
sparse interpolation. It works for polynomials with integer
coefficients.
\hyperref[chapter:sparsearithmetic]{Chapter~\ref*{chapter:sparsearithmetic}}
focuses on sparse polynomial arithmetic. We present fast algorithms for
multiplication, exact division and divisibility testing. As a necessary
tool, we develop fast algorithms for the verification of certain
polynomial products. Finally, we also study the factorization of sparse
polynomials.

This part is largely based on a series of works carried out as part of
Armelle Perret du Cray's PhD thesis, co-supervised with Pascal Giorgi
(U. Montpellier), some of which are also co-authored by Daniel S. Roche
(U.S. Naval
Academy)~\autocite{GiorgiGrenetPerretduCray2020,GiorgiGrenetPerretduCray2021,GiorgiGrenetPerretduCrayRoche2022,GiorgiGrenetPerretduCrayRoche2022a,GiorgiGrenetPerretduCrayRoche2022b,GiorgiGrenetPerretduCray2023,GiorgiGrenetPerretduCrayRoche2024}.
The root-finding algorithm is based on joint works with Grégoire Lecerf
and Joris van der Hoeven (CNRS, École
polytechnique)~\autocite{GrenetvanderHoevenLecerf2015,GrenetvanderHoevenLecerf2016}.
Results about sparse factorization are based on single-authored
works~\autocite{Grenet2014,Grenet2016,Grenet2016a}.

\section*{Notations and
conventions}\label{Section:notations-and-conventions-1}
\addcontentsline{toc}{section}{Notations and conventions}

This part deals with sparse polynomials of the form
\(f = \sum_{i=0}^{t-1} c_i x^{e_i}\). In such an expression, we shall
always assume that \(c_i \neq 0\) for all \(i\), and
\(e_0 < \dotsb < e_{t-1}\). The \emph{support} of \(f\) is the set
\(\{e_i:0\leq i<n\}\) of its exponents. The degree of \(f\) is
\(\deg(f) = \max_i e_i = e_{t-1}\). Its \emph{sparsity} \(\#f\) is the
number \(t\) of nonzero terms. The \emph{height} of a polynomial
\(f\in\mathbb{Z}[x]\) is \(\height(f) = \max_i |c_i|\) where \(|\cdot|\)
denote the absolute value. The bit size of the sparse representation of
\(f\) is denoted by \(\textsf{bitsize}(f)\) and bounded by
\(O(\#f(\log\deg f + \log b))\), where \(b = q\) if the base ring is
\(\mathbb{F}_q\), and \(b = \height(f)\) if the base ring is
\(\mathbb{Z}\).

For an integer \(p\) (usually prime), \(f^{[p]}\) denotes the polynomial
\(f\bmod x^p-1\) and is called a \emph{fold} of \(f\). We say that a
term \(cx^e\) belongs to some polynomial \(f\), written \(cx^e\in f\),
if \(c\) is the coefficient of degree \(e\) of \(f\).

\notation[degree, sparsity and height of $f$\nomrefpage]{\(\deg(f)\),
\(\#f\), \(\height f\)}(spp)
\notation[bit size of the sparse representation of $f$\nomrefpage]{\(\bitsize(f)\)}(sps)
\notation[fold $f\bmod x^p-1$ of a sparse polynomial $f$\nomrefpage]{\(f^{[p]}\)}(spf)
\notation[the coefficient of degree $e$ of $f$ is $c$\nomrefpage]{\(cx^e\in f\)}(spe)
\index{polynomial!sparse|see{sparse polynomial}}
\index{sparse polynomial!support} \index{sparse polynomial!sparsity}
\index{sparse polynomial!height} \index{sparse polynomial!bit size}

Some algorithms are given \emph{black box access} to a polynomial
\(f\in\mathsf{R}[x]\). This means that the algorithm is allowed to
evaluate \(f\) on any point \(\alpha\in\mathsf{R}\) at unit cost. The
model of black boxes is further discussed in
\hyperref[section:comparisons]{Section~\ref*{section:comparisons}}.

\chapter{Main approaches for sparse
interpolation}\label{chapter:sparseinterpolation}

In this chapter, we review the two main lines of work for sparse
interpolation: black box interpolation based on Prony's method
(\hyperref[section:SparseInterpolationBB]{Section~\ref*{section:SparseInterpolationBB}}),
and SLP interpolation based on reductions modulo \(x^p-1\)
(\hyperref[section:SparseInterpolationSLP]{Section~\ref*{section:SparseInterpolationSLP}}).
We then compare the two approaches, and the two models used as inputs in
\hyperref[section:comparisons]{Section~\ref*{section:comparisons}}.
While the focus of the first three sections is on univariate polynomial,
\hyperref[section:multivariate]{Section~\ref*{section:multivariate}}
presents with the multivariate case. Finally,
\hyperref[section:FFTfriendly]{Section~\ref*{section:FFTfriendly}} is
devoted to some special finite fields, which we will refer to as
\emph{FFT-friendly}, that allow for faster sparse interpolation
algorithms. A thorough study of the case of polynomials with integer
coefficients is deferred to the next chapter.

\section{Black box sparse
interpolation}\label{section:SparseInterpolationBB}

The starting point of this approach is Prony's
method~\autocite{deProny1795}. Given \(2t\) initial terms of a linear
recurrent sequence \((u_i)_{i\geq 0}\), the goal is to express each term
as an exponential sum \(u_i = \sum_{j=0}^t \lambda_j\rho_j^i\). By
\hyperref[proposition:linrec]{Proposition~\ref*{proposition:linrec}},
the \(\rho_j\)'s are the roots of the minimal polynomial of
\((u_i)_{i\geq 0}\). Recall that \(V_t(\vec \rho)\) denotes the
Vandermonde matrix with second column \(\vec \rho\in\mathsf{R}^t\).

\begin{algorithm}[Prony]

\begin{description}
\item[Input:]
\(2t\) terms \(u_0\), \ldots, \(u_{2t-1}\) such that
\(u_i = \sum_{j=0}^{t-1}\lambda_j\rho_j^i\) for \(0\leq i<2t\), where
\(\vec \lambda\), \(\vec \rho\in \mathsf{R}^t\)
\item[Output:]
The two vectors \(\vec \lambda\) and \(\vec \rho\)
\end{description}

\begin{enumerate}

\item

\(p \gets\) minimal polynomial of \(\vu=(u_0,\dotsc,u_{2t-1})\)
\comment{\hyperref[proposition:minpoly]{Proposition~\ref*{proposition:minpoly}}}
\label{step:minpoly}

\item

\((\rho_0, \dotsc, \rho_{t-1}) \gets\) roots of \(p\) \label{step:roots}

\item

\((\lambda_0, \dotsc, \lambda_{t-1}) \gets V_t^\transp(\vec \rho)^{-1}\cdot\vec u_{[0,t[}\)
\comment{transposed Vandermonde syst.} \label{step:trVdm}

\item

Return \(\vec \lambda\), \(\vec \rho\)

\end{enumerate}\gobblepar

\end{algorithm}

To fully specify the algorithm, an algorithm for
\hyperref[step:roots]{Step~\ref*{step:roots}} is required. Actually,
there is no generic root-finding algorithm for an abstract ring
\(\mathsf{R}\)~\autocite{FrohlichShepherdson1955,FrohlichShepherdson1956}.
We focus on the case where \(\mathsf{R} = \mathbb{F}_q\) is a finite
field. Let \(p \in \mathbb{F}_q[x]\). We assume, as it is the case in
\hyperref[algorithm:Prony]{Algorithm~\ref*{algorithm:Prony}}, that \(p\)
is split and square-free in \(\mathbb{F}_q\). In other words, the
polynomial \(p\) has as many distinct roots in \(\mathbb{F}_q\) as its
degree. In this case, Berlekamp-Rabin
algorithm~\autocite{Berlekamp1970,Rabin1980} can be used to compute its
roots, as follows.

Since \(\alpha^{q-1} = 1\) for any nonzero \(\alpha\in\mathbb{F}_q\),
the set of roots of the polynomial \(x^{q-1}-1\) is exactly
\(\mathbb{F}_q^\times\). Assume for simplicity that \(q\) is odd. Then
\(x^{q-1}-1 = (x^{\frac{q-1}{2}}-1)\cdot(x^{\frac{q-1}{2}}+1)\).
Therefore, the roots of \(g = \textsc{gcd}(p,x^{\frac{q-1}{2}}-1)\) are
exactly the roots \(\alpha\) of \(p\) such that
\(\alpha^{\frac{q-1}{2}} = 1\), in other words the roots of order at
most \((q-1)/2\). The roots of \(p/g\) are the roots of \(p\) of order
\(q-1\). More generally, the polynomial
\(g_\sigma = \textsc{gcd}(p,(x+\sigma)^{\frac{q-1}{2}}) = 1\) contains
the roots \(\alpha\) of \(p\) such that \(\alpha+\sigma\) has order at
most \((q-1)/2\), and \(p/g_\sigma\) contains the rest of the roots. For
any \(\sigma\), the polynomial \(g_\sigma\) can be computed by fast
exponentiation of \((x+\sigma)\) modulo \(p\). Berlekamp-Rabin algorithm
takes a random shift \(\sigma\), computes \(g_\sigma\) and
\(p/g_\sigma\), and recursively computes their roots.

\begin{theorem}[\autocite{Berlekamp1970,Rabin1980}]\label{theorem:rootfinding}

Let \(p\in\mathbb{F}_q[x]\) of degree \(d\) with \(d\) distinct roots in
\(\mathbb{F}_q\). Its roots can be computed in
\(O(\M(d)\log(d)\log(q))\) operations in \(\mathbb{F}_q\), or
\(O^{\widetilde{}}(d\log^{2}q)\) bit operations.\gobblepar

\end{theorem}

Using this result, we can analyze
\hyperref[algorithm:Prony]{Algorithm~\ref*{algorithm:Prony}} (\textsc{Prony})
in the case of finite fields. Note that
\hyperref[step:trVdm]{Step~\ref*{step:trVdm}} can be solved in
\(O^{\widetilde{}}(t)\) operations in \(\mathbb{F}_q\) by transposition
(\hyperref[proposition:transposition]{Proposition~\ref*{proposition:transposition}})
of dense interpolation
(\hyperref[proposition:evaluationinterpolation]{Proposition~\ref*{proposition:evaluationinterpolation}}).

\begin{theorem}\label{theorem:seq2expcomplexity}

If \(\vec u\), \(\vec \lambda\) and \(\vec \rho\) lie in the finite
field \(\mathbb{F}_q\),
\hyperref[algorithm:Prony]{Algorithm~\ref*{algorithm:Prony}} (\textsc{Prony})
requires \(O^{\widetilde{}}(t\log^{2}q)\) bit operations.\gobblepar

\end{theorem}

As a corollary of
\hyperref[proposition:linrec]{Proposition~\ref*{proposition:linrec}}, we
obtain the fundamental result that underlies black box interpolation.
Remark that
\(f(\omega^j) = \sum_{i=0}^{t-1} c_i \omega^{j\cdot e_i} = \sum_{i=0}^{t-1} c_i (\omega^{e_i})^j\).
The following result, sometimes known as Blahut's theorem, is the
implication \ref{proposition:linrec:minpoly} \(\Rightarrow\)
\ref{proposition:linrec:expsum} of
\hyperref[proposition:linrec]{Proposition~\ref*{proposition:linrec}}
where \(u_j = f(\omega^j)\), \(\rho_i = \omega^{e_i}\) and
\(\lambda_j = c_j\).

\begin{corollary}[\autocite{Blahut1979}]\label{corollary:Blahut}

Let \(f = \sum_{i=0}^{t-1} c_i x^{e_i}\in \mathsf{R}[x]\) and
\(\omega\in \mathsf{R}\) of multiplicative order \(\geq \deg f\). The
sequence \((f(\omega^j))_{j\geq 0}\) is linear recurrent with minimal
polynomial \(p = \prod_{j=0}^{k-1} (x-\omega^{e_j})\).\gobblepar

\end{corollary}

As a consequence, we obtain a black box sparse interpolation algorithm,
whose principle is originally due to Ben-Or and
Tiwari~\autocite{Ben-OrTiwari1988} in the context of multivariate sparse
interpolation of polynomials with integer coefficients.

\begin{algorithm}[SparseInterpolationBB]

\begin{description}
\item[Inputs:]
Black box access \(f = \sum_{i=0}^{t-1} c_i x^{e_i}\in \mathsf{R}[x]\)

Bound \(\tau\geq\sp f =t\)

Element \(\omega\in \mathsf{R}\) of multiplicative order \(\geq \deg f\)
\item[Output:]
The sparse representation of \(f\), viz the vectors \(\vec c\) and
\(\vec e\)
\end{description}

\begin{enumerate}

\item

\(\vec \alpha \gets (f(\omega^{0}), \dotsc, f(\omega^{2\tau-1}))\)
\comment{\(2\tau\) calls to the black box} \label{Step:eval}

\item

\((\vec c, \vec \rho) \gets \textsc{Prony}(\vec \alpha)\)
\comment{\(\alpha_i = \sum_j c_j\rho_j^i\)} \label{Step:Seq2Exp}

\item

for \(j=0\) to \(t-1\)\,: \(e_j \gets \log_\omega(\rho_j)\)
\comment{\(\rho_j = \omega^{e_j}\)} \label{Step:dlog}

\item

return \(\vec c\), \(\vec e\)

\end{enumerate}\gobblepar

\end{algorithm}

As for Prony's algorithm, one cannot provide a complete complexity
analysis of
\hyperref[algorithm:SparseInterpolationBB]{Algorithm~\ref*{algorithm:SparseInterpolationBB}} (\textsc{SparseInterpolationBB})
over an abstract ring so focus on the case of finite fields. The most
computationally expensive part is the discrete logarithm computations.
With no assumption of \(\mathbb{F}_q\), a discrete logarithm can be
computed with \(O(\sqrt q)\) operations in \(\mathbb{F}_q\) using the
\emph{baby steps / giant steps} algorithm~\autocite{Shanks1971}. The
computation of \(t\) discrete logarithms can be amortized. As long as
\(t \ll q\), they can be computed in \(O(\sqrt{tq})\) bit
operations~\autocite{KuhnStruik2001}. This dominates the cost of the
whole algorithm.

\begin{theorem}

If \(f\in \mathbb{F}_q[x]\),
\hyperref[algorithm:SparseInterpolationBB]{Algorithm~\ref*{algorithm:SparseInterpolationBB}} (\textsc{SparseInterpolationBB})
requires \(O^{\widetilde{}}(\sqrt{tq})\) bit operations.\gobblepar

\end{theorem}

\begin{remark}

The bound given in the theorem can be improved by using better discrete
logarithm algorithms. Over any finite field, there exist subexponential
algorithms~\autocite{AdlemanDeMarrais1993}.\footnote{Many algorithms
  have been described for computing discrete logarithms problems. Some
  of them are proven while others are heuristic. Also, the best
  complexities depend on the relative size of the field compared to its
  characteristic. We refer to~\autocite{Galbraith2012} for some more
  details.} Also, if a bound \(\delta\geq\deg f\) is given, the discrete
logarithm computations can be sped up to be performed in time
\(O^{\widetilde{}}(\sqrt{t\delta})\) instead of
\(O^{\widetilde{}}(\sqrt{tq})\)~\autocite{KuhnStruik2001,Pollard1978}.
We do not discuss into more details these faster algorithms that remain
too expensive in our case. In the following, we shall mainly explain how
to circumvent discrete logarithm computations. An exception is for
finite fields with multiplicative group of smooth cardinality, discussed
in
\hyperref[section:FFTfriendly]{Section~\ref*{section:FFTfriendly}}.\gobblepar

\end{remark}

\hyperref[algorithm:SparseInterpolationBB]{Algorithm~\ref*{algorithm:SparseInterpolationBB}} (\textsc{SparseInterpolationBB})
requires an upper bound \(\tau\) on the sparsity \(t\) of the output
polynomial. This is used for the computation of the minimal polynomial
of the sequence \((f(\omega^j))_{j\geq 0}\), that is known to have
degree at most \(\tau\). Hence, \(2\tau\) evaluations are sufficient to
compute it. In order the remove this upper bound, one can use a
randomized approach, coined \emph{early
termination}~\autocite{KaltofenLeeLobo2000,KaltofenLee2003}. Consider
the sequence of polynomials \((f(x^j))_{j\geq 0}\). It is linear
recurrent of order \(t\), and has the same minimal polynomial as
\((f(\omega^j))_{j\geq 0}\). Let \(y_j = f(x^j)\) for all \(i\). For any
\(\tau\), define the Hankel matrix \[H_\tau = \begin{bmatrix}
y_{2\tau-1} & y_{2\tau-2} & \dots & y_\tau\\
y_{2\tau-2} & y_{2\tau-3} & \dots & y_{\tau-1}\\
\vdots   & \vdots   & \ddots& \vdots\\
y_{\tau}    & y_{\tau-1}  & \dots & y_1
\end{bmatrix}.\] For any \(\tau > t\), \(H_\tau\) is singular since the
first column is, by definition of a linear recurrent sequence, a linear
combination of the other ones. But for \(\tau \leq t\), this matrix can
be shown to be full rank. The sparsity of \(f\) is therefore the largest
value \(\tau\) for which \(H_\tau\) is full rank, of equivalently for
which \(\det(H_\tau)\) does not vanish. Since \(\det(H_\tau)\) is a
polynomial, its vanishing can be tested by an evaluation on a random
point \(\alpha\in\mathbb{F}_q\). Algorithmically, this means that one
computes the sequence \((f(\alpha^i))_{1\leq i\leq 2\tau-1}\) and tests
whether the corresponding Hankel matrix is full rank. This can be done
in a quasi-linear number of operations in \(\mathbb{F}_q\). From this,
it is clear that one can compute the sparsity with the same approach.
Actually, this test can be made implicit within the computation of the
minimal polynomial~\autocite{KaltofenLeeLobo2000,KaltofenLee2003}.

Altogether, this provides a variant of
\hyperref[algorithm:SparseInterpolationBB]{Algorithm~\ref*{algorithm:SparseInterpolationBB}} (\textsc{SparseInterpolationBB})
that does not take as input a bound on the sparsity, if the base ring is
large enough. Also, no element of large order needs to be provided since
the evaluations are made on random elements of \(\mathbb{F}_q\). Since
the algorithm requires the field to be large enough, namely
\(q = \Omega(\deg(f)^{4})\), random elements have large order with high
probability.

\section{SLP-based sparse
interpolation}\label{section:SparseInterpolationSLP}

The second approach assumes a richer representation of the polynomial as
input, namely a straight-line program. The starting point is the obvious
inefficient algorithm. Given an SLP for
\(f = \sum_{j=0}^{t-1} c_j x^{e_j}\), one can explicitly compute the
result of each instruction. This approach is inefficient since
intermediate results may be dense polynomials much larger than the final
result. The solution is then to compute a \emph{fold}
\(f^{[p]} = f\bmod x^p-1 = \sum_{j=0}^{t-1} c_j x^{e_j\bmod p}\). This
is achieved by replacing each multiplication in the SLP by a
multiplication modulo \(x^p-1\).

\begin{definition}\label{definition:fold}

Let \(f\in\mathsf{R}[x]\) and \(p\in\mathbb{Z}_{>0}\). The \emph{fold of
\(f\) modulo \(p\)} is the polynomial
\(f^{[p]} = f\bmod x^p-1\).\gobblepar

\end{definition}

\index{sparse polynomial!fold}

There are two difficulties to overcome:

\begin{enumerate}[label=(\roman*)]

\item

The exponents of \(f\) are reduced modulo \(p\) in \(f^{[p]}\);

\item

There can be \emph{collisions}, that is two monomials whose degrees are
congruent modulo \(p\) are mapped to a same monomial in \(f^{[p]}\).

\end{enumerate}

\index{sparse polynomial!monomial collisions}

The first difficulty can be dealt with either using the Chinese
Remainder Theorem (using several primes) with some technicalities, or
using the derivative of \(f\) to \emph{embed the exponents into the
coefficients}. We only present the second technique which is more
efficient, although less general since it requires the characteristic of
\(\mathsf{R}\) to be zero or large enough. The idea is to compute not
only the fold of the polynomial \(f\), but also the fold of \(xf'\)
where \(f'\) is the derivative of \(f\). A monomial \(cx^e\) of \(f\)
becomes \(cex^e\) in \(xf'\), and they are mapped to \(cx^{e\bmod p}\)
and \(cex^{e\bmod p}\) after reduction modulo \(x^p-1\), respectively.
If there is no collision, from two monomials \(cx^d\) in \(f^{[p]}\) and
\(c'x^d\) in \((xf')^{[p]}\), one can reconstruct the original monomial
\(cx^{c'/c}\) from \(f\). \index{sparse polynomial!exponent embedding}

\begin{algorithm}[TentativeTerms]

\begin{description}
\item[Inputs:]
\(f^{[p]}\) and \((xf')^{[p]}\)

Bound \(\delta \geq \deg f\)
\item[Outputs:]
\(f_{\smash *}\) that contains all terms of \(f\) that do not collide
modulo \(p\), and \(g_{\smash *} = xf_{\smash *}'\)
\end{description}

\begin{enumerate}

\item

\(f_{\smash *} \gets 0\); \(g_{\smash *} \gets 0\)

\item

for each pair of terms \(cx^e\in f^{[p]}\) and \(c'x^e\in (xf')^{[p]}\):

\item

~~~~if \(c\) divides \(c'\) in \(\mathbb{Z}\) and
\(0 \leq c'/c \leq \delta\):

\item

~~~~~~~~\(f_{\smash *} \gets f_{\smash *} + c\cdot x^{c'/c}\)

\item

~~~~~~~~\(g_{\smash *} \gets g_{\smash *} + c'\cdot x^{c'/c}\)

\item

return \(f_{\smash *}\) and \(g_{\smash *}\)

\end{enumerate}\gobblepar

\end{algorithm}

Note that the previous algorithm correctly computes all the terms of
\(f\) if no collision occurs. Otherwise, some terms can be missed and
some spurious terms can be added to \(f_{\smash *}\). This is the second
difficulty mentioned above. A first approach is to probabilistically
avoid any collision, taking for \(p\) a random prime in a suitable
interval. This requires to have \(p = O(t^{2}\log d)\) where
\(t = \sp f\) and \(d = \deg f\)~\autocite{ArnoldRoche2015} and is too
costly. A second approach is to allow \emph{some}
collisions~\autocite{ArnoldGiesbrechtRoche2013}. A monomial \(cx^e\) of
\(f\) is said \emph{collision-free modulo \(p\)} if no other monomial of
\(f\) has a degree congruent to \(e\) modulo \(p\). Even in the presence
of collisions, \textsc{TentativeTerms} returns some polynomial
\(f_{\smash *}\) that \emph{approximates} \(f\). It contains all the
collision-free monomials of \(f\), as well as some spurious monomials
coming from collisions. For \(p = O(t\log d)\), \(\frac{2}{3}t\)
monomials are probably
collision-free~\autocite{ArnoldGiesbrechtRoche2013}. If this is the
case, at most \(\frac{1}{3}t\) terms provide collisions. Since each
collision involves at least two terms, the number of collisions is at
most \(\frac{1}{6}t\). If each collision creates a spurious term in
\(f_{\smash *}\), then \(f-f_{\smash *}\) has at most \(\frac{1}{3}t\)
original terms from \(f\) and \(\frac{1}{6}t\) spurious terms.
Therefore, \(f-f_{\smash *}\) has at most \(\frac{1}{2}t\) terms. This
means that \(\log t\) iterations are enough to fully reconstruct \(f\).

\index{sparse polynomial!collision-free monomial}

The algorithm requires an SLP for \(xf'\) in addition to an SLP for
\(f\). It is easily computed in linear time using the standard rules for
derivation: \((f_1+f_2)' = f_1'+f_2'\) and
\((f_1\times f_2)' = f_1'f_2+f_1f_2'\). The size of the new SLP is at
most thrice the size of original one. (The generalization to
multivariate SLPs is known as automatic differentiation and due to Baur
and Strassen~\autocite{BaurStrassen1983}.\footnote{This attribution
  makes sense in the computer algebra community. The same algorithm and
  its very close cousin named backpropagation are central to machine
  learning and have been independently described several times in that
  community~\autocite{BaydinPearlmutterRadulSiskind2018}.}) This
provides the following algorithm which is a slight modification of an
algorithm of Huang~\autocite{Huang2019}.

\begin{algorithm}[SparseInterpolationSLP]

\begin{description}
\item[Inputs:]
SLP \(\mathcal{F}\) for
\(f = \sum_{i=0}^{t-1} c_i x^{e_i} \in \mathsf{R}[x]\)

Bounds \(\delta > \deg f\), \(\tau > \sp f\)
\item[Output:]
The sparse representation of \(f\), with probability
\(\geq \frac{2}{3}\)
\item[Assume:]
\(\charac(\mathsf{R}) =0\) or \(\charac(\mathsf{R}) > \deg f\)
\item[Constants:]
\(k = \lceil\log \tau\rceil\);
\(\lambda = \lceil\frac{5}{3\varepsilon_k}(\tau-1)\ln \delta\rceil\)
\end{description}

\begin{enumerate}

\item

\(\mathcal{G} \gets\) SLP for \(xf'\)

\item

\(f_{\smash *} \gets 0\); \(g_{\smash *} \gets 0\)

\item

repeat \(k\) times:

\item

~~~~\(p \gets\) random prime in \([\lambda, 2\lambda]\)

\item

~~~~\(u \gets f^{[p]} - f_{\smash *}^{[p]}\) using \(\mathcal{F}\) and
\(f_{\smash *}\) \label{SpIntSLP:fp}

\item

~~~~\(v \gets (xf')^{[p]} - g_{\smash *}^{[p]}\) using \(\mathcal{G}\)
and \(g_{\smash *}\) \label{SpIntSLP:xfp}

\item

~~~~\((f_{\smash **}, g_{\smash **}) \gets \textsc{TentativeTerms}(u,v,\delta)\)

\item

~~~~\(f_{\smash *} \gets f_{\smash *} + f_{\smash **}\)

\item

~~~~\(g_{\smash *} \gets g_{\smash *} + g_{\smash **}\)

\item

~~~~if \(\#f_{\smash *} > 2\tau\): return \textsc{failure}

\item

return \(f_{\smash *}\)

\end{enumerate}\gobblepar

\end{algorithm}

\notation[characteristic of the ring $\mathsf{R}$\nomrefpage]{\(\charac(R)\)}(dringzj)

\begin{theorem}

\hyperref[algorithm:SparseInterpolationSLP]{Algorithm~\ref*{algorithm:SparseInterpolationSLP}} (\textsc{SparseInterpolationSLP})
requires \(O^{\widetilde{}}(\ell\tau\log(\delta))\) operations in
\(\mathsf{R}\) and \(O(\tau\log(\delta))\) operations on integers of
size \(O(\log(\tau\log(\delta)))\), where \(\ell\) is the size of
\(\mathcal{F}\).\gobblepar

\end{theorem}

\begin{remark}

To improve the dependency on the failure probability, a classical
solution is to repeat several times the algorithm and use a majority
vote. To have a success probability \(1-\varepsilon\),
\(O(\log(\frac{1}\varepsilon))\) repetitions are sufficient. In the
context of this algorithm, a finer approach exists. Since any failure
comes from too many collisions, one can also in each iteration sample a
set of \(O(\log(\frac{1}\varepsilon))\) random primes and keep the prime
that maximizes the sparsity of \(f^{[p]}-f_{\smash *}^{[p]}\). This is
the approach used in Huang's original
algorithm~\autocite{Huang2019}.\gobblepar

\end{remark}

Huang presented this algorithm in the context of a finite field
\(\mathbb{F}_q\) with large characteristic.

\begin{theorem}[\autocite{Huang2019}]\label{theorem:Huang}

Over a finite field \(\mathbb{F}_q\) such that
\(\charac(\mathbb{F}_q) \geq \deg f\),
\hyperref[algorithm:SparseInterpolationSLP]{Algorithm~\ref*{algorithm:SparseInterpolationSLP}} (\textsc{SparseInterpolationSLP})
returns the sparse representation of \(f\) with probability at least
\(1-\varepsilon\) in time
\(O^{\widetilde{}}(\ell\tau\log \delta \log q\log\frac{1}\varepsilon)\),
where \(\ell\) is the size of the SLP.\gobblepar

\end{theorem}

We note that a slightly faster algorithm has been proposed by
Huang~\autocite{Huang2023}, that uses a Prony-like approach combined
with \emph{exponent embedding}. This is presented in the next section.

\section{Comparisons}\label{section:comparisons}

In this section, we discuss alternative presentations of the classical
algorithms described above, and links with other sparse interpolation
algorithms.

First, note that
\hyperref[algorithm:SparseInterpolationBB]{Algorithm~\ref*{algorithm:SparseInterpolationBB}} (\textsc{SparseInterpolationBB})
actually evaluates the black box for \(f\) on powers of an element
\(\omega\) of order \(n \geq \deg(f)\). Another presentation of this
algorithm could consider that the input is the full vector
\((f(\omega^i))_{0\leq i<n}\). With random access to this vector, it is
possible to interpolate \(f\) within the complexity described earlier.
In other words, this algorithm can be seen as a(n inverse) \emph{sparse
Fast Fourier Transform}. This name originates in the field of signal
processing. There, the input is a vector \(\vec v\in \mathbb{C}^n\)
whose discrete Fourier transform is known to be \(t\)-sparse. The goal
is to compute it in time sublinear in \(n\), closer to linear in \(t\).
Actually, the more general problem is to compute, given
\(\vv\in\mathbb{C}^n\), a \(t\)-sparse \emph{approximation} of its DFT.
The algorithms for sparse FFT can be traced back to a sparse polynomial
interpolation algorithm of Mansour for polynomials over \(\mathbb{Z}\),
given a black box over
\(\mathbb{C}\)~\autocite{Mansour1995,AlonMansour1995}. The best sparse
FFT algorithms have complexity \(O(t\log n)\) for the exact case, and
\(O(t\log(n)\log(n/t))\) for the approximate
case~\autocite{HassaniehIndykKatabiPrice2012,Hassanieh2016}. Although
these algorithms use techniques that are specific to the complex
numbers, others are very close to the techniques present in sparse
interpolation. For instance, they use the idea of \emph{folding} the
polynomials modulo \(x^p-1\) as in
\hyperref[algorithm:SparseInterpolationSLP]{Algorithm~\ref*{algorithm:SparseInterpolationSLP}} (\textsc{SparseInterpolationSLP}).

This idea of \emph{folding} can also be used to provide an alternative
view of
\hyperref[algorithm:SparseInterpolationSLP]{Algorithm~\ref*{algorithm:SparseInterpolationSLP}} (\textsc{SparseInterpolationSLP}).
The SLP is indeed only used to compute \(f^{[p]}\) and \((xf')^{[p]}\)
for some primes \(p\). This means that it can be described as taking as
inputs \emph{folding black boxes} for \(f\) and \(xf'\) which, on input
\(p\), return \(f^{[p]}\) and \((xf')^{[p]}\) respectively. In other
words, the algorithm does not make full use of its input since it does
not exploit the \emph{structure} of the SLP.\footnote{This structure is
  better viewed in the arithmetic circuit representation.} To the best
of our knowledge, no sparse interpolation algorithm exploits this
structure.

To summarize the previous discussion, let us define several models of
black boxes for a polynomial.

\begin{definition}\label{definition:BlackBoxes}

Let \(f = \sum_{i=0}^{t-1} c_i x^{e_i} \in \mathsf{R}[x]\).

\begin{itemize}
\item
  A \emph{regular black box} takes as input \(r\in\mathsf{R}\), and
  returns \(f(r)\).
\item
  A \emph{geometric black box}, parameterized by some
  \(\omega\in\mathsf{R}\), takes as input \(i\) and returns
  \(f(\omega^i)\).
\item
  A \emph{folding black box} takes as input \(p\in\mathbb{Z}_{>0}\), and
  returns \(f^{[p]}\), that is
  \(f(\zeta)\in\mathsf{R}[\zeta]/\langle\zeta^p-1\rangle\).
\item
  A \emph{modular black box} takes as inputs \(\mathcal{I}\) and
  \(m\in\mathsf{R}/\mathcal{I}\) where \(\mathcal{I}\) is an ideal of
  \(\mathsf{R}\), and returns \(f(m)\in \mathsf{R}/\mathcal{I}\).
\item
  An \emph{extended black box} takes as inputs \(\mathsf{S}\) and
  \(s\in\mathsf{S}\) where \(\mathsf{S}\) is an extension ring of
  \(\mathsf{R}\), and returns \(f(s)\in\mathsf{S}\).
\end{itemize}

A black box \emph{with derivative} not only returns the evaluation of
\(f\), but also the evaluation of its derivative \(f'\) (or equivalently
of \(xf'\)).\gobblepar

\end{definition}

\index{black box}

As discussed earlier,
\hyperref[algorithm:SparseInterpolationBB]{Algorithm~\ref*{algorithm:SparseInterpolationBB}} (\textsc{SparseInterpolationBB})
only requires a geometric black box rather than a regular black box and
\hyperref[algorithm:SparseInterpolationSLP]{Algorithm~\ref*{algorithm:SparseInterpolationSLP}} (\textsc{SparseInterpolationSLP})
only requires access to a folding black box with derivative instead of
an SLP. A natural question is to understand which black box is needed is
which context.

\begin{figure}

\centering
\input{blackboxes.tikz}

\caption{Relation between the main black box models and SLPs. An arrow
indicates that the source is able to simulate the target.}
\label{figure:blackboxes}\gobblepar

\end{figure}
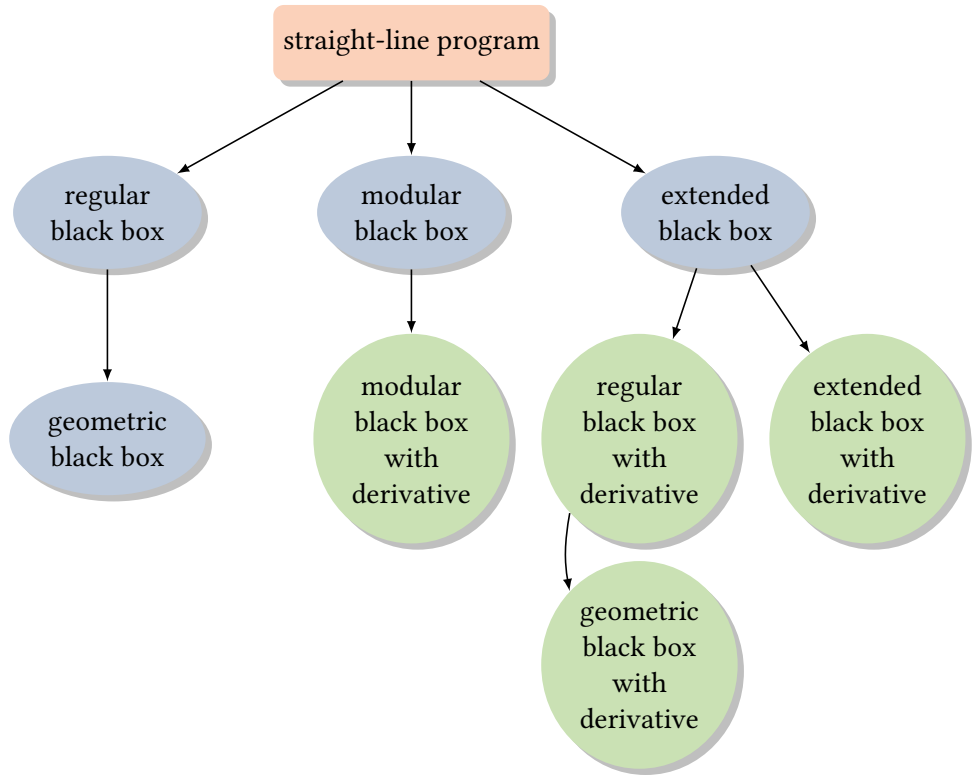

First, we can relate the power of some of these black boxes and of SLPs.
\hyperref[figure:blackboxes]{Figure~\ref*{figure:blackboxes}} summarizes
this paragraph. First, an SLP for \(f\) is enough to implement any type
of black box, with or without derivative. (For the geometric black box,
this assumes that \(\omega\) is given.) In this sense, the SLP model is
the strongest one. In terms of complexity, the simulation of each call
to a black box requires \(\ell\) operations in the ring where the black
box operates, where \(\ell\) is the size of the SLP. For instance, a
call to a folding black box requires \(O(\ell\M(p))\) operations in
\(\mathsf{R}\). The presentation in terms of black boxes requires some
care to avoid any \emph{cheating}. For instance, \(f\) can be read
directly from one evaluation of a folding black box, if \(p > \deg(f)\).
Associating a cost to each call to the black box prevents this problem.

A regular black box can implement a geometric black box, given access to
\(\omega\). A folding black box is a special case of an extended black
box. An extended black box can implement a regular black box with
derivative. Indeed, consider
\(\mathsf{R}_\varepsilon = \mathsf{R}[\varepsilon]/\langle\varepsilon^{2}\rangle\).
Then for any \(r\in\mathsf{R}\),
\(f(r+\varepsilon) = f(r)+\varepsilon f'(r)\) and a call to the black
box with inputs \(r\) and \(r+\varepsilon\) provides both \(f(r)\) and
\(f'(r)\). Note that if
\(\mathsf{R} = \mathbb{F}_p[\zeta]/\langle\psi\rangle\) for some prime
\(p\) and \(\psi\in\mathbb{F}_p[\zeta]\), the black box with derivative
can also be implemented using
\(\mathsf{R}_2 = (\mathbb{Z}/p^{2}\mathbb{Z})[\zeta]/\langle\psi\rangle\)
since \(f(r\cdot(1+p)) = f(r) + p\cdot(xf')(r) \in \mathsf{R}_2\). The
same idea proves that a modular black box can always evaluate \(f\) and
its derivative.

We are also interested in the limits of some of the models. First, for
\(\mathsf{R} = \mathbb{Z}\), it is quite clear that from the viewpoint
of bit complexity, the regular black box is not satisfactory. The output
\(f(r)\) has exponential size in the size of the sparse representation
of \(f\) in the worst case. Therefore, it is natural to use a modular
black box and to reconstruct \(f\) from evaluations \(f(\alpha)\bmod m\)
for some \(m\) and \(\alpha\in\mathbb{Z}/m\mathbb{Z}\). Note that a
modular black box for \(f\) also provides a modular black box with
derivative using the same technique as before to evaluate both \(f\) and
\(xf'\) over some ring \(\mathbb{Z}/m\mathbb{Z}\).

This difficulty disappears when \(\mathsf{R} = \mathbb{F}_q\) since
every element has the same size. But the complexity of
\hyperref[algorithm:SparseInterpolationBB]{Algorithm~\ref*{algorithm:SparseInterpolationBB}} (\textsc{SparseInterpolationBB})
is exponential because it requires some discrete logarithm computations.
It is a natural question to ask whether they can be avoided. The answer
is negative if the input is a geometric black box.

\begin{proposition}\label{proposition:sparseFFTlowerbound}

Consider an algorithm that, given access to a geometric black box for
\(f\in\mathbb{F}_q[x]\), computes the sparse representation of the
\(f\). This algorithm can be used to compute, given \(\omega\) and
\(\alpha\in \mathbb{F}_q\), the discrete logarithm \(e\) of \(\alpha\)
that satisfies \(\alpha^e = \omega\).\gobblepar

\end{proposition}

\begin{proof}

Assume we are given an algorithm \(\mathcal{A}\) that makes queries to a
geometric black box for \(f\) and outputs the sparse representation of
\(f\). Given \(\alpha\), we want to compute \(e\) such that
\(\alpha = \omega^e\), using \(\mathcal{A}\). To this end, define
\(f(x) = x^e\). We can apply \(\mathcal{A}\) on \(f\), simulating each
query to the geometric black box. Indeed,
\(f(\omega^i) = \omega^{e\cdot i} = \alpha^i\) can be computed from
\(\alpha\) and \(i\) without knowing \(e\). Therefore, \(\mathcal{A}\)
returns the sparse representation of \(f\), that is the value
\(e\).\gobblepar

\end{proof}

One could argue that we apply the algorithm in some very specific case
where \(f\) has sparsity one. There might exist an algorithm that does
not compute discrete logarithms but only works for polynomials of
sparsity at least \(2\), or at least some other constant. Actually, the
reduction still works in these more general settings. It is enough to
add to \(f\) any known sparse polynomial \(g\), and even multiply
\(x^e\) by some known coefficient. The queries to the geometric black
box can still be simulated and the result holds.

The lower bound does not hold for a geometric black box with derivative,
since there is no \emph{a priori} way to simulate the black box calls
for \(f'\). Actually, a fast algorithm exists when the input is a
geometric black box with derivative. The idea is to evaluate both \(f\)
and \(xf'\) on powers of \(\omega\) and to use
\hyperref[algorithm:Prony]{Algorithm~\ref*{algorithm:Prony}} (\textsc{Prony})
on both sequences. This provides in particular the coefficients of \(f\)
and of \(xf'\), from which the exponents of \(f\) can be computed. This
algorithm has been described in the case of finite fields by
Huang~\autocite{Huang2023}, using an SLP as input. We present his result
in more restricted settings.

\begin{theorem}[\autocite{Huang2023}]\label{theorem:Huang2023}

Given a geometric black box with derivative for
\(f = \sum_{i=0}^{t-1} c_ix^{e_i} \in \mathbb{F}_q[x]\) where
\(\charac(\mathbb{F}_q)\geq\deg(f)\), and bounds \(\tau \geq t\) and
\(\delta > \deg(f)\), one can compute the sparse representation of \(f\)
using \(O^{\widetilde{}}(\tau)\) calls to the black box and
\(O^{\widetilde{}}(t\log^{2} q)\) bit operations.\gobblepar

\end{theorem}

The most expensive part of this algorithm is actually the root finding
part of
\hyperref[algorithm:Prony]{Algorithm~\ref*{algorithm:Prony}} (\textsc{Prony}).
This is an intriguing open question whether this root finding step can
be bypassed to get a better complexity.

\section{Multivariate sparse interpolation}\label{section:multivariate}

We have presented univariate interpolation algorithms. Actually, these
algorithms were usually first presented in the multivariate settings. We
now present the two main generic solutions to turn a univariate
interpolation algorithm into a multivariate one.

The classical reduction from the multivariate case to the univariate
case in Kronecker substitution~\autocite{Kronecker1882}. Let
\(f\in \mathsf{R}[x_0, \dotsc, x_{n-1}]\) where the individual degree of
\(f\) in each variable is \(<d\). We define
\(f_u = f(x,x^d, x^{d^{2}}, \dotsc, x^{d^{n-1}}))\). The map
\(f\mapsto f_u\) is easily inverted by writing the exponents of \(f_u\)
in base \(d\). If \(\deg_{x_i} f < d_i\) for \(0\leq i<n\), this can be
improved by setting
\(f_u = f(x,x^{d_0}, x^{d_0d_1}, \dotsc, x^{d_0\dotsb d_{n-2}})\). The
resulting polynomial has the same number of terms as \(f\), and degree
\(< d_0\dotsb d_{n-1} < d^n\) if \(d_i < d\) for all \(i\). Therefore,
univariate algorithms can be used for the interpolation of multivariate
polynomials. In the complexity analyses, a term \(\log(d)\) is replaced
by \(\log(d^n) = n\log(d)\). This has two main drawbacks. First, a
univariate algorithm that is not quasi-linear in \(\log(d)\) will result
in a multivariate algorithm that is not quasi-linear in the number \(n\)
of variables. Second, if a univariate algorithm only works for rings of
characteristic larger than the degree, the condition becomes that the
characteristic be larger than \(d^n\) in the multivariate case.

A way to reduce the degree of the resulting univariate polynomial is to
use a \emph{randomized} Kronecker
substitution~\autocite{ArnoldRoche2014}. The idea is to define
\(f_{\va} = f(x^{a_0}, \dotsc, x^{a_{n-1}})\) for some randomly chosen
vector \(\va\) where \(a_i = O^{\widetilde{}}(nt\log d)\) for each
\(i\). The resulting degree is much smaller. But the difficulty is that
the map \(f\mapsto f_{\va}\) is not invertible anymore, since there may
be some \emph{collisions} between monomials. This can be dealt with by
using several vectors \(\va\). Many approaches, with sometimes
incomparable complexities, have been proposed. One can use a set of
random vector and rely on linear
algebra~\autocite{ArnoldRoche2014,ArnoldGiesbrechtRoche2016}, or the
columns of a random Hankel matrix and rely on structured linear algebra
(attributed to Pernet~\autocite{Arnold2016}), or a set made of \(\va\)
and perturbations of it~\autocite{Arnold2016,HuangGao2019,Huang2019}.

A slight issue with (randomized) Kronecker is that it requires to
evaluate the polynomial to interpolate (be it given as a black box or an
SLP) on several sets of inputs. The complexities therefore all contain a
hidden \(n^{2}\) factor~\autocite{PerretduCray2023}. Another approach
using partial derivatives can remove this hidden
factor~\autocite{Huang2021}.

\section{FFT-friendly finite fields}\label{section:FFTfriendly}

The complexity of sparse interpolation over finite fields does not make
any assumption on the field. In the case of black-box interpolation, the
bottleneck is discrete logarithm computations which is not even
polynomial in the output size. For specific finite fields where the
cardinality of the multiplicative group is smooth, one can use
Pohlig-Hellman algorithm~\autocite{PohligHellman1978} to fasten this
computation. This approach works nicely when one can choose the finite
field over which to work, for instance in a modular approach for sparse
interpolation over the integers or the rational numbers. This has been
proposed and analyzed by Kaltofen~\autocite{Kaltofen1988}.

For such an approach, as well as the fastest SLP interpolation
algorithm~\autocite{Huang2019,Huang2021}, polynomial root finding is a
costly step. The roots of a degree-\(d\) polynomial over
\(\mathbb{F}_q\) can be computed in \(O(\M(d)\log(d)\log(q))\)
operations in \(\mathbb{F}_q\) using Berlekamp-Rabin
algorithm~\autocite{Berlekamp1970,Rabin1980}. The bit complexity is not
quasi-linear in the output size due to the factor \(\log^{2}q\) that
appears. In practice, the root-finding step appears to be the bottleneck
too~\autocite{vanderHoevenLecerf2014}.

In this section, we investigate practical improvements when the finite
field has some special structure. We consider finite fields
\(\mathbb{F}_q\) such that \(q-1\) is a smooth integer. In particular,
we can focus on so-called \emph{FFT-friendly} finite fields, where
\(q-1 = m\cdot 2^e\) for some \(m = O(\log q)\). Example of primes \(q\)
that satisfy the condition are \(3\cdot 2^{12}+1\), \(7\cdot 2^{26}+1\)
or \(5\cdot 2^{55}+1\). While it is not known how to generate prime
numbers of this form within a given interval in general, examples are
quite abundant and used in practice. The typical situation is some
computation with integer polynomials with not-too-large coefficients,
that is performed modulo an FFT-friendly prime. Their name come from the
fact that \(\mathbb{F}_q\) has a primitive root of unity of order \(e\),
making FFT-based algorithms very efficient.

For FFT-friendly finite fields, and more generally finite fields for
which \(q-1\) is smooth, most classical root-finding algorithms can be
fastened~\autocite{GrenetvanderHoevenLecerf2015}. In the same article,
we also introduce a new algorithm based on the \emph{Graeffe transform},
a tool originating in numerical analysis.

\begin{definition}

Let \(f = \prod_{i=0}^{d-1} (x-\alpha_i)\in\mathbb{F}_q[x]\). Its
\emph{Graeffe transform} \(Gf\) is the degree-\(d\) polynomial defined
by either of two ways:

\begin{itemize}
\item
  \(Gf(x^{2}) = (-1)^df(x)f(-x)\), or
\item
  \(Gf(x) = \prod_{i=0}^{d-1} (x-\alpha_i^{2})\).
\end{itemize}

\end{definition}

\index{Graeffe transform}

The first definition provides a way to compute the Graeffe transform
from the dense representation of \(f\). The second definition shows that
the effect of the Graeffe transform is to map the roots of \(f\) to
their squares. If \(q = m\cdot 2^e+1\), the \(e\)th iterate of the
Graeffe transform \(G^ef\) has roots \(\alpha_i^e\), \(1\leq i\leq e\).
They have multiplicative order at most \(m\). If \(\omega\) is a
generator of \(\mathbb{F}_q^\times\), the roots of \(G^ef\) belong to
the set \(\{\omega^{s\cdot 2^e}:0\leq s<m\}\) and can be computed by
multipoint evaluation. A root \(\alpha = \omega^{s\cdot 2^e}\) of
\(G^ef\) corresponds to one or two roots of \(G^{e-1}f\). The two
possibilities are \(\omega^{s\cdot 2^{e-1}}\) and
\(\omega^{(s+m)\cdot 2^{e-1}}\). Given the roots of \(G^ef\), this
provides a superset of the roots of \(G^{e-1}f\). Using multipoint
evaluation again, we can filter out elements of the superset that are
not actual roots of \(G^{e-1}f\). An algorithm follows: Compute the
\(e\)th Graeffe transform of \(f\); Compute the roots of \(G^ef\) by
exhaustive search using multipoint evaluation; Iteratively obtain the
roots of each \(G^if\) until \(i=0\).

\begin{figure}

\centering
\input{Graeffe.tikz}

\caption{Illustration of root-finding using Graeffe transforms. In the
first stage, the polynomial \(G^ef\) is computed and its roots \(Z_e\)
are obtained by multipoint evaluation. From \(Z_e\), the sets
\(Z_{e-1}\), \ldots, \(Z_0\) are computed by multipoint evaluation too.}
\label{figure:Graeffe}\gobblepar

\end{figure}
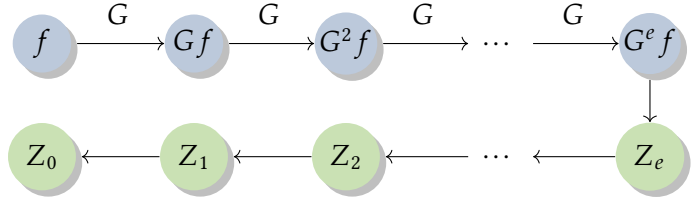

A generalization of this algorithm, together with some optimization,
provides the following deterministic algorithm.

\begin{theorem}[\autocite{GrenetvanderHoevenLecerf2016}]\label{theorem:detGraeffeRootFinding}

Let \(f \in \mathbb{F}_q[x]\) of degree \(d\) with \(d\) distinct roots
in \(\mathbb{F}_q\). Given the dense representation of \(f\), the
irreducible factorization of \(q-1\) and a generator of
\(\mathbb{F}_q^\times\), the roots of \(f\) can be computed in
\(O^{\widetilde{}}(\sqrt{P_1(q-1)}d\log^{2}q)+(d\log^{2}q)^{1+o(1)}\)
bit operations, where \(P_1(q-1)\) is the largest prime factor of
\(q-1\).\gobblepar

\end{theorem}

In particular, if \(q = m\cdot 2^e+1\) with \(m = O(\log q)\), the
algorithm runs in the same asymptotic complexity
\(O(\M(d)\log d\log q)\) operations in \(\mathbb{F}_q\) as the
randomized algorithm of Berlekamp and Rabin. In general, its complexity
slightly refines Shoup's complexity bounds~\autocite{Shoup1991}. An
adaptation of this method provides better randomized algorithms. To this
end, we introduce a generalization of the Graeffe transform.

\begin{definition}

Let \(f\in\mathbb{F}_q[x]\) and consider the extension ring
\(\mathbb{F}_q[\varepsilon]/\langle\varepsilon^{2}\rangle\). The
\emph{tangent Graeffe transform} of \(f\) is the Graeffe transform of
\(f_\varepsilon = f(x+\varepsilon) \in \mathbb{F}_q[\varepsilon]/\langle\varepsilon^{2}\rangle\).\gobblepar

\end{definition}

Note that \(f_\varepsilon = f(x) + \varepsilon f'(x)\) and
\(Gf_\varepsilon = Gf + \varepsilon g\) for some polynomial \(g\). The
roots of \(f_\varepsilon\) are of the form \(\alpha-\varepsilon\) where
\(\alpha\) is a root of \(f\), therefore they are mapped to
\((\alpha-\varepsilon)^{2} = \alpha^{2}-2\alpha\varepsilon\) in
\(Gf_\varepsilon\). Since
\(Gf_\varepsilon(\alpha^{2}-2\alpha\varepsilon) = Gf(\alpha^{2})-2\alpha\varepsilon(Gf)'(\alpha^{2}) + \varepsilon g(\alpha^{2})\)
and \(Gf(\alpha^{2}) = 0\), \(\alpha^{2}\) is a root of \(Gf'\) if and
only if \(g(\alpha^{2}) = 0\). This provides a way to detect simple
roots \(\alpha^{2}\) of \(Gf\), and for each of them to compute the
corresponding root \(\alpha\) of \(f\). Indeed, \(\alpha^{2}\) is a
simple root if \(g(\alpha^{2}) \neq 0\) and
\(\alpha = g(\alpha^{2})/2(Gf)'(\alpha^{2})\) in that case. The same
argument actually works more generally for the \(\ell\)th iterate
\(G^\ell f\). This translates into a variant of the previous algorithm.
Once the roots of some \(G^if\) are known, discard the multiple roots
and compute directly the corresponding roots of \(f\). (In
\hyperref[figure:Graeffe]{Figure~\ref*{figure:Graeffe}}, this this would
be a direct arc from \(Z_i\) to \(Z_0\).) Randomization, working with
\(f(x+\tau)\) instead of \(f\), ensures that for \(i \simeq q/d^{2}\)
all the roots of \(G^if\) are simple with good probability. With fast
computation of the tangent Graeffe transform, we obtain the following
complexity.

\begin{theorem}[\autocite{GrenetvanderHoevenLecerf2015}]\label{theorem:randGraeffeRootFinding}

Let \(f \in \mathbb{F}_q[x]\) of degree \(d\) with \(d\) distinct roots
in \(\mathbb{F}_q\), for \(q = m\cdot 2^e+1\) with \(m = O(\log q)\).
Given the dense representation of \(f\) and a generator of
\(\mathbb{F}_q^\times\), the roots of \(f\) can be computed in expected
\(O(\M(d)\log d\log q)\) operations in \(\mathbb{F}_q\).\gobblepar

\end{theorem}

The asymptotic complexity is the same as Berlekamp-Rabin algorithm.
Nevertheless, it is in practice faster due to the use of multipoint
evaluations instead of \textsc{gcd} computations. To make the algorithm
still faster, the goal would be to compute the roots of \(f\) from
\(G^if\) for a larger value of \(i\). The difficulty is that we are not
(yet) able to prove that randomization is sufficient. Anyway, we can
make the following heuristic.

\begin{quote}
\makebox[0pt][r]{\textbf{(H)}\hspace*{1em}}For any subset
\(\{\alpha_0,\dotsc,\alpha_{d-1}\} \subseteq \mathbb{F}_q\),
\(i = O(p/d)\),
\[\Pr_{\tau\in\mathbb{F}_q}\left[\#\left\{(\alpha_j-\tau)^i:0\leq j<d\right\} \leq 2d/3\right] \leq \frac{1}{2}.\]
\end{quote}

The heuristic states that at least \(\frac{1}{3}\) of the roots of
\(f(x+\tau)\) remain simple in the \(i\)th iterate of the Graeffe
transform with probability at least \(\frac{1}{2}\). It is supported by
a very similar result when instead of taking \(\tau\) at random, \(f\)
itself is chosen uniformly at random.

Using the heuristic, we obtain the following improved complexity.

\begin{theorem}[\autocite{GrenetvanderHoevenLecerf2015}]\label{theorem:heurGraeffeRootFinding}

Let \(f \in \mathbb{F}_q[x]\) of degree \(d\) with \(d\) distinct roots
in \(\mathbb{F}_q\), for \(q = m\cdot 2^e+1\) with \(m = O(\log q)\).
Given the dense representation of \(f\) and a generator \(\omega\) of
\(\mathbb{F}_q^\times\), the roots of \(f\) can be computed in expected
\(O(\M(d)(\log d+\log q))\) operations in \(\mathbb{F}_q\), assuming
heuristic \textbf{(H)}.\gobblepar

\end{theorem}

The complexity is this time lower than Berlekamp-Rabin's algorithm.
Further complexity analyzes and optimized implementations have
demonstrated the relevance of our algorithm in
practice~\autocite{vanderHoevenMonagan2020,vanderHoevenMonagan2020a}. We
also note that this root-finding method became popular in cryptographic
applications where the field can once again be
chosen~\autocite{SansoVitto2025,ZhaoDing2025}.

As a corollary, we obtain that for finite fields as in
\hyperref[theorem:heurGraeffeRootFinding]{Theorem~\ref*{theorem:heurGraeffeRootFinding}}
and under the same heuristic, it is possible to reconstruct a
\(t\)-sparse degree-\(d\) polynomial from a geometric black box with
derivative in \(O^{\widetilde{}}(\M(t)(\log t+\log q))\) operations in
\(\mathbb{F}_q\) (\emph{cf.}
\hyperref[theorem:Huang2023]{Theorem~\ref*{theorem:Huang2023}}).

The situation with polynomial root finding is still unsatisfactory. A
natural open question is to remove the heuristic assumption from
\hyperref[theorem:heurGraeffeRootFinding]{Theorem~\ref*{theorem:heurGraeffeRootFinding}}.
But more generally, the complexity of root finding is
\(O^{\widetilde{}}(d\log q)\) operations in \(\mathbb{F}_q\), that is
\(O^{\widetilde{}}(d\log^{2}q)\). This complexity is not quasi-linear.

\begin{openproblem}

Given a degree-\(d\) polynomial with coefficients in \(\mathbb{F}_q\)
with \(d\) distinct roots, is it possible to compute its roots in
\(O^{\widetilde{}}(d)\) operations in \(\mathbb{F}_q\), or in
\(O^{\widetilde{}}(d\log q)\) bit operations?\gobblepar

\end{openproblem}

\chapter{Quasi-linear sparse interpolation over the
integers}\label{chapter:sparseinterpolationZZ}

We describe our quasi-linear time algorithms for sparse interpolation
over the integers. The idea is to combine the techniques described so
far, taking advantage of the possibility to embed the ring of integers
into many modular rings. The algorithms are described in the model of
\emph{modular black box} described in
\hyperref[definition:BlackBoxes]{Definition~\ref*{definition:BlackBoxes}}:
Given \(m\) and \(\alpha\in \mathbb{Z}/m\mathbb{Z}\), the black box
returns \(f(\alpha)\in \mathbb{Z}/m\mathbb{Z}\). This model encompasses
the model of SLP since given a size-\(\ell\) SLP for
\(f\in \mathbb{Z}[x]\) and \(\alpha\in \mathbb{Z}/m\mathbb{Z}\), one can
compute \(f(\alpha)\) in \(O(\ell\Z(\log m))\) bit operations. Note that
the model of (standard) black box is not adapted to sparse integer
polynomials since \(f(\alpha)\) for
\(\alpha\in \mathbb{Z}\setminus\{-1,0,1\}\) is an integer of bit size
\(O(\deg(f)\log\height f)\), exponential in the size of the sparse
representation of \(f\).

Our approach is to follow Huang's algorithm and compute \(f\) from
several folds \(f^{[p]}\) and \((xf')^{[p]}\). But while Huang computes
these folds directly from an SLP for \(f\), we use the techniques of
\hyperref[algorithm:SparseInterpolationBB]{Algorithm~\ref*{algorithm:SparseInterpolationBB}} (\textsc{SparseInterpolationBB})
to compute them.

We distinguish two cases. The first and simpler one is the case of a
\emph{balanced} polynomial where the coefficients of the polynomial
\(f\) are assumed to be of similar bit sizes. The second one is the case
of an \emph{unbalanced} polynomial where the coefficients bit sizes may
vary a lot.

\index{balanced polynomial} \index{unbalanced polynomial}

\section{The balanced case}\label{Section:the-balanced-case}

As mentioned above, the goal is to reconstruct \(f\) from folds
\(f^{[p]}\) and \((xf')^{[p]}\). The only difference with Huang's
algorithm is that \(f^{[p]}\) and \((xf')^{[p]}\) are not directly
computed from an SLP since that would be too costly. Instead, they are
computed from a modular black box using the techniques of black box
interpolation.

The first remark is that
\hyperref[algorithm:SparseInterpolationBB]{Algorithm~\ref*{algorithm:SparseInterpolationBB}} (\textsc{SparseInterpolationBB})
evaluates \(f\) on powers of an element \(\omega\) of order \(>\deg(f)\)
in order to reconstruct \(f\). If we apply the algorithm with an element
\(\omega\) of order \(p\), then the output polynomial is actually
\(f^{[p]}\) since \(f(\omega^i) = f^{[p]}(\omega^i)\) in this case. If
we can find an element \(\omega\in \mathbb{Z}/m\mathbb{Z}\) of order
\(p\) where \(m > 2\height f\), we can use this algorithm in
\(\mathbb{Z}/m\mathbb{Z}\) with \(\omega\) to compute
\(f^{[p]}\bmod m\), and deduce \(f^{[p]}\).

The second aspect is to be able to compute \((xf')^{[p]}\). To apply the
same technique, one needs to be able to evaluate \((xf')\) on some
\(\omega\). If the input is actually an SLP, we can as in Huang's
algorithm compute an SLP for \(xf'\). As explained before, we can
evaluate \(f\) on \((1+m)\omega\) in \(\mathbb{Z}/m^{2}\mathbb{Z}\) to
obtain \(f(\omega)+m(xf')(\omega)\). Actually, \(\omega\) as an element
of \(\mathbb{Z}/m^{2}\mathbb{Z}\) may have a larger order than
\(\omega\in \mathbb{Z}/m\mathbb{Z}\). Hence, we need to use
\((1+m)\omega'\) where \(\omega'\in \mathbb{Z}/m^{2}\mathbb{Z}\) has
order \(p\) and \(\omega'\bmod m = \omega\).

The remaining question is the choice of \(m\) and the computation of
\(\omega\) and \(\omega'\). First, we need \(m > 2\height f\). The
complexity of
\hyperref[algorithm:SparseInterpolationBB]{Algorithm~\ref*{algorithm:SparseInterpolationBB}} (\textsc{SparseInterpolationBB})
in \(\mathbb{Z}/m\mathbb{Z}\) is then much too costly due to the
discrete logarithm computations. The solution is to remark that
\textsc{SparseInterpolationBB} can be (reorganized and) split into two
parts. The first part computes the support of \(f\) while the second one
uses the support of \(f\) to compute its coefficients. And only the
first part is very costly. Since the support of \(f^{[p]}\) contains the
support of \((xf')^{[p]}\), one can first compute the support of
\(f^{[p]}\) and then the coefficients of both \(f^{[p]}\) and
\((xf')^{[p]}\). For the computation of the support, the only condition
on \(m\) is that with high probability, no coefficient of \(f^{[p]}\)
vanishes. We can take for \(m\) a small prime \(q\). In a second stage,
the coefficients of \(f^{[p]}\) and \((xf')^{[p]}\) must be computed
exactly. This requires a larger \(m\), that we choose to be a power of
\(q\).

For this approach to work, we need a random prime \(p\) and prime \(q\)
together with an element \(\omega\in\mathbb{F}_q\) of order \(p\) (for
the computations modulo \(q\)), and an element
\(\omega_k\in\mathbb{Z}/q^k\mathbb{Z}\) of the same order for the
computations modulo \(q^k\). The strategy is to first sample \(p\). Then
using effective versions of Dirichlet's theorem on arithmetic
progressions, we can sample a prime number \(q\) of the form
\(q = ap+1\). Then for \(\zeta\in\mathbb{F}_q\), the order of
\(\zeta^a\) divides \((q-1)/a = p\). This means that an element
\(\omega\) of order \(p\) can be found by sampling
\(\zeta\in\mathbb{F}_q\) until \(\zeta^a \neq 1\). Finally, we can
\emph{lift} \(\omega\in \mathbb{F}_q\) to obtain an element
\(\omega_k\in\mathbb{Z}/q^k\mathbb{Z}\) of the same order by means of
Hensel lifting~\autocite{Hensel1918}. Indeed, if \(\omega_{2^i}\) has
order \(p\) in \(\mathbb{Z}/q^{2^i}\mathbb{Z}\), we can find
\(\omega^{2^{i+1}}\) of order \(p\) in
\(\mathbb{Z}/q^{2^{i+1}}\mathbb{Z}\) as
\(\omega_{2^{i+1}}  = \omega_{2^i} +\sigma q^{2^i}\) for some
\(\sigma\). Solving the equation
\((\omega_{2^i}+\sigma q^{2^i})^p\bmod q^{2^{i+1}} = 1\) provides a
formula for \(\sigma\).

\begin{algorithm}[PrincipalRootsOfUnity]

\begin{description}
\item[Inputs:]
Prime number \(p\)

Bound \(\gamma\)
\item[Outputs:]
Prime number \(q \leq p^{6}\) such that \(p\) divides \(q-1\)

Integer \(k\) such that \(q^k \geq \gamma\)

Principal \(p\)th root of unity \(\omega\in \mathbb{F}_q\)

Principal \(p\)th root of unity \(\omega_k\in \mathbb{Z}/q^k\mathbb{Z}\)
such that \(q^k \geq \gamma\)
\end{description}

\begin{enumerate}

\item

~~Sample odd integers \(q = a\cdot p+1\) for \(t\in[p^{5}]\) until \(q\)
is prime

\item

~~Sample \(\omega = \zeta^t\) for \(\zeta\in\mathbb{F}_q^\times\) until
\(\omega \neq 1\)

\item

~~\(k \gets \lceil\log_q \gamma\rceil\); \(\omega_k\gets \omega\)

\item

~~for \(i=0\) to \(\lfloor\log k\rfloor\):

\item

~~~~~~\(\sigma \gets (1-\omega_k^p)\bmod q^{2^{i+1}} / q^{2^i}\)

\item

~~~~~~\(\sigma \gets \sigma\cdot\omega_k\cdot p^{-1}\bmod q^{2^i}\)

\item

~~~~~~\(\omega_k \gets \omega_k + \sigma\cdot q^{2^i}\)

\item

~~return \((q,k,\omega,\omega_k\bmod q^k)\)

\end{enumerate}\gobblepar

\end{algorithm}

\begin{lemma}

\hyperref[algorithm:PrincipalRootsOfUnity]{Algorithm~\ref*{algorithm:PrincipalRootsOfUnity}} (\textsc{PrincipalRootsOfUnity})
is correct with high probability and uses
\(O^{\widetilde{}}(\log \gamma\log p) + \log(p)^{O(1)}\) bit
operations.\gobblepar

\end{lemma}

The exact complexity of the term \(\log(p)^{O(1)}\) and the exact
failure probability depend on the algorithm used to test the primality
of \(q\). The algorithm is either Las Vegas using a (slower)
deterministic test~\autocite{AgrawalKayalSaxena2004} or Monte Carlo
using a (faster) randomized test~\autocite{Rabin1980}.

Altogether, we can describe an algorithm which computes both \(f^{[p]}\)
and \((xf')^{[p]}\) for some prime \(p\).

\begin{algorithm}[SparseFoldedInterpolation]

\begin{description}
\item[Inputs:]
Black box for \(f\in \mathbb{Z}[x]\)

Bounds \(\delta > \deg f\), \(\tau \geq \sp f\) and
\(\gamma \geq \height f\)

Prime number \(p\)
\item[Outputs:]
The sparse representations of \(f^{[p]}\) and \((x\cdot f)^{[p]}\)
\end{description}

\begin{enumerate}

\item[]

\emph{Set up the rings}
\comment{\(\textsc{PrincipalRootsOfUnity}(p,2\gamma)\)}

\item

\((\omega,\omega_{2k}) \gets\) \(p\)th PRUs in \(\mathbb{F}_q\) and
\(\mathbb{Z}/q^{2k}\mathbb{Z}\) resp., where \(q^k \geq 2\gamma\)

\item[]

\item[]

\emph{Compute the support of \(f^{[p]}\) modulo \(q\)}

\item

\(\vec u \gets (f(\omega^i))_{0\leq i<2\tau}\) \comment{\(2\tau\) calls
to the black box}

\item

\((\vec \nu, \vec \sigma) \gets \textsc{Prony}(\vec u)\)
\comment{\(u_j = \sum_i \nu_i \sigma_i^j\)}

\item

for \(i=0\) to \(t-1\): \(a_i \gets \log_\omega \sigma_j\)
\comment{support of \(f^{[p]}\)}

\item[]

\item[]

\emph{Compute the coefficients of \(f^{[p]}\) and \((xf')^{[p]}\) modulo
\(q^k\)}

\item

\(\vec v \gets (f(\omega_{2k}^i))_{0\leq i<t}\) \comment{\(t\) calls to
the black box}

\item

\(\vec w \gets (f((1+q^k)\omega_{2k}^i))_{0\leq i<t}\) \comment{\(t\)
calls to the black box}

\item

\((\rho_0,\dotsc,\rho_{t-1}) \gets (\omega_{2k}^{a_0}, \dotsc, \omega_{2k}^{a_{t-1}})\)

\item

\((\lambda_0, \dotsc, \lambda_{t-1}) \gets V_t^\transp(\vec \rho)^{-1}\cdot\vv\)
\comment{transposed Vandermonde syst. solv.}

\item

\((\mu_0, \dotsc, \mu_{t-1}) \gets V_t^\transp(\vec \rho)^{-1}\cdot[(\vw-\vv)/q^k]\)

\item

Return \(f^{[p]} = \sum_{i=0}^{t-1} \lambda_i x^{a_i}\) and
\((x\cdot f')^{[p]} = \sum_{i=0}^{t-1} \mu_i x^{a_i}\)

\end{enumerate}\gobblepar

\end{algorithm}

This algorithm can replace to computation of \(f^{[p]}\) and
\((xf')^{[p]}\) in
\hyperref[algorithm:SparseInterpolationSLP]{Algorithm~\ref*{algorithm:SparseInterpolationSLP}} (\textsc{SparseInterpolationSLP})
(Lines \ref{SpIntSLP:fp} and \ref{SpIntSLP:xfp}). As a result, we obtain
a quasi-linear time sparse interpolation algorithm over \(\mathbb{Z}\).

\begin{theorem}[\autocite{GiorgiGrenetPerretduCrayRoche2022}]\label{theorem:SparseInterpolationZZ}

There is an algorithm \(\textsc{SparseInterpolationOver}\mathbb{Z}\)
that, given a modular black box for a polynomial \(f\in \mathbb{Z}[x]\),
bounds \(\delta\geq\deg(f)\), \(\tau\geq\#f\) and
\(\gamma\geq\height f\), and \(0<\varepsilon<1\), returns the sparse
representation of \(f\) with probability at least \(1-\varepsilon\)
using \(O(\tau)\) calls to the black box and
\(O^{\widetilde{}}(\tau(\log \delta+\log \gamma)\log\frac{1}\varepsilon)\)
bit operations.

If the input is an SLP of size \(\ell\), the calls to the black box have
total bit cost
\(O^{\widetilde{}}(\ell\tau(\log \delta+\log \gamma))\).\gobblepar

\end{theorem}

This algorithm is of Monte Carlo type. To get a Las Vegas algorithm, one
would need either a deterministic or a Las Vegas algorithm to verify if
a modular black box (resp. an SLP) represents a given sparse polynomial.
Such a deterministic algorithm is known, albeit not of quasi-linear
complexity~\autocite{BlaserHardtLiptonVishnoi2009}. The existence of a
quasi-linear Las Vegas algorithm is an open question.

As mentioned in
\hyperref[section:multivariate]{Section~\ref*{section:multivariate}}, it
is possible to turn a univariate algorithm into a multivariate one
through Kronecker substitution. For \(n\)-variate polynomials, the
resulting complexity is
\(O(\tau(n\log \delta+\log\gamma)\log\frac{1}\varepsilon)\).

Another question is to get rid of the bound \(\tau\) on the sparsity of
\(f\), by means of early termination as presented in
\hyperref[section:SparseInterpolationBB]{Section~\ref*{section:SparseInterpolationBB}}.
The difficulty for polynomials over \(\mathbb{Z}\) is that this requires
to work modulo some large prime number \(p\) of bit size
\(\Theta(\log \delta+\log\log \gamma)\). Producing a \(b\)-bit prime
number has cost \(\Omega(b^{3})\) which is far from quasi-linear in
\(b\)~\autocite{Shoup2008}. Therefore, we developed a technique of using
\emph{random primes without primality
testing}~\autocite{GiorgiGrenetPerretduCrayRoche2022a}. The idea is to
replace a random prime \(p\) with a random integer \(m\) of slightly
larger size. With good probability, \(m\) has a prime factor
\(\geq \sqrt m\). We show that given an algorithm designed to work
modulo a randomly chosen prime number, we can make it work modulo \(m\)
instead. The idea is to precede any inversion modulo \(p\) or zero-test
by a \textsc{gcd} computation with the current modulus \(m\), and update
if a common factor is found. The subtlety lies in the capacity for this
transformation to preserve the success probability when the original
algorithm is able to sample random elements modulo \(p\). Applying our
program transformation technique to early termination in sparse
interpolation algorithms provide the following result.

\begin{theorem}[\autocite{GiorgiGrenetPerretduCrayRoche2022a}]\label{theorem:SparseInterpolationZZnobound}

There is an algorithm \(\textsc{SparsityOver}\mathbb{Z}\) that, given a
modular black box for a polynomial \(f\in\mathbb{Z}[x]\), bounds
\(\delta\geq\deg(f)\) and \(\gamma \geq \height f\), and
\(0 < \varepsilon < 1\), returns an integer \(\tau \leq \#f\) that
equals \(\#f\) with probability at least \(1-\varepsilon\), using
\(O(\tau\log \frac{1}\varepsilon)\) calls to the black box and
\(O^{\widetilde{}}(\tau(\log \delta+\log\log \gamma)\log\frac{1}\varepsilon)\)
bit operations.\gobblepar

\end{theorem}

We remark that this sparsity estimation algorithm is one-sided error in
the sense that the approximation it returns is always at most \(\#f\).

\section{The unbalanced case}\label{Section:the-unbalanced-case}

The complexity of Algorithm
\(\textsc{SparseInterpolationOver}\mathbb{Z}\) given in
\hyperref[theorem:SparseInterpolationZZ]{Theorem~\ref*{theorem:SparseInterpolationZZ}}
is quasi-linear in the standard parameters \(\tau \geq \sp f\),
\(\delta \geq \deg f\) and \(\gamma \geq \height f\). But if
\(s = \bitsize f\) denotes the bit size of \(f\), the bound
\(s \leq \tau(\log \delta + \log \gamma)\) may be loose. If there is for
instance one very large coefficient with \(\Theta(\log \gamma)\) bits
and \(O(\tau)\) coefficients of \(O(1)\) bits,
\(s  = O(\tau\log \delta + \log \gamma)\). The term \(\tau\log \gamma\)
in the complexity of \(\textsc{SparseInterpolationOver}\mathbb{Z}\) can
therefore be quadratic in the bit size \(s\). The same remark applies on
the size of the exponents. There may be one term of degree \(\delta\)
and all the other exponents of degree \(O(\tau)\). In this case, the
representation of the support of \(f\) has size
\(O(\log \delta + \tau\log \tau)\). Combining both remarks, the size of
the representation can be as small as
\(O(\tau\log \tau + \log \delta+\log \gamma)\).

In this part, we investigate the first situation, that is when the
coefficient sizes are very unbalanced. Another work has investigated the
second situation when the exponent sizes are very
unbalanced~\autocite{vanderHoevenLecerf2025}. It remains an open problem
to combine both techniques to handle unbalancedness in both the
coefficients and the exponents.

To manage unbalanced coefficients, there are two natural approaches:
either a \emph{bottom-up} approach that recovers first the many terms
with small coefficients and then the few remaining ones that have large
coefficients; or a \emph{top-down} approach that first recovers the few
terms with large coefficients and then the many terms with small
coefficients.

\subsection{Bottom-up approach}\label{Section:bottom-up-approach}

The first approach starts by interpolating \(f_{\smash *} = f\bmod m\)
for some small \(m\). The many coefficients \(<m\) in \(f\) are
preserved in \(f_{\smash *}\), whence the sparsity of \(f-f_{\smash *}\)
is much less than the sparsity of \(f\). The process can be iterated to
interpolate \(f-f_{\smash *}\) modulo a larger integer \(m'\). Once
\(m > 2\height f\), \(f = f\bmod m\).

The heart of the algorithm is interpolating \(f-f_{\smash *}\) modulo
some integer \(m\) where \(f\) is given by a modular black box, and
\(f_{\smash *}\) as a sparse polynomial. The idea is to reuse the same
strategy as in the balanced case. The difference is in the computation
\(f^{[p]}\) and \((xf')^{[p]}\) using
\hyperref[algorithm:SparseFoldedInterpolation]{Algorithm~\ref*{algorithm:SparseFoldedInterpolation}} (\textsc{SparseFoldedInterpolation})
since we only need these polynomials modulo \(m\). We can simply give
the bound \(\gamma=m\) as input to the algorithm. It will return
\(f^{[p]}\bmod q^k\) and \((xf')^{[p]}\bmod q^k\) for some
\(q^k \geq 2\gamma\) even if \(\gamma < \height f\).

The problem of this approach is its complexity. From the beginning, the
sparsity of \(f_{\smash *}\) is of the same order as the sparsity of
\(f\). During the final step, the modulus must satisfy
\(m = \Theta(\height f)\) to recover \(f\) exactly. This means that
\(f_{\smash *}\) must be evaluated on a large input. The cost is
\(\Omega(t\log d\log\height f)\) bit operations where
\(d=  \deg(f_{\smash *})\). The term \(t\log \height f\) is exactly the
one we try to avoid. This is why we turn to the top-down approach.
Still, this method is adapted to unbalanced
exponents~\autocite{vanderHoevenLecerf2025}. The difficulty to combine
both kinds of unbalancedness precisely lies in the fact that the
unbalancedness of exponents and coefficients seem to need two opposite
approaches.

\subsection{Top-down approach}\label{Section:top-down-approach}

We turn to the second approach. The difficulty here is to only
interpolate the \emph{large} terms, that is with large coefficients. The
solution is to treat the other terms as \emph{noise}. Let us fix some
bound \(\beta\) and consider the polynomial \(f_{\smash\star}\) made of
the terms of \(f\) whose coefficients are larger than \(\beta\) (in
absolute value). Then the fold \(f_{\smash\star}^{[p]}\) is close to
\(f^{[p]}\) in the sense that \(f^{[p]}_{\smash\star}-f^{[p]}\) has only
small coefficients. If a large term \(cx^e\) does not collide modulo
\(p\), it can be reconstructed from the two corresponding terms
\(cx^{e\bmod p}\) and \(ce x^{e\bmod p}\) in \(f^{[p]}\) and
\((xf')^{[p]}\) as in
\hyperref[algorithm:TentativeTerms]{Algorithm~\ref*{algorithm:TentativeTerms}} (\textsc{TentativeTerms}).
If it collides with small terms, the corresponding terms in \(f^{[p]}\)
and \((xf')^{[p]}\) are \(\tilde cx^{e\bmod p}\) and
\(\tilde c'x^{e\bmod p}\) where \(\tilde c \approx c\) and
\(\tilde c' \approx ce\). As long as the noise is small enough, the
exponent \(e\) can still be recovered as the closest integer
\(\lfloor\tilde c'/\tilde c\rceil\) to \(\tilde c'/\tilde c\). This way,
we reconstruct an approximation \(\tilde cx^e\) of \(cx^e\). If all
large terms can be approximated this way, we get an approximation
\(f_{\smash *}\) of \(f_{\smash\star}\), such that
\(f_{\smash *}-f_{\smash\star}\) has only small coefficients. As a
result, since
\(f-f_{\smash *} = (f-f_{\smash\star}) - (f_{\smash *}-f_{\smash\star})\)
is a difference of two polynomials with small coefficients, it has
itself only small coefficients. Iterating this process with
\(f-f_{\smash *}\) provides better and better approximations
\(f_{\smash *}\) of \(f\).

An iteration as presented above should compute approximations of all and
only the largest terms. Since subsequent iterations only compute terms
with smaller coefficients, a missing or surplus term with large
coefficient cannot be corrected. There are two pitfalls. The first one
is missing a large term. This happens if a large term collides with many
small terms in \(f^{[p]}\) and if the resulting coefficient is too
small. To avoid this, we define a \emph{buffer zone} of \emph{medium}
coefficients such that a sum of a large coefficient and many small ones
remains larger than a sum of small coefficients. Ideally, we would
choose a large \(p\) to avoid any collisions between medium or large
terms, so that large terms only collide with small terms. But for
efficiency reasons, the size of \(p\) we can afford only ensures that a
constant fraction of the large terms only collide with small terms.
Using several values for \(p\) provides a tentative list of large terms,
that contains all actual large terms with high probability. The second
pitfall is that this list may contain spurious terms. Instead of
avoiding this situation, we only use this list to compute a superset of
the exponents that serves as a filter afterwards. We redo the same
process of computing large terms from several folds \(f^{[p]}\). This
time, we only reconstruct large terms of \(f\) that did not collide with
any other exponent in the superset modulo \(p\). Since there can still
be collisions with medium or small terms, a large term may become too
small because of collisions. We add another \emph{buffer zone} inside
the large terms. We distinguish between the mere \emph{large} terms and
the \emph{huge} terms. In one iteration, only the huge ones are
computed.

\begin{definition}

Let \(f\in \mathbb{Z}[x]\) and \(\gamma \geq \height f\). A term
\(cx^e\) of \(f\) is said

\begin{itemize}
\item
  \emph{small} if \(|c| < \gamma^\frac{1}{6}\),
\item
  \emph{medium} if \(|c| \geq \gamma^\frac{1}{6}\),
\item
  \emph{large} if \(|c| \geq \frac{1}{2} \gamma^{\frac{13}{30}}\), and
\item
  \emph{huge} if \(|c| \geq \gamma^\frac{1}{2}\).
\end{itemize}

\end{definition}

Note that every \emph{huge} term is also considered \emph{large}, and
every \emph{huge} or \emph{large} term is also considered \emph{medium}.
The algorithm can be summarized as follows. It makes use of
\hyperref[algorithm:SparseFoldedInterpolation]{Algorithm~\ref*{algorithm:SparseFoldedInterpolation}} (\textsc{SparseFoldedInterpolation}).

\begin{algorithm}[LargeSupportSuperset]

\begin{description}
\item[Inputs:]
Modular black box for \(f\)

Bounds \(s \geq \bitsize f\), \(\delta\geq\deg f\),
\(\gamma\geq\height f\)
\item[Output:]
Set \(\mathcal{T}\subset \mathbb{Z}_{\geq 0}\) that contains the
exponents of all large terms of \(f\)
\item[Constants:]
\(k = \lceil 2\log s\rceil+3\); \(N = \frac{60}{13} s\log_\gamma(4s)\);
\(\lambda = \max(21,18s\log_\gamma\delta)\);
\(m =4\gamma^{\frac{7}{6}}\)
\end{description}

\begin{enumerate}

\item

\(\mathcal{T} \gets \emptyset\)

\item

repeat \(k\) times:

\item

~~~~\(p \gets\) random prime in \([\lambda,2\lambda]\)
\comment{\(\textsc{SparseFoldedInterpolation}\)}

\item

~~~~\((g, h)\gets(f^{[p]},(xf')^{[p]})\) \comment{with \(\tau=p+1\) and
\(\gamma=m\)}

\item

~~~~for each pair \((ax^e, bx^e)\) of terms of \(g\) and \(h\):

\item

~~~~~~~~if \(ax^e\) is \emph{large} and
\(0 \leq \lfloor b/a\rceil \leq \delta\): add \(\lfloor b/a\rceil\) to
\(\mathcal{T}\) \comment{\(|a|\geq\frac{1}{2}\gamma^{\frac{13}{30}}\)}

\item

~~~~~~~~if \(\#\mathcal{T} > N\): return \textsc{failure}

\item

return \(\mathcal{T}\)

\end{enumerate}\gobblepar

\end{algorithm}

Given \(\mathcal{T}\), we can now compute all the huge coefficients of
\(f\). We use \(f^{[p]}\) for \(p = O(\#\mathcal{T}\log \delta)\) to
mostly avoid collisions between elements of \(\mathcal{T}\). Since they
still occur, \(\mathcal{T}\) is used to detect them and discard the
spurious terms they create. With \(\Theta(\log s)\) repetitions, each
exponent of a large term is non-colliding modulo \(p\) for at least one
\(p\) with high probability. If a huge term only collides with non-large
terms, it remains large enough (that is
\(\geq \frac{1}{2}\gamma^\frac{1}{2}\)) not to be missed.

\begin{algorithm}[HugeCoefficients]

\begin{description}
\item[Inputs:]
Modular black box \(\mathcal{B}\) for \(f\)

Bounds \(s \geq \bitsize f\), \(\delta\geq\deg f\),
\(\gamma\geq\height f\)
\item[Output:]
\(f_{\smash *}\in \mathbb{Z}[x]\) such that
\(\height(f-f_{\smash *}) \leq \sqrt\gamma\)
\item[Constants:]
\(k = \lceil 2\log s\rceil+3\);
\(\lambda = \max(21,3\#\mathcal{T}\log\delta)\); \(m=2\gamma\)
\end{description}

\begin{enumerate}

\item

\(\mathcal{T} \gets \textsc{LargeSupportSuperset}(\mathcal{B}, s, \delta, \gamma)\)

\item

\(f_{\smash *} \gets 0\)

\item

repeat \(k\) times:

\item

~~~~\(p \gets\) random prime in \([\lambda,2\lambda]\)

\item

~~~~\(g \gets f^{[p]} - f_{\smash *}^{[p]}\)
\comment{\textsc{SparseInterpolationFolded} with \(\tau=p\) and
\(\gamma=m\)}

\item

~~~~\(\mathcal{E} \gets\) array of \(p\) sets such that
\(\mathcal{E}_{[i]} = \{e \in \mathcal{T}: e\bmod p = i\}\)

\item

~~~~for each huge term \(cx^i\) of \(g\):
\comment{\(|c|\geq\frac{1}{2}\gamma^\frac{1}{2}\)}

\item

~~~~~~~~if \(\mathcal{E}_{[i]} = \{e\}\) and \(f_{\smash *}\) has no
term of degree \(e\): \comment{\(e\) does not collide mod \(p\)}

\item

~~~~~~~~~~\(f_{\smash *} \gets f_{\smash *} +cx^e\)

\item

~~~~~~~~~~if \(\bitsize(f_{\smash *}) > s\): return \textsc{failure}

\item

return \(f_{\smash *}\)

\end{enumerate}\gobblepar

\end{algorithm}

After a call to \textsc{HugeCoefficients}, we get an approximation
\(f_{\smash *}\) of \(f\) such that \(\height(f-f*) \leq \sqrt\gamma\).
We can update \(\gamma\) to its square root, until we reach a small
enough value that guarantees that the remaining polynomial
\(f-f_{\smash *}\) is actually balanced, and finish the computation with
\(\textsc{SparseInterpolationOver}\mathbb{Z}\)
(\hyperref[theorem:SparseInterpolationZZ]{Theorem~\ref*{theorem:SparseInterpolationZZ}}).
Since the algorithm may fail to return a correct answer, and since we do
not have a good verification algorithm in the unbalanced case, we repeat
several times the algorithm and use a majority vote. Finally, a bound on
the total bit size of \(f\) is enough to derive bounds on its sparsity
and height.

\begin{theorem}[\autocite{GiorgiGrenetPerretduCrayRoche2024}]\label{theorem:UnbalancedInterpolation}

There is an algorithm \textsc{UnbalancedInterpolation} that, given a
modular black box for \(f\in\mathbb{Z}[x]\) and bounds
\(s \geq \bitsize(f)\) and \(\delta \geq \deg(f)\), returns the sparse
representation of \(f\) with probability at least \(1-1/s\), using
\(O^{\widetilde{}}(s)\) calls to the modular black box and
\(O^{\widetilde{}}(s\log \delta)\) bit operations.\gobblepar

\end{theorem}

When implemented by an SLP, the calls to the modular black box have
total cost \(O^{\widetilde{}}(\ell s\log \delta)\). As mentioned
earlier, the complexity of \textsc{UnbalancedInterpolation} is not
quasi-linear in the output size due to the extra factor \(\log\delta\).
Removing it from the complexity is a challenging open problem.

\begin{openproblem}

Given a modular black box for \(f\in\mathbb{Z}[x]\) and a bound
\(s \geq \bitsize(f)\), is it possible to compute the sparse
representation of \(f\) with \(O(s)\) calls to the black box and
\(O^{\widetilde{}}(s)\) bit operations?\gobblepar

\end{openproblem}

\chapter{Sparse polynomial arithmetic}\label{chapter:sparsearithmetic}

In this chapter, we present algorithms for sparse polynomials. The case
of addition and subtraction is rather simple and the naive algorithm is
already optimal. The situation becomes different for multiplication and
division. In both cases, one specific difficulty of sparse polynomial
arithmetic is the output sensitivity. For instance, the product of two
polynomials of sparsity at most \(t\) may have between \(2\) and
\(t^{2}\) nonzero terms.

The strategy to compute a sparse polynomial product or quotient while
the size of the output is initially unknown is to \emph{guess} an output
size, perform the computation assuming the correctness of the size, and
\emph{a posteriori} check whether the result is correct. For the
strategy to be efficient, the verification must be extremely fast. The
problem at hand is, given three polynomials \(f\), \(g\) and \(h\), to
determine whether \(h = f\times g\).

We present in \hyperref[section:verif]{Section~\ref*{section:verif}} our
verification algorithms.
\hyperref[section:sparsemul]{Section~\ref*{section:sparsemul}} is about
multiplication and
\hyperref[section:sparsediv]{Section~\ref*{section:sparsediv}} about
division and divisibility testing. Finally,
\hyperref[section:sparsefact]{Section~\ref*{section:sparsefact}} is
independent from the preceding ones and focuses on computing low-degree
factors of sparse polynomials.

\section{Polynomial product verification}\label{section:verif}

The classical strategy to verify a polynomial product relies on
evaluation. Given \(f\), \(g\) and \(h\in \mathsf{R}[x]\), the equality
\(h = f\times g\) is tested by evaluation
\(h(\alpha) = f(\alpha)\times g(\alpha)\) where \(\alpha\) is chosen at
random in \(\mathsf{R}\). Beyond other issues such as the size of
\(\mathsf{R}\), this strategy does not work for high-degree sparse
polynomials since the evaluation itself is too costly. Since the
evaluation of a polynomial on \(\alpha\) corresponds to the computation
of the remainder of this polynomial modulo \(x-\alpha\), the equality
\(h(\alpha) = f(\alpha)\times g(\alpha)\) corresponds to
\(h\bmod x-\alpha = (f\times g)\bmod x-\alpha\). A generalization,
introduced by Kaminski~\autocite{Kaminski1989}, is to test
\(h\bmod \ell = f\times g\bmod \ell\) for some polynomial \(\ell\).
Kaminski uses polynomials \(\ell\) of the form \(x^i-1\). The reduction
\(h\bmod \ell\) is computed in linear time. Then, the modular equality
is verified by computing the product
\((f\bmod \ell)\times(g\bmod \ell)\) and then its reduction. Using a
small enough value for \(i\), this product can be computed in sublinear
time (in the input sizes).

To check a sparse polynomial product, we combine both methods. First we
use Kaminski's framework but with prime exponents, that is we test
whether \(h\bmod x^p-1 = f\times g\bmod x^p-1\) for some prime \(p\).
This allows us to consider a much smaller value \(p\) than with
Kaminski's approach. Furthermore, this equality is not verified by
performing a polynomial multiplication, but instead by relying on
evaluation. That is, we verify whether
\((h\bmod x^p-1)(\alpha) = (f\times g\bmod x^p-1)(\alpha)\). This
requires to be able to evaluate the right-hand side without computing
the modular product.

This approach, that we developed for sparse polynomial product
verification, is actually more general. It allows us to check modular
equalities such as \(h = f\times g\bmod \ell\) for a polynomial
\(\ell\), both when the inputs are in dense and in sparse
representation. We first present our method for the evaluation of a
modular product, and then its application to modular product
verification, and finally to (standard) product verification.

\subsection{Modular product
evaluation}\label{Section:modular-product-evaluation}

We first consider the evaluation of the (implicitly known) polynomial
\(fg\bmod x^p-1\) on \(\alpha\in \mathsf{R}\). Let us assume that \(f\)
and \(g\) have been reduced modulo \(x^p-1\), hence have size \(p\). The
idea here is to rely on the linear-algebraic interpretation of
polynomial arithmetic. The modular product \(fg\bmod x^p-1\) corresponds
to the matrix-vector product \(C_f\cdot\vg\) where
\[ C_f = \begin{bmatrix}
    f_0 & f_{p-1} & \dots & f_1 \\
    f_1 & f_0 & \dots & f_2\\
    \vdots & \vdots & & \vdots\\
    f_{p-1} & f_{p-2} & \dots & f_0
   \end{bmatrix}
\] is a circulant matrix. The evaluation of a size-\(p\) polynomial on
\(\alpha\) is given by the inner product with
\(\vec \alpha_n = (1, \alpha, \dotsc, \alpha^{n-1})\). Therefore, the
evaluation of \(fg\bmod x^p-1\) on \(\alpha\) is given by
\(\vec \alpha_n\cdot\left(C_f \cdot\vg\right)\). To perform the
evaluation without computing the polynomial \(fg\bmod x^p-1\), the
solution is to parenthesize the previous equation on the left:
\(\left(\vec \alpha_n\cdot C_f\right)\cdot\vg\).

By taking into account the circulant structure of \(C_f\) and the fact
that \(\vec \alpha_n\) is a geometric progression,
\(\vec \alpha_n\cdot C_f\) can be computed in linear time, as was
already noticed by Giorgi~\autocite{Giorgi2018}. This approach can be
extended to the case of sparse polynomial by leveraging the double
structure, circulant and sparse.

\begin{lemma}

Let \(f\), \(g\in \mathsf{R}[x]\) of size \(p\) and
\(\alpha\in \mathsf{R}\). The polynomial \(f\times g\bmod x^p-1\) can be
evaluated on \(\alpha\) in

\begin{itemize}
\item
  \(O(p)\) operations in \(\mathsf{R}\), or
\item
  \(O((\#f+\#g)\log p)\) operations in \(\mathsf{R}\) and \(O(\#f+\#g)\)
  operations on the exponents.
\end{itemize}

\end{lemma}

\begin{remark}

In the second complexity bound, the term \(O((\#f+\#g)\log p)\) can
actually be improved to \(\log p + O((\#f+\#g)\log p/\log\log p)\),
using Yao's simultaneous exponentiation algorithm~\autocite{Yao1976}.
The same remark holds for all the complexity bounds given in this
part.\gobblepar

\end{remark}

These results can be extended to any sparse monic modulus \(\ell\). The
idea is still to express \((fg\bmod \ell)(\alpha)\) as
\((\vec \alpha_n\cdot M_{f\bmod \ell})\cdot\vg\) where
\(M_{f\bmod \ell}\) is the matrix of the multiplication by \(f\) modulo
\(\ell\). This matrix is not circulant anymore but keeps some structure
which allows for a fast vector-matrix product.

If \(f\) and \(g\) are given in sparse representation, the complexity of
the evaluation of \(fg\bmod \ell\) depends on the ratio between the
degree and the \emph{second degree} of the monic polynomial \(\ell\),
defined as \(\deg_2(\ell) = \deg(\ell-x^d)\) where \(d = \deg(\ell)\).

\notation[second degree $\deg(f-x^{\deg f})$ of a monic polynomial\nomrefpage]{\(\deg_2(f)\)}(spps)

\begin{lemma}

Let \(f\), \(g\), \(\ell\in \mathsf{R}[x]\), \(\ell\) monic, and
\(\alpha\in \mathsf{R}\). Let \(d = \deg(\ell)\), \(d_2 = \deg_2(\ell)\)
and \(\gamma = \lceil d_2/(d-d_2)\rceil\). Assume that \(\deg(f)\),
\(\deg(g)<d\), and \(\#f \leq \#g\). Then \(f\times g\bmod \ell\) can be
evaluated on \(\alpha\) in

\begin{itemize}
\item
  \(O(\M(d))\) operations in \(\mathsf{R}\), or
\item
  \(O(d\cdot\#\ell)\) operations in \(\mathsf{R}\), or
\item
  \(O((\#f\#\ell^\gamma+\#g)\log d)\) operations in \(\mathsf{R}\).
\end{itemize}

\end{lemma}

\begin{remark}

The complexity \(O(\M(d))\) comes from computing \(fg\bmod \ell\) before
the evaluation. The complexity \(O(d\#\ell)\) becomes smaller as soon as
\(\#\ell < \M(d)/d\). One can think of the second complexity as adapted
for a modulus \(\ell\) with a constant number of terms and dense
polynomials \(f\) and \(g\). The third complexity is the case of sparse
polynomials \(f\) and \(g\). Note that for \(\deg_2(\ell) \leq d/2\),
\(\gamma \leq 1\) and the complexity simplifies to
\(O((\#f\#\ell+\#g)\log d)\).\gobblepar

\end{remark}

\begin{remark}

The evaluation of a polynomial on a random point is the basis for
polynomial product verification. If the base ring is too small, it is
customary to select a random point in a suitable extension of the base
ring. One can refine our complexity estimates to split between
operations in the base ring and operations in the
extension~\autocite{GiorgiGrenetPerretduCray2023}.\gobblepar

\end{remark}

\subsection{Modular product
verification}\label{Section:modular-product-verification}

The algorithm for modular product verification is directly based on
modular product evaluation. The idea is simply to choose a random point
\(\alpha\) in a large enough set, and perform the modular product
evaluation. We obtain the following result.

\begin{theorem}[\autocite{GiorgiGrenetPerretduCray2023}]\label{theorem:ModProdVerif}

Let \(f\), \(g\), \(h\), \(\ell\in \mathsf{R}[x]\), \(\ell\) monic of
degree \(d\), \(f\), \(g\), \(h\) of degrees \(< d\), and
\(0 < \varepsilon < 1\). Assume that
\(\#\mathsf{R} >\frac{1}\varepsilon(d-1)\). One can test whether
\(h = f\times g\bmod \ell\) in

\begin{itemize}
\item
  \(O(\M(d))\) operations in \(\mathsf{R}\), or
\item
  \(O(d\cdot\#\ell)\) operations in \(\mathsf{R}\), or
\item
  \(O((\#f\#\ell^\gamma+\#g+\#h)\log d)\) operations in \(\mathsf{R}\),
\end{itemize}

where \(\gamma = \lceil\deg_2(\ell)/(d-\deg_2(\ell))\rceil\), with a
failure probability at most \(\varepsilon\) when
\(h \neq fg\bmod \ell\).\gobblepar

\end{theorem}

This general result can be specialized to some rings of interests, such
as the integers or finite fields. For polynomials over the integers, the
idea is to perform the computation modulo some prime number, to avoid
coefficient swell.

\begin{corollary}

Let \(f\), \(g\), \(h\), \(\ell\) be as in
\hyperref[theorem:ModProdVerif]{Theorem~\ref*{theorem:ModProdVerif}},
with \(\mathsf{R} = \mathbb{Z}\). Let
\(H = \max(\height f, \height g, \height h, \height \ell)\),
\(t = \max(\#f, \#g, \#h)\) and
\(s = \log(\frac{1}\varepsilon d\log H)\). The algorithm of
\hyperref[theorem:ModProdVerif]{Theorem~\ref*{theorem:ModProdVerif}}
uses \(O^{\widetilde{}}(\log^{3} s)\) operations on \(s\)-bit integers
to get a prime number and

\begin{itemize}
\item
  \(O(d(\#\ell + \log_s H))\) operations on \(s\)-bit integers, or
\item
  \(O(\#f\#\ell^\gamma\log d + (t+\#\ell) \log_s H)\) operations on
  \(s\)-bit integers.
\end{itemize}

\end{corollary}

For polynomials over finite fields, the problem is to have enough points
in the finite field to ensure a good success probability. In the general
case, the idea is to go to a field extension. Note that in the next
statement, the value \(\delta\) is actually \(< 1\) for large fields.
This is the case where no field extension is required.

\begin{corollary}

Let \(f\), \(g\), \(h\), \(\ell\) be as in
\hyperref[theorem:ModProdVerif]{Theorem~\ref*{theorem:ModProdVerif}},
with \(\mathsf{R} = \mathbb{F}_q\). Let \(t = \max(\#f, \#g, \#h)\) and
\(\delta = \log_q(\frac{1}\varepsilon d)\). The algorithm of
\hyperref[theorem:ModProdVerif]{Theorem~\ref*{theorem:ModProdVerif}}
uses \(O^{\widetilde{}}(\delta^{3}\log q)\) operations in
\(\mathbb{F}_q\) to get a degree-\(\delta\) irreducible polynomial, and

\begin{itemize}
\item
  \(O(d(\#\ell +  \M(\delta)))\) operations in \(\mathbb{F}_q\), or
\item
  \(O(\#f\#\ell^\gamma\M(\delta)\log d)\) operations in
  \(\mathbb{F}_q\).
\end{itemize}

\end{corollary}

To improve the complexity for small finite fields in the dense case, one
can actually do without field extension. We reuse the idea of replacing
an evaluation on \(\alpha\) by a reduction modulo a random polynomial.
The goal becomes to test if a random polynomial \(r\) divides
\(h-fg\bmod \ell\). We can once again rely on evaluation, but now the
polynomials must be evaluated on a matrix. Indeed, a polynomial \(r\) is
always minimal polynomial of its companion matrix \(C_r\). This means
that for any polynomial \(b\), \(r\) divides \(b\) if and only if
\(b(C_r) = 0\). To get a fast algorithm, one cannot directly evaluate
\(h\) and \(fg\bmod \ell\) on \(C_r\) since this requires to perform
some matrix multiplications in dimensions \(k\) where \(k\) is the size
of \(r\). Yet \(h(C_r)\) and \((fg\bmod \ell)(C_r)\) are two (implicit)
matrices whose equality must be checked. The standard solution is
Freivalds' algorithm~\autocite{Freivalds1979}: Take a random vector
\(\vv\in\{0,1\}^k\) and test whether
\(\vv\cdot h(C_r) = \vv\cdot(fg\bmod \ell)(C_r)\). Using the same
algorithm as before to evaluate \(fg\bmod \ell\) on \(C_r\), the final
algorithm checks the equality
\[\vv\cdot\vec C_r\cdot\vh = \vv\cdot\vec C_r\cdot M_{f\bmod \ell}\cdot\vg.\]
For dense polynomials \(f\) and \(g\) and small fields, this technique
improves the complexity by replacing the term \(\M(\delta)\) by
\(\delta\) by using a random irreducible polynomial \(r\). The same
technique can also be used for another goal. When using field
extensions, the modular product is verified using an algorithm that
itself performs some modular products (for the computations in the
finite field extension). Although this is not a problem in theory, it
may be seen as unsatisfactory. The above algorithm does not rely on
finite fields extensions but still performs some polynomial arithmetic
to produce an irreducible polynomial \(r\). To get rid of this
dependency, we take for \(r\) any random polynomial without any
irreducibility test. The same algorithm works as well albeit with worse
success probability, and must therefore be repeated. The evaluation on a
random companion matrix provides the following results.

\begin{theorem}[\autocite{GiorgiGrenetPerretduCray2023}]

Let \(f\), \(g\), \(h\), \(\ell\in \mathbb{F}_q[x]\), \(\deg(f)\),
\(\deg(g)\), \(\deg(h) \leq \deg(\ell) = d\), \(0 < \varepsilon < 1\)
and \(\delta = \log_q(\frac{1}\varepsilon d)\). We can check whether
\(h = fg\bmod \ell\) in

\begin{itemize}
\item
  \(O(d(\#\ell+\delta))\) operations in \(\mathbb{F}_q\), or
\item
  \(O((\#\ell+\delta^{2})d)\) operations in \(\mathbb{F}_q\) without any
  polynomial multiplication, or
\item
  \(O(\#f\#\ell^\gamma \delta^{3}\log(d))\) operations in
  \(\mathbb{F}_q\) without any polynomial multiplication,
\end{itemize}

with a probability of error \(\leq \varepsilon\) if
\(h \neq fg\bmod \ell\).\gobblepar

\end{theorem}

\subsection{Dense and sparse polynomial product
verification}\label{Section:dense-and-sparse-polynomial-product-verification}

Let us come back to standard polynomial product verification. As
mentioned earlier, the standard technique to test whether
\(h = f\times g\) is to test \(h(\alpha) = f(\alpha)g(\alpha)\) for a
random point \(\alpha\). In the dense case, this is done in an optimal
linear number of operations in the base ring as long as it is large
enough. For small finite fields for instance, the complexity is not
optimal anymore since it requires an extension field. For polynomial
over \(\mathbb{Z}\), the difficulty is the possible growth of the
intermediate values, that requires to work modulo some prime number. And
finally, this approach is not suitable at all for sparse polynomials
where the evaluation is too costly.

To overcome these difficulties, the idea is to replace evaluation by
modular reduction. Test if \(h\bmod \ell = f\times g\bmod \ell\) for
some random polynomial \(\ell\). The idea of Kaminski is to use for
\(\ell\) a polynomial \(x^i-1\) for some random \(i\) between
\(d^{1-e}\) and \(2d^{1-e}\) for some \(0 < e < \frac{1}{2}\). Then,
\(f\), \(g\) and \(h\) are reduced modulo \(x^i-1\), and the product
\(f\times g\bmod x^i-1\) is computed explicitly.

\begin{theorem}[\autocite{Kaminski1989}]\label{theorem:Kaminski}

Let \(f\), \(g\), \(h\in \mathsf{R}[x]\) of degree \(< n\) and
\(0 < e < \frac{1}{2}\). One can test whether \(h = f\times g\) using
\(O(n)\) operations in \(\mathsf{R}\), with failure probability
\(O(\log\log n/n^{1-2e})\).\gobblepar

\end{theorem}

Kaminski's algorithm has one main drawback. It relies directly on a
polynomial multiplication algorithm to compute \(f\times g\bmod x^i-1\).
And since \(e < \frac{1}{2}\) the multiplication algorithm must have
subquadratic complexity. Also, analyzed in the bit complexity model for
\(\mathsf{R} =\mathbb{Z}\) or \(\mathsf{R} = \mathbb{F}_q\), the
resulting algorithm is not optimal anymore. Using modular product
verification instead, we can on the one hand get rid of any
multiplication algorithm and on the other hand obtain a linear bit
complexity. This fails to be true if the coefficients are extremely
large compared to the degree of the polynomial. But for this case, we
can rely on the integer product verification algorithm of Kaminski.
Altogether we obtain the following optimal results.

\begin{theorem}[\autocite{GiorgiGrenetPerretduCray2023}]

Let \(f\), \(g\), \(h\in \mathsf{R}[x]\) of degree \(< n\), where
\(\mathsf{R}\) is either \(\mathbb{Z}\) or \(\mathbb{F}_q\). For
\(\mathsf{R} = \mathbb{Z}\), let
\(q = \max(\height f, \height g,\height h)\). One can check if
\(h = f\times g\), with failure probability at most
\(\varepsilon = 1/n^{O(1)}\) if \(h \neq f\times g\), in \(O(n\log q)\)
bit operations and without performing any polynomial
multiplication.\gobblepar

\end{theorem}

For sparse polynomials, we also rely on testing if
\(h\bmod \ell = f\times g\bmod \ell\) but this time \(\ell = x^p-1\)
where \(p\) is a prime number. In this case, we are not able to get
optimal results, only quasi-linear complexities.

\begin{theorem}[\autocite{GiorgiGrenetPerretduCray2023}]

Let \(f\), \(g\), \(h\in \mathsf{R}[x]\) of degree \(< n\) and sparsity
at most \(t\), where \(\mathsf{R}\) is either \(\mathbb{Z}\) or
\(\mathbb{F}_q\). For \(\mathsf{R} = \mathbb{Z}\), let
\(q = \max(\height f, \height g,\height h)\). Let
\(s = t(\log n+\log q)\) be the input size.

One can check if \(h = f\times g\), with a failure probability at most
\(\varepsilon\) if \(h \neq f\times g\), in

\begin{itemize}
\item
  \(O(s\log s\log\log s)\) bit operations if \(\mathsf{R} =\mathbb{Z}\),
  or \(\mathsf{R} =\mathbb{F}_q\) with
  \(q < \frac{c}{\varepsilon}(\#f\#g+\#h)\log n)\), and
\item
  \(O(s\log^{2} s)\) bit operations if \(\mathsf{R} =\mathbb{F}_q\) and
  \(q \geq \frac{c}{\varepsilon}(\#f\#g+\#h)\log n)\),
\end{itemize}

where \(c\) is some constant. In the first case, the complexity becomes
\(O(s\log\log s)\) as soon as \(t = O(\log^k n)\) for some
\(k\).\gobblepar

\end{theorem}

\begin{remark}

The input size \(s = t(\log n+\log q)\) may be a pessimistic upper bound
for polynomials over \(\mathbb{Z}\), when the coefficients are very
unbalanced. If \(s\) is replaced by the actual bit size of the sparse
representations of \(f\), \(g\) and \(h\), the algorithm has complexity
\(O(s\log s\log\log s+s\log n\log\log n)\)~\autocite{GiorgiGrenetPerretduCrayRoche2024}.\gobblepar

\end{remark}

\section{Sparse polynomial multiplication}\label{section:sparsemul}

In this section, we investigate the question of sparse polynomial
multiplication. Given two sparse polynomials \(f\) and
\(g\in \mathsf{R}[x]\), how fast can we compute their product
\(h = f\times g\)? The naive algorithm requires \(O(\#f\cdot\#g)\)
products between the coefficients of \(f\) and \(g\). The exact
complexity depends on the data structures used by the algorithm to
represent the polynomials, and some work has been done to minimize
it~\autocite{Johnson1974,MonaganPearce2009}. Our goal is to provide an
asymptotic improvement.

The difficulty for such an improvement is that the algorithm has to be
\emph{output-sensitive}. The reason is that the size of the output is
not determined by the sole sizes of the inputs, contrary to, say, dense
polynomial multiplication. Even the support is not enough to determine
the size of the output, and one must also take into account the
coefficients.

\begin{example}

Let \(f = \sum_{i=0}^{t-1} x^i\),
\(g_1 = \sum_{i=0}^{t-1} (x^{ti+1}-x^{ti})\) and
\(g_2 = \sum_{i=0}^{t-1} (x^{ti+1}+x^{ti})\), in \(\mathbb{Z}[x]\).
While \(g_1\) and \(g_2\) have the same support,
\(f\times g_1 = x^{t^{2}-1}-1\) has only two monomials while
\(f\times g_2 = x^{t^{2}} + \sum_{i=1}^{t^{2}-1} 2x^i + 1\) has sparsity
\(t^{2}\).\gobblepar

\end{example}

With this example in mind, one cannot improve on the upper bound
\(O(\#f\cdot\#g)\) for the worst-case complexity. The goal is to provide
an output-sensitive bound that depends on \(t = \max(\#f,\#g,\#h)\)
where \(h = f\times g\).

The idea to perform fast sparse multiplication is to rely on sparse
interpolation. As noticed by several authors, one difficulty is due to
the cancellations between coefficients. These cancellations are the
reason why the support of the output cannot be computed from the sole
support of the inputs. This motivated Arnold and Roche to introduce the
\emph{structural support} of a sparse polynomial product, defined as the
sumset of the supports of the inputs. Formally, if \(S_f\) and \(S_g\)
denote the supports of \(f\) and \(g\), the structural support of
\(f\times g\) is the set \(S_f+S_g = \{a+b:a\in S_f, b\in S_g\}\). They
prove the following result.

\begin{theorem}[\autocite{ArnoldRoche2015}]\label{theorem:ArnoldRoche}

Let \(f\), \(g\in\mathbb{Z}[x]\) of size \(n\) and height \(\leq h\).
Let \(t = \max(\#f,\#g,\#(fg))\) and \(s\) be the size of the structural
support of \(f\times g\). The polynomial \(f\times g\) can be computed
by a randomized algorithm in \(O^{\widetilde{}}(t\log h+s\log n)\) bit
operations.\gobblepar

\end{theorem}

The previous example shows the shortcoming of this result. The
structural support may have size \(O(t^{2})\) while the actual output
size is constant. Therefore, this algorithm remains quadratic in \(t\)
in the worst case. We can remark that the structural support coincides
with the support when the input polynomials only have nonnegative
coefficients. In the discrete algorithms community, the problem is
studied under the name \emph{sparse convolution} and many results focus
on the case of nonnegative
coefficients~\autocite{ColeHariharan2002,BringmannFischerNakos2021}. In
particular, some efforts are made to make the algorithms
deterministic~\autocite{AmirKapahPorat2007,BringmannFischerNakos2022}.
The most recent results~\autocite{JinXu2024} provide very precise
complexity bounds in the word RAM model with \(O(\log t+\log d)\)-bit
words where \(d\) is the input degree. Roughly speaking, a
bit-complexity bound can be obtained by multiplying the result by
\(O^{\widetilde{}}(\log t+\log d)\).

\begin{theorem}[\autocite{JinXu2024}]

Let \(f\), \(g\in \mathbb{Z}_{\geq 0}[x]\) of size \(n\) and height
\(\leq h\). The polynomial \(f\times g\) can be computed by a Las Vegas
randomized algorithm that uses \(O(t\log t)\) word operations (with high
probability), where \(t = \#(fg)\).\gobblepar

\end{theorem}

Finally, these results have also been investigated in the real RAM. This
model allows for exact computations on real numbers, but is restricted
to purely algebraic computations. This forces to replace some crucial
steps of a Prony-like interpolation. Nevertheless, the obtained
complexity still remains similar as in the previous works.

\begin{theorem}[\autocite{Fischer2025}]

Let \(f\), \(g\in \mathbb{R}_{\geq 0}[x]\) of degree \(d\). The
polynomial \(f\times g\) can be computed by a Las Vegas algorithm in
\(O^{\widetilde{}}(t)\) operations on a real RAM (with high
probability), where \(t = \#(fg)\).\gobblepar

\end{theorem}

Our goal is to get rid of either the structural sparsity in the
complexity, or nonnegativity assumption. Since it is difficult to
predict the size of the output, our approach is to guess it and check
afterwards. More precisely, we assume an upper bound on the output
sparsity, compute a tentative result using this upper bound, and check
the result using our sparse polynomial product verification algorithm.
The initial upper bound is set to equal the size of the inputs, and is
doubled as long as the correct result is not found. Note that since the
verification algorithm is randomized with some false-positive
probability, this requires some careful probability analysis of the
overall algorithm. In particular, a failure probability of
\(\varepsilon\) in the verification algorithm does not imply the same
failure probability for the overall algorithm.

\begin{theorem}[\autocite{GiorgiGrenetPerretduCray2020}]

Let \(f\), \(g\in \mathbb{Z}[x]\) of degree \(\leq d\) and height
\(\leq h\). The polynomial \(f\times g\) can be computed by an Atlantic
City randomized algorithm that uses
\(O^{\widetilde{}}(t(\log d+\log h))\) bit operations where
\(t =\#(fg)\) (with high probability).\gobblepar

\end{theorem}

As mentioned earlier, the upper bound \(t(\log d+\log h)\) may be a
pessimistic upper bound. Using our unbalanced interpolation algorithm,
we can refine the bound.

\begin{theorem}[\autocite{GiorgiGrenetPerretduCrayRoche2024}]

Let \(f\), \(g\in\mathbb{Z}[x]\) of degree \(\leq d\). The polynomial
\(f\times g\) can be computed by an Atlantic City randomized algorithm
that uses \(O^{\widetilde{}}(\ell\log d)\) bit operations where
\(\ell = \max(\textsf{bitsize}(f),\textsf{bitsize}(g),\textsf{bitsize}(f\times g))\)
(with high probability).\gobblepar

\end{theorem}

Due to the extra factor \(\log(d)\), the algorithm is not quasi-linear
for sparse polynomials. Indeed, the bit size may already contain a
\(\log(d)\) factor. But for moderate degrees, in particular for dense
polynomials, the algorithm has quasi-linear complexity in the input and
output sizes. This provides the first multiplication algorithm for
(dense) polynomials over \(\mathbb{Z}\) with quasi-linear complexity in
the actual size of the input. All the previous approaches, which do not
take the unbalancedness into account, have a worst-case complexity
\(\Omega(\ell^{2})\).

To obtain a fully quasi-linear time algorithm, we would need to also
take into account the unbalancedness of the exponents, rather than using
a bound \(d\) on each of them. Some recent work investigates this
problem~\autocite{vanderHoevenLecerf2025}, but using the pessimistic
bound on the coefficients. Combining both results is a challenging open
problem.

\begin{openproblem}

Given \(f\), \(g\in\mathbb{Z}[x]\), can we compute their product
\(h = f\times g\) in quasi-linear time in the bit size of the sparse
representations of \(f\), \(g\) and \(h\)?\gobblepar

\end{openproblem}

Another question concerns sparse polynomials over finite fields.
Currently, the best known sparse interpolation algorithms over finite
fields do not have quasi-linear complexity. Therefore, using them for
sparse multiplication does not provide a quasi-linear algorithm. Some
partial results are known, for instance for not-too-sparse
polynomials~\autocite{JinXu2024}. Yet the general case remains open.

\begin{openproblem}

Given \(f\), \(g\in\mathbb{F}_q[x]\), can we compute their product
\(h = f\times g\) in quasi-linear time in the bit size of the sparse
representations of \(f\), \(g\) and \(h\)?\gobblepar

\end{openproblem}

A promising approach is to use some techniques developed for the real
RAM~\autocite{Fischer2025}. In particular, due to some limitations of
the real RAM, Fischer bypasses the root finding part of Prony's
algorithm, which is the only non-quasi-linear part of Huang's fastest
sparse interpolation algorithm over finite fields
(\hyperref[theorem:Huang2023]{Theorem~\ref*{theorem:Huang2023}}).

\section{Sparse polynomial division and
divisibility}\label{section:sparsediv}

Sparse polynomial division suffers from the same difficulty of
output-sensitivity as multiplication, but to a much larger scale.
Indeed, the output size of a division may vary from constant to
exponential in the size of the inputs. This is the case for the sparsity
and the height.

\begin{example}

Let \(f_1 = x^{2d+1}-x^{2d}+x^d\), \(f_2 = x^{2d+1}+x^{2d}+x^d\),
\(f_3=x^{2d+1}-x^{2d}\), and \(g = x^{d+1}-x^d+1\) in \(\mathbb{Z}[x]\).
Then

\begin{itemize}
\item
  \(g\) divides \(f_1\) and the quotient is \(q_1 = x^d\);
\item
  \(g\) does not divide \(f_2\) and the quotient is
  \(x^d + \sum_{i=0}^{d-1} 2x^i\) and the remainder is
  \(2x^d - \sum_{i=0}^{d-1} 2x^i\);
\item
  \(g\) does not divide \(f_3\) and the quotient is
  \(x^d +\sum_{i=0}^{d-1} 2^{d-i-1} x^i\) and the remainder is
  \((2^d-3)x^d -\sum_{i=0}^{d-1} 2^{d-i-1}x^i\).
\end{itemize}

\end{example}

In part due to the extreme variability of the output size, a fast
output-sensitive algorithm for Euclidean division of sparse polynomials
remains elusive. As a first step, we focus on the simpler problem of
exact division. There are two distinct problems. The first one is to
compute the quotient \(f/g\) when \(g\) is known to divide \(f\). The
second one is to decide \(g\) divides \(f\).

The first problem can be tackled with similar techniques as
multiplication. The idea is thus to interpolate \(h = f/g\) while the
division is exact. Nevertheless, two additional difficulties appear. The
first one is that the interpolation algorithms need to evaluate \(q\).
The evaluation points must not be roots of \(g\). Since the evaluations
are performed on \(p\)th roots of unity in \(\mathbb{F}_q\) or
\(\mathbb{Z}/q^{2k}\mathbb{Z}\) for some primes \(p\) and \(q\) and
integer \(k\), it is sufficient for \(g\) to be coprime with
\(\phi_p = \sum_{i=0}^{p-1}x^i\) in \(\mathbb{F}_q[x]\). A careful
choice of \(p\) and \(q\) makes this event most likely. The second
difficulty concerns the size growth of the quotient. For multiplication
algorithms, the sparsity can increase quadratically. In the case of
exact division, the sparsity can increase exponentially. Also, for
polynomials over \(\mathbb{Z}\) the size of the coefficients can
increase a lot. Although recent results have reduced the bounds from
exponential to polynomial when the division is
exact~\autocite{NahshonShpilka2024,NahshonShpilka2025}, quasi-linear
time algorithms require two \emph{a priori} bounds, one on the sparsity
and one on the height. If the result appears to be incorrect, one of the
two bounds has to be increased. The difficulty is to identify which
bound is too small. It happens that detecting if the bound on the height
is too small can actually be performed inside the interpolation
algorithm. As a result, there are two distinct tests at two distinct
steps to identify which bound(s) must be increased. For polynomials over
finite fields, no coefficient growth occurs. But no quasi-linear sparse
interpolation algorithm is known.

\begin{theorem}[\autocite{GiorgiGrenetPerretduCray2021,GiorgiGrenetPerretduCrayRoche2022}]

Let \(f\), \(g\in \mathsf{R}[x]\) such that \(g\) divides \(f\),
\(d = \deg(f)\) and \(t=\max(\#f,\#g,\#(f/g))\). The polynomial \(f/g\)
can be computed by an Atlantic City randomized algorithm that uses

\begin{itemize}
\item
  \(O^{\widetilde{}}(t(\log d+\log h))\) bit operations if
  \(\mathsf{R}=\mathbb{Z}\) and
  \(h = \max(\height f, \height g, \height(f/g))\);
\item
  \(O^{\widetilde{}}(t\log d\log q)\) bit operations if
  \(\mathsf{R}=\mathbb{F}_q\) has characteristic \(> d-\deg(g)\);
\item
  \(O^{\widetilde{}}(t\log^{2}(d)(\log d+\log q))\) bit operations if
  \(\mathsf{R}=\mathbb{F}_q\) has characteristic \(\leq d-\deg(g)\).
\end{itemize}

\end{theorem}

This theorem provides partial results. Beyond improving the complexities
in the case of finite field, it is not yet known how to adapt the
algorithm for polynomials over \(\mathbb{Z}\) with unbalanced
coefficients, as in the case of polynomial multiplication. A much more
ambitious goal is to extend the result to the general euclidean
division.

\begin{openproblem}

Given \(f\), \(g\in\mathsf{R}[x]\), can we compute the quotient \(q\)
and remainder \(r\) such that \(f = gq+r\) in time quasi-linear in the
sparse representations of \(f\), \(g\), \(q\) and \(r\)?\gobblepar

\end{openproblem}

The second problem of testing whether \(g\) divides \(f\) remains open
in full generality. Some special cases are easy though. When \(\deg(g)\)
or \(\deg(f)-\deg(g)\) are small, the long division algorithm has
polynomial running time, since in the first case the remainder cannot be
large, and in the second case the quotient cannot be large. We extend
these results to more general cases. In particular, we prove sparsity
bounds on the quotient based on the structure of the divisor \(g\). If
\(g = 1+x^kg_1\) for some sufficiently large \(k\), the quotient is
sparse. Using the fact that \(g\) divides \(f\) if and only if
\(g^{\shortleftarrow} = x^{\deg g}g(1/x)\) divides
\(f^{\shortleftarrow} = x^{\deg f}f(1/x)\), the same holds if
\(g = x^d + g_1\) for a low-degree \(g_1\). These bounds directly
translate into polynomial-time algorithms. The bounds fail for more
general divisors \(g = g_0+x^k g_1\). Nevertheless, using a
factorization \(g = g_0\cdot(1+x^k g_1/g_0)\), we prove that testing
divisibility by such divisors is still polynomial. The recursive use of
this idea gives our most general result. One can test divisibility by a
divisor \(g\) if there exists in \(g\) a low-degree polynomial \(g_1\)
surrounded by large \emph{gaps}.

\begin{theorem}[\autocite{GiorgiGrenetPerretduCray2021}]

Let \(f\) and \(g\in \mathsf{K}[x]\) be two polynomials of respective
degrees \(m+n\) and \(m\), and sparsity at most \(t\). One can check
whether \(g\) divides \(f\) in deterministic polynomial time if \(g\)
can be written as \(g_0+x^kg_1+x^\ell g_2\) where \(g_0\), \(g_1\),
\(g_2\in \mathsf{K}[x]\) satisfy \(\deg(g_1) = \poly(t\log(m+n))\),
\(k = \deg(g_0)+\Omega(n)\) and
\(\ell = \deg(x^kg_1) + \Omega(n)\).\gobblepar

\end{theorem}

A general result remains out of reach.

\begin{openproblem}

Given two sparse polynomials \(f\), \(g\in\mathsf{K}[x]\), what is the
complexity of testing whether \(g\) divides \(f\)? Is it possible to
perform the computation in polynomial time in the sparse representation
of \(f\) and \(g\)? Is the problem \(\NP\)-hard?\gobblepar

\end{openproblem}

\section{Factorization}\label{section:sparsefact}

In this section, we focus on a computationally more difficult problem,
namely the factorization of sparse polynomials. Many open questions
exist in this domain. In particular, it is not yet known how sparse or
dense are the irreducible factors of sparse polynomials. More generally,
understanding how the structure of a polynomial reflects in the
structure of its irreducible factors is a challenging question. A very
recent striking result provides some answers to this question in terms
of arithmetic circuit
depth~\autocite{BhattacharjeeKumarRaiRamanathanSaptharishiSaraf2025a}.

Over finite fields, it is \(\NP\)-hard (under randomized reductions) to
decide whether a given sparse polynomial has roots, that is degree-\(1\)
factors, in the base field~\autocite{BiChengRojas2016}. In
characteristic zero, more computational results are known. A first
algorithm was developed to compute the roots of sparse polynomials over
\(\mathbb{Z}\)~\autocite{CuckerKoiranSmale1999}. This result was
extended to \(\mathbb{Q}\) and number fields, and to low-degree
factors~\autocite{Lenstra1999a}. By generalizing the latter technique,
it was shown possible to extend these results to multivariate
polynomials as well~\autocite{KaltofenKoiran2005,KaltofenKoiran2006}.

We propose another approach for multivariate polynomials. It is
combinatorial in nature and applies to any field of characteristic zero.
We prove that computing low-degree factors of multivariate polynomials
\emph{reduces} to the univariate case and to low-degree factorization.
We distinguish two families of factors. A polynomial
\(f\in\mathsf{K}[x_1,\dotsc,x_n]\) is \emph{unidimensional} if it can be
written \(\vx^{\vec \mu}f_u(\vx^{\vec \nu})\) for some univariate
polynomial \(f_u\) and multivariate monomials \(\vx^{\vec \mu}\) and
\(\vx^{\vec \nu}\), and \emph{multidimensional} otherwise.

\index{polynomial!unidimensional} \index{polynomial!multidimensional}

The case of unidimensional factors is reduced to the univariate case.
The reduction is straightforward. Unidimensional factors of a polynomial
are factors of its \emph{unidimensional components}, that can be
computed easily from the support of the polynomial. Now a unidimensional
factorization problem is equivalent to its univariate counterpart. The
required algorithms for sparse univariate factorization are only known
for \(\mathbb{Z}\), \(\mathbb{Q}\) and more generally number fields.

The case of multidimensional factors is also a reduction, this time to
low-degree factorization. Algorithms for this problem are known for a
large variety of base rings: \(\mathbb{Z}\) or
\(\mathbb{Q}\)~\autocite{LenstraLenstraLovasz1982}, number
fields~\autocite{Lenstra1983,Landau1985,Lenstra1987}, real and complex
numbers~\autocite{Pan2002,KaltofenMayYangZhi2008,Moroz2022}, the
\(p\)-adic numbers~\autocite{CantorGordon2000} or the algebraic closure
of \(\mathbb{Q}\) (absolute
factorization)~\autocite{BostanLecerfSalvySchostWiebelt2004,ChezeGalligo2005}.

The approach for multidimensional factors can be seen as a
generalization of the first root-finding algorithm for sparse
polynomials~\autocite{CuckerKoiranSmale1999}, based on a so-called
\emph{Gap Theorem}. They consider a sparse univariate polynomial
\(f\in\mathbb{Z}[x]\). Assuming that \(f = f_1 + x^k f_2\) where
\(k \gg \deg(f_1)\), they prove that any root
\(\alpha\in\mathbb{Z}\setminus\{-1,0,1\}\) of \(f\) must be a root of
both \(f_1\) and \(f_2\). The argument is that whenever \(\alpha\) is
not a root of \(f_2\), the integer \(\alpha^kf_2(\alpha)\) is a much
larger integer than \(f_1(\alpha)\) in absolute value, whence their sum
cannot vanish. A recursive application of this result reduces the
computation of the roots of \(f\) to the computation of the roots of its
low-degree parts.

To mimic this technique to compute low-degree factors of multivariate
polynomials, we need to consider roots of such factors. Consider first a
bivariate polynomial \(f \in \mathsf{K}[x, y]\) over some field
\(\mathsf{K}\) of characteristic \(0\), and some irreducible factor
\(g\) of \(f\). Viewing \(g\) as a polynomial in \(\mathsf{K}[x][y]\),
it has a root \(\psi\) in the field of Puiseux series \(\puiseux\). This
root satisfy \(g(x,\psi(x)) = 0\), therefore \(f(x,\psi(x)) = 0\) as
well. The notion of size used in the integer-case argument is played
here by the valuation of the Puiseux series, that is the largest power
of \(x\) that divides the series. We have the following result.

\notation[field of Puiseux series over the algebraic closure $\overline{\mathsf{K}}$ of $\mathsf{K}$\nomrefpage]{\(\puiseux\)}(dringzRpolz)

\begin{theorem}[\autocite{Grenet2014,Grenet2016a}]

Let
\(f = \sum_{j=1}^\ell c_j x^{\alpha_j}y^{\beta_j} \in \mathsf{K}[x,y]\)
and \(g\in \mathsf{K}[x,y]\) of individual degree \(d\) that does not
divide \(f\). There exists an explicit bound
\(\gamma_v(d,\ell) = O(d^{2}\ell^{2})\) such that for any root
\(\psi\in\puiseux\) of \(g\), the valuation of \(f(x,\psi(x))\) is at
most
\(\min_{1\leq j\leq\ell} (\alpha_j+v\beta_j) + \gamma_v(d,\ell)\).\gobblepar

\end{theorem}

To use it for multivariate polynomial, we single out two variables from
\(n\) and work with the polynomial ring
\(\mathsf{K}(\vx\setminus\{x_i,x_j\})[x_i,x_j]\) where
\(\mathsf{K}(\vx\setminus\{x_i,x_j\})\) denotes the field of rational
functions in \(n-2\) variables. To handle all multidimensional factors,
that may depend on only two variables each, we actually consider all the
\(\binom{n}{2}\) possible pairs of variables. The theorem is only proved
in the case of a field of characteristic zero. For positive
characteristic, one needs to consider roots as more general power
series, namely Hahn series~\autocite{Kedlaya2001}. It is plausible that
a generalization of our techniques can prove the result for some finite
fields, at least when the characteristic is larger than the degree of
\(f\). This would provide a generalization of our previous results that
allow to compute multilinear factors of sparse polynomial over finite
field of large
characteristic~\autocite{ChattopadhyayGrenetKoiranPortierStrozecki2021}.

Combining both the unidimensional case and the multidimensional case, we
obtain the following algorithm.

\begin{theorem}[\autocite{Grenet2014,Grenet2016a}]

Let \(f\in\mathsf{K}[x_1,\dotsc,x_n]\) be a degree-\(D\) sparsity-\(t\)
polynomial, where \(\mathsf{K}\) is a field of characteristic zero.

\begin{enumerate}[label=\roman*.]

\item

Computing its irreducible degree-\(\leq d\) \emph{unidimensional}
factors can be reduced to computing the degree-\(\leq d\) irreducible
factors of \(\poly(t,n,\log D,d)\) univariate polynomials of degree
\(\leq D\) and sparsity \(\leq t\), plus \(\poly(t,n,\log D,d)\) bit
operations.

\item

Computing its irreducible degree-\(\leq d\) \emph{multidimensional}
factors can be reduced to the irreducible factorization of at most \(t\)
polynomials of degree \(\poly(t,d)\), plus \(\poly(t,n,\log D,d)\) bit
operations.

\end{enumerate}

As a consequence, if \(\mathsf{K}\) is some number field, one can
compute all the irreducible factors of \(f\) of degree \(\leq d\) in
time polynomial in the input size.\gobblepar

\end{theorem}

The natural generalization of the previous result is to put a bound on
the sparsity of the factors, rather than on their degree.

\begin{openproblem}

Given a polynomial \(f\in\mathsf{K}[x_1,\dotsc,x_n]\) of degree \(D\)
and sparsity \(t\), and a bound \(\tau\), can we compute all the
irreducible factors of \(f\) of sparsity at most \(\tau\) in time
polynomial in the input and output size?\gobblepar

\end{openproblem}

\chapter{Conclusions and
perspectives}\label{Section:conclusions-and-perspectives-1}

We have shown that one can sometimes provide fast algorithms for
computing with sparse polynomials. For basic arithmetic, one tool of
choice is sparse interpolation which plays to some extent the same role
for sparse polynomials as the FFT for dense polynomials. The comparison
ends quickly. On the one hand, quasi-linear sparse interpolation is
known only for polynomials over the integers, but not over finite fields
for instance. On the other hand, many computational problems about dense
polynomials reduce to polynomial multiplication, itself based on FFT in
the asymptotic regime. For sparse arithmetic, only multiplication and
exact division have been reduced to sparse interpolation for now.

\section{Sparse interpolation}\label{Section:sparse-interpolation}

The main challenge in sparse interpolation is to have an algorithm that
has quasi-linear complexity for polynomials over finite fields.

\begin{openproblem}

Is there a quasi-linear time algorithm for sparse interpolation over
(some, or all) finite fields? In which model(s)?\gobblepar

\end{openproblem}

\emph{A priori}, the algorithm is given access to a black box with
derivative to avoid the exponential lower bound of
\hyperref[proposition:sparseFFTlowerbound]{Proposition~\ref*{proposition:sparseFFTlowerbound}}.
Yet this lower bound does not apply directly to regular black box
interpolation. Indeed, it relies on the fact that the polynomial \(f\)
is evaluated on some values \(\omega^i\) where \(i\) is known. This
includes
\hyperref[algorithm:SparseInterpolationBB]{Algorithm~\ref*{algorithm:SparseInterpolationBB}} (\textsc{SparseInterpolationBB})
but also includes any algorithm that would for instance evaluate \(f\)
on random values. Indeed, sampling uniformly \(r\in\mathsf{R}\) can be
done by first sampling an integer \(i\) and setting \(r = \omega^i\).
But it does not rule out an algorithm that would for instance evaluate
\(f\) on an arithmetic progression. No such algorithm is known yet, and
it is not clear that few evaluations are sufficient.

\begin{openproblem}

Let \(f=\sum_{i=0}^{t-1} c_i x^{e_i} \in\mathsf{R}[x]\) and let
\((r_i)_i\) be a sequence of elements of \(\mathsf{R}\) in arithmetic
progression. What is the smallest possible \(s\) such that \(f\) can be
reconstructed uniquely from \((f(r_i))_{0\leq i<s}\)?\gobblepar

\end{openproblem}

This open question may receive different answers depending on the ring
\(\mathsf{R}\). It can also be generalized and phrased as a standard
interpolation problem.

\begin{openproblem}

Given \(s\) evaluations of a \(t\)-sparse degree-\(d\) polynomial
\(f\in\mathsf{R}[x]\), on what condition(s) on \(s\) and the evaluation
points can \(f\) be reconstructed?\gobblepar

\end{openproblem}

Geometric black box interpolation shows that it is possible if
\(s \geq 2t\) and the evaluation points are in geometric progression.
Lagrange interpolation theorem states that \(s > d\) is sufficient,
whatever the points are. Other cases are treated by Borodin and
Tiwari~\autocite{BorodinTiwari1991} although they do not solve the
problem in full generality. One intriguing special case arises from the
study of sparse polynomial interpolation
codes~\autocite{KaltofenPernet2014}. If \(\mathsf{R} = \mathbb{R}\),
Descartes' rule of signs~\autocite{Descartes1637} implies that \(f\) has
at most \(t-1\) positive real roots and at most \(t-1\) negative real
roots. This implies for instance that the value of \(f\) on \(2t\)
positive real numbers uniquely defines it. (If two \(t\)-sparse
polynomials \(f_1\) and \(f_2\) are consistent with the evaluations,
their difference \(f_1-f_2\) has \(2t\) positive real roots and at most
\(2t\) monomials, hence vanishes.) But it is not clear how to compute
\(f\) from its evaluations. To stick with exactly representable
polynomials, we phrase the question for polynomials with rational
coefficients. In this case, an exhaustive search is possible. Since
\(\mathbb{Q}[x]\) is countable, one can enumerate all \(t\)-sparse
polynomials in a certain order and recompute the evaluations to compare.
This algorithm is extremely inefficient.

\begin{openproblem}

What is the complexity of computing the sparse representation of a
\(t\)-sparse polynomial \(f\in\mathbb{Q}[x]\), given as inputs \(2t\)
evaluations of \(f\) on positive rational numbers?\gobblepar

\end{openproblem}

Similar bounds depending on the sparsity of the polynomials have been
proved when \(\mathsf{R}\) is a number field or a \(p\)-adic
field~\autocite{Lenstra1999}, and the same question of sparse
interpolation can be asked. On the other hand, no such bound is known,
nor even possible, in the case of finite
fields~\autocite{BiChengRojas2016,ChengGaoRojasWan2017}.

\section{Sparse polynomial
arithmetic}\label{Section:sparse-polynomial-arithmetic}

Computing with sparse polynomials remains a challenge. We can identify
at least two reasons. First the complexity of many basic operations is
still unclear. For instance, due to a lack of quasi-linear sparse
interpolation algorithm over finite fields, it is currently not known if
a quasi-linear multiplication algorithm exists for sparse polynomials
over finite fields. Second the structures of the problems are not well
understood. As mentioned earlier, one difficulty in sparse arithmetic is
the output sensitivity. Some bounds are known for the output size for
some problems (multiplication, exact division, \ldots) but their
tightness is not always known. For instance, some non-trivial bounds
exist on the bit size of the exact quotient of two sparse polynomials.
But the case of Euclidean division in much less understood and no
reduction of Euclidean division to sparse interpolation is currently
known. As for exact division, the goal would be to have a fast
output-sensitive algorithm. A related decision problem is wide open too:
Given two sparse polynomials \(f\) and \(g\), does \(g\) divide \(f\)?
We have given some partial results but the general case still seems out
of reach.

Structural questions also arise for polynomial factorization, or more
simply root finding. Giving tight upper bounds on the number of roots of
sparse polynomials is very challenging. The theory of fewnomials
attempts to provide such results for systems of
polynomials~\autocite{Khovanskii1991}. But even very basic questions
remain open. For instance, consider polynomials over the real numbers.
As a consequence of Descartes' rule of signs~\autocite{Descartes1637}, a
polynomial \(f\) with \(t\) nonzero terms has at most \(2t-2\) nonzero
roots, irrespective of its degree. The product \(fg\) of two such
polynomials has at most \(t^{2}\) nonzero terms, so Descartes' rule
implies that it has \(O(t^{2})\) nonzero roots. But of course, the roots
of \(fg\) are roots of \(f\) or \(g\), therefore a tighter (and
attained) bound is \(4t-4\). But the argument fails for the very close
polynomial \(fg+1\) for instance.

\begin{openproblem}

Let \(f\), \(g\in\mathbb{R}[x]\) with \(t\) nonzero terms each. What is
the maximal number of roots of the polynomial \(fg+1\)? In particular,
is it always linear? Or are there examples where the number of roots is
super-linear, even quadratic?\gobblepar

\end{openproblem}

This and similar questions actually have implications in algebraic
complexity theory~\autocite{Koiran2011}. The same holds for a variant
that asks for the number of edges in the Newton polygon of a bivariate
polynomial. For this variant, an upper bound \(O(t^{\frac{4}{3}})\) can
be proven instead of the trivial bound
\(O(t^{2})\)~\autocite{KoiranPortierTavenasThomasse2015}. Results are
also known for random polynomials~\autocite{BriquelBurgisser2020}. But
the original question remains completely open.

For polynomial factorization, it is a notoriously hard problem to bound
the sparsity of factors of sparse
polynomials~\autocite{vonzurGathenKaltofen1985}. From the perspective of
algorithms, no algorithm is known to output the sparse factors of a
sparse polynomial. Very recent results on the factorization of
low-degree polynomials represented by
circuits~\autocite{BhattacharjeeKumarRaiRamanathanSaptharishiSaraf2025a}
suggest a natural approach. Given a sparse polynomial, are there some
low-depth (but high-degree!) arithmetic circuits for its irreducible
factors? And can we compute them?

\section{Sparse interpolation, error-correcting codes, and
cryptography}\label{Section:sparse-interpolation-error-correcting-codes-and-cryptography}

Finally, outside polynomial arithmetic, sparse interpolation has strong
links with the field of error-correcting
codes~\autocite{ComerKaltofenPernet2012} and applications in
cryptography~\autocite{FleischhackerLarsenSimkin2023}. Sparse
interpolation may be seen as an instance of the \emph{Syndrome Decoding
Problem} (see for instance~\autocite{WegerGassnerRosenthal2024})
underlying Niederreiter cryptosystem~\autocite{Niederreiter1986}.

Consider a linear code \(\mathcal{C}\). A message \(\vm\) is encoded as
a codeword \(\vc\). The code has a parity-check matrix \(H\) such that
\(H\cdot\vv = \vec 0\) if and only if \(\vv\) is a codeword. The
codeword is sent through some channel and the received word is
\(\vr = \vc+\ve\) where \(\ve\) is an error vector. Given \(\vr\) and
\(H\), one can compute the \emph{syndrome}
\(\vs = H\cdot\vr = H\cdot\ve\) using a matrix-vector product. The
Syndrome Decoding Problem (SDP) asks to reconstruct the sparse vector
\(\ve\) from the syndrome \(\vs\), knowing the parity-check matrix.
Niederreiter cryptosystem uses this problem with two parity-check
matrices. The first one is the parity-check matrix of code for which SDP
has an efficient algorithm. It serves as the secret key. The second
parity-check matrix some randomly scrambled version of the first one
that (hopefully) has lost all its structure due to randomization. It
serves as the public key. To encrypt some data, it is first mapped to an
error vector \(\ve\), that is a sparse vector. No codeword is used, or
rather the only codeword is the all-zero vector. The encryption is the
computation of the syndrome, as a matrix-vector product using the
random-looking parity-check matrix. The decryption reduces to solving
SDP with the structured parity-check matrix. The efficiency of
decryption relies on a fast algorithm for SDP with this matrix, while
the security relies on the hardness of SDP for the random-looking parity
check matrix.

Sparse interpolation, and more precisely black box sparse interpolation,
can be seen as solving SDP where the parity check matrix is
\[H = \begin{pmatrix}
1 & 1 & \dots & 1 \\
1 & \omega & \dots & \omega^{n-1} \\
1 & \omega^{2}& \dots & \omega^{2(n-1)} \\
\vdots&\vdots&\ddots&\vdots\\
1 & \omega^{2t-1} & \dots & \omega^{(2t-1)(n-1)}
\end{pmatrix}.\] Indeed, viewing \(\ve\) as a sparse polynomial
\(f = \sum_{j=0}^{t-1} e_{i_j} x^{i_j}\), the syndrome
\(\vs = H\cdot\ve\) corresponds to the vector of evaluations
\((f(1), f(\omega), \dotsc, f(\omega^{2t-1}))\). Therefore, computing
sparse vector \(\ve\) from the syndrome \(\vs\), or computing the sparse
polynomial \(f\) from its evaluations on \(1\), \(\omega\), \ldots,
\(\omega^{2t-1}\), are exactly equivalent. Note that the parity-check
matrix \(H\) corresponds to a \emph{normalized Reed-Solomon
code}~\autocite{Roth2006}.

Beyond this formal equivalence, sparse interpolation can be used in
cryptography for \emph{oblivious ciphertext compression} and
\emph{decompression}~\autocite{FleischhackerLarsenSimkin2023,BienstockPatelSeoYeo2024}.
Ciphertext compression takes as input a sparse vector \(\vv\), encrypted
under a linearly homomorphic encryption scheme (LHE), and outputs a
compressed (still encrypted) version of \(\vv\). Upon reception, the
owner of the private key can decrypt and decompress to reconstruct
\(\vv\). We interpret \(\vv\) as an error vector, and compress it as the
syndrome. This requires a product between a clear matrix and an
encrypted vector, which is possible using an LHE. Once the syndrome is
decrypted, the reconstruction of the vector \(\vv\) is an instance of
SDP.

A ciphertext compression protocol can be used in a Private Information
Retrieval protocol (PIR)~\autocite{ChorKushilevitzGoldreichSudan1998}.
In such a protocol, a server stores a database and a client makes
queries to the database to retrieve some entries. But the server must
not learn which entries were queried. One inefficient solution would be
for the server to send the whole database. The objective of a PIR
protocol is to minimize the amount of the data that the server must send
to enable the client retrieve its entries. Without going into more
details, some PIR protocols have a first step where the server gets the
queries as an encrypted sparse vector, containing a \(1\) exactly for
those entries that have been queried~\autocite{OstrovskySkeith2007}.
Then, the server computes a pointwise (homomorphic) product between the
database and the query vector. This results in an encrypted sparse
database containing only the relevant entries. Ciphertext compression
allows then the server to send this sparse database with a low
communication cost. Similar techniques can also be used for searchable
encryption~\autocite{RenWang2023} where the database is now encrypted.

The interpretation of sparse interpolation as a special case of the
Syndrome Decoding Problem forms the basis of ongoing work with Pascal
Giorgi (U. Montpellier) and Mark Simkin (Aarhus U.) to improve the
current state-of-the-art in ciphertext compression and decompression,
with applications to PIR protocol and searchable encryption.

\backmatter
\bookmarksetup{startatroot}
\addtocontents{toc}{\bigskip}

\chapter*{Publication list}\label{Section:publication-list}
\addcontentsline{toc}{chapter}{Publication list}

\bibliopart{J}{journals}{Journal publications}
\bibliopart{C}{conferences}{Conference publications}

\bibliopart{M}{other}{Unpublished manuscripts}
\bibliopart{S}{software}{Software}

\chapter*{References}\label{Section:references}
\addcontentsline{toc}{chapter}{References}

\newrefcontext[labelprefix=] %,sorting=nty]
\printbibliography[notkeyword=perso,heading=none]

\renewcommand\printbibliography[1][]{}

\printindex
\printnomenclature

\backmatter
\printbibliography

\end{document}

%% file: linrec.tikz
\begin{tikzpicture}[x=6cm,y=4cm,%
    every edge/.style={draw,semithick,-latex,bend left=25,%
        every node/.style={fill=vert!30,align=center,drop shadow},%
    },%
    every node/.style={shape=ellipse,fill=bleu!30,minimum height=1.5cm,drop shadow},%
    every pin edge/.style={-,bend left=0},%
    every pin/.style={fill=none,reset preactions},%
]
\node[align=center] (init) at (0,0) {\ref{proposition:lrsrepr:init}~$2k$ initial terms};
\node[align=center] (minpoly) at (1,1) {\ref{proposition:lrsrepr:minpoly}~Minimal polynomial\\\phantom{\ref{proposition:lrsrepr:minpoly}~}and $k$ initial terms};
\node[align=center] (rational) at (1,-1) {\ref{proposition:lrsrepr:rat}~Rational function};
\node[align=center] (expsum) at (1.35,0) {\ref{proposition:lrsrepr:expsum}~Exponential sum};

    \path (init)    edge node[pin=100:{Berlekamp-Massey\\algorithm}] {LFSR\\synthesis} (minpoly);
    \path (minpoly) edge node[pin=15:{transposed\\Euclidean division}] {Recurrence\\extension} (init);
    \path (init)    edge node[pin=-15:{extended  Euclid's\\ algorithm}] {Padé\\approximation} (rational);
\path (rational)edge node {Taylor\\expansion} (init);
\path (init)    edge[bend left=0,dashed] node[solid] {Prony's\\method} (expsum);
\end{tikzpicture}

%% file: circuit.tikz
\begin{tikzpicture}[circuit,baseline=(F.north)]
  \node[state] (X) at (0,0) {$x$};
  \node[state] (I) at (2,0) {$1$};
  \node[state] (P) at (0,-2) {$+$};
  \node[state] (M) at (2,-2) {$-$};
  \node[state,accepting] (F) at (1,-4) {$\times$};

  \path 
    (X) edge (P)
        edge (M)
    (I) edge (P)
        edge (M)
    (P) edge (F)
    (M) edge (F)

;
\end{tikzpicture}

%% file: semicumulativeprod.tikz
\begin{tikzpicture}[matvec]
    \begin{scope}[local bounding box=T,shift={(3,0)}]
        \draw (2,4) pic {fullprod={toeplitz=bleute}{5.5}{$f_t$}};
        \foreach \x in {0,2,...,6}
            \draw (\x,\x) pic {fullprod={toeplitz=violette}{2}{$f_b$}};
        \foreach \y in {2, 4, 6}
            \draw (0,\y) pic {fullprod={toeplitz=orangerouge}{2}{}};
        \node at (1,6) {$f_t$};
        \draw (0,0) pic {fullprod={thick}{7.5}{}};
        \fill[white,opacity=.5] (7.52,0) rectangle (8.1,15);
        \fill[sharp corners,white,opacity=.5] (-.1,7.45) -- ++(2.2,2.2) -- ++(0,1) -- ++(-2.2,2.2) -- cycle;
        %\node at (-.5,1) {$f_b$};
    \end{scope}
    \scoped[local bounding box=f,shift={(12,4)}]
    {
        \draw[thick] (0,0) rectangle +(1,7);
        \draw[fill=orangerouge] (0,0) rectangle +(1,2);
        \draw[fill=bleute] (0,2) rectangle +(1,5);
        \node at (.5,1) {$g_b$};
        \node at (.5,4.5) {$g_t$};
    }
    \scoped[local bounding box=h]
    {
      \draw[thick] (0,0) rectangle +(1,15);
      \draw[fill=vertjaune] (0,0) rectangle +(1,7);
      \node at (.5, 9.5) {$h$};
    }
    \node at ($(h.east)!.5!(T.west)$) {$\pe$};
    \node at ($(f.west)!.5!(T.east)$) {$\times$};
    %\foreach \l in {f,h,T} \crochets{\l};

\end{tikzpicture}

%% file: lowerprod.tikz
\begin{tikzpicture}[matvec]
    \begin{scope}[local bounding box=T,shift={(3,0)}]
        \draw[toeplitz=vertjaune] (0,0) -- ++(0,5.5) -- ++(5.5,0) -- cycle;
        \node at (2,4) {$f_b$};
        \foreach \x in {0,...,3}
        {
            \draw[toeplitz=rougete] (5.5-2*\x,5.5) -- ++(0,2) -- ++(2,0) -- cycle;
            \node at (6.45-2*\x,7) {$f_{\x}$};
        }
        \foreach \y in {0,...,2}
        {
            \draw[toeplitz=bleute] (3.5-2*\y,5.5) -- ++(2,0) -- ++(0,2) -- cycle;
            \node at (4.75-2*\y,6) {$f_{\y}$};
        }
        \fill[sharp corners,white,opacity=.5] (0,5.5) rectangle +(-.52,2);
        \draw[thick] (0,0) -- ++(0,7.5) -- ++(7.5,0) -- cycle;
        %\node at (1,6) {$f_1$};
        %\node at (-.5,1) {$f_0$};
    \end{scope}
    \scoped[local bounding box=h]
    {
      \draw[thick] (0,0) rectangle +(1,7.5);
      \draw (0,0) rectangle +(1,5.5);
      \draw (0,5.5) rectangle +(1,2);
      \node at (.5,2.75) {$h_b$};
      \node at (.5,6.5) {$h_t$};
    }
    \scoped[local bounding box=f,shift={(12,0)}]
    {
        \foreach \y in {0,...,3}
        {
            \draw[fill=orangerouge] (0,2*\y-.5) rectangle +(1,2);
            \node at (.5,2*\y+.5) {$g_{\y}$};
        }
        \fill[sharp corners,white,opacity=.5] (0,0) rectangle +(1,-.51);
        \draw[thick] (0,0) rectangle +(1,7.5);
        %\node at (.5,1) {$g_0$};
        %\node at (.5,4.5) {$g_1$};
    }
    \node at ($(h.east)!.5!(T.west)$) {$\se$};
    \node at ($(f.west)!.5!(T.east)$) {$\times$};
    %\foreach \l in {f,h,T} \crochets{\l};

\end{tikzpicture}

%% file: middleprod.tikz
\begin{tikzpicture}[matvec]
    \begin{scope}[local bounding box=T,shift={(3,0)}]
        \draw[toeplitz=vertjaune] (0,2) rectangle +(9.5,5.5);
        %\node at (2,4) {$f_{[0,n-k[}$};
        %\foreach \x/\a/\b in {0/0/k,1/k/2k,2/2k/3k,3/3k/4k,4/4k/n}
        \foreach \x in {0,...,4}
        {
            \draw[toeplitz=rougete] (7.5-2*\x,0) rectangle ++(2,2);
            \node at (8.5-2*\x,1) {$f_{\x}$};
        }
        \draw[thick] (0,0) rectangle ++(9.5,7.5);
        \fill[sharp corners,white,opacity=.5] (0,0) rectangle +(-.52,2);
        \node at (4.75,4.75) {$f_t$};
        %\node at (-.5,1) {$f_0$};
    \end{scope}
    \scoped[local bounding box=h]
    {
      \draw[thick] (0,0) rectangle +(1,7.5);
      \draw (0,0) rectangle +(1,2);
      \draw (0,2) rectangle +(1,5.5);
      \node at (.5,1) {$h_b$};
      \node at (.5,4.75) {$h_t$};
    }
    \scoped[local bounding box=f,shift={(14,-1)}]
    {
        \foreach \y in {0,...,4}
        {
            \draw[fill=orangerouge] (0,2*\y-.5) rectangle +(1,2);
            \node at (.5,2*\y+.5) {$g_{\y}$};
        }
        \fill[sharp corners,white,opacity=.5] (0,0) rectangle +(1,-.52);
        \draw[thick] (0,0) rectangle +(1,9.5);
        %\node at (.5,4.75) {$g$};
        %\node at (.5,4.5) {$g_1$};
    }
    \node at ($(h.east)!.5!(T.west)$) {$\se$};
    \node at ($(f.west)!.5!(T.east)$) {$\times$};
    %\foreach \l in {f,h,T} \crochets{\l};

\end{tikzpicture}

%% file: euclideandiv1.tikz
\begin{tikzpicture}[matvec,x=.65cm,y=-.65cm]
  \begin{scope}[shift={(0,.25)},local bounding box=L]
    \scoped[local bounding box=h,shift={(1,0)}]
      \draw[fill=vertjaune] (0,0) rectangle node {$f$} +(1,8);
    \scoped[local bounding box=T,shift={(3,0)}]
        \draw (0,0) pic {fullprod={toeplitz=orangerouge}{4}{$g$}};
    \scoped[local bounding box=f,shift={(8,2)}]
        \draw[fill=bleute](0,0) rectangle node {$q$} +(1,4);
    \scoped[local bounding box=r,shift={(10,0)}]
    {
        \draw[fill=violette] (0,0) rectangle node {$r$} +(1,4);
        \draw (0,0) rectangle +(1,8);
    }
    \node at ($(h.east)!.5!(T.west)$) {$=$};
    \node at ($(f.west)!.5!(T.east)$) {$\times$};
    \node at ($(r.west)!.5!(f.east)$) {$+$};
  \end{scope}
  \path (L.west) -- +(-5,0) -- (L.east) -- +(5,0);
\end{tikzpicture}

%% file: euclideandiv2.tikz
\begin{tikzpicture}[matvec,x=.65cm,y=-.65cm]
  \begin{scope}[shift={(0,10)},local bounding box=C]
      \scoped[shift={(1,0)},local bounding box=htop] \draw[fill=vertjaune] (0,0) rectangle node {$f_0$} +(1,4);
      \scoped[shift={(1,0)},local bounding box=hbot] \draw[fill=vertjaune] (0,4.5) rectangle node {$f_1$} +(1,4);
      \scoped[shift={(3,0)},local bounding box=Ttop] {
          \draw[toeplitz=orangerouge] (0,0) -- ++(0,4) -- ++(4,0) -- cycle;
          \node at (1.5,3) {$g$};
      }
      \scoped[shift={(3,0)},local bounding box=Tbot] {
          \draw[toeplitz=orangerouge] (0,4.5) -- ++(4,0) -- ++(0,4) -- cycle;
          \node at (2.5,5.5) {$g$};
      }
      \scoped[shift={(8,0)},local bounding box=ftop] \draw[fill=bleute](0,0) rectangle node {$q$} +(1,4);
      \scoped[shift={(8,0)},local bounding box=fbot] \draw[fill=bleute](0,4.5) rectangle node {$q$} +(1,4);
      \scoped[local bounding box=r,shift={(10,0)}]   \draw[fill=violette] (0,0) rectangle node {$r$} +(1,4);

      \node at ($(htop.east)!.5!(Ttop.west)$) {$=$};
      \node at ($(hbot.east)!.5!(Tbot.west)$) {$=$};
      \node at ($(ftop.west)!.5!(Ttop.east)$) {$\times$};
      \node at ($(fbot.west)!.5!(Tbot.east)$) {$\times$};
      \node at ($(r.west)!.5!(ftop.east)$) {$+$};
  \end{scope}
  \path (C.west) -- +(-5,0) -- (C.east) -- +(5,0);
\end{tikzpicture}

%% file: euclideandiv3.tikz
\begin{tikzpicture}[matvec,x=.65cm,y=-.65cm]
  \begin{scope}[shift={(0,20)},local bounding box=R]
      \scoped[shift={(3,0)},local bounding box=htop] \draw[fill=vertjaune] (0,0) rectangle node {$f_0$} +(1,4);
      \scoped[shift={(5,0)},local bounding box=Ttop] {
          \draw[toeplitz=orangerouge] (0,0) -- ++(0,4) -- ++(4,0) -- cycle;
          \node at (1.5,3) {$g$};
      }
      \scoped[shift={(10,0)},local bounding box=ftop] \draw[fill=bleute](0,0) rectangle node {$q$} +(1,4);
      \scoped[shift={(1,0)},local bounding box=r]   \draw[fill=violette] (0,0) rectangle node {$r$} +(1,4);
      \scoped[shift={(8,0)},local bounding box=hbot] \draw[fill=vertjaune] (0,4.5) rectangle node {$f_1$} +(1,4);
      \scoped[shift={(3,0)},local bounding box=Tbot] {
          \draw[toeplitz=orangerouge] (0,4.5) -- ++(4,0) -- ++(0,4) -- cycle;
          \node at (2.5,5.5) {$g$};
      }
      \scoped[shift={(1,0)},local bounding box=fbot] \draw[fill=bleute](0,4.5) rectangle node {$q$} +(1,4);

      \node at ($(r.east)!.5!(htop.west)$) {$=$};
      \node at ($(ftop.west)!.5!(Ttop.east)$) {$\times$};
      \node at ($(Ttop.west)!.5!(htop.east)$) {$-$};
      \node at ($(fbot.east)!.5!(Tbot.west)$) {$=$};
      \node at ($(hbot.west)!.5!(Tbot.east)$) {$\times$};
      \node at ($(Tbot.north east)+(.3,-.15)$) {$-1$};
  \end{scope}
  \path (R.west) -- +(-5,0) -- (R.east) -- +(5,0);
\end{tikzpicture}

%% file: reductions.tikz
\begin{tikzpicture}[%
        common/.style={
            rounded corners,
            sharp corners,%
            %draw=#1,fill=bleu!30,%
            %fill=#1!30,
            align=center,%
            text width=22em,
        },%
        cadre/.style={rectangle,common,fill=#1!30,drop shadow},%
        double/.style={rectangle split,rectangle split parts=2,common,rectangle split part fill={#1!30,white}},%
        rightcol/.append style={text width=10em,xshift=.5cm},%
        node distance=.75cm,%
        mv/.style={anchor=north,align=center,text width=22em},%
        every edge/.style={vert,semithick,<-,draw},%
        sh/.style={shorten <= 2pt},%
    ]
    \node[cadre=vert] (prod) at (0,0)  {$h\pe f\times g$\\Algorithms~\ref{algorithm:CumulativeKaratsuba} and \ref{algorithm:CumulativeFFTMultiplication}};
    \node[cadre=vert,anchor=north west,rightcol](conv) at (prod.north east){%
            $h\pe f\times g\bmod x^n-\lambda$\\Algorithm~\ref{algorithm:CumulativeConvolution}};
    \node[mv,rightcol] at (conv.south) {$\lambda$-circulant matrix-vector product};
    \node[cadre=vert,below=of prod] (sp) {%
        $h\pe f\times g\bmod x^n$ and $h\pe f\times g\bquo x^n$\\Algorithm~\ref{algorithm:CumulativeLowerProduct}};%
    \node[mv] (spmv) at (sp.south) {triangular Toeplitz matrix-vector product};
    \node[cadre=vert,below=of spmv]  (mp) {
        $h\pe f\times g\bmod x^{m+n-1}\bquo x^n$\\Corollary~\ref{corollary:CumulativeSlice}};
    \node[mv] (mpmv) at (mp.south) {rectangular Toeplitz matrix-vector product};
    \node[cadre=orange,below=of mpmv,text width=22em]  (overmul) {
        $g \fe f\bmod x^n$, $g \fe f\bquo x^n$ and $g \de f\bmod x^n$\\Algorithms~\ref{algorithm:InPlaceLowerProduct} and \ref{algorithm:InPlaceDivision}};
    \node[mv] (overmv) at (overmul.south) {triangular Toeplitz matrix-vector product and system solving};
    \node[cadre=orange,rightcol,right=0cm of overmul] (mod) {$r\se f\bmod g$\\Algorithm~\ref{algorithm:Remainder}};
    \node[cadre=orange,below=of overmv] (overmod) {$(f,g) \longleftrightarrow (q|r,g)$\\Algorithm~\ref{algorithm:InPlaceEuclideanDivision}};
    \node[cadre=orange,right=0cm of overmod,rightcol] (modacc) {$r\pe f\bmod g$\\Corollary~\ref{corollary:CumulativeRemainder}};
    \node[cadre=orange,below=of overmod] (modmul) {$r\pe fg\bmod p$\\Algorithms~\ref{algorithm:ModularMultiplication} and \ref{algorithm:ModularMultiplicationAllSizes}};
    \draw (prod) edge[sh] (conv)
          (prod) edge[sh] (sp)
          (spmv) edge (mp)
          (mpmv) edge[<-,orange] node[right] {$\log n$} (overmul)
          %(mp) edge[<-,orange] node[right] {$\log n$} (overdiv)
          %(overdiv) edge (mod)
          (overmul) edge[sh] (mod)
          (overmv) edge (overmod)
          (overmod) edge[sh] (modmul)
          (overmod) edge[sh] (modacc)
    ;
    \node[cadre=vert,text width=,minimum width=.5cm, minimum height=.5cm,anchor=west,below left=.75cm and -1cm of modmul] (carrevert) {};
    \node[right=0pt of carrevert,align=left] (legend) {Time: $O(\M(n))$\\Space: $O(1)$};
    \node[cadre=orange,text width=,minimum width=.5cm, minimum height=.5cm,anchor=west,right=of legend] (carreorange) {};
    \node[right=0pt of carreorange,align=left] {Time: $O(\M^*(n))$\\Space: $O(1)$ algebraic registers, $O(\log n)$ pointers};
    \node[below left=.75cm and 0cm of carrevert] (g) {};
    \node[below=.5em of g] (r) {};
    \draw[semithick, vert,->] (g) -- +(.75,0);
    \draw[semithick, orange,->] (r) -- +(.75,0);
    \node[right=of g] {space- and time-preserving reduction};
    \node[right=of r] {reduction with a call stack and an extra logarithmic factor in the time};

\end{tikzpicture}

%% file: blackboxes.tikz
\begin{tikzpicture}[x=4cm,y=-3cm,%
    every edge/.style={draw,-latex,semithick},%
    bb/.style={shape=ellipse,fill=bleu!30,minimum height=1.5cm, align=center,drop shadow},%
    deriv/.style={fill=vert!30},%
]
    \node[shape=rectangle,fill=orange!30,minimum height=1cm,rounded corners,drop shadow] (slp) at (0,0.25) {straight-line program};
    \node[bb] (reg) at (-1,1) {regular\\black box};
    \node[bb] (mod) at (0,1) {modular\\black box};
    \node[bb] (ext) at (1,1) {extended\\black box};
    \node[bb] (geo) at (-1,2) {geometric\\black box};
    \node[bb,deriv] (dmod) at (0,2) {modular\\black box\\with\\ derivative};
    \node[bb,deriv] (dext) at (1.5,2) {extended\\black box\\with\\derivative};
    \node[bb,deriv] (dreg) at (0.75,2) {regular\\black box\\with\\derivative};
    \node[bb,deriv] (dgeo) at (0.75,3) {geometric\\black box\\with\\derivative};

    \path
        (slp)   edge    (reg)
                edge    (mod)
                edge    (ext)
        (reg)   edge    (geo)
        (mod)   edge    (dmod)
        (ext)   edge    (dext)
                edge    (dreg)
        (dreg.south west)  edge[bend right=10]    (dgeo.north west)
    ;

\end{tikzpicture}

%% file: Graeffe.tikz
\begin{tikzpicture}[%
    x=2cm, y=1.5cm,%
    every node/.style={minimum size=2em},%
    poly/.style={fill=bleu!30,circle,drop shadow,inner sep=0pt},%
    roots/.style={fill=vert!30,circle,drop shadow},%
    every edge/.style={semithick},% shorten >= 2pt},%
    epol/.style={shorten <= 2pt},%
    erts/.style={shorten >= 2pt},%
]
\node[poly] (f) at (0,1) {$f$};
\node[poly] (h0) at (1,1) {$Gf$};
\node[poly] (h1) at (2,1) {$G^2f$};
\node (dots) at (3,1) {\dots};
\node[poly] (hm1) at (4,1) {$G^ef$};

\node[roots] (zf) at (0,0) {$Z_0$};
\node[roots] (z0) at (1,0) {$Z_1$};
\node[roots] (z1) at (2,0) {$Z_2$};
\node (zdots) at (3,0) {\dots};
\node[roots] (zm1) at (4,0) {$Z_e$};

\draw[->,shorten <= 2pt] (hm1) -- (zm1);

\draw[epol,->] (f)  -- node[above] {$G$} (h0);
\draw[epol,->] (h0) -- node[above] {$G$} (h1);
\draw[epol,->] (h1) -- node[above] {$G$} (dots);
\draw[epol,->] (dots) -- node[above] {$G$} (hm1);

\draw[erts,->] (zm1) -- (zdots);
\draw[erts,->] (zdots) -- (z1);
\draw[erts,->] (z1) -- (z0);
\draw[erts,->] (z0) -- (zf); 
\end{tikzpicture}